\documentclass{jpp}

\usepackage{amsmath,amsfonts,amssymb}
\usepackage{authblk}
\usepackage{graphicx}
\usepackage{color}
\usepackage{bm}
\usepackage{import}

\numberwithin{equation}{section}

\newcommand{\Mu}{M}
\newcommand{\Apa}{A_{1,\shortparallel}}
\newcommand{\Ad}{\vc{A}_1}
\newcommand{\Adperp}{\vc{A}_{1,\perp}}
\newcommand{\phid}{\phi_1}

\newcommand{\htau}{\vc{\hat{\tau}}}
\newcommand{\hrho}{\vc{\hat{\rho}}}
\newcommand{\hb}{\vc{\hat{b}}_0}

\DeclareMathOperator{\sign}{sgn}

\newcommand\burby{\varsigma}
\newcommand\gc[1]{\bar{#1}}
\newcommand\gcburby[1]{{\bar{#1}}^{\dagger,\burby}}
\newcommand\gyFIXED[1]{\bar{\bar{#1}}}
\newcommand\gy[1]{\gyFIXED{#1}}

\newcommand\jacobiangy{\mathfrak{J}_s}
\newcommand\dotjacobiangy{\dot{\mathfrak{J}}_s}
\newcommand\jacobiangcslab{\mathfrak{J}_{0,s}}

\newcommand{\vc}[1]{{\boldsymbol{#1}}}
\newcommand{\pard}[2]{\frac{\partial #1}{\partial #2}}
\newcommand{\pardd}[2]{\frac{\partial^2 #1}{\partial #2^2}}
\newcommand{\dd}{\mathrm{d}}
\newcommand{\wt}[1]{\widetilde{#1}}
\newcommand{\totd}[2]{\frac{\dd #1}{\dd #2}}
\newcommand{\hatdd}[1]{\partial_{\hat{#1}}}

\newcommand\bigO{\mathcal{O}}

\usepackage{scalerel,stackengine}
\stackMath
\newcommand\reallywidecheck[1]{%
\savestack{\tmpbox}{\stretchto{%
  \scaleto{%
    \scalerel*[\widthof{\ensuremath{#1}}]{\kern-.6pt\bigwedge\kern-.6pt}%
    {\rule[-\textheight/2]{1ex}{\textheight}}
  }{\textheight}%
}{0.5ex}}%
\stackon[1pt]{#1}{\scalebox{-1}{\tmpbox}}%
}
\newcommand{\evalperturbedgc}[1]{\mathring{#1}}
\newcommand{\evalperturbedgcadjoint}[1]{\reallywidecheck{#1}}
\newcommand{\evalperturbedgcadjointalt}[1]{\reallywidecheck{#1}}
\newcommand{\evalperturbedgcburby}[1]{\mathring{#1}^\burby}

\newcommand{\gav}[1]{\langle #1 \rangle}
\newcommand{\sgav}[1]{\left\langle #1 \right\rangle}
\newcommand{\dgav}[1]{\langle\!\langle #1 \rangle\!\rangle}
\newcommand{\sdgav}[1]{\left\langle\!\!\left\langle #1 \right\rangle\!\!\right\rangle}
\newcommand{\radgav}[1]{\langle\!| #1 |\!\rangle}

\newcommand\effective{{\star}} 
\newcommand\bracketgc[2]{\{#1, #2\}_{0}}
\newcommand\bracketgy[2]{\{#1, #2\}}

\newcommand\bcddot{
  \mathbin{\vcenter{%
    {\hbox{$\bcdot$}}\vspace{-10pt}{\hbox{$\bcdot$}}}%
}}%

\newcommand{\evalAt}[2]{\left. #1 \right|_{#2}}
\newcommand{\cp}{\Lambda}
\newcommand{\ca}{\vc{\cp}}
\newcommand{\capa}{\cp_\shortparallel}

\newcommand{\Vtau}{\gy{U}_\tau}
\newcommand{\Vtausqr}{\gy{U}_\tau^2}

\newcommand{\zlr}{\mathrm{ZLR}}
\newcommand{\flr}{\mathrm{FLR}}

\newcommand{\darwin}{\mathrm{Dar}}
\newcommand{\bh}{\mathrm{BH}}
\newcommand{\reduced}{\shortparallel}
\newcommand{\cm}{\mathrm{CM}}
\newcommand{\free}{\mathrm{f}}
\newcommand{\bound}{\mathrm{b}}
\newcommand{\brizard}{\mathrm{B}}

\newcommand{\gparam}{\xi}
\newcommand{\gparamr}{\gparam_R}
\newcommand{\gparamtheta}{\gparam_\Theta}
\newcommand{\temperature}{T}

\usepackage[hidelinks]{hyperref}
\usepackage[capitalize]{cleveref}
\crefname{section}{Sec.}{Secs.}
\crefname{equation}{Eq.}{Eqs.}

\usepackage{booktabs}

\newcommand\curl{\nabla \times}

\newcommand\defeq{\mathrel{\mathop:}=}

\newcommand{\kpa}{k_{\shortparallel}}

\newcommand{\kperp}{k_{\perp}}

\newcommand\uth{u_\mathrm{th}}
\newcommand\uths{u_{\mathrm{th},s}}

\newcommand\uthi{u_{\mathrm{th},\mathrm{i}}}
\newcommand{\ii}{\mathrm{i}}
\newcommand{\euler}{\mathrm{e}}
\newcommand\diff[1]{\mathop{}\! \mathrm{d} #1}

\newcommand\variation[3]{
  \frac{\delta #3}{\delta #1}\squarepar{#2}
}

\newcommand{\kboltz}{k_\mathrm{B}}

\newcommand\roundpar[1]{\left( #1 \right)}
\newcommand\squarepar[1]{\left[ #1 \right]}

\newcommand\abs[1]{\lvert #1 \rvert}
\newcommand\sabs[1]{\left\lvert #1 \right\rvert}

\newcommand\gyfzero{\gy{f}^0_{s}}
\newcommand\gyfzerocm{\gy{f}^{0,\cm}_{s}}
\newcommand\gyfone{\delta\gy{f}_{s}}

\newcommand\gyzzero{\gy{\vc{z}}^0}
\newcommand\gyzzerod{\gy{z}^0}
\newcommand\tzero{t^0}
\newcommand\tone{t^1}

\usepackage{tikz,pgfplots}
\usepackage{subcaption}

\def\twofigwidth{0.475\textwidth}

\newcommand*\circled[1]{\tikz[baseline=(char.base)]{
            \node[shape=circle, draw, inner sep=0.5pt, line width=0.5pt] (char) {\footnotesize\textrm{#1}};}}

\makeatletter
\newcommand*{\transpose}{%
  {\mathpalette\@transpose{}}%
}
\newcommand*{\@transpose}[2]{%
  \raisebox{\depth}{$\m@th#1\intercal$}%
}
\makeatother

\DeclareFontFamily{U}{mathx}{\hyphenchar\font45}
\DeclareFontShape{U}{mathx}{m}{n}{
      <5> <6> <7> <8> <9> <10>
      <10.95> <12> <14.4> <17.28> <20.74> <24.88>
      mathx10
      }{}
\DeclareSymbolFont{mathx}{U}{mathx}{m}{n}
\DeclareMathAccent{\widecheck}    {0}{mathx}{"71}

\newcommand\mat[1]{\mathsfbi{#1}}
\newcommand\matcmp[1]{\mathsfi{#1}}
\newcommand\fourier{\widehat} 

\newcommand\nondim[1]{\widecheck{#1}}
\newcommand\units[1]{[#1]}

\newcommand\coeffcone{\frac{\sum_s m_s \gy{n}_{0,s}}{B_0^2}}
\newcommand\coeffcupar{\frac{\sum_s m_s \gy{n}_{0,s} \gy{u}_{0,\shortparallel,s}}{B_0^2}}

\definecolor{matlablines1}{rgb}{0, 0.4470, 0.7410}
\definecolor{matlablines2}{rgb}{0.8500, 0.3250, 0.0980}
\definecolor{matlablines3}{rgb}{0.9290, 0.6940, 0.1250}
\definecolor{matlablines4}{rgb}{0.4940, 0.1840, 0.5560}
\definecolor{matlablines5}{rgb}{0.4660, 0.6740, 0.1880}
\definecolor{matlablines6}{rgb}{0.3010, 0.7450, 0.9330}
\definecolor{matlablines7}{rgb}{0.6350, 0.0780, 0.1840}

\newcommand\parallelcolor{matlablines1}

\newcommand\darwincolor{matlablines2}
\newcommand\gaugeinvariantcolor{matlablines3}

\def\twofigwidth{0.475\textwidth}

\newcommand\axheight{0.4\textwidth}

\def\twoaxwidth{0.44\textwidth}

\newcommand\tikzlinewidth{1.5pt}
\tikzset{every picture/.style={line width=\tikzlinewidth}}
\pgfplotsset{every axis/.style={axis lines=left, axis line style = thin}}

\usepackage{titlesec}
\titleformat{\paragraph}
{\normalfont\normalsize\itshape}{\theparagraph}{1em}{}
\titlespacing*{\paragraph}
{0pt}{3.25ex plus 1ex minus .2ex}{1.5ex plus .2ex}

\newtheorem{theorem}{Theorem}

\newtheorem{remark}{Remark}

\makeatletter
\def\oversortoftilde#1{\mathop{\vbox{\m@th\ialign{##\crcr\noalign{\kern3\p@}%
      \sortoftildefill\crcr\noalign{\kern3\p@\nointerlineskip}%
      $\hfil\displaystyle{#1}\hfil$\crcr}}}\limits}

\def\sortoftildefill{$\m@th \setbox\z@\hbox{$\braceld$}%
  \braceld\leaders\vrule \@height\ht\z@ \@depth\z@\hfill\braceru$}

\makeatother

\newcommand\drho{\mathfrak{r}}
\usepackage[sort]{natbib}

\begin{document}

  \title{Gauge-invariant variational formulations of electromagnetic gyrokinetic theory}

\author[1]{Ronald Remmerswaal\corresp{Email address for correspondence: \href{mailto:ronald.remmerswaal@ipp.mpg.de}{ronald.remmerswaal@ipp.mpg.de}}}
\author[1]{Roman Hatzky}
\author[1,2]{Eric Sonnendr\"ucker}

\affil[1]{Max Planck Institute for Plasma Physics, D-85748 Garching, Germany}
\affil[2]{Technical University of Munich, Department of Mathematics, D-85748 Garching, Germany}


\maketitle

  \begin{abstract}
    The use of gyrokinetics, wherein phase-space coordinate transformations result in a phase-space dimensionality reduction as well as the removal of fast time scales, has enabled the simulation of microturbulence in fusion devices.
The state-of-the-art gyrokinetic models used in practice are parallel-only models wherein the perpendicular part of the vector potential is neglected.
Such models are inherently not gauge invariant.
We generalise the work of [Burby, Brizard. \emph{Physics Letters A}, 383(18):2172–2175] by deriving a sufficient condition on the gyrocentre coordinate transformation which ensures gauge invariance.
This leads to a parametrized family of gyrokinetic models for which we motivate a specific choice of parameters that results in the smallest gyrocentre coordinate transformation for which the resulting gyrokinetic model is consistent, gyro-phase independent, gauge invariant and has an invariant magnetic moment.
Due to gauge invariance this model can be expressed directly in terms of the electromagnetic fields, rather than the potentials, and the gyrokinetic model thereby results in the macroscopic Maxwell's equations.
For the linearised model, it is demonstrated that the shear and compressional Alfv\'en waves are present with the correct frequencies.
The fast compressional Alfv\'en wave can be removed by making use of a Darwin-like approximation.
This approximation retains the gauge invariance of the proposed model.
  
  \end{abstract}


  \section{Introduction}
The role of numerical modelling is prevalent not only in understanding the physics of fusion plasmas, but also in the design and optimization of magnetic fusion devices.
The collisionless Vlasov--Maxwell model is found to be appropriate when the collision frequency of the charged particles is much lower than the frequencies that are of interest, e.g.\ when studying microturbulence~\citep{garbet2010gyrokinetic}.
Nonetheless, such a model is still very challenging to use, not only because of its six dimensional phase-space, but also because of the large range of length (four orders of magnitude between the plasma size and the Debye length) and time (seven orders of magnitude between the ion-ion collision frequency and the electron plasma frequency) scales.

Two of the aforementioned challenges are, at least partially, addressed by gyrokinetic theory~\citep{Frieman1982,littlejohn1983,sugama,brizard2007,burby2019gauge} wherein a sequence of phase-space coordinate transformations is used to decouple the fast gyration of a charged particle from its otherwise slower motion along the magnetic field.
This thereby results in the removal of the (high) cyclotron frequency, while at the same time reducing the phase-space dimensionality by one.
When moreover considering the quasi-neutral limit, it is found that also the light wave and the Langmuir wave (or plasma oscillation) are removed from the model.
The use of gyrokinetic theory thereby permits the numerical modelling of turbulent transport in tokamaks, as discussed in the well written review paper~\citep{garbet2010gyrokinetic} and was more recently applied to study electromagnetic turbulence in stellarator plasmas~\citep{mishchenko2023global}.

The gyrokinetic model results from a sequence of -- mostly near-identity -- phase-space coordinate transformations which are applied to the collisionless Vlasov--Maxwell model, and therefore \emph{in theory} there is no approximation error.
In practice, however, one must always truncate the near-identity phase-space coordinate transformation to some order of the small expansion parameter, resulting in an unavoidable modelling error.
The more recently developed gyrokinetic models are based on a variational principle, which thereby, despite this modelling error, still preserve essential structures of the original Vlasov--Maxwell model.
For instance the total (free) charge, momentum and energy should be conserved~\citep{brizard2021exact,hirvijoki2020energy,sugama2018eulerian,peifeng2021discovering}, while the choice of gauge condition on the vector potential should leave the model invariant~\citep{burby2019gauge}, resulting in so-called gauge invariance.

However, to our knowledge all global gyrokinetic \emph{simulations} either neglect the part of the vector potential that is perpendicular to the background magnetic field and thereby result in a `parallel-only' gyrokinetic model~\citep{Kleiber_pullback}, or use a simplified model for the parallel component of the perturbed magnetic field~\citep{Chen_Zonka_2016}.
Both approximations are irreconcilable with gauge invariance and lead to intermediate wavelength models wherein perpendicular system-scale effects are incorrectly modelled.
There are, however, numerous gyrokinetic \emph{theories} and \emph{models} which include the perpendicular part of the vector potential \citep{brizard2007,qin1999gyrokinetic,qin2000gyrocenter,qin2005short}.
Furthermore, in \citep{burby2019gauge} a novel gauge-invariant gyrokinetic model is introduced for which exact conservation laws are derived in \citep{brizard2021hamiltonian,brizard2021exact}.

We extend the approach followed in \citep{burby2019gauge} by proposing a parametrized family of gyrocentre coordinate transformations, resulting in a family of gauge-invariant gyrokinetic models where a specific choice of parameters yields the model of \citep{burby2019gauge}.
A different choice of parameters is motivated in detail in this paper, resulting in the smallest gyrocentre coordinate transformation for which the resulting gyrokinetic model is consistent, gyro-phase independent, gauge invariant and has an invariant magnetic moment.


The proposed gyrokinetic model is derived in detail using the language of vector calculus in favour of the customarily used language of differential geometry and Lie transform methods.
Such a derivation is equivalent to the more traditionally used techniques, as found for example in \citep{hahm1988nonlinear,brizard_PhD1990,qin1999gyrokinetic}, but we have opted for the vector calculus approach as it reduces the required prerequisite knowledge of our readers.


Our paper starts with a brief overview of its main results in \cref{sec:highlights}.
In \cref{sec:notation} we discuss the preliminary phase-space coordinate transformation leading to the guiding-centre single-particle phase-space Lagrangian, wherein only the stationary background magnetic field is considered.
The perturbed time-dependent electromagnetic fields are included as a perturbation to the guiding-centre Lagrangian in \cref{sec:single_particle}, followed by a detailed description of the proposed gyrocentre coordinate transformation, resulting eventually in the gyrocentre single-particle phase-space Lagrangian.
We then combine the gyrocentre single-particle Lagrangian for each species with Maxwell's equations in \cref{sec:vlasov_maxwell}, eventually resulting in the gyrocentre equations of motion, Gauss's law as well as the Amp\`ere--Maxwell law.
A low-frequency quasi-neutral Darwin approximation of the proposed model is considered in \cref{sec:darwin}.
A comparison with reduced models is made in \cref{sec:reduced}, where we compare the two proposed models to several models from the literature.
We conclude with a discussion in \cref{sec:conclusion}.
  \renewcommand\gy[1]{#1}
\section{A brief overview of the main results}\label{sec:highlights}
To compensate the length and complexity of this paper, we provide a brief overview of the main results in this section. 
This is of particular use to those readers who are not necessarily interested in a detailed derivation of the model, but who are instead interested in (the implementation of) the resulting models and their key properties.

In essence, two gauge invariant models are derived, analysed and proposed in this paper: the first is referred to as the `gyrokinetic Maxwell model' (see \cref{sec:vlasov_maxwell}), and the second is referred to as the `quasi-neutral gyrokinetic Darwin model' (see \cref{sec:darwin}).
The former yields a model in which fast waves (such as the light wave, Langmuir wave and compressional Alfv\'en wave) are retained, whereas such fast waves are eliminated in the latter model.
Each model is derived from an action principle which can be found in \cref{eq:low_split_explicit,eq:low_split_explicit_darwin} respectively, where the action is based on the gyrocentre single-particle Lagrangian (which is derived in detail in \cref{sec:single_particle}) and is a function of the gyrocentre coordinate $\vc{Z}(t) = (\vc{R}(t), U_\shortparallel(t), \Mu, \Theta(t))$ (where $\vc{R}$ denotes the gyrocentre position, $U_\shortparallel$ denotes the velocity component parallel to the background magnetic field $\vc{B}_0$, $\Mu$ denotes the invariant magnetic moment and $\Theta$ denotes the gyro-phase), the perturbed scalar potential~$\phi_1$ and the perturbed vector potential~$\vc{A}_1$.



\subsection{The gyrocentre equations of motion}\label{sec:highlight:gyrocentre_eoms}
Imposing the principle of least action (see \cref{sec:pola}) w.r.t. the gyrocentre coordinate yields the gyrocentre equations of motion, which are discussed in detail in \cref{sec:vlasov_maxwell_eoms} and are presented here for convenience
\begin{subequations}
  \label{eq:highlight:gyrocentre_eoms}
  \begin{align}
    \dot{\gy{\vc{R}}} &= 
        \gy{U}_\shortparallel \vc{b}_s^\effective 
      - \frac{1}{q_s B_{s,\shortparallel}^\effective} \hb \times \squarepar{
          q_s \vc{E}^\effective_1
        - \Mu \nabla (B_0 + \dgav{\evalperturbedgcburby{B}_{1,\shortparallel}})
      },\label{eq:highlight:gyrocentre_eoms_R}
      \\
    \dot{\gy{U}}_\shortparallel &= 
        \frac{1}{m_s} \vc{b}_s^\effective \bcdot \squarepar{
            q_s \vc{E}^\effective_1
          - \Mu \nabla (B_0 + \dgav{\evalperturbedgcburby{B}_{1,\shortparallel}})}
      ,\label{eq:highlight:gyrocentre_eoms_Vpa}
  \end{align}
\end{subequations}
where we note that the magnetic moment is invariant $\dot{\Mu} = 0$ and the gyro-phase $\Theta(t)$ is an ignorable coordinate as none of the other equations depend on it.
Here $\hb = \vc{B}_0 / B_0$ is the normalized background magnetic field, $B_0 = \abs{\vc{B}_0}$ is the Euclidean norm of the background magnetic field, $\vc{b}_s^\effective$ is the effective magnetic field direction defined in \cref{eq:gy_bstarstar}, $B_{s,\shortparallel}^\effective$ is the effective parallel magnetic field defined in \cref{eq:gy_Bpastar}, $q_s$ and $m_s$ denote the charge and mass of the species $s$, $\vc{E}^\effective_1$ is the effective electric field defined in \cref{eq:gy_flr_electricfield} (and is approximately equal to the gyro-averaged electric field $\vc{E}^\effective_1 \approx \gav{\evalperturbedgc{\vc{E}}_1}$), and $\dgav{\evalperturbedgcburby{B}_{1,\shortparallel}}$ denotes the disc average of the parallel component of the perturbed magnetic field as defined in \cref{eq:disc_average}.
We note that the equations of motion are identical for the two models proposed in this paper.

\subsection{The gyrokinetic Maxwell model}\label{eq:highlight:gyrokinetic_maxwell}
Imposing the principle of least action w.r.t. the perturbed vector potential $\vc{A}_1$, where the action is given by \cref{eq:low_split_explicit}, yields the following Amp\`ere--Maxwell law together with Faraday's law (we refer to \cref{sec:vlasov_maxwell_fieldeqs} for corresponding weak formulation and to \cref{sec:macromaxwell_formulation,sec:initial_value_problem} for a discussion on the strong formulation)
\begin{subequations}
  \begin{align}
    \biggl(\epsilon_0 \mat{I}_3 + \underbrace{\coeffcone \mat{\Pi}_\perp\biggr) \pard{\vc{E}}{t}}_{\text{polarisation current}} = {} &
    \curl {\biggl(
      \frac{1}{\mu_0} \vc{B} 
      - \underbrace{\frac{p_{0,\shortparallel} - p_{0,\perp}}{B_0^2} \vc{B}_{\perp}
      + \coeffcupar \hb \times \vc{E}}_{\text{magnetisation}}
    \biggr)}
    \nonumber \\
    & + \underbrace{\coeffcupar \hb \times (\curl \vc{E})}_{\text{polarisation current}}\label{eq:highlight:maxwell_ampere}
    - \evalperturbedgcadjointalt{\vc{\mathcal{J}}}{}^\free  
    ,\\
    \pard{\vc{B}}{t} = {} & - \curl \vc{E},
  \end{align}
\end{subequations}
where we note that both Gauss's law (cf.\ \cref{eq:maxwell_gauss}) and the magnetic Gauss's law (cf.\ \cref{eq:maxwell_maggauss}) are satisfied provided that they are satisfied initially.
Here $\epsilon_0$ denotes the vacuum permittivity, $\mu_0$ denotes the magnetic vacuum permeability, $n_{0,s}$ and $\gy{u}_{0,\shortparallel,s}$ denote the background particle density and parallel velocity of species $s$ (as defined in \cref{eq:background_density_velocity}), $\mat{\Pi}_\perp = \mat{I}_3 - \hb \otimes \hb$ is the perpendicular projection matrix and $\mat{I}_3$ denotes the $3 \times 3$ identity matrix, $\vc{E} = \vc{E}_1$ and $\vc{B} = \vc{B}_0 + \vc{B}_1$ denote the electric and magnetic field, $p_{0,\perp}, p_{0,\shortparallel}$ denote the perpendicular and parallel background pressure as defined in \cref{eq:par_perp_pressure} and $\evalperturbedgcadjointalt{\vc{\mathcal{J}}}{}^\free$ denotes the gyrocentre free-current density defined weakly in \cref{eq:free_current_gavadjoint}.

The evolution equation for the electric field $\vc{E}$ can be obtained from \cref{eq:highlight:maxwell_ampere} by multiplying by the inverse of the $3 \times 3$ matrix shown on the left-hand side.
In this formulation it is therefore advantageous to let the vacuum permittivity $\epsilon_0$ be finite, such that this matrix is invertible, resulting in field equations which are entirely local and thereby result in a local gyrokinetic model which can be integrated explicitly in time.
The fast compressional Alfv\'en wave is present in this model as we demonstrate in \cref{sec:slab_susceptibility}.

Conditions on the background magnetic field $\vc{B}_0$ and the initial distribution function~$\gyfzero$ are derived in \cref{sec:field_equilibria}, ultimately leading to the MHD equilibrium condition~\eqref{eq:mhd_equilibrium} for the correct balance in the perpendicular part of the Amp\`ere--Maxwell law and \cref{eq:mhd_parallel_ampere} for the remaining parallel component.
Moreover, a local energy conservation law for this model in terms of the kinetic and potential energy densities is derived in \cref{sec:energy}.

\subsection{The quasi-neutral  gyrokinetic Darwin approximation}\label{eq:highlight:gyrokinetic_darwin}
The gyrokinetic Maxwell model enjoys a favourable local structure of the equations, but it contains the fast compressional Alfv\'en wave, fast light wave as well as the Langmuir wave, which are often undesirable.
To this end, we propose a quasi-neutral gyrokinetic Darwin model in \cref{sec:darwin}, wherein these fast waves are removed, ultimately resulting in the following field equations in the perpendicular Coulomb gauge~\eqref{eq:highlight:darwin_constraint}
\begin{subequations}\label{eq:highlight:darwin}
  \begin{align}
    -\nabla \bcdot \roundpar{
    \sum_s \frac{ m_s \gy{n}_{0,s}}{B_0^2} \squarepar{\nabla_\perp \phi_1 - \gy{u}_{0,\shortparallel,s} \hb \times (\curl \Ad)}
  } &= \evalperturbedgcadjoint{\mathcal{R}}{}^\free,\label{eq:highlight:quasineutrality}\\
    \curl \biggl[
      \frac{1}{\mu_0} \curl \Ad 
    - \frac{p_{0,\shortparallel} - p_{0,\perp}}{B_0^2} (\curl \Ad)_\perp 
    -  \coeffcupar &\hb \times \nabla_\perp \phi_1
    \biggr] \nonumber \\ + \, \coeffcone \nabla_\perp \lambda &= \evalperturbedgcadjointalt{\vc{\mathcal{J}}}{}^\free 
    - \frac{1}{\mu_0} \curl \vc{B}_0,\label{eq:highlight:darwin_ampere}\\
    \nabla \bcdot \roundpar{\coeffcone \Adperp} &= 0,\label{eq:highlight:darwin_constraint}
  \end{align}
\end{subequations}
where $\evalperturbedgcadjoint{\mathcal{R}}{}^\free$ denotes the gyrocentre free-charge density defined weakly in \cref{eq:free_charge_gavadjoint} and $\lambda$ is the Lagrange multiplier associated with the constraint~\eqref{eq:highlight:darwin_constraint}.
Due to the quasi-neutral Darwin approximation the field equations are no longer local, but we note that the quasi-neutrality equation~\eqref{eq:highlight:quasineutrality} is entirely decoupled from the Amp\`ere--Maxwell law~\eqref{eq:highlight:darwin_ampere} (together with its constraint~\eqref{eq:highlight:darwin_constraint}) if the background distribution function is symmetric ($\gy{u}_{0,\shortparallel,s} = 0$) and can therefore be solved for independently.

In \cref{sec:slab_susceptibility} we demonstrate that in this model the fast compressional Alfv\'en wave is removed.
Moreover, the local energy conservation law from \cref{sec:energy} also holds for this model, except that the energy flux vector is altered slightly as discussed in \cref{sec:darwin_energy}.

\renewcommand\gy[1]{\gyFIXED{#1}}
  \section{Preliminary transformations}\label{sec:notation}
In this section we establish most of our notation and apply preliminary coordinate transformations which result in the guiding-centre single-particle phase-space Lagrangian wherein only a background magnetic field is considered.

\subsection{Motivation}\label{sec:motivation_physcoords}
We start by considering the model for the motion of a charged particle, of charge $q$ and mass~$m$, in the presence of a stationary background magnetic field $\vc{B}_0 = \curl \vc{A}_0$, with magnitude $B_0 = \abs{\vc{B}_0}$, where $\vc{A}_0$ denotes the background vector potential and $\abs{\cdot}$ denotes the Euclidean norm.
The background magnetic field yields a coordinate system whose orthogonal basis vectors are denoted by $(\hb, \vc{\hat{e}}_1, \vc{\hat{e}}_2)$, where $\hb = \vc{B}_0 / B_0$, and $\vc{\hat{e}}_1, \vc{\hat{e}}_2$ are unit vectors orthogonal to $\hb$ for which $\hb = \vc{\hat{e}}_1 \times \vc{\hat{e}}_2$.
For any vector $\vc{S}$ (such as the velocity $\vc{U}$), we denote the component parallel to the background magnetic field as
\begin{equation}
  S_\shortparallel \defeq \vc{S} \bcdot \hb.
\end{equation}
The resulting parallel and perpendicular parts of a vector are denoted by
\begin{equation}
  \vc{S}_\shortparallel \defeq S_\shortparallel \hb, \quad
  \vc{S}_\perp \defeq \vc{S} - \vc{S}_\shortparallel.
\end{equation}
Equivalent notation is used for the perpendicular gradient operator
\begin{equation}
  \nabla_\perp Q \defeq \nabla Q - (\nabla Q \bcdot \hb) \hb.
\end{equation}

It is well-known that the following single-particle phase-space Lagrangian $L_0$ is a model for the motion of a charged particle in physical coordinates
\begin{equation}\label{eq:part_lag_phys}
  L_0 \defeq \Gamma_0 - H_0, \quad
  \Gamma_0 \defeq \roundpar{q \vc{A}_0 + m \vc{U}} \bcdot \dot{\vc{R}}, \quad
  H_0 \defeq K_0,
\end{equation}
where $\Gamma_0$ and $H_0$ denote the symplectic and Hamiltonian part of the Lagrangian $L_0$, respectively.
The kinetic energy per particle is given by
\begin{equation}\label{eq:energy_kin0}
  K_0 = \frac{m}{2} \abs{\vc{U}}^2.
\end{equation}
The model is expressed in terms of the phase-space coordinates $\tilde{\vc{Z}} = (\vc{R}, \vc{U}) \in \mathbb{R}^6$, where $\vc{R}$ and $\vc{U}$ denote the particle position and velocity, respectively.

Imposing the principle of least action (this is explained in more detail in \cref{sec:gc_least_action}) on the single-particle phase-space Lagrangian~\eqref{eq:part_lag_phys} results in the well-known Euler--Lagrange equations, which in turn yield the equations of motion (EOMs) for a charged particle in the presence of a stationary background magnetic field
\begin{equation}\label{eq:physical_eoms}
  \dot{\vc{R}} = \vc{U}
  , \quad
  \dot{\vc{U}} = \frac{q}{m} \vc{U} \times \vc{B}_0
  .
\end{equation}
If $\vc{B}_0$ is constant then it can easily be verified that the following is the solution to the EOMs
\begin{equation}\label{eq:physical_solution}
  \vc{U}(t) = 
    U_\shortparallel(0) \hb 
  + \squarepar{\cos(\omega_\mathrm{c} t) (\vc{\hat{e}}_1 \vc{\hat{e}}_1^\transpose + \vc{\hat{e}}_2 \vc{\hat{e}}_2^\transpose) + \sin(\omega_\mathrm{c} t) (\vc{\hat{e}}_1 \vc{\hat{e}}_2^\transpose - \vc{\hat{e}}_2 \vc{\hat{e}}_1^\transpose)} \vc{U}_\perp(0),
\end{equation}
where we have defined the cyclotron frequency as
\begin{equation}\label{eq:cyclotron_frequency}
  \omega_\mathrm{c} \defeq \frac{q B_0}{m}.
\end{equation}

In many applications the frequency of interest is much smaller than the cyclotron frequency and therefore the aim is to decouple this fast gyrating motion by applying coordinate transformations to the single-particle phase-space Lagrangian.

\subsection{Field aligned velocity coordinates}
\begin{figure}
  \centering
  \def\svgwidth{\textwidth}
  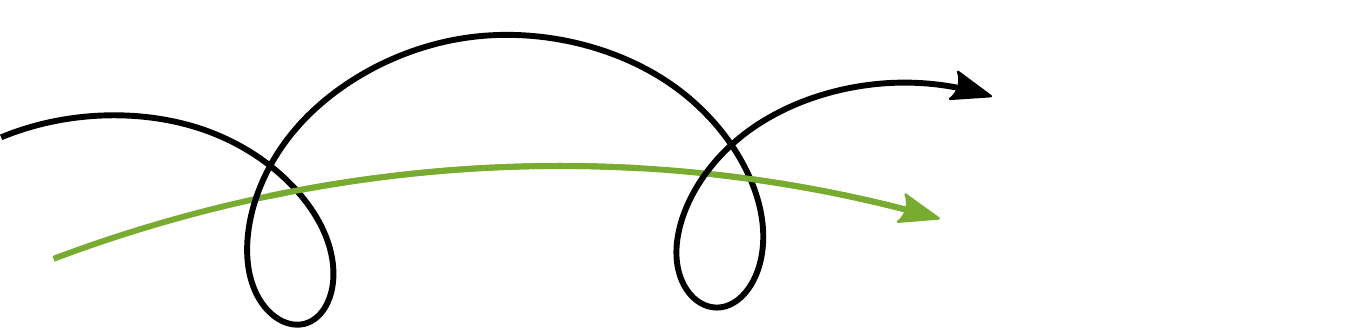
  \caption{Illustration of the guiding-centre coordinate system.
  We denote the physical particle position in black and the guiding-centre position in green.
  The particle moves along the background magnetic field in the (blue) $\hb$ direction, while gyrating in the (red) plane perpendicular to the background magnetic field, in the direction of the (red) arrow $\htau$.
  The extremal values of the $\burby$ parameter (introduced in \cref{sec:gauge_invariance}) are indicated in grey.}\label{fig:illustrate_gc}
\end{figure}
The first coordinate transformation that we consider results in coordinates which are field aligned in velocity space $\vc{Z} = (\vc{R}, U_\shortparallel, \Mu, \Theta)$, where the parallel velocity component~$U_\shortparallel$, magnetic moment $\Mu$, and gyro-phase $\Theta$ are defined as
\begin{equation}\label{eq:field_aligned_velocity_coords}
  U_\shortparallel \defeq \vc{U} \bcdot \hb, \quad
  \Mu \defeq \frac{m U_\tau^2}{2 B_0}, \quad
  \Theta \defeq \arctan \roundpar{\frac{\vc{U} \bcdot \vc{\hat{e}}_1}{\vc{U} \bcdot \vc{\hat{e}}_2}},
\end{equation}
where $U_\tau$ is defined below. 
Using the gyro-phase we define the new coordinate system $(\hb, \htau, \hrho)$, where $\htau, \hrho$ are given by
\begin{equation}
  \htau \defeq - \vc{\hat{e}}_1 \sin\Theta - \vc{\hat{e}}_2 \cos\Theta, \quad
  \hrho \defeq \vc{\hat{e}}_1 \cos\Theta - \vc{\hat{e}}_2 \sin\Theta
\end{equation}
and are such that $\hb = \htau \times \hrho$.
See also \cref{fig:illustrate_gc}.
We denote the tangential and radial components of a vector field $\vc{S}$ by
\begin{equation}
  S_\tau \defeq \vc{S} \bcdot \htau, \quad
  S_\rho \defeq \vc{S} \bcdot \hrho.
\end{equation}

Note that
\begin{equation}
  \tan\Theta 
  = \frac{\vc{U} \bcdot \vc{\hat{e}}_1}{\vc{U} \bcdot \vc{\hat{e}}_2} 
  = \frac{U_\tau \tan\Theta - U_\rho}{U_\tau + U_\rho \tan\Theta}
  \quad \implies \quad
  U_\rho = 0
\end{equation}
and therefore the velocity can be expressed in terms of the field aligned velocity coordinates as
\begin{equation}\label{eq:field_aligned_velocity}
  \vc{U} = U_\shortparallel \hb + U_\tau \htau,
\end{equation}
where the signed tangential velocity can be obtained from the magnetic moment as follows
\begin{equation}\label{eq:utau_def}
  U_\tau = \sign(q) \sqrt{\frac{2 \Mu B_0}{m}}.
\end{equation}
Thus, the kinetic energy can be written as $K_0 = m (U_\shortparallel^2 + U_\tau^2)/2 = m U_\shortparallel^2/2 + \Mu B_0$.

The single-particle phase-space Lagrangian in field aligned velocity coordinates is expressed as
\begin{equation}\label{eq:part_lag_field_aligned}
  L_0 = \vc{\gamma}_0 \bcdot \dot{\vc{Z}} - H_0, \quad
  \vc{\gamma}_{0,\vc{R}} = q \vc{A}_0 + m U_\shortparallel \hb + m U_\tau \htau, \quad
  H_0 = \frac{m U_\shortparallel^2}{2} + \Mu B_0,
\end{equation}
where the remaining components of $\vc{\gamma}_{0}$ are zero.
Note that the symplectic part of the Lagrangian is now written as $\vc{\gamma}_0 \bcdot \dot{\vc{Z}}$, where we interchangeably refer to $\vc{\gamma}_0$ as the symplectic part of the Lagrangian.

\subsection{Small parameters}
Before we discuss near-identity phase-space coordinate transformations, we discuss the corresponding small parameters in which the coordinate transformation is expanded.

We let $L_B$ be the length scale on which the background magnetic field varies and let $\varrho$ denote the Larmor radius
\begin{equation}
  L_B \defeq \frac{\units{B_0}}{\units{\nabla B_0}}
  , \quad
  \varrho \defeq \frac{m \uth}{q \units{B_0}}
  , \quad
  \uth \defeq \sqrt{\frac{2 \kboltz \temperature}{m}}
  ,
\end{equation}
where $\units{Q}$ is the constant dimensional part of $Q$ and $\uth, \kboltz, \temperature$ denote the thermal velocity, the Boltzmann constant and the temperature, respectively.
The ratio of the two length scales is denoted by $\varepsilon_B$
\begin{equation}
  \varepsilon_B \defeq \frac{\varrho}{L_B}, 
\end{equation}
which is much smaller than one when the background magnetic field has a weak inhomogeneity~\citep{brizard2007}.
This is the parameter that is used in the guiding-centre coordinate transformation.

We let $\varepsilon_\delta$ denote the size of the perturbed magnetic field, which is introduced in \cref{sec:single_particle}, relative to the background magnetic field,
and we use the first subscript of any function or vector field to indicate the magnitude in terms of $\varepsilon_\delta$:
\begin{equation}
  \varepsilon_\delta \defeq \frac{\units{\vc{B}_1}}{\units{\vc{B}_0}}, \quad
  Q_l = \bigO(\varepsilon_\delta^l).
\end{equation}
It is assumed that the perturbed electric field~$\vc{E}_1$ scales identically such that any function linear in $\vc{E}_1$ is $\bigO(\varepsilon_\delta)$.
This is the parameter that is used in the gyrocentre coordinate transformation, which is discussed in \cref{sec:single_particle}.
The smallness of this parameter is motivated in \cref{sec:field_equilibria}.

Frequencies are non-dimensionalised using the cyclotron frequency resulting in the non-dimensional frequency $\varepsilon_\omega$
\begin{equation}\label{eq:eps_omega}
  \varepsilon_\omega \defeq \frac{\omega}{\units{\omega_\mathrm{c}}},
\end{equation}
which is a small parameter in the magnetic fusion devices that we consider \citep{Zoni2021}.
The assumed smallness of this parameter plays a crucial role in the approximation of the perturbed gyrocentre Lagrangian, which is discussed in \cref{sec:gy_H1_gauge_invariant}.

Finally, we non-dimensionalise the perpendicular length scale $2\upi / \kperp$ (that is, the typical length scale in the plane perpendicular to $\hb$) by the Larmor radius, which results in the non-dimensional wavenumber $\varepsilon_\perp$
\begin{equation}\label{eq:eps_perp}
  \varepsilon_\perp \defeq \kperp \varrho.
\end{equation}
We emphasize that this last parameter is not necessarily small; in particular when turbulence is considered we find that $\varepsilon_\perp \sim 1$.
This parameter is used to approximate the second-order (in $\varepsilon_\delta$) gyrocentre Hamiltonian in \cref{sec:H2}.

\subsection{Guiding-centre coordinates}\label{sec:guiding_center}
The second coordinate transformation that we consider results in the guiding-centre coordinates $\gc{\vc{Z}} = (\gc{\vc{R}}, \gc{U}_\shortparallel, \gc{\Mu}, \gc{\Theta})$. 
This transformation is aimed specifically at removing the gyro-phase dependence of $\vc{\gamma}_0$ (which depends on the gyro-phase \textit{via} the coordinate vector $\htau(\vc{R}, \Theta)$).
It results in the desired decoupling of the EOM for $\Theta$ from the other EOMs and thereby also decouples the fast gyrating motion of the particle.

The leading-order (in $\varepsilon_B$) contribution to the near-identity coordinate transformation of the particle position is given by (see \cref{fig:illustrate_gc})
\begin{align}\label{eq:gc_transformR}
  \gc{\vc{R}} = \vc{R} - \vc\rho,
\end{align}
where the particle radial vector and gyroradius are defined as
\begin{equation}\label{eq:gyroradius}
  \vc\rho \defeq \rho \hrho, \quad
  \rho \defeq \frac{m \gc{U}_\tau}{q B_0},
\end{equation}
and we note that a derivation of this well-known result can be found in \citet[Eq.~(2.58)]{brizard_PhD1990}.
The transformation of the remaining phase-space coordinates is not of interest to us here and therefore we do not list them.
The resulting guiding-centre single-particle phase-space Lagrangian is given by \citep[Eq.~(2.57)]{brizard_PhD1990}
\begin{equation}\label{eq:gc_lag}
  \gc{L}_0 \defeq \gc{\vc{\gamma}}_0 \bcdot \dot{\gc{\vc{Z}}} - \gc{H}_0, \quad
  \gc{\vc{\gamma}}_{0,\vc{R}} \defeq q \vc{A}_0^\effective, \quad
  \gc{\gamma}_{0,\Theta} \defeq \frac{m \gc{\Mu}}{q}, \quad
  \gc{H}_0 \defeq \gc{K}_0,
\end{equation}
where the effective guiding-centre vector potential is defined as
\begin{equation}\label{eq:gc_genvecpot}
  \vc{A}_0^\effective \defeq \vc{A}_0 + \frac{m \gc{U}_\shortparallel}{q} \hb - \frac{m \gc{\Mu}}{q^2} \vc{w}_0
  , \quad
  \vc{w}_0 \defeq (\nabla \htau) \hrho + \frac{1}{2} (\curl \hb)_\shortparallel \hb,
\end{equation} 
and the guiding-centre kinetic energy per particle is given by
\begin{equation}\label{eq:gc_energy_kin0}
  \gc{K}_0 = \frac{m \gc{U}_\shortparallel^2}{2} + \gc{\Mu} B_0.
\end{equation}

Here, the gradient of a vector field $\vc{S}$ is defined component wise as
\begin{equation}\label{eq:jacobian_matrix_def}
  (\nabla \vc{S})_{ij} \defeq \frac{\partial S_j}{\partial R_i}
\end{equation}
such that, for example, the components of $\vc{w}_0$ are given by the matrix-vector product
\begin{equation}
  (\vc{w}_0)_i = \sum_{j=1}^3 \frac{\partial \hat\tau_j}{\partial R_i} \hat{\rho}_j + \frac{1}{2} (\curl \hb)_\shortparallel (\hb)_i.
\end{equation}

The relevance of the first contribution to the $\vc{w}_0$ term becomes apparent when considering the transformation
\begin{equation}\label{eq:gg_transformation}
  \Theta \mapsto \Theta + \Psi(\gc{\vc{R}}), \quad
  \vc{\hat{e}}_i \mapsto \mat{T}(-\Psi) \vc{\hat{e}}_i,
\end{equation}
where the rotation matrix $\mat{T}$ is such that
\begin{equation}
  \mat{T}(\psi) \htau(\vc{R}, \Theta) = \htau(\vc{R}, \Theta + \psi), \quad
  \mat{T}(\psi) \hb = \hb.
\end{equation}
Invariance of the Lagrangian under this transformation reflects that we should be free to choose the coordinate vectors $\vc{\hat{e}}_i$. This is referred to as gyro-gauge invariance.
Indeed, we find that two terms of the symplectic part of the Lagrangian now depend on the gyro-phase $\gc{\Theta}$ or the coordinate vectors $\vc{\hat{e}}_i$, and their sum is given by
\begin{equation}
  - [(\nabla \htau) \hrho] \bcdot \dot{\gc{\vc{R}}} + \dot{\gc{\Theta}} 
  = \squarepar{- (\nabla \htau)^\transpose \dot{\gc{\vc{R}}}  + \dot{\gc{\Theta}} \hrho } \bcdot \hrho 
  = -\totd{\htau}{t} \bcdot \hrho,
\end{equation}
which is invariant under the transformation given by \cref{eq:gg_transformation}.

Note that we can furthermore show that
\begin{equation}
  \pard{}{\Theta} \squarepar{(\nabla \htau) \hrho} 
  = (\nabla \hrho) \hrho - (\nabla \htau) \htau 
  = \frac{1}{2} \nabla (\hrho \bcdot \hrho) - \frac{1}{2} \nabla (\htau \bcdot \htau)
  = \vc{0}_3,
\end{equation}
from which it follows that $\vc{w}_0$, and therefore also $\gc{\vc{\gamma}}_0$ is gyro-phase independent.
This implies that we can select the value $\Theta = \upi/2$ resulting in
\begin{equation}
  \vc{w}_0 
  = 
  \evalAt{\squarepar{(\nabla \htau) \hrho}}{\Theta = \upi/2}  + \frac{1}{2} (\curl \hb)_\shortparallel \hb
  = 
  (\nabla \vc{\hat{e}}_1) \vc{\hat{e}}_2  + \frac{1}{2} (\curl \hb)_\shortparallel \hb.
\end{equation}

\subsection{Principle of least action}\label{sec:gc_least_action}
Provided with the guiding-centre single-particle phase-space Lagrangian, we impose the principle of least action to obtain the corresponding EOMs.
That is, we impose
\begin{equation}\label{eq:action_principle_spl}
  \evalAt{\totd{}{\varepsilon}}{\varepsilon = 0} \int_{\tzero}^{\tone} \gc{L}_0(\gc{\vc{Z}} + \varepsilon \vc{\delta}, \dot{\gc{\vc{Z}}} + \varepsilon \dot{\vc{\delta}}) \diff t = 0,
\end{equation}
where $\vc{\delta}$ is arbitrary with $\vc{\delta}(\tzero) = \vc{\delta}(\tone) = \vc{0}_6$.
This results in the well-known Euler--Lagrange equations, which are given by
\begin{equation}\label{eq:hamilton_J0}
  \totd{}{t} \pard{\gc{L}_0}{\dot{\gc{\vc{Z}}}} = \pard{\gc{L}_0}{\gc{\vc{Z}}}
  \quad \iff \quad
  \dot{\gc{\vc{Z}}} = \gc{\mat{J}}_0 \roundpar{\pard{\gc{\vc{\gamma}}_0}{t} + \pard{\gc{H}_0}{\gc{\vc{Z}}}},
\end{equation}
where we have defined the Lagrange and Poisson matrices as
\begin{equation}\label{eq:gc_matrixdef}
  \gc{\mat{W}}_0 \defeq \roundpar{\pard{\gc{\vc{\gamma}}_0}{\gc{\vc{Z}}}}^\transpose - \pard{\gc{\vc{\gamma}}_0}{\gc{\vc{Z}}}, \quad
  \gc{\mat{J}}_0 \defeq (\gc{\mat{W}}_0)^{-1},
\end{equation}
respectively.
Here the components of the Jacobian matrix are given by (cf.\ \cref{eq:jacobian_matrix_def})
\begin{equation}
  \roundpar{\pard{\gc{\vc{\gamma}}_0}{\gc{\vc{Z}}}}_{ij} = \pard{\gc{\vc{\gamma}}_{0,i}}{\gc{Z}_j}
\end{equation}
and $^\transpose$ denotes the transpose of a matrix.

Provided with the Poisson matrix $\gc{\mat{J}}_0$, we can define the guiding-centre Poisson bracket as
\begin{equation}\label{eq:gc_bracket_def}
  \bracketgc{\mathcal{F}}{\mathcal{G}} \defeq \pard{\mathcal{F}}{\gc{\vc{Z}}} \bcdot \roundpar{\gc{\mat{J}}_{0} \pard{\mathcal{G}}{\gc{\vc{Z}}}},
\end{equation}
which allows the EOMs as given by \cref{eq:hamilton_J0} to be expressed as
\begin{equation}\label{eq:gc_eoms_as_bracket}
  \dot{\gc{\vc{Z}}} = \bracketgc{\gc{\vc{Z}}}{\gc{H}_0},
\end{equation}
where we have made use of the time-independence of $\gc{\vc{\gamma}}_0$, and we evaluate the bracket component-wise: $(\bracketgc{\gc{\vc{Z}}}{\gc{H}_0})_i = \bracketgc{\gc{Z}_i}{\gc{H}_0}$.

When using our expression for the symplectic part of the Lagrangian $\gc{\vc{\gamma}}_0$, as given by \cref{eq:gc_lag}, we find that the Lagrange matrix is given by
\begin{equation}
  \label{eq:gc_Lagrangematrix}
  \gc{\mat{W}}_0 = \begin{pmatrix}
    q \mat{B}_0^\effective & -m \hb & \frac{m}{q} \vc{w}_0 & \vc{0}_3\\
    m \hb^\transpose & 0 & 0 & 0\\
    - \frac{m}{q} \vc{w}_0^\transpose  & 0 & 0 & \frac{m}{q} \\
    \vc{0}_3^\transpose & 0 & - \frac{m}{q}& 0\\
  \end{pmatrix},
\end{equation}
where we have defined the matrix
\begin{equation}
  \mat{B}_0^\effective \defeq  \nabla \vc{A}_0^\effective - (\nabla \vc{A}_0^\effective)^\transpose
\end{equation}
for which
\begin{equation}\label{eq:matvec_is_timescurl}
  \mat{B}_0^\effective \vc{S} = \vc{S} \times \vc{B}_0^\effective, \quad
  \vc{B}_0^\effective \defeq \curl \vc{A}_0^\effective.
\end{equation}
Inversion of the Lagrange matrix, resulting in the Poisson matrix, is somewhat tedious and is therefore described in detail in \cref{app:matrix_inverse} (this coincides with the result of \citet[Appendix E]{parra2011phase}, except that therein the derivation is absent).
The result is given by
\begin{equation}\label{eq:gc_Poissonmatrix}
  \gc{\mat{J}}_0 = \begin{pmatrix}
    -\frac{\mat{B}_0}{q B_0 B_{0,\shortparallel}^\effective} & \frac{\vc{b}_{0}^\effective}{m} & \vc{0}_3 & - \frac{\vc{w}_0 \times \hb}{q B_{0,\shortparallel}^\effective}\\
    -\frac{(\vc{b}_{0}^\effective)^\transpose}{m} & 0 & 0 & -\frac{\vc{b}_{0}^\effective \bcdot \vc{w}_0}{m}\\
    \vc{0}_3^\transpose & 0 & 0 & -\frac{q}{m}\\
    \frac{(\vc{w}_0 \times \hb)^\transpose}{q B_{0,\shortparallel}^\effective} & \frac{\vc{b}_{0}^\effective \bcdot \vc{w}_0}{m} & \frac{q}{m}& 0\\
  \end{pmatrix},
\end{equation}
where we have defined
\begin{equation}\label{eq:gc_b0starstar}
  \vc{b}_0^\effective \defeq \frac{\vc{B}_{0}^\effective}{B_{0,\shortparallel}^\effective} = \hb + \frac{1}{q B_{0,\shortparallel}^\effective} \squarepar{
      m \gc{U}_\shortparallel \hb \times \vc{\kappa}
    - \frac{m \gc{\Mu}}{q} (\curl \vc{w}_0)_\perp
  },
\end{equation}
the parallel component of $\vc{B}_{0}^\effective$ is given by
\begin{equation}\label{eq:gy_Bpastar0}
  B_{0,\shortparallel}^\effective 
  =   B_0 
    + \frac{m \gc{U}_\shortparallel}{q} (\curl \hb)_\shortparallel 
    - \frac{m \gc{\Mu}}{q^2} (\curl \vc{w}_0)_\shortparallel,
\end{equation}
and the curvature vector $\vc{\kappa}$ is defined as
\begin{equation}\label{eq:curvature}
  \vc{\kappa} \defeq (\curl \hb)\times \hb.
\end{equation}
The matrix $\mat{B}_0$ is defined analogously to \cref{eq:matvec_is_timescurl} and therefore is given by
\begin{equation}
  \mat{B}_0 \defeq \nabla \vc{A}_0 - (\nabla \vc{A}_0)^\transpose
  \quad \implies \quad
  \mat{B}_0 \vc{S} = \vc{S} \times \vc{B}_0.
\end{equation}

This results in the following guiding-centre Poisson bracket
\begin{align}\label{eq:gc_bracket}
  \bracketgc{\mathcal{F}}{\mathcal{G}} = {} &
  - \frac{\hb }{q B_{0,\shortparallel}^\effective} \bcdot (\nabla \mathcal{F} \times \nabla \mathcal{G})
  + \frac{\vc{b}_{0}^\effective}{m} \bcdot \roundpar{\nabla \mathcal{F} \pard{\mathcal{G}}{\gc{U}_\shortparallel}
    - \pard{\mathcal{F}}{\gc{U}_\shortparallel} \nabla \mathcal{G}}\nonumber \\ &
  + \frac{q}{m} \roundpar{\pard{\mathcal{F}}{\gc{\Theta}} \pard{\mathcal{G}}{\gc{\Mu}}
    - \pard{\mathcal{F}}{\gc{\Mu}} \pard{\mathcal{G}}{\gc{\Theta}}} 
    + \frac{\vc{w}_0 \times \hb}{q B_{0,\shortparallel}^\effective} \bcdot \roundpar{\pard{\mathcal{F}}{\gc{\Theta}} \nabla \mathcal{G} 
  - \nabla \mathcal{F} \pard{\mathcal{G}}{\gc{\Theta}}}\nonumber \\ &
  + \frac{\vc{b}_{0}^\effective \bcdot \vc{w}_0}{m} \roundpar{\pard{\mathcal{F}}{\gc{\Theta}} \pard{\mathcal{G}}{\gc{U}_\shortparallel}
    - \pard{\mathcal{F}}{\gc{U}_\shortparallel} \pard{\mathcal{G}}{\gc{\Theta}}}
\end{align}
by substituting \cref{eq:gc_Poissonmatrix} into \cref{eq:gc_bracket_def}.
Substitution of \cref{eq:gc_bracket,eq:gc_lag} in \cref{eq:gc_bracket_def} yields the following guiding-centre EOMs
\begin{subequations}
  \label{eq:gc_eoms}
  \begin{align}
    \dot{\gc{\vc{R}}}           
      &= \gc{U}_\shortparallel \vc{b}_{0}^\effective + \frac{\gc{\Mu}}{q B_{0,\shortparallel}^\effective} \hb \times \nabla B_0
      ,\label{eq:gc_eoms_R}\\
    \dot{\gc{U}}_\shortparallel 
      &= -\frac{\gc{\Mu}}{m} \vc{b}_{0}^\effective \bcdot \nabla B_0,\\
    \dot{\gc{\Mu}}              
      &= 0,\\
    \dot{\gc{\Theta}}           
      &= \omega_\mathrm{c} + \vc{w}_0 \bcdot \dot{\gc{\vc{R}}} \,.
  \end{align}
\end{subequations}

Note that whereas the guiding-centre EOMs still contain the fast gyrating motion for which the frequency is given by the cyclotron frequency $\omega_\mathrm{c}$, this motion has been decoupled from the EOMs for the guiding-centre position and parallel velocity.
This means that if one is not interested in the gyro-phase $\gc{\Theta}$, then the corresponding EOM can be omitted entirely, thereby resulting in a phase-space dimensionality reduction.

\subsection{Discussion on guiding-centre coordinates}
We compare the guiding-centre EOMs given by \cref{eq:gc_eoms} to the EOMs in physical coordinates as given by \cref{eq:physical_eoms}. 
When integrating \cref{eq:physical_solution} in time, we find that the physical particle position is given by
\begin{align}\label{eq:physical_solutionR}
  \vc{R}(t) = {} &
  \vc{R}(0) 
  + U_\shortparallel(0) \hb t \nonumber \\ &
  + \frac{1}{\omega_{\mathrm{c}}} \squarepar{\sin(\omega_\mathrm{c} t) (\vc{\hat{e}}_1 \vc{\hat{e}}_1^\transpose + \vc{\hat{e}}_2 \vc{\hat{e}}_2^\transpose) - (\cos(\omega_\mathrm{c} t) - 1) (\vc{\hat{e}}_1 \vc{\hat{e}}_2^\transpose - \vc{\hat{e}}_2 \vc{\hat{e}}_1^\transpose)} \vc{U}_\perp(0),
\end{align}
where we recall that this result holds only if $\vc{B}_0$ is constant.
Under the same assumption, we find that the guiding-centre EOMs result in
\begin{equation}
  \dot{\gc{\vc{R}}}           
  = \gc{U}_\shortparallel \hb
  , \quad
  \dot{\gc{U}}_\shortparallel 
  = 0
  , \quad
  \dot{\gc{\Mu}} = 0
  , \quad
  \dot{\gc{\Theta}} = \omega_\mathrm{c},
\end{equation}
which upon integration in time yields
\begin{equation}\label{eq:gc_solution}
  \gc{\vc{R}}(t) = \gc{\vc{R}}(0) + \gc{U}_\shortparallel(0) \hb t
  , \quad
  \gc{\Mu}(t) = \gc{\Mu}(0)
  , \quad
  \gc{\Theta}(t) = \gc{\Theta}(0) + \omega_\mathrm{c} t.
\end{equation}

According to \citet[Eq.~(2.58)]{brizard_PhD1990} the velocity coordinates $(U_\shortparallel, \Mu, \Theta)$ transform trivially under the guiding-centre coordinate transformation whenever $\vc{B}_0$ is constant and therefore \cref{eq:gc_solution} can be written in physical coordinates as
\begin{equation}\label{eq:gc_solutionR}
  \vc{R}(t) = \vc{R}(0) + U_\shortparallel(0) \hb t + \frac{1}{\omega_{\mathrm{c}}}\squarepar{\hrho(\Theta^0 + \omega_\mathrm{c} t)- \hrho(\Theta^0)} \gc{U}_\tau(0),
\end{equation}
upon substitution of \cref{eq:gc_transformR,eq:gyroradius,eq:cyclotron_frequency} and letting $\Theta(0) = \Theta^0$.
Here we use the notational convention, as we do throughout this paper, that a superscripted~`$0$' indicates the initial value.
By making use of $\vc{U}_\perp(0) = U_\tau(0) \htau(\Theta^0)$, which follows from \cref{eq:field_aligned_velocity}, it can be shown that the solutions given by \cref{eq:physical_solutionR,eq:gc_solutionR} are identical, thereby confirming that we have consistently decoupled the fast gyrating motion using the guiding-centre coordinate transformation.
  \section{Gyrocentre single-particle phase-space Lagrangian}\label{sec:single_particle}
Thus far, we have discussed a model for the motion of a charged particle in the presence of a stationary background magnetic field $\vc{B}_0$, where the introduction of the guiding-centre coordinates has resulted in decoupling the fast gyration and has furthermore resulted in a phase-space dimensionality reduction.
However, the moving charged particle itself deposits a charge and current and thereby generates an electromagnetic field, which in turn affects the motion of the particle.
In this section we introduce a `perturbation' to the guiding-centre single-particle phase-space Lagrangian in the form of time-varying electromagnetic potentials, which in \cref{sec:vlasov_maxwell} allows us to derive a self-consistent formulation of the proposed gyrokinetic model.

\subsection{Perturbed guiding-centre Lagrangian}
In physical coordinates, the perturbed guiding-centre Lagrangian is given by
\begin{equation}\label{eq:gc_lagrangian_perturbation_tmp}
  \gc{L}_1^\dagger \defeq q \vc{A}_1(\vc{R}, t) \bcdot \dot{\vc{R}} - q \phi_1(\vc{R}, t),
\end{equation}
where $ \vc{A}_1$ and $\phi_1$ are the perturbed vector and scalar potentials resulting in the perturbed electric and magnetic field which are assumed to be small compared to $\vc{B}_0$, i.e.\ $\varepsilon_\delta \ll 1$.
Note that we have added a superscripted~${}^\dagger$ which we have introduced to distinguish this Lagrangian from the final perturbed guiding-centre Lagrangian in which we have subtracted the total derivative of some function.

Using \cref{eq:gc_transformR} we find that
\begin{equation}
  \vc{R} 
  = \gc{\vc{R}} + \vc\rho,
\end{equation}
which expresses the particle position $\vc{R}$ in terms of the guiding-centre position $\gc{\vc{R}}$ and the radial vector $\vc\rho$.
We introduce the following compact notation to indicate the evaluation of a scalar function, the gradient of a scalar function, or a vector field at the particle position
\begin{equation}\label{eq:eval_at_particle}
  \evalperturbedgc{Q} \defeq Q(\gc{\vc{R}} + \vc{\rho}), \quad
  \evalperturbedgc{\nabla} Q \defeq (\nabla Q)(\gc{\vc{R}} + \vc{\rho}), \quad
  \evalperturbedgc{S}_\tau \defeq \vc{S}(\gc{\vc{R}} + \vc{\rho}) \bcdot \htau(\gc{\vc{R}}).
\end{equation}
When considering \cref{fig:illustrate_gc} one might expect that the coordinate vector $\htau$ should be evaluated at the particle position $\gc{\vc{R}} + \vc{\rho}$.
However, from the derivation of the model it turns out that the evaluation is always done at the guiding-centre position $\gc{\vc{R}}$, which is equivalent to the evaluation at the particle position up to an $\bigO(\varepsilon_B)$ contribution.

When making use of \cref{eq:eval_at_particle} it follows that \cref{eq:gc_lagrangian_perturbation_tmp} can be rewritten as
\begin{equation}
  \label{eq:gc_lagrangian_perturbation}
  \gc{L}_1^\dagger = q \evalperturbedgc{\vc{A}}_1 \bcdot (\dot{\gc{\vc{R}}} + \dot{\vc{\rho}}) - q \evalperturbedgc{\phi}_1.
\end{equation}
Note that both $\vc\rho$ and $\evalperturbedgc{\vc{A}}_1$ (\textit{via} $\vc\rho$) depend on the gyro-phase $\gc{\Theta}$, and therefore we are in need of a third coordinate transformation which is aimed at removing the gyro-phase dependence of the perturbed guiding-centre Lagrangian and results in the gyrocentre phase-space coordinates $\gy{\vc{Z}} = (\gy{\vc{R}}, \gy{U}_\shortparallel, \gy{\Mu}, \gy{\Theta})$.

\subsection{Gyrocentre coordinate transformation}
Before we can perform the gyrocentre coordinate transformation, however, we must briefly discuss Lie transformations \citep{littlejohn1982,cary1983,dragtfinn1976}, which is used for this purpose.
We consider second-order Lie transformations, which are phase-space coordinate transformations of the form
\begin{equation}\label{eq:lie_coord_transform}
  \gc{\vc{Z}}
  = \gy{\vc{Z}} - \gy{\vc{G}}_1 + \frac{1}{2} \pard{\gy{\vc{G}}_1}{\gy{\vc{Z}}} \gy{\vc{G}}_1 - \gy{\vc{G}}_2,
\end{equation}
where $\gy{\vc{G}}_1$ and $\gy{\vc{G}}_2$ are the first- and second-order generating vectors.

The resulting gyrocentre Lagrangian is defined such that
\begin{equation}\label{eq:lie_lagrangian_def}
  \gy{L}(\gy{\vc{Z}}, \dot{\gy{\vc{Z}}}) = \gc{L}(\gc{\vc{Z}}, \dot{\gc{\vc{Z}}}) + \totd{\gy{S}}{t} + \bigO(\varepsilon_\delta^3),
\end{equation}
where we have added the total derivative of a generating function $\gy{S} = \gy{S}_1 + \gy{S}_2$ to the gyrocentre Lagrangian
\begin{equation}
  \gy{L} \mapsto \gy{L} + \totd{\gy{S}}{t},
\end{equation}
resulting in the following additions to the Hamiltonian and symplectic part
\begin{equation}\label{eq:total_derivative_update}
  \gy{H} \mapsto \gy{H} - \pard{\gy{S}}{t}
  , \quad
  \gy{\vc{\gamma}} \mapsto \gy{\vc{\gamma}} + \pard{\gy{S}}{\gy{\vc{Z}}}.
\end{equation}

This results in the following gyrocentre Hamiltonian $\gy{H} = \gy{H}_0 + \gy{H}_1 + \gy{H}_2$
\begin{subequations}
  \label{eq:lie2_F_transform}
  \begin{align}
    \gy{H}_0 &= \gc{H}_0, \\
    \gy{H}_1 &= \gc{H}_1 - \pard{\gc{H}_0}{\gy{\vc{Z}}} \bcdot \gy{\vc{G}}_1 - \pard{\gy{S}_1}{t}, \label{eq:lie2_F_transform1}\\
    \gy{H}_2 &= \gc{H}_2 - \pard{\gc{H}_0}{\gy{\vc{Z}}} \bcdot \gy{\vc{G}}_2 - \squarepar{\pard{}{t} \roundpar{\gc{\vc{\gamma}}_1
    - \frac{1}{2} \pard{\gc{\vc{\gamma}}_0}{\gy{\vc{Z}}} \gy{\vc{G}}_1} + \pard{}{\gy{\vc{Z}}}\roundpar{\gc{H}_1 - \frac{1}{2} \pard{\gc{H}_0}{\gy{\vc{Z}}} \bcdot \gy{\vc{G}}_1}} \bcdot \gy{\vc{G}}_1 - \pard{\gy{S}_2}{t},\label{eq:lie2_F_transform2}
  \end{align}
\end{subequations}
whereas the symplectic part $\gy{\vc{\gamma}} = \gy{\vc{\gamma}}_0 + \gy{\vc{\gamma}}_1 + \gy{\vc{\gamma}}_2$ is given by
\begin{subequations}
  \label{eq:lie2_gamma_transform}
  \begin{align}
    \gy{\vc{\gamma}}_0 &= \gc{\vc{\gamma}}_0 ,\\
    \gy{\vc{\gamma}}_1 &= \gc{\vc{\gamma}}_1 + \gc{\mat{W}}_0 \gy{\vc{G}}_1 + \pard{\gy{S}_1}{\gy{\vc{Z}}},\label{eq:lie2_gamma_transform1}\\
    \gy{\vc{\gamma}}_2 &= \gc{\vc{\gamma}}_2 + \gc{\mat{W}}_0 \gy{\vc{G}}_2 + \frac{1}{2}(\gc{\mat{W}}_1 + \gy{\mat{W}}_1) \gy{\vc{G}}_1 + \pard{\gy{S}_2}{\gy{\vc{Z}}}.\label{eq:lie2_gamma_transform2}
  \end{align}
\end{subequations}
We use the Lagrange matrix $\gc{\mat{W}}_0$ as given by \cref{eq:gc_matrixdef,eq:gc_Lagrangematrix} and have equivalently defined the perturbed Lagrange matrices as
\begin{subequations}\label{eq:perturbed_Lagrange}
  \begin{align}
    \gc{\mat{W}}_1 &\defeq \roundpar{\pard{\gc{\vc{\gamma}}_1}{\gy{\vc{Z}}}}^\transpose - \pard{\gc{\vc{\gamma}}_1}{\gy{\vc{Z}}}, \\
    \gy{\mat{W}}_1 
    &\defeq 
    \roundpar{\pard{\gy{\vc{\gamma}}_1}{\gy{\vc{Z}}}}^\transpose - \pard{\gy{\vc{\gamma}}_1}{\gy{\vc{Z}}}
    = 
    \gc{\mat{W}}_1 + \squarepar{\pard{}{\gy{\vc{Z}}} \roundpar{\gc{\mat{W}}_0 \gy{\vc{G}}_1} }^\transpose - \pard{}{\gy{\vc{Z}}} \roundpar{\gc{\mat{W}}_0 \gy{\vc{G}}_1} 
    .\label{eq:perturbed_Lagrange_gy}
  \end{align}
\end{subequations}
These transformation rules are classical results which can be obtained using Lie transform methods \citep{cary1983}, but can also be derived using Taylor series expansions, as shown in \cref{app:transform}.
Note that the contribution due to the first-order generating function vanishes in \cref{eq:perturbed_Lagrange_gy} because the skew-symmetric part of the Hessian matrix of $\gy{S}_1$ vanishes.


\subsection{The general form of the gyrocentre coordinate transformation}
The generating vectors $\gy{\vc{G}}_1$ and $\gy{\vc{G}}_2$, which are used to define the gyrocentre coordinate transformation, are chosen to satisfy some desired form of the symplectic part $\gy{\vc{\gamma}}_1, \gy{\vc{\gamma}}_2$ of the Lagrangian, by inverting \cref{eq:lie2_gamma_transform1,eq:lie2_gamma_transform2}, respectively.
This yields a transformation of the Hamiltonian part of the Lagrangian, as given by \cref{eq:lie2_F_transform}, where the generating vectors re-introduce gyro-phase dependence in the gyrocentre Hamiltonian.
The role of the generating functions $\gy{S}_1$ and $\gy{S}_2$ is to absorb the gyro-phase dependent part of the resulting gyrocentre Hamiltonian.

\subsubsection{The first-order transformation}
Without specifying the desired form of $\gy{\vc{\gamma}}_1$, we find that this approach results in the following first-order generating vector field
\begin{equation}\label{eq:gy_genvector1}
  \gy{\vc{G}}_1 = \gc{\mat{J}}_0 \roundpar{\gy{\vc{\gamma}}_1 - \gc{\vc{\gamma}}_1 - \pard{\gy{S}_1}{\gy{\vc{Z}}}},
\end{equation}
which, upon substitution in \cref{eq:lie2_F_transform1}, results in the following first-order Hamiltonian
\begin{equation}\label{eq:gy_H1_withbracket}
  \gy{H}_1 
  = \gc{H}_1 +  \roundpar{\gy{\vc{\gamma}}_1 - \gc{\vc{\gamma}}_1} \bcdot \dot{\gc{\vc{Z}}} - \pard{\gy{S}_1}{t} - \bracketgc{\gy{S}_1}{\gy{H}_0} 
  = q \psi_1 - \pard{\gy{S}_1}{t} - \bracketgc{\gy{S}_1}{\gy{H}_0},
\end{equation}
where we have used \cref{eq:hamilton_J0,eq:gc_bracket_def}, denote by $\dot{\gc{\vc{Z}}}$ the unperturbed guiding-centre EOMs~\eqref{eq:gc_eoms} evaluated at the gyrocentre coordinate $\gy{\vc{Z}}$, and we have defined the effective potential as
\begin{equation}\label{eq:gy_effpot}
  q \psi_1 \defeq \gc{H}_1 + \roundpar{\gy{\vc{\gamma}}_1 - \gc{\vc{\gamma}}_1} \bcdot \dot{\gc{\vc{Z}}}.
\end{equation}

We let the generating function $\gy{S}_1$ absorb the gyro-phase dependent part of $\psi_1$ such that \cref{eq:gy_H1_withbracket} results in
\begin{equation}\label{eq:gy_S1_def}
  \pard{\gy{S}_1}{t} + \bracketgc{\gy{S}_1}{\gy{H}_0} = q \wt{\psi}_1
  \quad\implies\quad
  \gy{H}_1 
  = q \gav{\psi_1},
\end{equation}
where we define the gyro-average and the resulting gyro-phase dependent part of some function~$Q$ as
\begin{equation}\label{eq:gyro_average}
  \gav{Q} \defeq \frac{1}{2\upi} \int_0^{2\upi} Q \diff \gy{\Theta},
  \quad
  \wt{Q} \defeq Q - \gav{Q},
\end{equation}
which is defined component-wise for vector fields.
It follows that the first-order part of the gyrocentre single-particle phase-space Lagrangian is given by
\begin{equation}\label{eq:gy_L1_general}
  \gy{L}_1 
  = \gy{\vc{\gamma}}_1 \bcdot \dot{\gy{\vc{Z}}} - \gy{H}_1
  = 
    \gy{\vc{\gamma}}_1 \bcdot \dot{\gy{\vc{Z}}} 
  + \gav{\gc{\vc{\gamma}}_1 - \gy{\vc{\gamma}}_1} \bcdot \dot{\gc{\vc{Z}}}
  - \gav{\gc{H}_1}.
\end{equation}

It is insightful to consider the two limiting cases of \cref{eq:gy_L1_general}: if $\gy{\vc{\gamma}}_1 = \vc{0}_6$ then the gyrocentre coordinate transformation transforms the entire symplectic part of the first-order guiding-centre Lagrangian to the Hamiltonian part of the Lagrangian (this is referred to as the Hamiltonian formulation)
\begin{subequations}
  \begin{equation}
    \gy{\vc{\gamma}}_1 = \vc{0}_6
    \quad \implies \quad
    \gy{H}_1 
    =
      \gav{\gc{H}_1}
    - \gav{\gc{\vc{\gamma}}_1} \bcdot \dot{\gc{\vc{Z}}}
  \end{equation}
  and conversely if $\gy{\vc{\gamma}}_1 = \gav{\gc{\vc{\gamma}}_1}$ then the symplectic and Hamiltonian parts of the first-order guiding-centre Lagrangian simply end up being gyro-averaged
  \begin{equation}
    \gy{\vc{\gamma}}_1 = \gav{\gc{\vc{\gamma}}_1}
    \quad \implies \quad
    \gy{H}_1 
    = 
     \gav{\gc{H}_1}.
  \end{equation}
\end{subequations}

\subsubsection{The second-order transformation}
We follow the same approach for deriving the second-order Hamiltonian.
That is, we solve \cref{eq:lie2_gamma_transform2} for $\gy{\vc{G}}_2$ resulting in
\begin{equation}
  \gy{\vc{G}}_2 = \gc{\mat{J}}_0 \squarepar{\gy{\vc{\gamma}}_2 - \frac{1}{2}(\gc{\mat{W}}_1 + \gy{\mat{W}}_1) \gy{\vc{G}}_1 - \pard{\gy{S}_2}{\gy{\vc{Z}}}},
\end{equation}
without specifying $\gy{\vc{\gamma}}_2$ and by making use of $\gc{\vc{\gamma}}_2 = \vc{0}_6$.
This allows us to express the second-order Hamiltonian~\eqref{eq:lie2_F_transform2} in the following way
\begin{equation}
  \gy{H}_2 = 
      \gy{\vc{\gamma}}_2 \bcdot \dot{\gc{\vc{Z}}}
    + \vc{T}_1 \bcdot \gy{\vc{G}}_1 
    - \pard{\gy{S}_2}{t}
    - \bracketgc{\gy{S}_2}{\gc{H}_0},
\end{equation}
where we have made use of $\gc{H}_2 = 0$, and we have defined
\begin{equation}\label{eq:gy_Tvector}
  \vc{T}_1 \defeq 
    \frac{1}{2} (\gc{\mat{W}}_1 + \gy{\mat{W}}_1) \dot{\gc{\vc{Z}}} 
  - \pard{}{t} \roundpar{\gc{\vc{\gamma}}_1
  - \frac{1}{2} \pard{\gc{\vc{\gamma}}_0}{\gy{\vc{Z}}} \gy{\vc{G}}_1} 
  - \pard{}{\gy{\vc{Z}}}\roundpar{\gc{H}_1 - \frac{1}{2} \pard{\gc{H}_0}{\gy{\vc{Z}}} \bcdot \gy{\vc{G}}_1}
\end{equation}
by making use of \cref{eq:hamilton_J0}.

As with the first-order generating function, the second-order generating function $\gy{S}_2$ is defined such that it absorbs the gyro-phase dependent part of $\gy{H}_2$ resulting in
\begin{equation}\label{eq:gy_S2_def}
  \pard{\gy{S}_2}{t} + \bracketgc{\gy{S}_2}{\gc{H}_0} = 
     \wt{\gy{\vc{\gamma}}_2} \bcdot \dot{\gc{\vc{Z}}}
  + \wt{\vc{T}_1 \bcdot \gy{\vc{G}}_1}
\end{equation}
and therefore
\begin{equation}\label{eq:gy_H2_general}
  \gy{H}_2 = 
      \gav{\gy{\vc{\gamma}}_2} \bcdot \dot{\gc{\vc{Z}}}
    + \gav{\vc{T}_1 \bcdot \gy{\vc{G}}_1}.
\end{equation}

To summarise, we have thus far considered a general gyrocentre coordinate transformation, where we are still free to choose the symplectic parts $\gy{\vc{\gamma}}_1$ and $\gy{\vc{\gamma}}_2$. 
The resulting first- and second-order Hamiltonians are given by
\begin{equation}\label{eq:gy_H1_general}
  \gy{H}_1 = \gav{\gc{H}_1} + \gav{\gy{\vc{\gamma}}_1 - \gc{\vc{\gamma}}_1} \bcdot \dot{\gc{\vc{Z}}}
\end{equation}
as well as \cref{eq:gy_H2_general}, respectively as follows from \cref{eq:gy_effpot,eq:gy_S1_def}.
For consistency, we require that $\gy{\vc{\gamma}}_1$ and $\gy{\vc{\gamma}}_2$ are $\bigO(\varepsilon_\delta)$ and $\bigO(\varepsilon_\delta^2)$, respectively.
As the purpose of the gyrocentre coordinate transformation is to decouple the gyro-phase from the perturbed Lagrangian, we must also require $\wt{\gy{\vc{\gamma}}_1} = \wt{\gy{\vc{\gamma}}_2} = \vc{0}_6$.
Moreover, we require the magnetic moment to remain an invariant in gyrocentre coordinates.
The requirement on the coordinate transformation to obtain invariance of the magnetic moment can be found by considering the Euler--Lagrange equation for $\gy{\Theta}$
\begin{equation}\label{eq:invariance_mu_condition}
  \totd{}{t} \pard{\gy{L}}{\dot{\gy{\Theta}}} = \pard{\gy{L}}{\gy{\Theta}}
  \quad \implies \quad
  \totd{\gy{\gamma}_{\Theta}}{t} = 0
  \quad \implies \quad
  \dot{\gy{\Mu}} = - \frac{q}{m} \totd{}{t} (\gy{\gamma}_{1,\Theta} + \gy{\gamma}_{2,\Theta})
  ,
\end{equation}
which shows that $\gy{\gamma}_{1,\Theta} + \gy{\gamma}_{2,\Theta} = 0$ is sufficient for obtaining invariance of $\gy{\Mu}$.
In what follows, we discuss a fourth requirement on $\gy{\vc{\gamma}}_1$ and $\gy{\vc{\gamma}}_2$, which ensures that the resulting gyrocentre single-particle phase-space Lagrangian is gauge invariant.

\subsection{Gauge invariance}\label{sec:gauge_invariance}
Gauge invariance refers to invariance under the gauge transformation
\begin{equation}\label{eq:gauge_transform}
  \phi_1 \mapsto \phi_1 - \pard{\eta}{t}, \quad
  \vc{A}_1 \mapsto \vc{A}_1 + \nabla \eta
\end{equation}
for some scalar function $\eta$.
The electromagnetic fields, as given by
\begin{subequations}
  \label{eq:electromagnetic_fields}
  \begin{align}
  \vc{E}_1 &\defeq - \nabla \phi_1 -\pard{\vc{A}_1}{t}, \\
  \vc{B}_1 &\defeq \curl \vc{A}_1
  \end{align}
\end{subequations}
are invariant under the gauge transformation~\eqref{eq:gauge_transform}, from which it follows that any (part of a) model which is expressed in terms of the electromagnetic fields is automatically gauge invariant.
If a model is gauge invariant, it means that it does not matter which gauge condition is used to fix the function $\eta$, which is what we would expect from a physical point of view.

Following the discussion in \citet{burby2019gauge}, we introduce the following parametrized perturbed Lagrangian
\begin{equation}
  \gcburby{L}_1 \defeq q \evalperturbedgcburby{\vc{A}}_1 \bcdot \roundpar{\dot{\gc{\vc{R}}} + \burby \dot{\vc\rho}} - q \evalperturbedgcburby{\phi}_1,
\end{equation}
where we have defined
\begin{equation}\label{eq:eval_pert_burby}
  \evalperturbedgcburby{Q} \defeq Q(\gc{\vc{R}} + \burby \vc{\rho}).
\end{equation}
The $\burby$ parameter therefore interpolates from the guiding-centre position ($\burby = 0$) to the particle position ($\burby = 1$), see also \cref{fig:illustrate_gc}.
It follows that the perturbed guiding-centre Lagrangian, as given by \cref{eq:gc_lagrangian_perturbation_tmp}, coincides with $\burby = 1$ and can therefore be written as
\begin{equation}\label{eq:gc_lag_pert_burby}
  \gc{L}_1^\dagger  
  = \gc{L}_1^{\dagger,\burby = 1}
  = \gc{L}_1^{\dagger,\burby = 0} + (\gc{L}_1^{\dagger,\burby = 1} - \gc{L}_1^{\dagger,\burby = 0})
  = \underbrace{\gc{L}_1^{\dagger,\burby = 0}}_{\gc{L}_1^{\dagger,\zlr}} + \underbrace{\int_0^1 \totd{\gcburby{L}_1}{\burby} \diff \burby}_{\gc{L}_1^{\dagger,\flr}},
\end{equation}
which can therefore be written as the sum of a zero Larmor radius (ZLR) contribution and a finite Larmor radius (FLR) contribution.

Computation of the $\burby$ derivative of the parametrized Lagrangian yields
\begin{equation}
  \totd{\gcburby{L}_1}{\burby} = 
      q \squarepar{(\evalperturbedgcburby{\nabla} \vc{A}_1)^\transpose \vc\rho} \bcdot (\dot{\gc{\vc{R}}} + \burby \dot{\vc\rho})
      + q \evalperturbedgcburby{\vc{A}}_1 \bcdot \dot{\vc\rho}
      - q \evalperturbedgcburby{\nabla} \phi_1 \bcdot \vc\rho.
\end{equation}
Furthermore, we note that
\begin{equation}\label{eq:gc_gaugeinvariance_totalder}
  \totd{}{t} (\vc\rho \bcdot \evalperturbedgcburby{\vc{A}}_1) 
  =   \dot{\vc{\rho}} \bcdot \evalperturbedgcburby{\vc{A}}_1 + \vc\rho \bcdot \totd{}{t} \vc{A}(\gc{\vc{R}} + \burby \vc\rho, t)
  =   \dot{\vc{\rho}} \bcdot \evalperturbedgcburby{\vc{A}}_1 
    + \vc{\rho} \bcdot \squarepar{\pard{\evalperturbedgcburby{\vc{A}}_1}{t} + (\evalperturbedgcburby{\nabla} \vc{A}_1)^\transpose (\dot{\gc{\vc{R}}} + \burby \dot{\vc\rho})},
\end{equation}
from which it follows that
\begin{equation}\label{eq:gc_lag_minus_term}
  \totd{\gcburby{L}_1}{\burby} - q\totd{}{t} (\vc\rho \bcdot \evalperturbedgcburby{\vc{A}}_1) = 
    q \vc\rho \bcdot \squarepar{(\dot{\gc{\vc{R}}} + \burby \dot{\vc\rho}) \times \evalperturbedgcburby{\vc{B}}_1 + \evalperturbedgcburby{\vc{E}}_1},
\end{equation}
and therefore the FLR part of the perturbed guiding-centre Lagrangian can, up to a total derivative, be expressed in terms of the gauge-invariant electromagnetic fields.

In what follows, we omit the contribution by the total derivative, as this does not alter the resulting EOMs after imposing the principle of least action.
We denote the resulting perturbed guiding-centre Lagrangian by $\gc{L}_1$ for which
\begin{equation}\label{eq:gc_L1_totalder}
  \gc{L}_1 \defeq \gc{L}_1^\dagger - q \totd{}{t} \int_0^1 \vc\rho \bcdot \evalperturbedgcburby{\vc{A}}_1 \diff\burby.
\end{equation}
And therefore
\begin{subequations}
  \label{eq:gc_lag_pert_burby_split}
  \begin{equation}
    \gc{L}_1 = (\gc{\vc{\gamma}}_{1}^\zlr + \gc{\vc{\gamma}}_{1}^\flr) \bcdot \dot{\gc{\vc{Z}}} - (\gc{H}_1^\zlr + \gc{H}_1^\flr),
  \end{equation}
  where the symplectic part $\gc{\vc{\gamma}}_{1} \defeq \gc{\vc{\gamma}}_{1}^\zlr + \gc{\vc{\gamma}}_{1}^\flr$ is given by
  \begin{equation}\label{eq:gc_lag_pert_burby_split_gamma10}
    \gc{\vc{\gamma}}_{1,\vc{R}}^\zlr \defeq q\vc{A}_1
    , \quad 
    \gc{\vc{\gamma}}_{1,\vc{R}}^\flr \defeq q \int_0^1  \evalperturbedgcburby{\vc{B}}_1 \diff \burby \times \vc\rho
    , \quad 
    \gc{\gamma}_{1,\Theta}^\flr \defeq - q \rho^2 \int_0^1 \burby \evalperturbedgcburby{\vc{B}}_1 \diff \burby \bcdot \hb,
  \end{equation}
  and the Hamiltonian part $\gc{H}_1 \defeq \gc{H}_1^\zlr + \gc{H}_1^\flr$ is given by
  \begin{equation}\label{eq:gc_lag_pert_burby_split_hamiltonian}
    \gc{H}_1^\zlr \defeq q \phi_1
    , \quad
    \gc{H}_1^\flr \defeq - q \int_0^1  \evalperturbedgcburby{\vc{E}}_1 \diff \burby \bcdot \vc\rho.
  \end{equation}
\end{subequations}
We distinguish the ZLR contributions from the FLR contributions.
Note that each of the FLR contributions is gauge invariant, as they are expressed in terms of the electromagnetic fields.

When considering the Lie coordinate transformation given by \cref{eq:lie2_gamma_transform,eq:lie2_F_transform}, we find that the following yields a sufficient condition for gauge invariance of the resulting gyrokinetic model.
This is a new result which provides a general approach for the development of gauge-invariant gyrokinetic models.
A proof can be found in \cref{sec:gauge_invariance_proof}.

\begin{theorem}[Sufficient condition for gauge invariance]\label{thm:gauge_invariance}
  The gyrocentre single-particle phase-space Lagrangian (to second-order) is gauge invariant up to a total derivative
  \begin{equation}\label{eq:gauge_invariance_thm}
    \gy{L} 
    \overset{\text{\cref{eq:gauge_transform}}}{\mapsto}
    \gy{L} + q \roundpar{\nabla \eta \bcdot \dot{\gy{\vc{R}}} + \pard{\eta}{t}}
    = 
    \gy{L} + q \totd{\eta}{t}.
  \end{equation}  
  provided that $\gy{\vc{\gamma}}_1 - \gc{\vc{\gamma}}_1$ and $\gy{\vc{\gamma}}_2$ are gauge invariant.
\end{theorem}

\begin{remark}[Cross terms of $\bigO(\varepsilon_\delta \varepsilon_B)$]
  In the expression for $\gc{\vc{\gamma}}_{1,\vc{R}}^\flr$ in \cref{eq:gc_lag_pert_burby_split_gamma10} we have neglected the $\bigO(\varepsilon_\delta \varepsilon_B)$ contribution.
  Neglecting this term is consistent with the leading-order (in $\bigO(\varepsilon_B)$) approximation of the particle position in terms of the guiding-centre coordinates as given in \cref{eq:gc_transformR}.
  
  When a conventional gyrokinetic ordering is used \citep{parra2011phase}, where in particular it is assumed that $\varepsilon_\delta = \varepsilon_B$, we find that the neglected cross terms are of the same order as terms that eventually end up in the second-order gyrocentre Hamiltonian $\gy{H}_2 = \bigO(\varepsilon_\delta^2)$ (see also \cref{sec:H2}).
  Hence, when a conventional gyrokinetic ordering is used one should retain these terms, as is done in \citet{parra2011phase}.
  In the present work such terms are not retained in favour of clarity and simplicity of the resulting model, and we note that such cross terms are also not included in the state-of-the-art parallel-only gyrokinetic models used in practice \citep{Kleiber_pullback,brizard2007}.
  However, the cross terms can easily be included without breaking the structure of the proposed model, as we now demonstrate.
  
  A more accurate approximation of the guiding-centre coordinate transformation is considered, which is given by
  \begin{equation}\label{eq:gc_transformR_parra}
    \vc{R} = \gc{\vc{R}} + \vc\rho + \vc\drho,
  \end{equation}
  where $\vc\drho$ is an $\bigO(\varepsilon_B)$ correction (i.e.\ $\vc\drho = -\vc\rho_1$ in \citet[Eq.~(2.58)]{brizard_PhD1990}).
  When taking this additional correction into account, the FLR part of the perturbed guiding-centre Lagrangian becomes (the ZLR part is unchanged)
  \begin{equation}
    q \int_0^1 (\vc\rho + \vc{\drho}) \bcdot \squarepar{(\dot{\gc{\vc{R}}} + \burby \dot{\vc\rho} + \burby \dot{\vc\drho}) \times \evalperturbedgcburby{\vc{B}}_1 + \evalperturbedgcburby{\vc{E}}_1} \diff\burby,
  \end{equation}
  which, when neglecting $\bigO(\varepsilon_\delta \varepsilon_B^2)$ terms, results in (cf.\ \cref{eq:gc_lag_minus_term})
  \begin{align}
    \gc{L}_1^\flr
    &= q\int_0^1 \roundpar{\evalperturbedgcburby{\vc{B}}_1 \bcdot \squarepar{\vc\rho \times \roundpar{\dot{\gc{\vc{R}}} + \burby \rho \htau \dot{\gc{\Theta}}}} + \vc\rho \bcdot \evalperturbedgcburby{\vc{E}}_1} \diff\burby \nonumber\\
    &+ q\int_0^1 \squarepar{\evalperturbedgcburby{\vc{B}}_1 \bcdot \roundpar{\vc{\drho} \times \squarepar{\dot{\gc{\vc{R}}} + \burby \rho \htau \dot{\gc{\Theta}}} + \vc\rho \times \squarepar{\burby (\nabla \vc\rho)^\transpose \dot{\gc{\vc{R}}} + \burby \pard{\vc\drho}{\Theta} \dot{\gc{\Theta}}}} + \vc{\drho} \bcdot \evalperturbedgcburby{\vc{E}}_1} \diff\burby,
  \end{align}
  where the second row contains all the cross terms which we do not include in the present work.
  We note that the definition of $\evalperturbedgcburby{Q}$ (cf.\ \cref{eq:eval_pert_burby}) is altered according to \cref{eq:gc_transformR_parra} when this more accurate approximation to the perturbed guiding-centre Lagrangian is considered.
  Moreover, care should be taken that $\bigO(\varepsilon_B^2)$ terms are also included in the guiding-centre Lagrangian $\gc{L}_0$ when such an approach is followed.
\end{remark}%

\subsection{A family of gauge-invariant gyrocentre coordinate transformations}
Thus far, we have considered a general gyrocentre coordinate transformation, which is defined by the symplectic part of the gyrocentre Lagrangian: $\gy{\vc{\gamma}}_1$ and $\gy{\vc{\gamma}}_2$.
In what follows, we let $\gy{\vc{\gamma}}_2 = \vc{0}_6$.
Therefore, we find that consistency (w.r.t.\ the Lie transformation), gyro-phase independence, invariance of the gyrocentre magnetic moment (cf.\ \cref{eq:invariance_mu_condition}) and gauge invariance of the resulting gyrocentre Lagrangian requires the following four conditions to be satisfied
\begin{equation}
  \gy{\vc{\gamma}}_1 = \bigO(\varepsilon_\delta)
  , \quad
  \wt{\gy{\vc{\gamma}}_1} = \vc{0}_6
  , \quad
  \gy{\gamma}_{1,\Theta} = 0
  , \quad
  \gy{\vc{\gamma}}_1 - \gc{\vc{\gamma}}_1 \overset{\text{\cref{eq:gauge_transform}}}{\mapsto} \gy{\vc{\gamma}}_1 - \gc{\vc{\gamma}}_1
  ,
\end{equation}
respectively.

\subsubsection{Overview}
Traditionally \citep{brizard_PhD1990,brizard2007}, the following choice was made
\begin{equation}\label{eq:gy_gamma_brizard}
  \gy{\vc{\gamma}}_{1,\vc{R}} = q \gav{\evalperturbedgc{\vc{A}}_1},
\end{equation}
with all other components equal to zero.
Note that this corresponds to $\gy{\vc{\gamma}}_{1,\vc{R}} = \gav{\gc{\vc{\gamma}}_{1,\vc{R}}^\dagger}$, where $\gc{\vc{\gamma}}_{1}^\dagger$ denotes the symplectic part of the perturbed guiding-centre Lagrangian before the total derivative has been omitted, see also \cref{eq:gc_L1_totalder}.
This choice satisfies the first, second and third of our requirements, but it does not lead to a gauge invariant model:
\begin{equation}
  \gy{\vc{\gamma}}_{1} - \gc{\vc{\gamma}}_{1}^\dagger 
    \overset{\text{\cref{eq:gauge_transform}}}{\mapsto}
  \gy{\vc{\gamma}}_{1} - \gc{\vc{\gamma}}_{1}^\dagger -
  \begin{pmatrix}
    q \wt{\evalperturbedgc{\nabla} \eta}\\
    0\\
    \frac{q \rho}{2\gc{\Mu}} \hrho \bcdot \evalperturbedgc{\nabla} \eta\\
    q \rho \htau \bcdot \evalperturbedgc{\nabla} \eta
  \end{pmatrix}.
\end{equation}
This approach leads to a gyrokinetic model in which the compressional Alfv\'en wave can be included by considering a high-frequency approximation for the first-order generating function $\gy{S}_1$, as proposed by \citet{qin1999gyrokinetic}.

More recently, the following gyrocentre coordinate transformation was proposed by \citet{burby2019gauge}
\begin{equation}\label{eq:gy_gamma_burby}
  \gy{\vc\gamma}_1 = \gc{\vc\gamma}_1^\zlr
  \quad\implies\quad 
  \gy{\vc\gamma}_{1,\vc{R}} = q \vc{A}_1,
\end{equation}
which satisfies all of our requirements since
\begin{equation}
  \gy{\vc{\gamma}}_1 - \gc{\vc{\gamma}}_1  = - \gc{\vc\gamma}_1^\flr,
\end{equation}
which is gauge invariant as the FLR parts of the perturbed guiding-centre Lagrangian are gauge invariant.
Rather than keeping only the ZLR part of the symplectic part of the perturbed guiding-centre Lagrangian, we can also include the FLR effects resulting in
\begin{equation}\label{eq:gy_gamma_minimal}
  \gy{\vc\gamma}_1 = \gc{\vc\gamma}_1^\zlr + \gav{\gc{\vc\gamma}_1^\flr} = \gav{\gc{\vc\gamma}_1}.
\end{equation}
Gauge invariance follows from the gauge invariance of $\gc{\vc\gamma}_1^\flr$, and we have gyro-averaged the FLR contribution to ensure that our second requirement is satisfied.
We note that \cref{eq:gy_gamma_minimal} results in a transformation for which the first-order generating vector is, in some sense, smallest.
This is of interest because the gyrokinetic model results from a \emph{truncated} coordinate transformation, where the truncation error is smaller if the coordinate transformation is smaller.
In particular, using \cref{eq:gy_genvector1} we find that
\begin{equation}\label{eq:gy_trans_trivial}
  \gy{\vc\gamma}_1 = \gav{\gc{\vc\gamma}_1}
  \quad \implies \quad
  \gav{\gy{\vc{G}}_1} = \gc{\mat{J}}_0 \sgav{\gy{\vc{\gamma}}_1 - \gc{\vc{\gamma}}_1 - \pard{\gy{S}_1}{\gy{\vc{Z}}}} = \vc{0}_6,
\end{equation}
and therefore the coordinate transformation given by \cref{eq:gy_gamma_minimal} contains, to first-order, only a fluctuating gyro-phase dependent part.

Henceforth, we consider the following parametrized form of the symplectic part of the gyrocentre Lagrangian
\begin{equation}\label{eq:gy_gamma_param}
  \gy{\vc\gamma}_1 \defeq \begin{pmatrix}
    q \vc{A}_1 + \gparamr q \radgav{\evalperturbedgcburby{\vc{B}}_1 \times \vc\rho}\\
    0\\
    0\\
    - \gparamtheta \frac{q \rho^2}{2} \dgav{\evalperturbedgcburby{B}_{1,\shortparallel}}
  \end{pmatrix},
\end{equation}
where $\gparam_R, \gparamtheta$ are real-valued parameters that define the coordinate transformation. 
The choice $(\gparamr, \gparamtheta) = (0, 0)$ yields the model proposed in \citep{burby2019gauge}, $(\gparamr, \gparamtheta) = (1, 1)$ results in \cref{eq:gy_gamma_minimal}, and we note that this general form is gauge invariant regardless of the value of $\gparamr, \gparamtheta$.
Here we have defined 
the radially averaged gyro-average as
\begin{equation}\label{eq:radial_average}
  \radgav{\evalperturbedgcburby{Q}} 
  \defeq \int_0^1 \gav{\evalperturbedgcburby{Q}} \diff \burby
  = \frac{1}{2\upi} \int_0^1 \int_0^{2\upi} Q(\gy{\vc{r}} + \burby \vc{\rho}) \diff \gy{\Theta} \diff \burby
\end{equation}
and the disc average as (cf.\ \cref{eq:eval_pert_burby})
\begin{equation}\label{eq:disc_average}
  \dgav{\evalperturbedgcburby{Q}} 
  \defeq 2 \radgav{\burby \evalperturbedgcburby{Q}}
  = \frac{1}{\upi} \int_0^1 \int_0^{2\upi} \burby Q(\gy{\vc{r}} + \burby \vc{\rho}) \diff \gy{\Theta} \diff \burby,
\end{equation}
which is defined component-wise for vector fields.
The latter operator is referred to as the disc average \citep{porazik2011gyrokinetic} because it exactly yields the average value of the `gyro-disc' shown in \cref{fig:illustrate_gc}.

The parametrized coordinate transformation results in the following first-order gyrocentre Lagrangian
\begin{align}
  \gy{L}_1 = {} &
      \gc{\vc\gamma}_1^\zlr \bcdot \dot{\gy{\vc{Z}}} 
    + \gav{\gc{\vc\gamma}_{1,\vc{R}}^\flr} \bcdot \squarepar{\gparamr \dot{\gy{\vc{R}}} + (1 - \gparamr) \dot{\gc{\vc{R}}}}
    + \gav{\gc{\gamma}_{1,\Theta}^\flr} \squarepar{\gparamtheta \dot{\gy{\Theta}} + (1 - \gparamtheta) \dot{\gc{\Theta}}} \nonumber\\ &
    - q \phi_1 
    + q \rho \int_0^1 \gav{\evalperturbedgcburby{E}_{1,\rho}} \diff\burby
    ,\label{eq:gy_L1}
\end{align}
where we have substituted \cref{eq:gy_gamma_param,eq:gc_lag_pert_burby_split_hamiltonian} into \cref{eq:gy_H1_general}.
This shows that the parameters $\gparam_R, \gparamtheta$ put the symplectic FLR part of the perturbed guiding-centre Lagrangian either in the symplectic ($(\gparam_R, \gparamtheta) = (1, 1)$) or in the Hamiltonian ($(\gparam_R, \gparamtheta) = (0, 0)$) part of the first-order gyrocentre Lagrangian.

We can already ensure that the gyrocentre magnetic moment is an invariant by imposing our third condition, where we note that
\begin{equation}\label{eq:eom_mu_gparamtheta}
  \dot{\gy{\Mu}} 
  = - \frac{q}{m} \totd{}{t} (\gy{\gamma}_{1,\Theta} + \gy{\gamma}_{2,\Theta}) 
  = \gparamtheta \totd{}{t} \roundpar{\gy{\Mu} \frac{\dgav{\evalperturbedgcburby{B}_{1,\shortparallel}}}{B_0}}
\end{equation}
by substituting \cref{eq:gy_gamma_param} into \cref{eq:invariance_mu_condition}.
Hence, requiring $\gy{\Mu}$ to remain invariant in gyrocentre coordinates implies that $\gparamtheta = 0$, which we use from here on out.

\subsubsection{The first-order transformation}\label{sec:gy_H1_gauge_invariant}
The first-order gyrocentre Hamiltonian is found by substituting \cref{eq:gc_eoms,eq:gc_lag_pert_burby_split_gamma10} into \cref{eq:gy_L1}
\begin{equation}\label{eq:gy_H1}
  \gy{H}_1 \defeq  
      q \phi_1 
    - q \rho \radgav{\evalperturbedgcburby{E}_{1,\rho}} 
    - (1 - \gparamr) q \rho \gy{U}_\shortparallel \radgav{\evalperturbedgcburby{B}_{1,\tau}}
    + 
    \gy{\Mu} \dgav{\evalperturbedgcburby{B}_{1,\shortparallel}}
    ,
\end{equation}
where we have neglected the $\bigO(\varepsilon_B)$ contributions due to $\dot{\gc{\vc{R}}}$ if $\gparamr \neq 1$.

We need an explicit expression for the first-order generating vector $\gy{\vc{G}}_1$ for the computation of the second-order Hamiltonian as follows from \cref{eq:gy_H2_general}. 
Recall that the first-order generating vector $\gy{\vc{G}}_1$ is given by \cref{eq:gy_genvector1}, which itself requires an expression for the first-order generating function $\gy{S}_1$.
From \cref{eq:gy_S1_def} it follows that
\begin{equation}\label{eq:gy_S1_def_approx}
  \underbrace{\frac{1}{\omega_\mathrm{c}}\pard{\gy{S}_{1}}{t}}_{\bigO(\varepsilon_\omega)}
  + \underbrace{\frac{\gy{U}_\shortparallel}{\omega_\mathrm{c}} \hb \bcdot \nabla \gy{S}_{1}}_{\bigO(\varepsilon_\omega)}
  + \underbrace{\pard{\gy{S}_{1}}{\gy{\Theta}}}_{\bigO(1)}
= \frac{q}{\omega_\mathrm{c}} \wt{\psi}_{1},
\end{equation}
where we have substituted the zeroth-order Hamiltonian, as given by \cref{eq:gc_lag} and have neglected the $\bigO(\varepsilon_B)$ contributions from the guiding-centre Poisson bracket~\eqref{eq:gc_bracket}.
Furthermore, we have indicated the magnitude of each of the terms, which is a result from non-dimensionalisation using
\begin{equation}\label{eq:non_dimensionalisation_S}
  \units{t} = \frac{1}{\omega}, \quad
  \units{U_\shortparallel} = \frac{\omega}{\kpa},
\end{equation}
where we recall that the non-dimensional frequency $\varepsilon_\omega$ is defined in \cref{eq:eps_omega}.
Using \cref{eq:gy_effpot,eq:gc_lag_pert_burby_split} we find that 
\begin{equation}\label{eq:gy_wtpsi}
  q \wt{\psi}_{1} 
  = \wt{\gc{H}_{1}^\flr} - \dot{\gc{\vc{Z}}} \bcdot \wt{\gc{\vc{\gamma}}_{1}^\flr}
  = - \rho \int_0^1 \wt{\evalperturbedgcburby{F}_{1,\rho}} \diff \burby 
    + \omega_{\mathrm{c}} q \rho^2 \int_0^1 \burby \wt{\evalperturbedgcburby{B}_{1,\shortparallel}} \diff \burby,
\end{equation}
where we have introduced the Lorentz force 
\begin{equation}\label{eq:gc_Lorentz}
  \vc{F}_1 \defeq q \roundpar{\vc{E}_1 + \gy{U}_\shortparallel \hb \times \vc{B}_1}
  , \quad
  \evalperturbedgcburby{\vc{F}}_1 \defeq q \roundpar{\evalperturbedgcburby{\vc{E}}_1 + \gy{U}_\shortparallel \hb \times \evalperturbedgcburby{\vc{B}}_1}.
\end{equation}

When considering approximations of \cref{eq:gy_S1_def_approx} it is important to keep gauge invariance of the resulting model in mind.
In particular, when considering the proof of \cref{thm:gauge_invariance} as given in \cref{sec:gauge_invariance_proof}, we find that gauge invariance of the first-order generating function $\gy{S}_1$ is needed, and this is proven by observing that $\gy{S}_1$ is the solution of a linear PDE~\eqref{eq:gy_S1_def} with a gauge-invariant right-hand side given by \cref{eq:gy_wtpsi}.
When obtaining an approximation to $\gy{S}_1$ it is therefore essential that we preserve its gauge invariance, which can rather easily be achieved by simply keeping the gauge invariant parts of the right-hand side $\wt{\psi}_1$ together.
The consequence of preserving gauge invariance is that the high-frequency contribution from $\partial \vc{A}_1 / \partial t$, which itself comes from the $\vc{E}_1$ term in the Lorentz force~\eqref{eq:gc_Lorentz}, is kept on the right-hand side of \cref{eq:gy_S1_def_approx}.
Keeping this term results in a high-frequency compressional Alf\'en wave as discussed in \cref{sec:darwin}.

We make several long wavelength approximations to \cref{eq:gy_wtpsi}, starting with
\begin{equation}
  \evalperturbedgcburby{F}_{1,\rho} = F_{1,\rho} + \bigO(\varepsilon_\perp)
  \quad \implies \quad
  \wt{\evalperturbedgcburby{F}_{1,\rho}} 
  = \evalperturbedgcburby{F}_{1,\rho} - \gav{\evalperturbedgcburby{F}_{1,\rho}}
  = F_{1,\rho} + \bigO(\varepsilon_\perp)
\end{equation}
as follows from a Taylor series expansion of $\vc{F}_1$ centred around the gyrocentre position~$\gy{\vc{R}}$ (see the discussion in \cref{app:gyro_identities}), where we recall that the non-dimensional perpendicular wave number $\varepsilon_\perp$ is as defined in \cref{eq:eps_perp}.
Similarly, we find that
\begin{equation}
  \wt{\evalperturbedgcburby{B}_{1,\shortparallel}} 
  = \evalperturbedgcburby{B}_{1,\shortparallel} - \gav{\evalperturbedgcburby{B}_{1,\shortparallel}}
  = \bigO(\varepsilon_\perp)
\end{equation}
and therefore neglecting $\bigO(\varepsilon_\perp)$ contributions to the right-hand side of \cref{eq:gy_wtpsi} and neglecting the $\bigO(\varepsilon_\omega)$ part of the left-hand side of \cref{eq:gy_S1_def_approx}, we find that the first-order generating function can be approximated by
\begin{equation}\label{eq:gy_S1_approx}
  \pard{\gy{S}_{1}}{\gy{\Theta}} = - \frac{\rho}{\omega_\mathrm{c}} F_{1,\rho}
  \quad \iff \quad
  \gy{S}_{1} = \frac{\rho}{\omega_\mathrm{c}} F_{1,\tau}.
\end{equation}
This allows us to approximate the first-order generating vector explicitly as
\begin{equation}\label{eq:gy_G10}
  \gy{\vc{G}}_{1} 
   = \gc{\mat{J}}_0 \roundpar{\gy{\vc{\gamma}}_{1} - \gc{\vc{\gamma}}_{1} - \pard{\gy{S}_{1}}{\gy{\vc{Z}}}}
   = \begin{pmatrix}
    \frac{1}{B_{0,\shortparallel}^\effective} \hb \times \int_0^1 \roundpar{\gparamr\gav{\evalperturbedgcburby{\vc{B}}_1 \times \vc\rho} - \evalperturbedgcburby{\vc{B}}_1 \times \vc\rho} \diff\burby + \frac{\rho B_{1,\rho}}{B_0} \hb\\
    - \frac{q}{m} \hb \bcdot \int_0^1 \roundpar{\gparamr\gav{\evalperturbedgcburby{\vc{B}}_1 \times \vc\rho} - \evalperturbedgcburby{\vc{B}}_1 \times \vc\rho} \diff\burby\\
    - \frac{q^2 \rho^2}{m} \int_0^1 \burby \evalperturbedgcburby{B}_{1,\shortparallel} \diff\burby - \frac{\rho}{B_0} F_{1,\rho}\\
    -\frac{\rho}{2 \gy{\Mu} B_0} F_{1,\tau}
  \end{pmatrix},
\end{equation}
where we have substituted \cref{eq:gy_S1_approx,eq:gy_gamma_param,eq:gc_lag_pert_burby_split_gamma10} into \cref{eq:gy_genvector1} and have furthermore neglected all spatial derivatives of the perturbed electromagnetic fields for the evaluation of the Poisson bracket $\bracketgc{\gy{\vc{Z}}}{\gy{S}_{1}}$.


From \cref{eq:gy_G10} it follows that the gyro-average all except one of the components of the first-order generating vector vanishes if we choose $\gparamr = 1$
\begin{equation}\label{eq:gy_G10_gav}
  \gav{\gy{\vc{G}}_{1}}^{\gparamr = 1} = \begin{pmatrix}
    \vc{0}_3\\
    0\\
    - \frac{\gy{\Mu}}{B_0} \dgav{\evalperturbedgcburby{B}_{1,\shortparallel}}\\
    0
  \end{pmatrix}.
\end{equation}
This choice, therefore, in some sense yields the smallest transformation to first-order in~$\varepsilon_\delta$, which is relevant as the magnitude of the coordinate transformation determines the magnitude of the truncation error of the resulting gyrokinetic model.
More specifically, in \cref{app:parameter_motivation} we show that this choice minimises the Euclidean norm of the gyro-average of the first-order generating vector, resulting in the minimisation of the truncation error of the gyrocentre coordinate transformation.

When considering \cref{eq:gy_G10_gav}, we find that only the magnetic moment is transformed non-trivially, which is a consequence of choosing $\gparamtheta = 0$ which in turn was required to ensure that the magnetic moment remains invariant in gyrocentre coordinates, as is shown in \cref{eq:eom_mu_gparamtheta}.
It follows that the gyrocentre coordinate transformation is smallest for $\gparamr = 1$, which, as discussed previously, is of interest as it affects the accuracy of the resulting model.
For this reason we choose the parameter value $\gparamr = 1$ resulting in the following symplectic part of the gyrocentre Lagrangian
\begin{equation}\label{eq:choose_gparams_show_gamma}
  (\gparamr, \gparamtheta) = (1, 0)
  \quad \implies \quad
  \gyFIXED{\vc{\gamma}}_{1,\vc{R}} 
  = \gav{\gc{\vc{\gamma}}_{1,\vc{R}}} 
  = q \vc{A}_1 + q \radgav{\evalperturbedgcburby{\vc{B}}_1\times \vc\rho},
\end{equation}
with all other components equal to zero.

\begin{remark}[Interpreting the gyrocentre magnetic moment]
  From \cref{eq:gy_G10_gav} it follows that the gyrocentre magnetic moment can be interpreted as follows
  \begin{equation}
    \gav{\gc{\Mu}(\gyFIXED{\vc{Z}})} 
    \defeq \gyFIXED{\Mu} - \gav{\gy{G}_{1,\Mu}} + \bigO(\varepsilon_\delta^2)
    \approx \gyFIXED{\Mu} \roundpar{1 + 
    \frac{\dgav{\evalperturbedgcburby{B}_{1,\shortparallel}}}{B_0}}
    \quad \implies \quad
    \gyFIXED{\Mu} \approx \frac{m_s \gc{U}_\tau^2}{2 (B_0 + \dgav{\evalperturbedgcburby{B}_{1,\shortparallel}})},
  \end{equation}
  where we find that the parallel component of the full magnetic field appears in the denominator, rather than just $B_0$ (cf.\ \cref{eq:field_aligned_velocity_coords}):
  \begin{equation}
    B_0 + \dgav{\evalperturbedgcburby{B}_{1,\shortparallel}} = \hb \bcdot (\vc{B}_0 + \dgav{\evalperturbedgcburby{\vc{B}}_1}).
  \end{equation}
  Thus, it is crucial to include the contribution of the perturbed magnetic field to
  make $\gyFIXED{\Mu}$ an invariant of motion.
\end{remark}

\subsubsection{The second-order transformation}\label{sec:H2}
When assuming $\gy{\vc{\gamma}}_2 = \vc{0}_6$, as we do throughout this discussion, we find that \cref{eq:gy_H2_general} results in
\begin{equation}\label{eq:gy_H2_general_}
  \gy{H}_{2} = \gav{\vc{T}_{1} \bcdot \gy{\vc{G}}_{1}},
\end{equation}
where $\vc{T}_{1}$ is as defined in \cref{eq:gy_Tvector}.
The approximation of $\vc{T}_{1}$, followed by substitution of the approximated first-order generating vector $\gy{\vc{G}}_{1}$ and subsequent gyro-averaging results in the following second-order gyrocentre Hamiltonian
\begin{equation}\label{eq:gy_H2}
  \gy{H}_{2} \defeq 
      \frac{\gy{\Mu}}{2 B_0} \abs{\vc{B}_{1,\perp}}^2
    - \frac{m}{2 q^2 B_0^2} \abs{\vc{F}_{1,\perp}}^2
    ,
\end{equation}
which agrees with the result from \citet{burby2019gauge} (hence with $(\gparamr, \gparamtheta) = (0, 0)$) upon substitution of the expression for the Lorentz force~\eqref{eq:gc_Lorentz}. 
The derivation of \cref{eq:gy_H2} is rather tedious and can be found in \cref{app:H20}.

It should be noted that many terms have been neglected in the derivation of the second-order Hamiltonian $\gy{H}_2$. 
In particular, only the terms of leading order in $\varepsilon_B$ and~$\varepsilon_\perp$ have been kept, resulting in a ZLR approximation of $\gy{H}_2$ wherein $\bigO(\varepsilon_B)$ terms have been neglected.
Even though there is no fundamental limitation that keeps us from including such higher-order terms, we have thus far opted not to do so, thereby keeping the resulting equations somewhat tractable, more easily interpretable as well as more suitable for discretisation.
We view the proposed model as a pragmatic first step towards a `fully gyrokinetic' gauge invariant model wherein such terms are kept also at second-order in~$\varepsilon_\delta$, i.e.\ in the second-order gyrocentre Hamiltonian.

\subsubsection{The gyrocentre single-particle phase-space Lagrangian}
When combining the symplectic and Hamiltonian parts of the zeroth-order Lagrangian defined by \cref{eq:gc_lag}, the first-order gyrocentre Lagrangian defined by \cref{eq:gy_gamma_param,eq:gy_H1} (with $(\gparamr, \gparamtheta) = (1, 0)$), as well as the second-order gyrocentre Lagrangian defined by $\gy{\vc{\gamma}}_2 = \vc{0}_6$ and \cref{eq:gy_H2}, we find that the total gyrocentre single-particle phase-space Lagrangian is given by
\begin{align}
  \gy{L} = {} & q (\vc{A}^\effective_0 + \vc{A}^\effective_1) \bcdot \dot{\gy{\vc{R}}}
  + \frac{m \gy{\Mu}}{q} \dot{\gy{\Theta}}
  - \frac{m \gy{U}_\shortparallel^2}{2} 
  - q \phi_1 
  + q \rho \radgav{\evalperturbedgcburby{E}_{1,\rho}} 
  - \gy{\Mu} (B_0 + \dgav{\evalperturbedgcburby{B}_{1,\shortparallel}})\nonumber \\ &
\label{eq:total_gy_lagrangian}
  - \frac{\gy{\Mu}}{2 B_0} \abs{\vc{B}_{1,\perp}}^2
  + \frac{m}{2 q^2 B_0^2} \abs{\vc{F}_{1,\perp}}^2
  ,
\end{align}
where we have defined the FLR corrected vector potential as
\begin{equation}\label{eq:gy_pert_effpot}
  \vc{A}^\effective_1 \defeq \vc{A}_1 + \radgav{(\evalperturbedgcburby{\nabla} \times \vc{A}_1) \times \vc\rho}.
\end{equation}

\subsection{Principle of least action}\label{sec:gy_least_action}
For the derivation of the EOMs we follow the same approach as followed in \cref{sec:gc_least_action}, and therefore we must compute the perturbed gyrocentre Lagrange matrix $\gy{\mat{W}}_1$, which follows from the skew-symmetric part of the Jacobian matrix of $\gy{\vc{\gamma}}_{1}$ as defined in \cref{eq:gy_gamma_param}.
We note that the unperturbed gyrocentre Lagrange matrix coincides with the unperturbed guiding-centre Lagrange matrix $\gc{\mat{W}}_0$, given by \cref{eq:gc_Lagrangematrix}, except that it is evaluated at gyrocentre coordinates.
From \cref{eq:gy_gamma_param} it follows that
\begin{equation}
  \pard{\gy{\vc{\gamma}}_{1}}{\gy{\vc{Z}}} = \begin{pmatrix}
    q (\nabla \vc{A}^\effective_1)^\transpose & \vc{0}_3 & -\vc{w}_1 & \vc{0}_3\\
    \vc{0}_3^\transpose & 0 & 0 & 0\\
    \vc{0}_3^\transpose & 0 & 0 & 0\\
    \vc{0}_3^\transpose & 0 & 0 & 0\\
  \end{pmatrix},
\end{equation}
where we let $\vc{w}_1$ be defined as
\begin{equation}
  \vc{w}_1 \defeq
  - \frac{q}{2 \gy{\Mu}} \roundpar{\radgav{\evalperturbedgcburby{\vc{B}}_1\times \vc\rho} + \frac{1}{2} \sdgav{\squarepar{(\evalperturbedgcburby{\nabla} \vc{B}_1)^\transpose \vc\rho} \times \vc\rho}}.
\end{equation}

In addition, we define the FLR corrected electromagnetic fields by
\begin{subequations}
  \label{eq:gy_flr_electomagfields}
  \begin{align}
    \label{eq:gy_flr_electricfield}
    \vc{E}^\effective_1 
    &\defeq \vc{E}_1 + \nabla\radgav{\rho \evalperturbedgcburby{E}_{1,\rho}} + \radgav{(\evalperturbedgcburby{\nabla} \times \vc{E}_1) \times \vc\rho}
    , \\
    \vc{B}^\effective_1 &\defeq \curl \vc{A}^\effective_1
    = \vc{B}_1 + \curl \radgav{\evalperturbedgcburby{\vc{B}}_1\times \vc\rho},
  \end{align}
\end{subequations}
where we recall that the radially averaged gyro-average $\radgav{\cdot}$ is defined in \cref{eq:radial_average}.
These fields, which end up being used in the EOMs in \cref{eq:gy_split_eoms}, are referred to as `FLR corrected' electromagnetic fields because they can be approximated as (cf.\ \cref{eq:electromagnetic_fields})
\begin{subequations}
  \label{eq:gy_flr_electomagfields_appro}
  \begin{align}
    \vc{E}^\effective_1 
    &= \gav{\evalperturbedgc{\vc{E}}_1} + \bigO(\varepsilon_B),\\
    \label{eq:flrcorrected_B1}
    \vc{B}^\effective_1
    &= 
    \gav{\evalperturbedgc{\vc{B}}_1} + \bigO(\varepsilon_B)
  \end{align}
\end{subequations}
by making use of \cref{eq:gradient_theorem_vector_approx,eq:gradient_theorem_curlvector_approx}.
We note that \cref{eq:gy_flr_electomagfields_appro} only holds due to our choice of the parameter value $\gparamr = 1$.

The resulting Lagrange matrix is given by (cf.\ \cref{eq:gc_Lagrangematrix})
\begin{equation}
  \label{eq:gy_Lagrangematrix}
  \gy{\mat{W}} = \begin{pmatrix}
    q \mat{B}^\effective & -m \hb & \frac{m}{q} \vc{w} & \vc{0}_3\\
    m \hb^\transpose & 0 & 0 & 0\\
    - \frac{m}{q} \vc{w}^\transpose  & 0 & 0 & \frac{m}{q} \\
    \vc{0}_3^\transpose & 0 & - \frac{m}{q}& 0\\
  \end{pmatrix},
\end{equation}
where we have defined the effective gyrocentre vector potential as (see also \cref{eq:gc_genvecpot})
\begin{equation}\label{eq:gy_genvecpot}
  \vc{A}^\effective 
  \defeq \vc{A}_0^\effective + \vc{A}^\effective_1 
  = \vc{A}_0 
  + \frac{m \gy{U}_\shortparallel}{q} \hb 
  - \frac{m \gy{\Mu}}{q^2} \vc{w}_0 
  + \vc{A}_1 
  + \radgav{\evalperturbedgcburby{\vc{B}}_1\times \vc\rho}
  , 
  \quad
  \vc{B}^\effective \defeq \curl \vc{A}^\effective
\end{equation}
as well as $\vc{w} = \vc{w}_0 + \vc{w}_1$.
The matrix $\mat{B}^\effective$ is defined analogously to \cref{eq:matvec_is_timescurl}.

For the computation of the gyrocentre Poisson bracket we must invert the gyrocentre Lagrange matrix.
Using the result of \cref{app:matrix_inverse} we find that the gyrocentre Poisson matrix is given by
\begin{equation}\label{eq:gy_Poissonmatrix}
  \gy{\mat{J}} = \begin{pmatrix}
    -\frac{\mat{B}_0}{q B_0 B_\shortparallel^\effective} & \frac{\vc{b}^\effective}{m} & \vc{0}_3 & - \frac{\vc{w} \times \hb}{q B_\shortparallel^\effective}\\
    -\frac{(\vc{b}^\effective)^\transpose}{m} & 0 & 0 & -\frac{\vc{b}^\effective \bcdot \vc{w}}{m}\\
    \vc{0}_3^\transpose & 0 & 0 & -\frac{q}{m}\\
    \frac{(\vc{w} \times \hb)^\transpose }{q B_\shortparallel^\effective} & \frac{\vc{b}^\effective \bcdot \vc{w}}{m} & \frac{q}{m}& 0\\
  \end{pmatrix},
\end{equation}
where we have defined (cf.\ \cref{eq:gc_b0starstar})
\begin{equation}\label{eq:gy_bstarstar}
  \vc{b}^\effective 
  \defeq \frac{\vc{B}^\effective}{B_\shortparallel^\effective} 
  = \hb + \frac{1}{B_{\shortparallel}^\effective} \squarepar{
      \frac{m \gy{U}_\shortparallel}{q} \hb \times \vc{\kappa}
    - \frac{m \gy{\Mu}}{q^2} (\curl \vc{w}_0)_\perp
    + \vc{B}^\effective_{1,\perp}
  },
\end{equation}
and note that $B_\shortparallel^\effective$ can be written explicitly as (cf.\ \cref{eq:gy_Bpastar0})
\begin{equation}\label{eq:gy_Bpastar}
  B_\shortparallel^\effective
  =   B^\effective_{0,\shortparallel} + B^\effective_{1,\shortparallel}
  =   B_0 
  + \frac{m \gy{U}_\shortparallel}{q} (\curl \hb)_\shortparallel 
  - \frac{m \gy{\Mu}}{q^2} (\curl \vc{w}_0)_\shortparallel
  + B^\effective_{1,\shortparallel}
  .
\end{equation}

Analogous to \cref{eq:gc_bracket_def}, we define the gyrocentre Poisson bracket as
\begin{equation}\label{eq:gy_bracket_def}
  \bracketgy{\mathcal{F}}{\mathcal{G}} \defeq \pard{\mathcal{F}}{\gy{\vc{Z}}} \bcdot \roundpar{\gy{\mat{J}} \pard{\mathcal{G}}{\gy{\vc{Z}}}}
\end{equation}
such that the EOMs, similar to \cref{eq:gc_eoms_as_bracket}, are given by
\begin{equation}\label{eq:gy_eoms_as_bracket}
  \dot{\gy{\vc{Z}}} = \gy{\mat{J}} \pard{\gy{\vc{\gamma}}}{t} + \bracketgy{\gy{\vc{Z}}}{\gy{H}},
\end{equation}
where the zeroth-order term $\gy{H}_0$ of the Hamiltonian is defined in \cref{eq:gc_lag}, the first-order term $\gy{H}_1$ is defined in \cref{eq:gy_H1}, and the second-order term is given by \cref{eq:gy_H2}.
Substitution of \cref{eq:gy_Poissonmatrix} in \cref{eq:gy_eoms_as_bracket} results in
\begin{subequations}
  \label{eq:gy_eoms}
  \begin{align}
    \dot{\gy{\vc{R}}} &= \frac{1}{m} \pard{\gy{H}}{\gy{U}_\shortparallel} \vc{b}^\effective + \frac{1}{q B_\shortparallel^\effective} \hb \times \roundpar{\nabla \gy{H} + q \pard{\vc{A}^\effective_1}{t}},\label{eq:gy_eoms_R}\\
    \dot{\gy{U}}_\shortparallel &= -\frac{1}{m} \vc{b}^\effective \bcdot \roundpar{\nabla \gy{H} + q\pard{\vc{A}^\effective_1}{t}},\label{eq:gy_eoms_Vpa}\\
    \dot{\gy{\Mu}} &= 0,\\
    \dot{\gy{\Theta}} &= \frac{q}{m} \pard{\gy{H}}{\gy{\Mu}} + \vc{w} \bcdot \dot{\gy{\vc{R}}} \, .
  \end{align}
\end{subequations}
Here we note that, even though the EOM for the gyro-phase $\gy{\Theta}$ is non-trivial, it does not have to be solved as none of the other terms on the right-hand side of the EOMs depend on the gyro-phase.

  \section{Gyrokinetic Maxwell model}\label{sec:vlasov_maxwell}
Thus far, we have derived a gyrokinetic model for single-particle motion for a given electromagnetic field, which includes a time-dependent perturbation.
This forms the basis for a gyrokinetic approximation of the coupled and self-consistent Vlasov--Maxwell system of equations, wherein the time-dependent perturbation of the electromagnetic field results from the motion of the charged particles themselves.

In this section we first introduce the particle distribution function for each of the species, which we then use to formulate the self-consistent action principle, following the work of e.g.\ \citet{sugama}.
Provided with this action principle we then derive the resulting EOMs for the particles as well as the corresponding field equations for the electromagnetic field.
The field equations are considered in a strong formulation wherein we recognise the macroscopic Maxwell's equations.
We discuss equilibrium solutions as well as well-posedness of the field equations.
The section is concluded with a discussion on energy conservation.

As we exclusively discuss the gyrokinetic model, which is expressed in gyrocentre coordinates, we drop the $\gy{\cdot}$~notation and simply write $\vc{Z}$ rather than $\gy{\vc{Z}}$.
\renewcommand\gy[1]{#1}

\subsection{Particle distribution function}
Several particle species are considered, which we denote by the subscript `$s$', where usually $s \in \{\ii, \euler\}$ for the ion species~`$\ii$' and the electron species~`$\euler$'.
Each species has its own particle mass $m_s$ and charge $q_s$.
A particle distribution function is considered, for each particle species $s$, which is denoted by $\gy{f}_s(\gy{\vc{r}}, \gy{u}_\shortparallel, \gy{\mu}, \gy{t})$ and coincides with the number of particles per unit phase-space volume.
The particle distribution function is split into its initial background part and time-dependent part
\begin{equation}
  \gy{f}_s(\gy{\vc{r}}, \gy{u}_\shortparallel, \gy{\mu}, \gy{t}) = \gyfzero(\gy{\vc{r}}, \gy{u}_\shortparallel, \gy{\mu}) + \gyfone(\gy{\vc{r}}, \gy{u}_\shortparallel, \gy{\mu}, \gy{t})
\end{equation}
with $\gyfone(\gy{\vc{r}}, \gy{u}_\shortparallel, \gy{\mu}, \tzero) = 0$.
Note that the particle distribution function is gyrotropic, i.e.\ it does not depend on the gyro-phase, which is a consequence of assuming that the initial particle distribution function $\gyfzero$ is gyrotropic.
This, in turn, can be justified by noting that the non-dimensionalisation of the Vlasov equation for $\gyfzero$ implies that ${\partial \gyfzero}/{\partial\gy{\theta}} = \bigO(\varepsilon_\omega)$ by making use of \cref{eq:non_dimensionalisation_S}.

We use the small case letter $\gy{\vc{z}} = (\gy{\vc{r}}, \gy{u}_\shortparallel, \gy{\mu}, \gy{\theta})$ to refer to the Eulerian equivalent of the Lagrangian phase-space coordinate $\gy{\vc{Z}}$.
The dependence of a particle's Lagrangian characteristic on the initial phase-space coordinate $\gyzzero = (\gy{\vc{R}}(\tzero), \gy{U}_\shortparallel(\tzero), \gy{\Mu}(\tzero), \gy{\Theta}(\tzero))$ is denoted in the following way (in the absence of collisions)
\begin{equation}
  \gy{\vc{Z}}(t; \gyzzero, \tzero) = \gy{\vc{Z}}(t) \quad \text{with} \quad \gy{\vc{Z}}(\tzero) = \gyzzero.
\end{equation}
The particle distribution function then satisfies (by definition)
\begin{equation}\label{eq:gy_distfun_def}
  \gy{f}_s(\gy{\vc{Z}}(t; \gyzzero, \tzero), t) = \gyfzero(\gyzzero),
\end{equation}
where we denote by $\gyfzero$ the particle distribution function at $\tzero$.
Hence, the following Vlasov equation is satisfied
\begin{equation}\label{eq:gy_vlasov_bracket}
  \totd{\gy{f}_s}{t} = 0 
  \quad \iff \quad
  \pard{\gy{f}_s}{t} + \dot{\gy{\vc{R}}} \bcdot \nabla \gy{f}_s + \dot{\gy{U}}_\shortparallel \pard{\gy{f}_s}{\gy{u}_\shortparallel} = 0,
\end{equation}
where we consider the EOMs $\dot{\gy{\vc{Z}}}$ given by \cref{eq:gy_eoms} to be evaluated at the Eulerian phase-space coordinate $\gy{\vc{z}}$.

Recall that the physical coordinates, as defined in \cref{sec:motivation_physcoords}, were denoted by $\tilde{\vc{Z}}$.
The field-theoretic Lagrangian, which is discussed in \cref{sec:low}, is formulated using integrals over physical space, which has to be transformed to integrals over the gyrocentre coordinates.
For instance, we consider the integral of a function $\tilde{\mathcal{F}}(\vc{x}, \vc{v}, t) = \mathcal{F}(\gy{\vc{r}}, \gy{u}_\shortparallel, \gy{\mu}, \gy{\theta}, \gy{t})$ (note that we now write $\vc{x}, \vc{v}$ for the physical position and velocity, to distinguish them from the gyrocentre position and velocity, which are now denoted by $\vc{r}, \vc{u}$ since we have omitted the $\gyFIXED{\cdot}$ notation)
\begin{equation}\label{eq:integral_physical_to_gyrocentre}
  \int_{\mathbb{R}^3} \int_\Omega \tilde{\mathcal{F}} \diff^3 x \diff^3 v = \int \mathcal{F} \jacobiangy \diff^6 \gy{z},
\end{equation}
where the integration limits and differentials are defined as
\begin{equation}
  \int \diff^6 \gy{z} \defeq \int \int_\Omega \diff^3 \gy{r} \diff^3 \gy{u}, \quad
  \int \diff^3 \gy{u} \defeq \int_0^{2 \upi} \int_0^\infty \int_{-\infty}^\infty \diff \gy{u}_\shortparallel \diff \gy{\mu} \diff \gy{\theta},
\end{equation}
and $\jacobiangy$ denotes the Jacobian of the coordinate transformation from physical to gyrocentre coordinates
\begin{equation}\label{eq:gy_jacobian}
  \jacobiangy  
  \defeq \det \pard{\tilde{\vc{z}}}{\gy{\vc{z}}}
  = \frac{B_{s,\shortparallel}^\effective}{m_s},
\end{equation}
which is derived in \cref{app:jacobian} (a proof can also be found in \citet[Appendix F]{parra2011phase}) and can be written explicitly by making use of \cref{eq:gy_Bpastar}.

We find that the gyrocentre EOMs, as given by \cref{eq:gy_eoms}, imply that phase-space volume is conserved. 
A proof is given in \cref{app:liouville}, and can also be found in \citet[Appendices G \& H]{parra2011phase}.
\begin{theorem}[Gyrocentre Liouville theorem]\label{thm:gy_liouville}
  The phase-space volume is conserved:
  \begin{equation}\label{eq:gy_liouville}
    \pard{\jacobiangy}{t} + \nabla \bcdot (\jacobiangy \dot{\gy{\vc{R}}}) + \pard{}{\gy{u}_\shortparallel} (\jacobiangy \dot{\gy{U}}_\shortparallel) 
    = 0.
  \end{equation}
  Furthermore, integrals of the form \cref{eq:integral_physical_to_gyrocentre} can be expressed in terms of the initial phase-space coordinates in the following way
  \begin{equation}\label{eq:gy_liouville_integral_equality}
    \int \gy{f}_s \mathcal{F} \jacobiangy \diff^6 \gy{z}
    =
    \int \gyfzero(\gyzzero) \mathcal{F}(\gy{\vc{Z}}(t; \gyzzero, \tzero)) \jacobiangy(\gyzzero, \tzero) \diff^6 \gyzzerod
    ,
  \end{equation}
  where an absence of arguments implies evaluation at ($\gy{\vc{z}}, t$).
\end{theorem}

By combining \cref{eq:gy_vlasov_bracket,eq:gy_liouville} we find the conservative form of the Vlasov equation
\begin{equation}\label{eq:gy_vlasov_conservative}
  \pard{}{t} (\gy{f}_s \jacobiangy) + \nabla \bcdot (\gy{f}_s \jacobiangy \dot{\gy{\vc{R}}}) + \pard{}{\gy{u}_\shortparallel} (\gy{f}_s \jacobiangy \dot{\gy{U}}_\shortparallel) 
  = 0.
\end{equation}
Note that integration of the conservative form of the Vlasov equation over velocity space, multiplication by $q_s$, and subsequent summation over the species $s$, results in the free-charge continuity equation
\begin{equation}\label{eq:gy_continuity}
  \pard{\mathcal{R}^\free}{t} + \nabla \bcdot \vc{\mathcal{J}}^\free = 0,
\end{equation}
where the gyrocentre free charge and current density are defined as
\begin{subequations}
  \label{eq:gy_freecharge_and_current}
  \begin{align}
    \mathcal{R}^\free &\defeq \sum_s q_s \int \gy{f}_s \jacobiangy \diff^3 \gy{u},\label{eq:gy_freecharge}\\
    \vc{\mathcal{J}}^\free &\defeq \sum_s q_s \int \gy{f}_s \dot{\gy{\vc{R}}} \, \jacobiangy \diff^3 \gy{u},\label{eq:gy_freecurrent}
  \end{align}
\end{subequations}
respectively. It implies local conservation of the free charge.

\subsection{Low's action}\label{sec:low}
We use a variational formulation in order to obtain a structure preserving self-consistent Vlasov--Maxwell system of equations.
Such a variational formulation is in particular suitable for our foreseen structure preserving discretisation using finite element exterior calculus (FEEC)~\citep{gempic}.

The starting point is Low's action \citep{low} in gyrocentre coordinates
\begin{subequations}
  \label{eq:low}
  \begin{equation}\label{eq:low_action}
    \mathfrak{A}(\gy{\vc{Z}}, \phi_1, \Ad) \defeq \int \mathfrak{L}(\gy{\vc{Z}}, \phi_1, \Ad) \diff t,
  \end{equation}
  where the field-theoretic Lagrangian is given by
  \begin{align}
      \mathfrak{L}(\gy{\vc{Z}}, \phi_1, \Ad) = {} & 
      \sum_s \int \gyfzero(\gyzzero) \gy{L}_s(\gy{\vc{Z}}(t; \gyzzero, \tzero), \dot{\gy{\vc{Z}}}(t; \gyzzero, \tzero)) \jacobiangy(\gyzzero, \tzero) \diff^6 \gyzzerod\nonumber \\ &
           + \frac{\epsilon_0}{2} \int \abs{\vc{E}_1}^2 \diff^3 x
           - \frac{1}{2 \mu_0} \int \abs{\vc{B}_0 + \vc{B}_1}^2 \diff^3 x
      .\label{eq:low_lagrangian}
  \end{align}
\end{subequations}
Here integration over the time coordinate $t$ is done over the interval $[\tzero, \tone]$, where $\tone$ denotes the final time, $\gy{L}_s$ denotes the gyrocentre Lagrangian corresponding to the species~$s$ and $\mu_0$ denotes the magnetic permeability in vacuum.
Note that we keep a finite value of the vacuum permittivity $\epsilon_0$ as this favourably yields field equations which can be integrated explicitly in time.
This is discussed in more detail in \cref{sec:initial_value_problem}.
In \cref{sec:darwin} a low-frequency approximation of this model is proposed, wherein the limit of quasi-neutrality $\epsilon_0 \rightarrow 0$ is considered, thereby eliminating the fast waves which would otherwise be present.

We may transform the first integral in \cref{eq:low_lagrangian} by making use of \cref{thm:gy_liouville}.
Rather than transforming the integral resulting from each of the contributions of the gyrocentre Lagrangian, we split the gyrocentre Lagrangian in two parts $\gy{L}_s = \gy{L}_s^\mathrm{part} + \gy{L}_s^\mathrm{field}$ referred to as the particle and field part, respectively.
Subsequently, we transform and linearize the contribution from the field part
\begin{multline}
  \int \gyfzero(\gyzzero) \gy{L}_s(\gy{\vc{Z}}(t; \gyzzero, \tzero), \dot{\gy{\vc{Z}}}(t; \gyzzero, \tzero)) \jacobiangy(\gyzzero, \tzero) \diff^6 \gyzzerod \\
  \approx 
  \int \gyfzero(\gyzzero) \gy{L}_s^\mathrm{part}(\gy{\vc{Z}}(t; \gyzzero, \tzero), \dot{\gy{\vc{Z}}}(t; \gyzzero, \tzero)) \jacobiangy(\gyzzero, \tzero) \diff^6 \gyzzerod
  + \int \gyfzero \gy{L}_s^\mathrm{field} \jacobiangcslab \diff^6 \gy{z}.
\end{multline}
Furthermore, we have defined the (unperturbed) guiding-centre Jacobian as (cf.\ \cref{eq:gy_jacobian})
\begin{equation}\label{eq:gc_jacobian}
  \jacobiangcslab \defeq \frac{B_0}{m_s},
\end{equation}
where we have neglected the $\bigO(\varepsilon_\delta)$ and $\bigO(\varepsilon_B)$ terms from the Jacobian.
Note that this is a modelling choice which does not break the structure of the resulting equations.
The field-theoretic Lagrangian as given by \cref{eq:low_lagrangian} is now approximated by
\begin{equation}\label{eq:low_split}
  \begin{aligned}
    \mathfrak{L}(\gy{\vc{Z}}, \phi_1, \Ad)
    \defeq {} & \sum_s \int \gyfzero(\gyzzero) \gy{L}_s^\mathrm{part}(\gy{\vc{Z}}(t; \gyzzero, \tzero), \dot{\gy{\vc{Z}}}(t; \gyzzero, \tzero)) \jacobiangy(\gyzzero, \tzero) \diff^6 \gyzzerod\\
    & + \sum_s \int \gyfzero \gy{L}_s^\mathrm{field} \jacobiangcslab \diff^6 \gy{z}
    + \frac{\epsilon_0}{2} \int \abs{\vc{E}_1}^2 \diff^3 x
    - \frac{1}{2 \mu_0} \int \abs{\vc{B}_0 + \vc{B}_1}^2 \diff^3 x
    .
  \end{aligned}
\end{equation}

The field part of the Lagrangian does not affect the EOMs of the particles directly, but only affects the potentials \textit{via} the field equations, which are derived in \cref{sec:vlasov_maxwell_fieldeqs}.
The reason for splitting the field-theoretic Lagrangian in this way, is to simplify the resulting discretised model.
For instance, we want to obtain linear field equations, and thus have linearised the corresponding field part of the Lagrangian.
Note that neglecting the time-dependent part~$\gyfone$ of the particle distribution function is justified only if $\gyfone$ is small compared to $\gyfzero$, which is the case e.g.\ when studying microturbulence in the core of fusion devices~\citep{garbet2010gyrokinetic}.

We recall that the gyrocentre single-particle phase-space Lagrangian is given by \cref{eq:total_gy_lagrangian}; the following splitting is considered
\begin{subequations}
  \label{eq:Hsplitting}
  \begin{equation}\label{eq:Lsplit}
    \gy{L}_s^\mathrm{part}(\gy{\vc{Z}}, \dot{\gy{\vc{Z}}}) \defeq q_s \vc{A}_s^\effective \bcdot \dot{\gy{\vc{R}}} + \frac{m_s \gy{\Mu}}{q_s} \dot{\gy{\Theta}} - \gy{H}_s^\mathrm{part}(\gy{\vc{Z}}), \quad
    \gy{L}_s^\mathrm{field}(\gy{\vc{z}}) \defeq - \gy{H}_s^\mathrm{field}(\gy{\vc{z}})
  \end{equation}
  for 
  \begin{equation}\label{eq:Hpart}
    \gy{H}_s^\mathrm{part}(\gy{\vc{Z}}) \defeq 
      \frac{m_s}{2} U_\shortparallel^2
    + \Mu (B_0 + \dgav{\evalperturbedgcburby{B}_{1,\shortparallel}})
    + q_s \phi_1 
    - q_s \rho \radgav{\evalperturbedgcburby{E}_{1,\rho}}
  \end{equation}
  and
  \begin{equation}\label{eq:Hfield}
    \gy{H}_s^\mathrm{field}(\gy{\vc{z}}) \defeq
      \frac{\gy{\mu}}{2 B_0} \abs{\vc{B}_{1,\perp}}^2
    - \frac{m_s}{2 q_s^2 B_0^2} \abs{\vc{F}_{1,\perp}}^2
    .
  \end{equation}
\end{subequations}

\subsection{Principle of least action}\label{sec:pola}
In what follows, we compute the EOMs for the gyrocentre characteristic $\gy{\vc{Z}}$ as well as the field equations for the potentials $\phi_1, \vc{A}_1$.
We follow the principle of least action, which states that the EOMs and field equations are stationary points of the action.
The following notation is introduced for computing the variation of the action w.r.t.\ the gyrocentre coordinate
\begin{equation}
  \variation{\gy{\vc{Z}}}{\vc{\delta}}{\mathfrak{A}} \defeq \evalAt{\totd{}{\varepsilon}}{\varepsilon = 0} \mathfrak{A}(\gy{\vc{Z}} + \varepsilon \vc{\delta}, \phi_1, \Ad),
\end{equation}
which we define analogously for the other arguments of the action.

\subsubsection{Equations of motion}\label{sec:vlasov_maxwell_eoms}
The EOMs are defined by setting the variation of the action w.r.t.\ the gyrocentre coordinate to zero, for all suitable trajectories $\vc{\delta}$ with $\vc{\delta}(t^0) = \vc{\delta}(t^1) = \vc{0}_6$.
It follows that the trajectories satisfy
\begin{equation}\label{eq:eoms_partial_integration}
  \sum_s \int \gyfzero(\gyzzero) \Biggl[\roundpar{\pard{\gy{L}_s^\mathrm{part}}{\gy{\vc{Z}}} - \totd{}{t} \pard{\gy{L}_s^\mathrm{part}}{\dot{\gy{\vc{Z}}}}} \bcdot \vc{\delta} + \underbrace{\totd{}{t} \roundpar{\pard{\gy{L}_s^\mathrm{part}}{\dot{\gy{\vc{Z}}}} \bcdot \vc{\delta}}}_{ = 0}\Biggr] \jacobiangy(\gyzzero, \tzero) \diff^6 \gyzzerod \diff t = 0
\end{equation}
by making use of partial integration in time.
As this must hold for all trajectories $\vc{\delta}$ it follows that the EOMs satisfy the Euler--Lagrange equations (see e.g.\ \cref{eq:hamilton_J0}), and therefore the EOMs are of the form given by \cref{eq:gy_eoms}, except that only the particle part of the Hamiltonian, as defined in \cref{eq:Hpart}, is used on the right-hand side.

We compute the required partial derivatives of $\gy{H}_s^\mathrm{part}$.
The term appearing in the EOMs given by \cref{eq:gy_eoms} can be written as
\begin{equation}\label{eq:gy_gradHpart}
    \nabla \gy{H}_s^\mathrm{part} + q_s \pard{\vc{A}^\effective_1}{t}
    = \Mu \nabla (B_0 + \dgav{\evalperturbedgcburby{B}_{1,\shortparallel}})
      - q_s \vc{E}^\effective_1
\end{equation}
by substituting \cref{eq:Hpart,eq:gy_pert_effpot,eq:electromagnetic_fields,eq:gy_flr_electricfield}.
Furthermore, we made use of Faraday's law
\begin{equation}\label{eq:Faraday}
  \pard{\vc{B}_1}{t} 
  = -\curl \vc{E}_1, 
\end{equation}
which follows from the definition of the electromagnetic fields~\eqref{eq:electromagnetic_fields}.

The second partial derivative that is required for the EOMs is the one w.r.t.\ the parallel velocity and is given by
\begin{equation}
  \pard{\gy{H}_s^\mathrm{part}}{\gy{U}_\shortparallel} 
  = m_s \gy{U}_\shortparallel.
\end{equation}

Substitution of these results in \cref{eq:gy_eoms} yields (we only show the relevant and non-trivial EOMs)
\begin{subequations}
  \label{eq:gy_split_eoms}
  \begin{align}
    \dot{\gy{\vc{R}}} &= 
        \gy{U}_\shortparallel \vc{b}_s^\effective 
      - \frac{1}{q_s B_{s,\shortparallel}^\effective} \hb \times \squarepar{
          q_s \vc{E}^\effective_1
        - \Mu \nabla (B_0 + \dgav{\evalperturbedgcburby{B}_{1,\shortparallel}})
      },\label{eq:gy_split_eoms_R}
      \\
    \dot{\gy{U}}_\shortparallel &= 
        \frac{1}{m_s} \vc{b}_s^\effective \bcdot \squarepar{
            q_s \vc{E}^\effective_1
          - \Mu \nabla (B_0 + \dgav{\evalperturbedgcburby{B}_{1,\shortparallel}})}
      ,\label{eq:gy_split_eoms_Vpa}
  \end{align}
\end{subequations}
where we recall that $\vc{b}_s^\effective$ is defined in \cref{eq:gy_bstarstar}
and $B_{s,\shortparallel}^\effective$ is given by \cref{eq:gy_Bpastar}.
When substituting \cref{eq:gy_bstarstar} we find that the EOM for the gyrocentre position $\gy{\vc{R}}$ can be written as
\begin{equation}\label{eq:gy_split_eoms_R_subbstarstar}
  \begin{aligned}
    \dot{\gy{\vc{R}}} = {} & 
          \gy{U}_\shortparallel \biggl(\hb + \overbrace{\frac{\vc{B}^\effective_{1,\perp}}{B_{s,\shortparallel}^\effective}}^\text{\parbox{1.4cm}{\centering magnetic flutter}}\biggr)
        + \frac{1}{q_s B_{s,\shortparallel}^\effective} \biggl[
            \overbrace{q_s \vc{E}^\effective_1}^\text{ExB drift}
          - \overbrace{\Mu \nabla (B_0 + \dgav{\evalperturbedgcburby{B}_{1,\shortparallel}})}^\text{grad-B drift}
          + \overbrace{m_s \gy{U}_\shortparallel^2 \vc{\kappa}}^\text{curvature drift}\\ 
          & - \underbrace{\frac{m_s \gy{\Mu} \gy{U}_\shortparallel}{q_s} (\curl \vc{w}_0) \times \hb}_\text{gyro-gauge invariance}
        \biggr] \times \hb,
  \end{aligned}
\end{equation}
where we have indicated the physical meaning of each of the terms.
The EOM for the gyrocentre parallel velocity $\gy{U}_\shortparallel$ as given by \cref{eq:gy_split_eoms_Vpa} contains two contributions: an acceleration due to the perturbed parallel component of the FLR corrected electric field~$\vc{E}^\effective_1$ as well as the contribution due to the magnetic mirror force.

We recall that the FLR corrected electromagnetic fields~$\vc{E}^\effective_1$ and $\vc{B}^\effective_1$ are approximations of their respective gyro-averaged counterparts~$\gav{\evalperturbedgc{\vc{E}}_1}$ and $\gav{\evalperturbedgc{\vc{B}}_1}$ according to \cref{eq:gy_flr_electomagfields_appro}.
It should be noted that letting $\vc{B}^\effective_1 = \gav{\evalperturbedgc{\vc{B}}_1}$ and/or $\vc{E}^\effective_1 = \gav{\evalperturbedgc{\vc{E}}_1}$ implies that the model no longer results from an action principle, thereby resulting in a loss of energy conservation.
Even if $\varepsilon_B = 0$ one must be aware that the identities given by \cref{eq:gy_flr_electomagfields_appro} result from application of the gradient theorem~\eqref{eq:gradient_theorem_vector_approx}, which is not likely to hold numerically.


\subsubsection{Field equations}\label{sec:vlasov_maxwell_fieldeqs}
We give Low's action explicitly for our parameter choice $(\gparamr, \gparamtheta) = (1, 0)$ such that we can find the field equations by computing the appropriate variations
\begin{multline}\label{eq:low_split_explicit}
  \mathfrak{A}(\gy{\vc{Z}}, \phi_1, \Ad)
  = \sum_s \int \gyfzero(\gyzzero) \biggl[
      q_s \roundpar{
          \vc{A}_{0,s}^\effective 
        + \vc{A}_1 
        + \radgav{\evalperturbedgcburby{\vc{B}}_1\times \vc\rho}} \bcdot \dot{\gy{\vc{R}}} 
      + \frac{m_s \gy{\Mu}}{q_s} \dot{\gy{\Theta}}
      \\
      - \frac{m_s}{2} \gy{U}_\shortparallel^2
      - \gy{\Mu} (B_0 + \dgav{\evalperturbedgcburby{B}_{1,\shortparallel}})
      - q_s \phi_1 
      + q_s \rho \radgav{\evalperturbedgcburby{E}_{1,\rho}}
  \biggr] \jacobiangy(\gyzzero, \tzero) \diff^6 \gyzzerod \diff t\\
  + \sum_s \int \gyfzero \biggl[
      \frac{m_s}{2 B_0^2} \abs{\vc{E}_{1,\perp}}^2
    - \frac{m_s \gy{u}_\shortparallel}{B_0^2} \hb \bcdot (\vc{E}_{1} \times \vc{B}_{1})
    - \roundpar{\gy{\mu} B_0 - m_s \gy{u}_\shortparallel^2} \frac{\abs{\vc{B}_{1,\perp}}^2}{2 B_0^2}
  \biggr] \jacobiangcslab \diff^6 \gy{z} \diff t\\
  + \frac{\epsilon_0}{2} \int \abs{\vc{E}_1}^2 \diff^3 x \diff t
  - \frac{1}{2 \mu_0} \int \abs{\vc{B}_0 + \vc{B}_1}^2 \diff^3 x \diff t
  ,
\end{multline}
where we have substituted \cref{eq:Hsplitting,eq:gy_pert_effpot,eq:low_split} into \cref{eq:low_action}.
Recall that the electromagnetic fields are defined in \cref{eq:electromagnetic_fields}.

Each of the field equations can be derived by setting the corresponding variation w.r.t.\ the function to zero.
We start by computing Gauss's law, which results from setting the variation of Low's action~\eqref{eq:low_split_explicit} w.r.t.\ the scalar potential $\phi_1$ to zero.
That is, Gauss's law is derived from
\begin{equation}
  \variation{\phi_1}{\cp}{\mathfrak{A}} = 0,
\end{equation}
where $\cp$ is a scalar test function.
We note that the substitution $\phi_1 \mapsto \phi_1 + \varepsilon \cp$ results in
\begin{equation}
  \vc{E}_1 \mapsto \vc{E}_1 - \varepsilon \nabla \cp.
\end{equation}
This results in the following Gauss's law
\begin{equation}\label{eq:quasi_neutrality}
  - \int \roundpar{\epsilon_0 \vc{E}_1 + \vc{\mathcal{P}}_1} \bcdot \nabla \cp \diff^3 \gy{r}
    =
  \sum_s q_s \int \gy{f}_s \gav{\evalperturbedgc{\cp}} \jacobiangy \diff^6 \gy{z}
  ,
\end{equation}
where we have used Liouville's theorem to transform the integral over $\gy{f}_s$, assumed the test function to be independent of time, and have defined the electric polarisation as 
\begin{equation}\label{eq:polarisation}
  \vc{\mathcal{P}}_1 \defeq \sum_s \int \gyfzero \vc{P}_{1,s} \jacobiangcslab \diff^3 \gy{u}
  , \quad
  \vc{P}_{1,s} \defeq \frac{m_s}{q_s B_0^2} \vc{F}_{1,\perp},
\end{equation}
where we recall that $\vc{F}_{1}$ denotes the Lorentz force as defined in \cref{eq:gc_Lorentz}.
In simplifying the right-hand side of \cref{eq:quasi_neutrality} we have made use of the gradient theorem~\eqref{eq:gradient_theorem_scalar}
\begin{equation}\label{eq:gradient_theorem_scalar_intext}
  \cp + \rho \radgav{\hrho \bcdot \evalperturbedgcburby{\nabla} \cp} = \gav{\evalperturbedgc{\cp}}.
\end{equation}
We note that, because the gradient theorem in this form does not hold numerically, it is important to discretise the gyro-average using the left-hand side of \cref{eq:gradient_theorem_scalar_intext}, whenever gauge invariance is to be preserved numerically.

The Amp\`ere--Maxwell law is derived by imposing
\begin{equation}
  \variation{\vc{A}_1}{\ca}{\mathfrak{A}} = 0,
\end{equation}
where $\ca$ is a vector-valued test function, and we note that the substitution $\vc{A}_1 \mapsto \vc{A}_1 + \varepsilon \ca$ results in
\begin{equation}
  \vc{E}_1 \mapsto \vc{E}_1 - \varepsilon \pard{\ca}{t}
  , \quad
  \vc{B}_1 \mapsto \vc{B}_1 + \varepsilon \curl \ca.
\end{equation}
This results in
\begin{multline}\label{eq:ampere}
  \frac{1}{\mu_0} \int (\vc{B}_0 + \vc{B}_1) \bcdot (\curl \ca) \diff^3 x 
  = 
  \int \squarepar{\pard{}{t} \roundpar{\epsilon_0 \vc{E}_1 + \vc{\mathcal{P}}_1} + \curl \vc{\mathcal{M}}_1} \bcdot \ca \diff^3 \gy{r} \\
  + \sum_s \int \gy{f}_s \biggl[
      q_s \dot{\gy{\vc{R}}} \bcdot \roundpar{ 
          \ca
        + \nabla \radgav{\rho \evalperturbedgcburby{\cp}_\rho}
        + \radgav{(\evalperturbedgcburby{\nabla} \times \ca) \times \vc\rho}
      }
      - \gy{\mu} \dgav{(\evalperturbedgcburby{\nabla} \times \ca)_\shortparallel}
      \biggr]
       \jacobiangy \diff^6 \gy{z},
\end{multline}
where we made use of partial integration in time, substituted \cref{eq:gc_jacobian,eq:gy_vlasov_conservative},
and have defined the magnetisation as
\begin{equation}\label{eq:magnetisation}
  \vc{\mathcal{M}}_1 \defeq \sum_s \int \gyfzero \vc{M}_{1,s} \jacobiangcslab \diff^3 \gy{u}
  , \quad
  \vc{M}_{1,s} \defeq
  - \gy{u}_\shortparallel \hb \times \vc{P}_{1,s} 
  - \frac{\gy{\mu}}{B_0} \vc{B}_{1,\perp}
  .
\end{equation}



We define the rest-frame magnetic and electric dipole moments per particle as
\begin{subequations}
  \begin{align}
    \vc{\mathfrak{m}}_s &\defeq - \gy{\mu} \roundpar{\hb + \frac{\vc{B}_{1,\perp}}{B_0}},\label{eq:rest_magnetic_moment}\\
    \vc{\mathfrak{p}}_{1,s} &\defeq \frac{m_s}{q_s B_0^2} \vc{F}_{1,\perp} = \frac{m_s}{B_0^2} \roundpar{\vc{E}_{1,\perp} + \gy{u}_\shortparallel \hb \times \vc{B}_1},
  \end{align}
\end{subequations}
where we have included the intrinsic guiding-centre magnetic moment $-\gy{\mu} \hb$~\cite[Eq.~(4.35)]{bittencourt2004} coming from the ZLR part of $- \gy{\mu} \dgav{(\evalperturbedgcburby{\nabla} \times \ca)_\shortparallel}$ (the last term on the right-hand side of \cref{eq:ampere}). The minus sign in \cref{eq:rest_magnetic_moment} reflects the fact that the plasma is diamagnetic as the magnetic dipole moment points in the opposite direction to the magnetic field.
It follows that the magnetic and electric dipole moments per particle are given by
\begin{subequations}
  \begin{align}
    \vc{\mathfrak{M}}_s  &= \vc{\mathfrak{m}}_s - \vc{v}_0 \times \vc{\mathfrak{p}}_{1,s},\\
    \vc{P}_{1,s} &= \vc{\mathfrak{p}}_{1,s} + \frac{1}{c^2} \underbrace{\vc{v} \times \vc{\mathfrak{m}}_s}_{= 0}
  \end{align}
\end{subequations}
by making use of \cref{eq:magnetisation,eq:polarisation}.
Here $c = 1 / \sqrt{\epsilon_0 \mu_0}$ denotes the speed of light, and we have defined the velocity $\vc{v}$ as
\begin{equation}
  \vc{v} \defeq \roundpar{\hb + \frac{\vc{B}_{1,\perp}}{B_0}} \gy{u}_\shortparallel,
\end{equation}
which includes the contribution from the magnetic flutter as found in \cref{eq:gy_split_eoms_R_subbstarstar}.
The expressions for the magnetic and electric dipole moments per particle can now directly be compared to those found in~\citet[Eqs.~(34) and (35)]{brizard2007} as well as to the expressions of a moving electric and magnetic dipole described in~\citet{Hnizdo_2012,fisher1971_dipoles}.

\begin{remark}
  In obtaining \cref{eq:ampere} we have made use of partial integration in time, thereby omitting the following term from the variation of the action
  \begin{align}\label{eq:ampere_partial_integration}
    \variation{\vc{A}_1}{\ca}{\mathfrak{A}}
    = {} & 
    \int (\text{the Amp\`ere--Maxwell law}) \diff t
    -\sum_s q_s \int \pard{}{t} \roundpar{\gy{f}_s \jacobiangy \rho \radgav{\evalperturbedgcburby{\cp}_\rho}} \diff^6 \gy{z} \diff t
    \nonumber\\ &
    - \sum_s \int \gyfzero \frac{m_s}{q_s B_0^2} \pard{}{t} (\vc{F}_{1,\perp} \bcdot \ca) \jacobiangcslab \diff^6 \gy{z} \diff t 
    .
  \end{align}
  We explicitly state this term as it plays a crucial role in the derivation of the conserved energy in \cref{sec:energy}.
\end{remark}

\subsection{Strong formulation of the field equations}\label{sec:macromaxwell_formulation}
The previously discussed field equations were given in a weak formulation, which is how they naturally arise from the variational formulation.
The weak formulation is exactly what we need for a future FEEC~\citep{gempic} discretisation of the field equations, however, when it comes to physical interpretation it is not the most convenient way to present the equations.
To this end, we consider the strong formulation of the field equations, where we moreover highlight the macroscopic Maxwell structure of the equations.

In essence, the strong formulation of Gauss's law is the equation which, once multiplied by the scalar test function $\cp$ and integrated over the spatial domain, results in Gauss's law~\eqref{eq:quasi_neutrality} after partial integration.
Here we note that the right-hand side of Gauss's law~\eqref{eq:quasi_neutrality} contains the gyro-average of the test function.
Hence, to find the strong formulation, we must define the gyro-average adjoint of the free charge density~$\mathcal{R}^\free$, which is defined such that
\begin{subequations}\label{eq:free_gavadjoints}
  \begin{equation}\label{eq:free_charge_gavadjoint}
    \int \evalperturbedgcadjoint{\mathcal{R}}{}^\free \cp \diff^3 r \defeq \sum_s q_s \int \gy{f}_s \gav{\evalperturbedgc{\cp}} \, \jacobiangy \diff^6 \gy{z}
  \end{equation}
  for all suitable test functions $\cp$.

A similar definition for the gyro-average adjoint of the free current density holds
  \begin{align}
    \int \evalperturbedgcadjointalt{\vc{\mathcal{J}}}{}^\free \bcdot \ca \diff^3 \gy{r}
    \defeq {} &
    \sum_s \int \gy{f}_s \Bigl[q_s \dot{\gy{\vc{R}}} \bcdot
        \roundpar{
            \ca
          + \nabla \radgav{\rho \evalperturbedgcburby{\cp}_\rho}
          + \radgav{(\evalperturbedgcburby{\nabla} \times \ca) \times \vc\rho}
        }\nonumber\\ &
        - \gy{\mu} \dgav{(\evalperturbedgcburby{\nabla} \times \ca)_\shortparallel}\Bigr]
        \jacobiangy \diff^6 \gy{z}
        \label{eq:free_current_gavadjoint}
  \end{align}
\end{subequations}
for which
\begin{equation}
  \int \evalperturbedgcadjointalt{\vc{\mathcal{J}}}{}^\free \bcdot \ca \diff^3 \gy{r} = \sum_s q_s \int \gy{f}_s \gav{\gy{\vc{U}}^\effective \bcdot \evalperturbedgc{\ca}} \, \jacobiangy \diff^6 \gy{z} + \bigO(\varepsilon_B)
\end{equation}
by making use of Eqs.~\eqref{eq:gradient_theorem_vector_approx} and \eqref{eq:gyro_identity_gav_tau}.
Here we have defined the effective gyrocentre velocity as~(cf.\ \cref{eq:field_aligned_velocity})
\begin{equation}\label{eq:gy_effective_velocity}
  \gy{\vc{U}}^\effective \defeq \dot{\gy{\vc{R}}} + \gy{u}_\tau \htau
\end{equation}
for which $\gav{\gy{\vc{U}}^\effective} = \dot{\gy{\vc{R}}}$.
The tangential velocity component~$\gy{u}_\tau$ used in \cref{eq:gy_effective_velocity} is in gyrocentre coordinates, i.e.\ it is defined according to \cref{eq:utau_def} evaluated at gyrocentre coordinates.
Similarly, the gyroradius~$\rho$ is defined according to \cref{eq:gyroradius} evaluated at gyrocentre coordinates.

We recall that the contribution to the free current density given by $-\gy{\mu} \dgav{(\evalperturbedgcburby{\nabla} \times \ca)_\shortparallel}$ on the right-hand side of \cref{eq:free_current_gavadjoint} results in the intrinsic guiding-centre magnetic moment and was included in \cref{eq:rest_magnetic_moment} in order to define the rest-frame magnetic moment.
The complicated term on the right-hand side of \cref{eq:free_current_gavadjoint} that multiplies $\dot{\gy{\vc{R}}}$ is essential in \cref{sec:proposed_wellposed} for showing that the field equations are compatible.
The key property of this term is found by letting $\ca = \nabla \cp$, as one does when computing the divergence of the adjoint of the free current density. 
This results in
\begin{equation}
  \int \evalperturbedgcadjointalt{\vc{\mathcal{J}}}{}^\free \bcdot \nabla \cp \diff^3 \gy{r} = \sum_s q_s \int \gy{f}_s \dot{\gy{\vc{R}}} \bcdot
      \nabla \gav{\evalperturbedgc{\cp}}
  \jacobiangy \diff^6 \gy{z},
\end{equation}
where we have made use of the gradient theorem~\eqref{eq:gradient_theorem_scalar} and shows that the gradient of the gyro-average of the test function is found, rather than the gyro-average of the gradient.
This equality is essential in showing that the gyro-average adjoint of the free-charge continuity equation also holds
\begin{equation}\label{eq:gy_continuity_gavadjoint}
  \int \roundpar{\pard{\evalperturbedgcadjoint{\mathcal{R}}{}^\free}{t} + \nabla \bcdot \evalperturbedgcadjointalt{\vc{\mathcal{J}}}{}^\free} \cp \diff^3 \gy{r} = 0
  \qquad \implies \qquad
                 \pard{\evalperturbedgcadjoint{\mathcal{R}}{}^\free}{t} + \nabla \bcdot \evalperturbedgcadjointalt{\vc{\mathcal{J}}}{}^\free = 0
\end{equation}
as follows from multiplying the conservative form of the Vlasov equation~\eqref{eq:gy_vlasov_conservative} by $q_s$ and the gyro-averaged scalar test function $\gav{\evalperturbedgc{\cp}}$, integrating over phase-space, by making use of partial integration, and by substituting the gyro-average adjoints defined in \cref{eq:free_gavadjoints}.

We can write the strong formulation of the field equations as
\begin{subequations}\label{eq:maxwell}
  \begin{align}
    \nabla \bcdot \vc{\mathcal{D}} &= \evalperturbedgcadjoint{\mathcal{R}}{}^\free,\label{eq:maxwell_gauss}\\
    \nabla \bcdot \vc{B} &= 0,\label{eq:maxwell_maggauss} \\
    \curl \vc{E} &= - \pard{\vc{B}}{t},\label{eq:maxwell_faraday}\\
    \curl \vc{\mathcal{H}} &= \pard{\vc{\mathcal{D}}}{t} + \evalperturbedgcadjointalt{\vc{\mathcal{J}}}{}^\free,\label{eq:maxwell_ampere}
  \end{align}
  where the constitutive relations defining the displacement and magnetising field are given by
  \begin{align}
    \vc{\mathcal{D}} &\defeq \epsilon_0 \vc{E}_1 + \vc{\mathcal{P}}_1,\label{eq:maxwell_displacement}\\
    \vc{\mathcal{H}} &\defeq \frac{1}{\mu_0} \vc{B} - \vc{\mathcal{M}}_1.\label{eq:maxwell_magfield}
  \end{align}
\end{subequations}
The displacement current density is given by $\partial {\vc{\mathcal{D}}} / \partial t$.
We recall that the polarisation $\vc{\mathcal{P}}_1$ and magnetisation $\vc{\mathcal{M}}_1$ are defined in \cref{eq:polarisation,eq:magnetisation}, respectively, and we note that $\vc{B} = \vc{B}_0 + \vc{B}_1$ as well as $\vc{E} = \vc{E}_1$.
Besides Gauss's law~\eqref{eq:maxwell_gauss} and the Amp\`ere--Maxwell law~\eqref{eq:maxwell_ampere}, we have additionally included Faraday's law~\eqref{eq:maxwell_faraday} as well as the magnetic Gauss's law~\eqref{eq:maxwell_maggauss}.
The latter two equations are satisfied automatically when a potential formulation is used, but due to the gauge invariance of the proposed model, we are able to express the proposed model entirely in terms of the electromagnetic fields, which thereby requires \cref{eq:maxwell_faraday,eq:maxwell_maggauss}.

Writing the field equations in this way shows that the proposed gauge-invariant gyrokinetic model can in fact be interpreted as a material property in the macroscopic Maxwell's equations.
As with the vacuum Maxwell's equations, we find that 
substituting the partial time derivative of Gauss's law~\eqref{eq:maxwell_gauss} in the divergence of the Amp\`ere--Maxwell law~\eqref{eq:maxwell_ampere}
yields the free-charge continuity equation~\eqref{eq:gy_continuity_gavadjoint}. 
Hence, the field equations possess a constraint which is automatically satisfied as a consequence of the particle EOMs.
The fact that precisely this constraint arises in the field equations is a consequence of gauge invariance of the gyrocentre single-particle phase-space Lagrangian (cf.\ \cref{eq:gauge_invariance_thm}) as discussed in \cref{sec:proposed_wellposed} and in particular \cref{rem:compatibility}. 

\subsection{Structure of the initial value problem}\label{sec:initial_value_problem}
The Amp\`ere--Maxwell law~\eqref{eq:maxwell_ampere} can be written as (upon substitution of Faraday's law~\eqref{eq:maxwell_faraday})
\begin{equation}\label{eq:maxwell_ampere_explicit}
  \roundpar{\epsilon_0 \mat{I}_3 + \mathcal{C}(1) \mat{\Pi}_\perp} \pard{\vc{E}}{t} = \curl \vc{\mathcal{H}} - \evalperturbedgcadjointalt{\vc{\mathcal{J}}}{}^\free + \mathcal{C}(\gy{u}_\shortparallel) \hb \times (\curl \vc{E})
\end{equation}
where the perpendicular projection matrix is defined as
\begin{equation}\label{eq:perpendicular_projection}
  \mat{\Pi}_\perp \defeq \mat{I}_3 - \hb \otimes \hb,
\end{equation}
and we have defined the spatially varying functions $\mathcal{C}(\zeta)$ as
\begin{equation}\label{eq:coeffC}
  \mathcal{C}(\zeta) \defeq \frac{1}{B_0^2} \sum_s m_s \int \gyfzero \zeta \jacobiangcslab \diff^3 \gy{u}
  .
\end{equation}
We note that the positivity of $\mathcal{C}(1)$ implies that the matrix on the left-hand side can be trivially inverted, provided that the vacuum permittivity $\epsilon_0$ is positive.
This results in an evolution equation for the electric field $\vc{E}$, which combined with the evolution equation for the magnetic field $\vc{B}$ (i.e.\ Faraday's law~\eqref{eq:maxwell_faraday}) as well as the particle EOMs~\eqref{eq:gy_split_eoms} yields an initial value problem (IVP) for the unknowns $(\vc{E}, \vc{B}, \gy{f}_s)$ (where the solution of the characteristics $\vc{Z}(t; \vc{z}^0, t^0)$ define the distribution function $\gy{f}_s$).
We note that solving this IVP requires an initial particle distribution function $\gyfzero$ which has to be compatible with the background magnetic field $\vc{B}_0$ (as discussed in \cref{sec:field_equilibria}), an initial electric field $\vc{E}^0$ which satisfies Gauss's law~\eqref{eq:maxwell_gauss}, and an initial magnetic field $\vc{B}^0$ which satisfies the magnetic Gauss's law~\eqref{eq:maxwell_maggauss}.

Moreover, it is worth noting that having a positive vacuum permittivity introduces light wave as well as the Langmuir wave into the proposed model, which sounds problematic due to their high velocity.
However, the light wave does not travel at the vacuum speed of light, but rather at the speed of light in the gyrokinetic plasma, which is much lower than the vacuum speed of light \citep{burby2015hamiltonian}.
The presence of such fast waves (including the compressional Alfv\'en wave, as we demonstrate in \cref{sec:slab_susceptibility}), however, imply that explicit time integration yields a stringent time step constraint, and to this end implicit time-integration methods might be of interest.
A quasi-neutral Darwin-like approximation to the gyrokinetic model can be considered when such fast waves are not of interest, 
as discussed in \cref{sec:darwin}.

In the limit of quasi-neutrality (i.e.\ $\epsilon_0 = 0$) the light wave as well as the Langmuir wave are removed from the model, while the compressional Alfv\'en wave remains.
In this limit we find that the displacement field is perpendicular to the background magnetic field $\vc{\mathcal{D}} \perp \hb$ (by substituting \cref{eq:polarisation,eq:gc_Lorentz})
\begin{equation}
  \epsilon_0 = 0
  \quad \implies \quad
  \vc{\mathcal{D}} = \mathcal{C}(1) \vc{E}_{\perp} + \mathcal{C}(\gy{u}_\shortparallel) \hb \times \vc{B}.
\end{equation}
It follows that \cref{eq:maxwell_ampere_explicit} yields an evolution equation for $\vc{E}_{\perp}$ only and not for $E_{\shortparallel}$.
This means that, upon discretizing \cref{eq:maxwell} in space, we find a differential algebraic system of equations (DAEs), rather than a system of ordinary differential equations (ODEs).

Here we follow the works of \citet{chen2021gyrokinetic,mcmillan2023relationship} and compute the time derivative of the parallel component of the Amp\`ere--Maxwell law~\eqref{eq:maxwell_ampere}, followed by substituting Faraday's law~\eqref{eq:maxwell_faraday}. 
This results in the following constraint equation for $E_{1,\shortparallel}$ (i.e.\ not an evolution equation)
\begin{equation}\label{eq:Eparallel_constraint}
  \frac{1}{\mu_0} \hb \bcdot \roundpar{ \curl\squarepar{\curl \roundpar{E_{1,\shortparallel} \hb}}} = -\pard{\evalperturbedgcadjointalt{\mathcal{J}_\shortparallel}{}^\free}{t} - \hb \bcdot \squarepar{\curl \roundpar{\frac{1}{\mu_0}\curl \vc{E}_{1,\perp} + \pard{\vc{\mathcal{M}}_1}{t}}}.
\end{equation}
In general, the magnetisation $\vc{\mathcal{M}}_1$ also depends on $\vc{B}_{1,\perp}$, and therefore its time derivative depends on $\vc{E}_1$.
However, this dependency vanishes when an isotropic background pressure is considered (see also \cref{eq:centeredMax_integrate0}), as is often the case.
It is worth noting that, for a constant background magnetic field, the operator on the left-hand side of \cref{eq:Eparallel_constraint} reduces to the perpendicular Laplacian $\nabla \bcdot \nabla_\perp E_{1,\shortparallel}$.
Hence, if so desired an equation for~$E_{1,\shortparallel}$ can be obtained, but we leave the details of a corresponding numerical solution strategy for a future paper.

\subsection{Equilibrium solutions of the field equations}\label{sec:field_equilibria}
The gyrocentre coordinate transformation discussed in \cref{sec:single_particle} is based on the \emph{assumption} that $\varepsilon_\delta \ll 1$, which we have not yet justified.
In order for this assumption to hold, we require that the initial particle distribution function $\gyfzero$ and background magnetic field $\vc{B}_0$ are close to equilibrium.
That is, at $t = t^0$ we require that the field equations approximately hold true to leading order in $\varepsilon_\delta$, i.e.\ when setting the perturbed fields to zero.
This results in equilibrium solutions of the field equations.

We assume that the background distribution function is nearly symmetric in $\gy{u}_\shortparallel$, that is, it is of the form
\begin{equation}\label{eq:shifted_background}
  \gyfzero(\gy{\vc{r}}, \gy{u}_\shortparallel, \gy{\mu}) = \gyfzero{}^{,\mathrm{S}}(\gy{\vc{r}}, \gy{u}_\shortparallel - \delta \gy{u}_{s}, \gy{\mu})
  \quad \text{with} \quad
  \gyfzero{}^{,\mathrm{S}}(\gy{\vc{r}}, \gy{u}_\shortparallel, \gy{\mu}) = \gyfzero{}^{,\mathrm{S}}(\gy{\vc{r}}, -\gy{u}_\shortparallel, \gy{\mu}),
\end{equation}
where $\varepsilon_{U,s} \defeq \delta \gy{u}_{s} / \uths$ is assumed to be small.
This results in
\begin{equation}\label{eq:shifted_integral}
  \gy{n}_{0,s} \gy{u}_{0,\shortparallel,s}
  = \int \gyfzero{}^{,\mathrm{S}} \delta \gy{u}_{s} \jacobiangcslab \diff^3 \gy{u}
  = \gy{n}_{0,s} \delta \gy{u}_{s},
\end{equation}
where have defined the background particle density and parallel velocity as
\begin{subequations}\label{eq:background_density_velocity}
  \begin{align}
    \gy{n}_{0,s} &\defeq \int \gyfzero \jacobiangcslab \diff^3 \gy{u},\label{eq:background_density}\\
    \gy{u}_{0,\shortparallel,s} &\defeq \frac{1}{\gy{n}_{0,s}} \int \gyfzero \gy{u}_\shortparallel \jacobiangcslab \diff^3 \gy{u}.\label{eq:background_velocity}
  \end{align}
\end{subequations}
Throughout the discussion on the equilibrium solutions we neglect $\bigO(\varepsilon_{B,s}^2)$ and $\bigO(\varepsilon_{B,s} \varepsilon_{U,s})$ terms, and we consider the ZLR limit $\varepsilon_\perp \rightarrow 0$.

For Gauss's law~\eqref{eq:maxwell_gauss} we find that the leading order part is given by
\begin{equation}
  0 
  = \sum_s q_s \gy{n}_{0,s}
  ,
  \label{eq:leading_order_constraint_qn}
\end{equation}
where we have made use of \cref{eq:gy_freecharge}.
The background distributions must result in an (approximately) quasi-neutral plasma in order to justify $\varepsilon_\delta \ll 1$.

The leading order part of the ZLR limit of the Amp\`ere--Maxwell law~\eqref{eq:maxwell_ampere} is given by
\begin{equation}
  \frac{1}{\mu_0} \curl \vc{B}_0 = \evalperturbedgcadjointalt{\vc{\mathcal{J}}}{}^{\free,\zlr}_0,
\end{equation}
where the ZLR limit of the background gyrocentre free-current density results in
\begin{equation}\label{eq:current_intermsof_pressure}
  \evalperturbedgcadjointalt{\vc{\mathcal{J}}}{}^{\free,\zlr}_0 = 
      \sum_s q_s \gy{n}_{0,s} \delta \gy{u}_{s} \hb
    + \frac{p_{0,\shortparallel}^\mathrm{S}}{B_0} \curl \hb 
    + \frac{p_{0,\perp}^\mathrm{S}}{B_0^2} \hb \times \nabla B_0
    -  \curl \roundpar{\frac{p_{0,\perp}^\mathrm{S}}{B_0} \hb}
\end{equation}
by making use of \cref{eq:free_current_gavadjoint,eq:gc_eoms_R,eq:shifted_integral}.
The background pressures are defined as
\begin{subequations}\label{eq:par_perp_pressure}
  \begin{align}
    p_{0,\shortparallel} &\defeq \sum_s m_s \int \gyfzero \gy{u}_\shortparallel^2 \jacobiangcslab \diff^3 \gy{u},\\
    p_{0,\perp} &\defeq \sum_s \frac{m_s}{2} \int \gyfzero \gy{u}_\tau^2 \jacobiangcslab \diff^3 \gy{u} = B_0 \sum_s \int \gyfzero \gy{\mu} \jacobiangcslab \diff^3 \gy{u}
  \end{align}
\end{subequations}
with equivalent definitions for the pressures resulting from the symmetric distribution function: $p_{0,\shortparallel}^\mathrm{S}, p_{0,\perp}^\mathrm{S}$.
Moreover, we have made use of \cref{eq:shifted_integral} as well as
\begin{subequations}
  \begin{align}
    \int \gyfzero \gy{u}_\shortparallel^2 \jacobiangcslab \diff^3 \gy{u}
    &= \int \gyfzero{}^{,\mathrm{S}} \gy{u}_\shortparallel^2 \jacobiangcslab \diff^3 \gy{u} + \bigO(\varepsilon_{U,s}^2),\\
    \int \gyfzero \gy{\mu} \jacobiangcslab \diff^3 \gy{u}
    &= \int \gyfzero{}^{,\mathrm{S}} \gy{\mu} \jacobiangcslab \diff^3 \gy{u}
    + \bigO(\varepsilon_{B,s} \varepsilon_{U,s}) 
    ,
  \end{align}
\end{subequations}
to conclude that $p_{0,\shortparallel} = p_{0,\shortparallel}^\mathrm{S}$ and $p_{0,\perp} = p_{0,\perp}^\mathrm{S}$ up to $\bigO(\varepsilon^2)$.

We find that the perpendicular part of the ZLR limit of the background gyrocentre free-current density can alternatively be written as
\begin{equation}\label{eq:perpcurrent_intermsof_pressuretensor}
  \evalperturbedgcadjointalt{\vc{\mathcal{J}}}{}^{\free,\zlr}_{0,\perp} = \frac{1}{B_0}\hb \times (\nabla \bcdot \mat{P}_{0})
  , \quad
  \mat{P}_{0} \defeq p_{0,\shortparallel}^\mathrm{S} \hb \otimes \hb + p_{0,\perp}^\mathrm{S} (\mat{I}_3 - \hb \otimes \hb),
\end{equation}
where we have made use of 
\begin{equation}\label{eq:curlcross_equals_perpdiv}
  Q (\curl \hb)_\perp = \hb \times \nabla \bcdot (Q \hb \otimes \hb).
\end{equation}
When combined with the Amp\`ere--Maxwell law this results in the equilibrium condition (as can also be found in e.g.\ \citet{grad1966})
\begin{equation}\label{eq:perpendicular_equilibrium}
  (\curl \vc{B}_0) \times \vc{B}_0 = \mu_0 (\nabla \bcdot \mat{P}_{0})_\perp,
\end{equation}
which, for an isotropic background distribution with $p_0 = p_{0,\shortparallel}^\mathrm{S} = p_{0,\perp}^\mathrm{S}$, results in the MHD equilibrium condition \citep{grad1966,grad1967}
\begin{equation}\label{eq:mhd_equilibrium}
  (\curl \vc{B}_0) \times \vc{B}_0 = \mu_0 \nabla p_0,
\end{equation}
wherein we have imposed $\hb \bcdot \nabla p_0 = 0$, as usually required by the tools for computing MHD equilibria.
The condition on the parallel derivative of the background pressure implies that the background particle density and temperature must be a function of the flux surface label.

MHD equilibria can be computed using software tools such as VMEC~\citep{VMEC1983} and GVEC~\citep{hindenlang2019gvec}, which for a given geometry and pressure find a magnetic field $\vc{B}_0$ such that \cref{eq:mhd_equilibrium} holds.
However, we note that \cref{eq:mhd_equilibrium,eq:perpendicular_equilibrium} only ensure that the perpendicular part of the Amp\`ere--Maxwell law is satisfied, and to this end we consider the parallel component of the Amp\`ere--Maxwell law
\begin{equation}\label{eq:mhd_parallel_ampere}
  \frac{1}{\mu_0} \hb \bcdot (\curl \vc{B}_0) 
  = \sum_s q_s \gy{n}_{0,s} \delta \gy{u}_{s}
  + \frac{p_{0,\shortparallel}^\mathrm{S} - p_{0,\perp}^\mathrm{S}}{B_0^2} \hb \bcdot (\curl \vc{B}_0)
\end{equation}
as follows substitution of the parallel component of \cref{eq:current_intermsof_pressure}.
Correctly satisfying the parallel component of the Amp\`ere--Maxwell law is crucial for the modelling of, for example, kink modes \citep{Dudkovskaia_2023}.
To this end, we note that the shift $\delta \gy{u}_{s}(\gy{\vc{r}})$ (which may be a function of the gyrocentre position) is the only unknown in \cref{eq:mhd_parallel_ampere}, and therefore this equation can be used to impose a constraint on the shift in order to satisfy the parallel component of the Amp\`ere--Maxwell law.
Moreover, it shows that the background distribution function must be strongly anisotropic if it is unshifted, unless $\hb \bcdot (\curl \hb) = 0$.
The smallness of the non-dimensional shift $\varepsilon_{U,s}$ per species can be deduced from the non-dimensionalisation of \cref{eq:mhd_parallel_ampere}
\begin{equation}\label{eq:mhd_parallel_ampere_nondim}
  \bigO(1)
  = \sum_s \frac{\beta_{0,s}}{1 - \frac{\mu_0 (p_{0,\shortparallel}^\mathrm{S} - p_{0,\perp}^\mathrm{S})}{B_0^2}} \frac{\varepsilon_{U,s}}{\varepsilon_{B,s}}
  ,
\end{equation}
where the plasma-$\beta$ is defined per species as
\begin{equation}\label{eq:beta}
  \beta_{0,s} 
  \defeq 
  \frac{2 \mu_0 \gy{n}_{0,s} \kboltz \temperature_s}{B_0^2}.
\end{equation}

This approach is comparable to the work presented in \citet{mcmillan2023relationship}. 
Therein a global Maxwellian particle distribution function is considered, for which it is shown that only part of the parallel component of the Amp\`ere--Maxwell law is correctly satisfied for a (toroidally symmetric) Grad--Shafranov equilibrium \citep{grad1958hydromagnetic}.
This issue is then resolved by introducing a slight modification (in particular, \citep[Eqs.~(3.5) and (3.6)]{mcmillan2023relationship}) to the global Maxwellian particle distribution function.
We similarly modify the originally symmetric particle distribution function by a shift $\delta \gy{u}_s$ to correctly satisfy the parallel component of the Amp\`ere--Maxwell law.
Our approach, however, does not require toroidal symmetry of the background magnetic field and can therefore also be applied to 3D MHD equilibria in stellarator devices.

To summarise, we can choose any symmetric background distribution function $\gyfzero{}^{,\mathrm{S}}$ from which we compute the parallel and perpendicular pressure according to \cref{eq:par_perp_pressure}.
We then solve the equilibrium equation~\eqref{eq:perpendicular_equilibrium} (or \cref{eq:mhd_equilibrium} if the background distribution function is isotropic), resulting in the background magnetic field $\vc{B}_0$.
Finally, we use \cref{eq:mhd_parallel_ampere} to find the shifts $\delta \gy{u}_{s}(\gy{\vc{r}})$, thereby adjusting the background distribution function~$\gyfzero$ according to \cref{eq:shifted_background}.
Constructing the background distribution function and magnetic field in this way ensures that we are near equilibrium, which thereby ensures that the underlying assumption of the gyrocentre coordinate transformation $\varepsilon_\delta \ll 1$ is justified.

\subsection{Well-posedness of the field equations}\label{sec:proposed_wellposed}
Well-posedness of the field equations is non-trivial.
With well-posedness we refer to the existence and uniqueness of solutions to the field equations.
This is a rather mathematical topic and therefore falls outside the scope of this paper when it is considered fully rigorously.
However, as we can see, a necessary condition for well-posedness is related to the bound-charge continuity equation and thereby allows for a physical interpretation.

To illustrate that well-posedness is a non-trivial property, we consider the following general form of Gauss's law and the Amp\`ere--Maxwell law 
\begin{subequations}\label{eq:hahm}
  \begin{align}
     \epsilon_0 \nabla \bcdot \vc{E}_1 &= \evalperturbedgcadjoint{\mathcal{R}}{}^\free + \mathcal{R}^\bound,\label{eq:general_gauss} \\
    \frac{1}{\mu_0}\curl \vc{B} &= \epsilon_0 \pard{\vc{E}_1}{t} + \evalperturbedgcadjointalt{\vc{\mathcal{J}}}{}^\free + \vc{\mathcal{J}}^\bound,\label{eq:general_ampere_strong}
  \end{align}
\end{subequations}
where we have defined the bound charge density as
\begin{equation}\label{eq:bound_charge}
  \mathcal{R}^\bound \defeq - \nabla \bcdot \vc{\mathcal{P}}_1
\end{equation}
and consider some unspecified bound current density denoted by $\vc{\mathcal{J}}^\bound$.
Computing the divergence of the Amp\`ere--Maxwell law~\eqref{eq:general_ampere_strong} results in
\begin{equation}\label{eq:darwin_divampere}
  0 
  =  \epsilon_0 \nabla \bcdot \pard{\vc{E}_1}{t} + \nabla \bcdot \evalperturbedgcadjointalt{\vc{\mathcal{J}}}{}^\free + \nabla \bcdot \vc{\mathcal{J}}^\bound
  \quad \iff \quad
  \underbrace{\pard{\mathcal{R}^\bound}{t} + \nabla \bcdot \vc{\mathcal{J}}^\bound = 0}_{\substack{\text{bound-charge} \\ \text{continuity equation}}},
\end{equation}
where we have substituted the free-charge continuity equation~\eqref{eq:gy_continuity_gavadjoint} as well as Gauss's law~\eqref{eq:general_gauss}.
That is, we find that the bound-charge continuity equation~\eqref{eq:darwin_divampere} must hold for the two field equations to be compatible.
Note that compatibility here means that the divergence of the Amp\`ere--Maxwell law should coincide with the time derivative of Gauss's law.
For the strong formulation we find
\begin{equation}
  \vc{\mathcal{J}}^\bound = \pard{\vc{\mathcal{P}}_1}{t} + \curl \vc{\mathcal{M}}_1
\end{equation}
for which it can easily be shown that \cref{eq:darwin_divampere} is satisfied.

\begin{remark}\label{rem:compatibility}
  Compatibility of the field equations can also directly be deduced from the action principle by computing the following gauge invariant variation
  \begin{equation}\label{eq:gaugeinvariant_variation}
    \evalAt{\totd{}{\varepsilon}}{\varepsilon = 0} \mathfrak{A}(\gy{\vc{Z}}, \phi_1 - \varepsilon \partial \eta / \partial t, \Ad + \varepsilon \nabla \eta) = 0
  \end{equation}
  for some scalar test function $\eta$.
  When substituting \cref{eq:low_split_explicit} we find that this results in
  \begin{equation}
    0
    = \sum_s q_s \int \gyfzero(\gyzzero) \roundpar{
        \nabla \eta \bcdot \dot{\gy{\vc{R}}} 
        + \pard{\eta}{t}
    } \jacobiangy(\gyzzero, \tzero) \diff^6 \gyzzerod \diff t
    ,
  \end{equation}
where we note that this is direct consequence of the gauge invariance of the gyrocentre single-particle phase-space Lagrangian, see \cref{eq:gauge_invariance_thm}.
  By making use of \cref{thm:gy_liouville} as well as partial integration this results in
  \begin{equation}
    0
    = - \int \underbrace{\roundpar{\pard{\mathcal{R}^\free}{t} + \nabla \bcdot \vc{\mathcal{J}}^\free}}_{\overset{\eqref{eq:gy_continuity}}{=} 0}
      \eta \diff^3 \gy{r} \diff t
      ,
  \end{equation}
  where we have moreover substituted the definitions of free charge and current density as given by \cref{eq:gy_freecharge_and_current} and thereby results in a constraint which is automatically satisfied as a consequence of the free-charge continuity equation~\eqref{eq:gy_continuity}.
  It follows that this gauge invariant variation does not give an additional constraint as it is consistent with the free-charge continuity equation~\eqref{eq:gy_continuity} derived from the conservative form of the Vlasov equation~\eqref{eq:gy_vlasov_conservative}. We note that this specific variation coincides with computing the difference between the divergence of the Amp\`ere--Maxwell law (i.e.\ letting $\ca = \nabla \eta$ in \cref{eq:ampere}) and the time derivative of Gauss's law (i.e.\ letting $\cp = \partial \eta / \partial t$ in \cref{eq:quasi_neutrality}).
\end{remark}

\subsection{Energy conservation}\label{sec:energy}
Typically, the derivation of conservation laws for quantities such as energy and momentum is achieved by making use of Noether's method \citep{Noether1918} wherein symmetries of the Lagrangian result in conserved quantities.
The derivation of exact local conservation laws for electromagnetic gyrokinetic systems has been elusive \citep{peifeng2021discovering}, however, due to the presence of the integrals over time only in the Lagrangian owing to the particles.
In \citet{peifeng2021discovering} a procedure is proposed to overcome this difficulty, resulting in exact local conservation laws for energy and momentum of arbitrary-order gyrokinetic models.
Alternatively, one can switch to Eulerian variables (i.e.\ transforming from the Lagrangian EOMs $\gy{\vc{Z}}$ to the particle distribution function $\gy{f}_s$) as it is done in \citet{hirvijoki2020energy} for models based on an Euler--Poincar\'e variational formulation.
The conservation laws derived in \citet{hirvijoki2020energy} are rederived in \citet{brizard2021exact} for models that are based on an Eulerian variational principle.

\subsubsection{Derivation}
In our work we follow a more direct and simple approach, as we do not aim to derive conservation laws for arbitrary-order gyrokinetic models.
We first derive the evolution of the kinetic energy per particle $\gy{K}_s$, which then leads to the evolution equation of the kinetic energy density $\mathcal{K}$ upon integration over the particles. 
Next, we derive the evolution equation of the potential energy density $\mathcal{U}$.
When combined, the two evolution equations result in the local energy conservation law, which, upon integration over the spatial coordinate then leads to the global energy conservation law.

The gyrocentre kinetic energy per particle is defined as
\begin{equation}\label{eq:energy_kin}
  \gy{K}_s \defeq \frac{m_s \gy{U}_\shortparallel^2 }{2} + \gy{\Mu} (B_0 + \dgav{\evalperturbedgcburby{B}_{1,\shortparallel}}),
\end{equation}
which is derived from applying the gyrocentre coordinate transformation~\eqref{eq:gy_G10_gav} to the guiding-centre kinetic energy $\gc{K}_0$ (as defined in \cref{eq:gc_energy_kin0}).
The kinetic energy of a particle evolves as
\begin{equation}
  \totd{\gy{K}_s}{t} 
  =   q_s \dot{\gy{\vc{R}}} \bcdot \vc{E}^\effective_1
    - \gy{\Mu} \dgav{(\evalperturbedgcburby{\nabla} \times \vc{E}_{1})_\shortparallel} 
\end{equation}
by making use of \cref{eq:gy_split_eoms} as well as Faraday's law~\eqref{eq:Faraday}.
Hence, if the perturbed electric field vanishes, we find that the kinetic energy per particle is conserved, as can be expected because static magnetic fields do no work.

The kinetic energy per particle can be integrated over all particles resulting in the kinetic energy density
\begin{subequations}
  \begin{equation}\label{eq:energy_kin_density}
    \gy{\mathcal{K}} \defeq \sum_s \int \gy{f}_s \gy{K}_s \jacobiangy \diff^3 \gy{u},
  \end{equation}
  whereas the displacement and magnetising field result in the potential energy density
  \begin{equation}
    \gy{\mathcal{U}} \defeq
    \frac{1}{2} (
      \vc{\mathcal{D}} \bcdot \vc{E}
      + 
      \vc{\mathcal{H}} \bcdot \vc{B}
    ) 
  .\label{eq:energy_pot}
  \end{equation}
\end{subequations}
This results in the following local energy conservation law for which a proof can be found in \cref{app:poynting}.
\begin{theorem}[Local energy conservation]\label{thm:poynting}
  The kinetic energy density~\eqref{eq:energy_kin_density} satisfies
  \begin{equation}\label{eq:energy_kin_evolution}
    \pard{\gy{\mathcal{K}}}{t} + \nabla \bcdot \roundpar{\sum_s \int \gy{f}_s \dot{\gy{\vc{R}}} \gy{K}_s \jacobiangy \diff^3 \gy{u}} = \evalperturbedgcadjointalt{\vc{\mathcal{J}}}{}^\free \bcdot \vc{E},
  \end{equation}
  whereas the potential energy density~\eqref{eq:energy_pot} satisfies Poynting's theorem
  \begin{equation}\label{eq:energy_pot_evolution}
    \pard{\gy{\mathcal{U}}}{t} + \nabla \bcdot (\vc{E} \times \vc{\mathcal{H}}) = - \evalperturbedgcadjointalt{\vc{\mathcal{J}}}{}^\free \bcdot \vc{E}.
  \end{equation}
  The magnetizing field $\vc{\mathcal{H}}$ and free current density $\evalperturbedgcadjointalt{\vc{\mathcal{J}}}{}^\free$ are defined in Eqs.~\eqref{eq:maxwell_magfield} and \eqref{eq:free_current_gavadjoint}, respectively.
  It follows that the following local energy conservation law holds
  \begin{equation}\label{eq:energy_tot_evolution}
    \pard{}{t} (\gy{\mathcal{K}} + \gy{\mathcal{U}}) + \nabla \bcdot \roundpar{\vc{E} \times \vc{\mathcal{H}} + \sum_s \int \gy{f}_s \dot{\gy{\vc{R}}} \gy{K}_s \jacobiangy \diff^3 \gy{u}} = 0.
  \end{equation}
\end{theorem}
On the right-hand side of the evolution equation for the energy densities, i.e. \cref{eq:energy_kin_evolution,eq:energy_pot_evolution}, we recognise the $\evalperturbedgcadjointalt{\vc{\mathcal{J}}}{}^\free \bcdot \vc{E}$ source term which is often used for diagnostic purposes \citep{bottino,NOVIKAU2021107032,kleiber2024euterpe}.

When integrating the sum of the kinetic and potential energy densities over the spatial domain we find the total energy
\begin{equation}\label{eq:energy_sum}
  \gy{\mathfrak{E}} \defeq \int (\gy{\mathcal{K}} + \gy{\mathcal{U}}) \diff^3 \gy{r},
\end{equation}
which is conserved as a consequence of \cref{eq:energy_tot_evolution}.

\begin{remark}[Total energy conservation in weak form]
  The derivation of the total energy presented above is based on the strong formulation of the equations, but we note that the conserved energy can also be derived directly from the field-theoretic Lagrangian.
  This is of interest as this implies that a numerical model based on this weak formulation can also be exactly energy conserving, provided that it is properly discretised.

  We start by considering the total time derivative of the field-theoretic Lagrangian~\eqref{eq:low_split}
  \begin{equation}\label{eq:energy_dldt}
    \totd{\mathfrak{L}}{t} = 
        \variation{\gy{\vc{Z}}}{\dot{\gy{\vc{Z}}}}{\mathfrak{L}} 
      + \variation{\phi_1}{\pard{\phi_1}{t}}{\mathfrak{L}} 
      + \variation{\vc{A}_1}{\pard{\vc{A}_1}{t}}{\mathfrak{L}} 
      .
  \end{equation}
  We then assume that the characteristics and fields satisfy the EOMs and field equations, respectively, and make use of the fact that these equations are found by setting the respective variation of the action to zero, up to partial integration in time.
  For instance, for the gyrocentre characteristics we find that
  \begin{equation}
    \variation{\gy{\vc{Z}}}{\dot{\gy{\vc{Z}}}}{\mathfrak{L}} = \totd{}{t} \sum_s \int \gyfzero(\gyzzero) {\pard{\gy{L}_s^\mathrm{part}}{\dot{\gy{\vc{Z}}}} \bcdot \dot{\gy{\vc{Z}}}} \jacobiangy(\gyzzero, \tzero) \diff^6 \gyzzerod
  \end{equation}
  by making use of \cref{eq:eoms_partial_integration}.
  Similarly, we find that \cref{eq:ampere_partial_integration} becomes 
  \begin{equation}
    \variation{\vc{A}_1}{\pard{\vc{A}_1}{t}}{\mathfrak{L}}
    = - \totd{}{t} \sum_s \int \roundpar{ q_s \gy{f}_s \jacobiangy \rho \pard{}{t}\radgav{\evalperturbedgcburby{A}_{1,\rho}}
    + \gyfzero \vc{P}_{1,s} \bcdot \pard{\vc{A}_1}{t} \jacobiangcslab} \diff^6 \gy{z}
  \end{equation}
  upon substitution of the Amp\`ere--Maxwell law.
  The variation for $\phi_1$ did not require any partial integration in time, and therefore we find that \cref{eq:energy_dldt} results in
  \begin{equation}\label{eq:global_energy_cons}
    \totd{\gy{\mathfrak{E}}}{t} = 0,
  \end{equation}
  where the conserved energy is given by
  \begin{equation}\label{eq:energy_sum_variational}
    \gy{\mathfrak{E}} = 
      \sum_s \int \roundpar{
      \gy{f}_s {\pard{\gy{L}_s^\mathrm{part}}{\dot{\gy{\vc{Z}}}} \bcdot \dot{\gy{\vc{Z}}}} \jacobiangy
    - q_s \gy{f}_s \rho \pard{}{t}\radgav{\evalperturbedgcburby{A}_{1,\rho}} \jacobiangy
    - \gyfzero \frac{m_s}{q_s B_0^2} \vc{F}_{1,\perp} \bcdot \pard{\vc{A}_1}{t} \jacobiangcslab
    } \diff^6 \gy{z}  
    - \mathfrak{L}.
  \end{equation}
  It can be shown that \cref{eq:energy_sum_variational} reduces to \cref{eq:energy_sum} by substituting the definitions of the particle~\eqref{eq:Hpart} and field~\eqref{eq:Hfield} Hamiltonian in Low's action~\eqref{eq:low_split} and by making use of Gauss's law as well as the gradient theorem~\eqref{eq:gradient_theorem_scalar}.
\end{remark}

\subsubsection{Comparison with results from literature}
Despite using a more direct approach in deriving the local energy conservation law, the resulting conserved energy density should agree with other results from literature.
To facilitate this comparison, we consider the work of \citet{brizard2021exact} wherein the ZLR limit of the model proposed in \citet{burby2019gauge} is considered, which conveniently coincides with the ZLR limit of the proposed model.

The conserved energy density of the proposed model in the ZLR limit is given by
\begin{equation}
  \mathcal{K}^\zlr + \mathcal{U}^\zlr
  = \sum_s \int \gy{f}_s \gy{K}_s^\zlr \jacobiangy \diff^3 \gy{u} 
  + \frac{1}{2} (
    \vc{\mathcal{D}} \bcdot \vc{E}
    + 
    \vc{\mathcal{H}} \bcdot \vc{B}
  ),
\end{equation}
where the ZLR limit of the kinetic energy per particle is given by
\begin{equation}
    \gy{K}_s^\zlr = \frac{m_s \gy{U}_\shortparallel^2 }{2} + \gy{\Mu} (B_0 + {B}_{1,\shortparallel})
\end{equation}
by making use of \cref{eq:energy_kin_density} as well as the observation that the displacement and magnetising field do not contain any FLR contributions.

The kinetic energy per particle in \citet[Eq.~(3.4)]{brizard2021exact} is defined differently from ours \eqref{eq:energy_kin} and is given by
\begin{equation}
  K_s^\brizard \defeq \gy{K}_s^\zlr + \gy{H}_{2,s},
\end{equation}
where we have neglected the contribution from the guiding-centre electric dipole moment as a result of neglecting the mixed $\bigO(\varepsilon_\delta \varepsilon_B)$ terms in the proposed model.
Note that the second-order gyrocentre Hamiltonian can be expressed in terms of the magnetisation and polarisation per particle as (cf.\ \cref{eq:magnetisation,eq:polarisation,eq:gy_H2})
\begin{equation}
  \gy{H}_{2,s} = -\frac{1}{2} (\vc{P}_{1,s} \bcdot \vc{E} + \vc{M}_{1,s} \bcdot \vc{B})
\end{equation}
such that the kinetic energy density can be written as
\begin{equation}
  \mathcal{K}^\brizard 
  = 
  \mathcal{K}^\zlr
  + \sum_s \int \gy{f}_s \gy{H}_{2,s} \jacobiangy \diff^3 \gy{u}
  = 
  \mathcal{K}^\zlr
  + \frac{1}{2} \squarepar{
    - \roundpar{\vc{\mathcal{D}} - \epsilon_0 \vc{E}} \bcdot \vc{E} 
    + \roundpar{\vc{\mathcal{H}} - \frac{1}{\mu_0} \vc{B}} \bcdot \vc{B}
  }
\end{equation}
by substituting \cref{eq:maxwell_displacement,eq:maxwell_magfield}.
Furthermore, we have ignored the linearization of the particle part of the Hamiltonian introduced in \cref{eq:low_split} as no such linearization is applied in \citet{brizard2021exact}.
Finally, we note that the potential energy density in \citet[Eq.~(5.11)]{brizard2021exact} is given by 
\begin{equation}
  \mathcal{U}^\brizard = 
    \vc{\mathcal{D}} \bcdot \vc{E}
    - \frac{\epsilon_0}{2} \abs{\vc{E}}^2 
    + \frac{1}{2\mu_0} \abs{\vc{B}}^2
\end{equation}
such that the conserved energy densities are indeed found to be the same: $\mathcal{K}^\zlr + \mathcal{U}^\zlr = \mathcal{K}^\brizard + \mathcal{U}^\brizard$.
We note that the conserved energy density also agrees with that found in \citet[Eq.~(82)]{hirvijoki2020energy}.

  \section{Quasi-neutral gyrokinetic Darwin model}\label{sec:darwin}
The proposed gyrokinetic Maxwell model keeps more physics than the popular reduced parallel-only model (discussed in \cref{sec:reduced_parallel}) and has more structure than the symplectic Brizard--Hahm model~\citep{brizard2007} (discussed in \cref{sec:brizardhahm}).
In particular, we find that the Amp\`ere--Maxwell law~\eqref{eq:ampere} contains a displacement current density, which is not present in either of the two other models.
Part of this displacement current density, however, gives rise to fast waves such as the light wave, the Langmuir wave and the compressional Alfv\'en wave (as demonstrated in \cref{sec:slab_susceptibility}), and in most situations such waves are undesired due to their high frequency.

In this section a quasi-neutral Darwin approximation is proposed, which removes the fast waves from the model.
This approximation consists of two steps: first the limit of quasi-neutrality is considered thereby removing the light wave as well as the Langmuir wave from the model, and second the compressional Alfv\'en wave is removed by considering a Darwin-like approximation wherein the transversal part of the displacement current density is removed from the Amp\`ere--Maxwell law.
The resulting model is gauge invariant, is obtained \textit{via} an action principle, has compatible field equations and still possesses a local energy conservation law.

\subsection{The Darwin approximation to Maxwell's equations}
One way to deal with such high frequency components is to damp them numerically using implicit time integration methods.
Another option is to remove such waves from the underlying model, i.e.\ from the Lagrangian.
For instance, to remove the light wave from Maxwell's equations, rather than considering the limit of quasi-neutrality, one can consider the Darwin approximation, wherein only the longitudinal (i.e.\ irrotational or curl free) part is kept in the $\epsilon_0 \partial \vc{E}_1 / \partial t$ term in the Amp\`ere--Maxwell law.
Thus, the transversal (i.e.\ solenoidal or divergence free), part of the displacement current density is neglected 
\begin{equation}
  \epsilon_0 \pard{\vc{E}_1}{t} \approx \epsilon_0 \pard{(\Pi_\mathrm{L} \vc{E}_1)}{t},
\end{equation}
where $\Pi_\mathrm{L}$ is the longitudinal projection operator which is defined as (with appropriate boundary conditions on the inverse Laplace operator)
\begin{equation}\label{eqn:ordinary_longitudinal_projection_operator}
  \Pi_\mathrm{L} \vc{S} \defeq \nabla [\nabla^{-2} (\nabla \bcdot \vc{S})].
\end{equation}
This results in a model which yields a second- and third-order accurate electric and magnetic field, respectively, in the small parameter $\varepsilon_c = v / c$~\citep{DegondRaviart} and restricts the dynamics of the Vlasov--Maxwell system to an invariant slow manifold of the Vlasov--Maxwell phase space~\citep{Miloshevich_Burby_2021}.

Note that the Darwin approximation is a gauge-invariant approximation, as the projection operator acts on the electric field directly.
If the Coulomb gauge is used then the vector potential is transversal, and we find that the Darwin approximation simply neglects the vector potential contribution to the electric field
\begin{equation}
  \nabla \bcdot \vc{A}_{1} = 0 
  \quad \implies \quad 
  \Pi_\mathrm{L} \vc{E}_1 = -\nabla \phi_1.
\end{equation} 

\subsection{The Darwin approximation of the gyrocentre Hamiltonian}
We follow an approach similar to the Darwin approximation to remove the fast compressional Alfv\'en wave.
As the proposed model is defined in terms of an action principle, we propose a modification to the action~\eqref{eq:low_split_explicit} which corresponds to the removal of the compressional Alfv\'en wave.
Recall that the second-order gyrocentre Hamiltonian is given by (cf.\ \cref{eq:gy_H2})
\begin{equation}
  \gy{H}_{2,s} = 
    - \frac{m_s}{2 B_0^2} \abs{\vc{E}_{1,\perp}}^2
    + \frac{m_s \gy{U}_\shortparallel}{B_0^2} \hb \bcdot (\vc{E}_{1,\perp} \times \vc{B}_1)
    + \frac{
          \gy{\Mu} B_0
        - m_s \gy{U}_\shortparallel^2
      }{2 B_0^2} {\abs{\vc{B}_{1,\perp}}^2},
\end{equation}
where we have substituted the definition of the Lorentz force~\eqref{eq:gc_Lorentz}, and thus agrees exactly with the result found in \citet[Eq.~(14)]{burby2019gauge}.
We note that the compressional Alfv\'en wave comes from the transversal contribution to the $\abs{\partial \Adperp / \partial t}^2$ term (which itself comes from the $\abs{\vc{E}_{1,\perp}}^2$ term), and it should therefore be sufficient to remove exactly this term from $\gy{H}_{2,s}$.

When keeping only the contribution from the longitudinal part of the electric field in the second-order Hamiltonian, we find the following contribution to the action from the second-order Hamiltonian
\begin{align}\label{eq:gy_darwin_H2}
  \sum_s \int \gyfzero \gy{H}_{2,s}^\darwin \jacobiangcslab \diff^3 \gy{u} \defeq {} &
    - \frac{1}{2 \mathcal{C}(1)} \sabs{\Pi_{\mathrm{L},\perp} \roundpar{\mathcal{C}(1) \vc{E}_{1}}}^2
    + \frac{\mathcal{C}(\gy{u}_\shortparallel)}{\mathcal{C}(1)} \hb \bcdot \squarepar{\Pi_{\mathrm{L},\perp}\roundpar{\mathcal{C}(1)\vc{E}_{1}} \times \vc{B}_1} 
    \nonumber \\
    &+ \frac{
      p_{0,\perp} - p_{0,\shortparallel}
      }{2 B_0^2} {\abs{\vc{B}_{1,\perp}}^2},
\end{align}
where the gyrokinetic longitudinal projection operator is defined as (cf.\ \cref{eqn:ordinary_longitudinal_projection_operator})
\begin{equation}
  \Pi_{\mathrm{L},\perp} \vc{S} \defeq \mathcal{C}(1) \nabla_\perp \roundpar{[\nabla \bcdot (\mathcal{C}(1) \nabla_\perp)]^{-1} \squarepar{\nabla \bcdot \vc{S}_\perp}},
\end{equation}
and we recall that $\mathcal{C}(\zeta)$ is defined in \cref{eq:coeffC}.

\subsection{Principle of least action} 
As the particle part of the Hamiltonian is still given by \cref{eq:Hpart}, and the symplectic part of the Lagrangian is unchanged when compared to the proposed gyrokinetic Maxwell model, we find that the EOMs are still given by \cref{eq:gy_split_eoms}.

The explicit expression of Low's action, while using \cref{eq:gy_darwin_H2}, is given by (cf. \cref{eq:low_split_explicit})
\begin{multline}\label{eq:low_split_explicit_darwin}
  \mathfrak{A}^\darwin(\gy{\vc{Z}}, \phi_1, \Ad)
  = \sum_s \int \gyfzero(\gyzzero) \biggl[
      q_s \roundpar{
          \vc{A}_{0,s}^\effective 
        + \vc{A}_1 
        + \radgav{\evalperturbedgcburby{\vc{B}}_1\times \vc\rho}} \bcdot \dot{\gy{\vc{R}}} 
        \\
      + \frac{m_s \gy{\Mu}}{q_s} \dot{\gy{\Theta}}
      - \gy{\Mu} (B_0 + \dgav{\evalperturbedgcburby{B}_{1,\shortparallel}})
      - \frac{m_s}{2} \gy{U}_\shortparallel^2 
      - q_s \phi_1 
      + q_s \rho \radgav{\evalperturbedgcburby{E}_{1,\rho}}
  \biggr] \jacobiangy(\gyzzero, \tzero) \diff^6 \gyzzerod \diff t\\
  + \int \Biggl( \frac{1}{2\mathcal{C}(1)} \sabs{\Pi_{\mathrm{L},\perp} \roundpar{\mathcal{C}(1) \vc{E}_{1}}}^2
    - \frac{\mathcal{C}(\gy{u}_\shortparallel)}{{\mathcal{C}(1)}} \hb \bcdot \squarepar{\Pi_{\mathrm{L},\perp}\roundpar{{\mathcal{C}(1)}\vc{E}_{1}} \times \vc{B}_1}\\
     - \frac{p_{0,\perp} - p_{0,\shortparallel}}{2B_0^2} {\abs{\vc{B}_{1,\perp}}^2} 
    \Biggr)\diff^3 \gy{r} \diff t
  - \frac{1}{2 \mu_0} \int \abs{\vc{B}_0 + \vc{B}_1}^2 \diff^3 x \diff t
  .
\end{multline}
Compared to the action of the proposed gyrokinetic Maxwell model, as given in \cref{eq:low_split_explicit}, we additionally consider the limit of quasi-neutrality (i.e.\ $\epsilon_0 = 0$) in order to eliminate the fast light wave as well as the Langmuir wave from the proposed model.

Setting the variations of the action~\eqref{eq:low_split_explicit_darwin} w.r.t.\ the scalar and vector potential to zero, results in the following quasi-neutrality equation (cf.\ \cref{eq:quasi_neutrality}) and the Amp\`ere--Maxwell law (cf.\ \cref{eq:ampere})
\begin{subequations}
  \begin{align}
    - \int \vc{\mathcal{P}}_{1}^\darwin \bcdot \nabla_\perp \cp \diff^3 \gy{r}
      = {} &
    \sum_s q_s \int \gy{f}_s \gav{\evalperturbedgc{\cp}} \jacobiangy \diff^6 \gy{z}
    ,\label{eq:darwin_qn}\\
    \frac{1}{\mu_0} \int (\vc{B}_0 + \vc{B}_1) \bcdot (\curl \ca) \diff^3 x 
    = {} &
    \int \roundpar{\pard{\vc{\mathcal{P}}_{1}^\darwin}{t} + \curl \vc{\mathcal{M}}_1^\darwin + \evalperturbedgcadjointalt{\vc{\mathcal{J}}}{}^\free} \bcdot \ca \diff^3 \gy{r} 
    ,\label{eq:darwin_ampere}
  \end{align}
\end{subequations}
respectively, by making use of the self-adjointness of the projection operator $\Pi_{\mathrm{L},\perp}$.
The Darwin polarisation (cf.\ \cref{eq:polarisation}) and magnetisation (cf.\ \cref{eq:magnetisation}) are given by
\begin{subequations}
  \begin{align}
    \vc{\mathcal{P}}_{1}^\darwin 
    &\defeq \Pi_{\mathrm{L},\perp} \vc{\mathcal{P}}_{1}
    = \Pi_{\mathrm{L},\perp} \roundpar{\mathcal{C}(1) \vc{E}_{1} + \mathcal{C}(\gy{u}_\shortparallel) \hb \times \vc{B}_1}
    , \label{eq:darwin_polarisation} \\
    \vc{\mathcal{M}}_{1}^\darwin &\defeq
    \frac{p_{0,\shortparallel} - p_{0,\perp}}{B_0^2} \vc{B}_{1,\perp}
    - \frac{\mathcal{C}(\gy{u}_\shortparallel)}{{\mathcal{C}(1)}} \hb \times \Pi_{\mathrm{L},\perp} \roundpar{{\mathcal{C}(1)} \vc{E}_{1}}
    ,\label{eq:darwin_magnetisation}
  \end{align}
\end{subequations}
where we note that the Darwin polarisation is entirely longitudinal, as desired.

\subsection{Strong formulation of the field equations}\label{sec:darwin_wellposed}

\subsubsection{Gauge invariant formulation}
We find that the resulting model can be written in strong formulation as
\begin{subequations}\label{eq:darwin}
  \begin{align}
    \nabla \bcdot \vc{\mathcal{D}}^\darwin &= \evalperturbedgcadjoint{\mathcal{R}}{}^\free,\label{eq:darwin_gauss} \\
    \curl \vc{\mathcal{H}}^\darwin &= \pard{\vc{\mathcal{D}}^\darwin}{t} + \evalperturbedgcadjointalt{\vc{\mathcal{J}}}{}^\free 
    , \label{eq:darwin_ampere_strong}
  \end{align}
\end{subequations}
  where the Darwin displacement and magnetizing field are defined as
  \begin{subequations}   
    \begin{align}
      \vc{\mathcal{D}}^\darwin &\defeq \vc{\mathcal{P}}_{1}^\darwin
      \label{eq:darwin_displacement},\\
      \vc{\mathcal{H}}^\darwin &\defeq \frac{1}{\mu_0} \vc{B} - \vc{\mathcal{M}}_{1}^\darwin.\label{eq:darwin_magfield}
    \end{align}
\end{subequations}
It follows that the field equations are compatible in the sense that the divergence of the Amp\`ere--Maxwell law yields the time derivative of the quasi-neutrality equation.
That is, the bound-charge continuity equation is satisfied for the gyrokinetic Darwin model analogous to the discussion from \cref{sec:proposed_wellposed}.

\subsubsection{The perpendicular Coulomb gauge}
Despite the favourable structure of the resulting field equations, we note that the explicit presence of the gyrokinetic longitudinal projection operator is not desirable because it is not a local operator and in particular involves the inversion of a perpendicular Laplace operator.
Specific choices of the gauge condition can be made to avoid this complication, and in particular we consider the perpendicular Coulomb gauge given by
\begin{equation}\label{eq:darwin_perp_gauge}
  \int \mathcal{C}(1) \nabla_\perp \cp \bcdot \Ad \diff^3 x = 0.
\end{equation}
This gauge condition implies that the longitudinal part of the scaled vector potential vanishes
\begin{equation}
  0 
  = \int \nabla_\perp \cp \bcdot \Pi_{\mathrm{L},\perp} \roundpar{{\mathcal{C}(1)} \Ad} \diff^3 x
  \quad \implies \quad
  \Pi_{\mathrm{L},\perp} \roundpar{{\mathcal{C}(1)}\Ad} = \vc{0}_3
\end{equation}
as follows from the adjoint of the projection operator as well as
\begin{equation}\label{eq:projector_property}
  \Pi_{\mathrm{L},\perp} \roundpar{{\mathcal{C}(1)} \nabla_\perp \cp} = {\mathcal{C}(1)} \nabla_\perp \cp.
\end{equation}

Within the perpendicular Coulomb gauge~\eqref{eq:darwin_perp_gauge} we find that the quasi-neutrality equation~\eqref{eq:darwin_gauss} reduces to
\begin{equation}\label{eqn:darwin_quasi_neutrality}
  -\nabla \bcdot \roundpar{
    \sum_s \frac{ m_s \gy{n}_{0,s}}{B_0^2} \squarepar{\nabla_\perp \phi_1 - \gy{u}_{0,\shortparallel,s} \hb \times (\curl \Ad)}
  } = \evalperturbedgcadjoint{\mathcal{R}}{}^\free,
\end{equation}
where we have substituted the value of $\mathcal{C}(1)$ and $\mathcal{C}(\gy{u}_\shortparallel)$ from \cref{eq:coeffC} and have made use of the adjoint of the longitudinal projection operator and \cref{eq:projector_property}.
Note that this equation is decoupled from the Amp\`ere--Maxwell law if the background distribution function is symmetric (i.e.\ $\gy{u}_{0,\shortparallel,s} = 0$) and thereby reduces to the well-known (and well-posed) perpendicular Laplace equation for $\phi_1$.
We note that using a perpendicular Lorenz-type gauge results in a similar simplification, but we do not consider this here.

For the vector potential $\vc{A}_1$ we have to solve the following system of equations (by substituting the definition of the magnetizing field \cref{eq:darwin_magfield} and magnetisation \cref{eq:darwin_magnetisation} into \cref{eq:darwin_ampere_strong})
\begin{subequations}\label{eq:darwin_saddle}
  \begin{align}
    \curl \biggl[
      \frac{1}{\mu_0} \curl \Ad 
    - \frac{p_{0,\shortparallel} - p_{0,\perp}}{B_0^2} (\curl \Ad)_\perp 
    -  \coeffcupar &\hb \times \nabla_\perp \phi_1
    \biggr] \nonumber \\ \quad + \coeffcone \nabla_\perp \lambda &= \evalperturbedgcadjointalt{\vc{\mathcal{J}}}{}^\free 
    - \frac{1}{\mu_0} \curl \vc{B}_0,\label{eq:darwin_ampere_saddle}\\
    \nabla \bcdot \roundpar{\coeffcone \Adperp} &= 0,\label{eq:darwin_constraint_saddle}
  \end{align}
\end{subequations}
where we have intentionally dropped the contribution from the \emph{longitudinal} displacement current density and have instead introduced a Lagrange multiplier $\lambda$ for which
\begin{equation}
  \pard{\vc{\mathcal{D}}^\darwin}{t} = -\mathcal{C}(1) \nabla_\perp \lambda
  \quad \implies \quad
  \lambda \defeq 
  \pard{\phi_1}{t} - [\nabla \bcdot (\mathcal{C}(1) \nabla_\perp)]^{-1} \squarepar{\nabla \bcdot \roundpar{
    \mathcal{C}(\gy{u}_\shortparallel) \hb \times \pard{\vc{B}_1}{t}
  }},
\end{equation}
which thereby replaces the displacement current density and simultaneously enforces the gauge condition~\eqref{eq:darwin_constraint_saddle}.
We note that keeping the contribution from the displacement current density would yield $\lambda = 0$ as a consequence of the compatibility of the field equations, which in turn is a consequence of the gauge invariance of the proposed model.
Hence, the Lagrange multiplier $\lambda$ is non-zero only because we have chosen to drop the contribution from the displacement current density for numerical reasons.

If the background particle distribution functions are symmetric, i.e.\ $\gy{u}_{0,\shortparallel,s} = 0$, it results in a full decoupling of the field equations for the potentials: \cref{eqn:darwin_quasi_neutrality} yields the scalar potential and \cref{eq:darwin_saddle} yields the vector potential.
The key property of \cref{eq:darwin_ampere_saddle} is that it is invariant under $\vc{A}_1 \mapsto \vc{A}_1 + \nabla \eta$ and a gauge condition is therefore needed to fix this freedom.
Such a gauge condition is exactly provided by the constraint~\eqref{eq:darwin_constraint_saddle}.
The well-posedness of \cref{eq:darwin_saddle} for an isotropic pressure and a symmetric background distribution is discussed in \cref{app:saddle_point}.

\subsection{Energy conservation}\label{sec:darwin_energy}
For the gyrokinetic Darwin model we find that the evolution equation for the kinetic energy density~\eqref{eq:energy_kin_evolution} is unchanged.
The equivalent to Poynting's theorem can also be shown, except that the potential energy density
\begin{equation}\label{eq:dar_energy_pot}
  \mathcal{U}^\darwin \defeq
  \frac{1}{2} (
    \vc{\mathcal{D}}^\darwin \bcdot \vc{E}
    + 
    \vc{\mathcal{H}}^\darwin \bcdot \vc{B}
  )
\end{equation}
evolves according to (cf.\ \cref{eq:energy_pot_evolution})
\begin{equation}
  \pard{\mathcal{U}^\darwin}{t} + \nabla \bcdot \roundpar{\vc{E} \times \vc{\mathcal{H}}^\darwin + \vc{\mathcal{S}}} = - \evalperturbedgcadjointalt{\vc{\mathcal{J}}}{}^\free \bcdot \vc{E},
\end{equation}
where the additional potential energy flux is given by
\begin{equation}
  \vc{\mathcal{S}} 
  = \frac{1}{2} \Biggl[
    \phi^E  \Pi_{\mathrm{T},\perp} \pard{\vc{\mathcal{P}}_1}{t}
    - \pard{\phi^E}{t} \Pi_{\mathrm{T},\perp} \vc{\mathcal{P}}_1
    + \phi^B \Pi_{\mathrm{T},\perp} \roundpar{{\mathcal{C}(1)} \pard{\vc{E}_\perp}{t}}
    - \pard{\phi^B}{t} \Pi_{\mathrm{T},\perp} \roundpar{{\mathcal{C}(1)} \vc{E}_\perp} 
  \Biggr]
  ,
\end{equation}
the gyrokinetic transversal projection operator $\Pi_{\mathrm{T},\perp}$ is defined as
\begin{equation}
  \Pi_{\mathrm{T},\perp} \vc{S} \defeq \vc{S} - \Pi_{\mathrm{L},\perp} \vc{S}
  ,
\end{equation}
and $\phi^E, \phi^B$ are the scalar potential parts of the (longitudinal) displacement
\begin{align}
  \phi^E &\defeq -[\nabla \bcdot (\mathcal{C}(1) \nabla_\perp)]^{-1} \squarepar{\nabla \bcdot \roundpar{\mathcal{C}(1) \vc{E}_\perp}},\\
  \phi^B &\defeq -[\nabla \bcdot (\mathcal{C}(1) \nabla_\perp)]^{-1} \squarepar{\nabla \bcdot \roundpar{\mathcal{C}(\gy{u}_\shortparallel)  \hb \times \vc{B}}},
\end{align}
which in the perpendicular Coulomb gauge yields $\phi^E = \phi_1$ and the potentials relate to the Lagrange multiplier~$\lambda$ from \cref{eq:darwin_saddle} \textit{via} $\lambda = \partial(\phi^E + \phi^B)/\partial t$.
 
This results in the following conserved energy
\begin{equation}
  \mathfrak{E}^\darwin \defeq \int (\mathcal{K} + \mathcal{U}^\darwin) \diff^3 \gy{r},
\end{equation}
where the kinetic energy density is still defined by \cref{eq:energy_kin_density}.

  \section{Comparison with some models from literature}\label{sec:reduced}
The two proposed gyrokinetic models which have been derived thus far are more comprehensive and possess more structure than the models usually found in literature~\citep{Kleiber_pullback,brizard2007,qin1999gyrokinetic}.
In this section, we compare the proposed gyrokinetic models to several models from the literature.
This comparison is not intended to be exhaustive: we compare the proposed models to the parallel-only model \citep{Kleiber_pullback} as it is the `working horse' of gyrokinetic simulations, the symplectic gyrokinetic model from \citet{brizard2007} as this is a frequently cited paper which presents a novel gyrokinetic model which includes $\Adperp$ but is not gauge invariant, and finally we compare the proposed models to the gauge-invariant gyrokinetic model from \citet{burby2019gauge} as the proposed models are a generalization thereof and are largely inspired by it.

For each model under consideration, we discuss the gyrocentre single-particle Lagrangian, the resulting EOMs and field equations as well as their corresponding strong form.
Furthermore, the well-posedness of the models is discussed, and dispersion relations are derived and used to compare the models in terms of the presence of shear and/or compressional Alfv\'en waves.

\subsection{Parallel-only model}\label{sec:reduced_parallel}
The parallel-only model, as is discussed for example in \citet{Kleiber_pullback}, is based on the assumption that the perpendicular part of the vector potential can be neglected: $\Adperp = \vc{0}_3$.
This assumption is combined with the following approximation in the derivation of the Hamiltonian
\begin{equation}\label{eq:reduced_assumption}
  \vc{B}_{1} 
  = \curl (\Apa \hb) 
  = \underbrace{\nabla_\perp \Apa \times \hb}_{\bigO(\varepsilon_\perp)}
  + \underbrace{\Apa \curl \hb}_{\bigO(\varepsilon_B)}
  \approx 
  \nabla_\perp \Apa \times \hb,
\end{equation}
where it is assumed that
\begin{equation}\label{eq:parallel_nosystemscale}
  \varepsilon_B \ll \varepsilon_\perp
  \quad \iff \quad
  \frac{1}{k_\perp} \ll L_B.
\end{equation}
That is, it is assumed that no system-scale effects are present in the perpendicular direction.
When considering that FLR effects are already neglected in the second-order gyrocentre Hamiltonian $\gy{H}_2$, it follows that the reduced parallel-only model is valid for intermediate wavelengths only: $\varrho \ll 1 / k_\perp \ll L_B$.

\subsubsection{The gyrocentre single-particle Lagrangian}
When neglecting the perpendicular part of the vector potential and by making use of the approximation given by \cref{eq:reduced_assumption}, we find that the symplectic part is given by 
\begin{equation}
  \gy{\vc{\gamma}}_{1,\vc{R},s}^{\reduced} 
  \defeq \gav{(\hb \bcdot \gc{\vc{\gamma}}_{1,s,\vc{R}}^\dagger) \hb} 
  = q_s \gav{\evalperturbedgc{A}_{1,\shortparallel}} \hb,
\end{equation}
whereas the first- and second-order Hamiltonian are reduced to (cf.\ \cref{eq:gy_H1})
\begin{subequations}
  \label{eq:H_parallel}
  \begin{equation}\label{eq:H1_parallel}
    \gy{H}_{1,s}^\reduced \defeq 
    q_s \gav{\evalperturbedgc{\phi}_1}
  \end{equation}
  and (cf.\ \cref{eq:gy_H2})
  \begin{equation}\label{eq:H2_parallel}
    \gy{H}_{2,s}^\reduced \defeq
    - \frac{m_s}{2 B_0^2} \abs{\nabla_\perp (\phi_1 - \gy{U}_\shortparallel \Apa)}^2
    + \frac{\gy{\Mu}}{2 B_0} \abs{\nabla_\perp \Apa}^2,
  \end{equation}
\end{subequations}
respectively, by making use of the gradient theorem~\eqref{eq:gradient_theorem_scalar}.

\subsubsection{Principle of least action}
We find that the reduced EOMs are given by (cf.\ \cref{eq:gy_split_eoms})
\begin{subequations}
  \label{eq:reduced_split_eoms}
  \begin{align}
    \dot{\gy{\vc{R}}} &= 
        \gy{U}_\shortparallel \vc{b}_s^{\effective,\reduced} 
      - \frac{1}{q_s B_{s,\shortparallel}^{\effective,\reduced}} \hb \times \roundpar{
          q_s \vc{E}_1^{\effective,\reduced}
        - \Mu \nabla B_0
      },\label{eq:reduced_split_eoms_R}
      \\
    \dot{\gy{U}}_\shortparallel &= 
        \frac{1}{m_s} \vc{b}_s^{\effective,\reduced} \bcdot \roundpar{
            q_s \vc{E}_1^{\effective,\reduced}
          - \Mu \nabla B_0
        }
      ,\label{eq:reduced_split_eoms_Vpa}
  \end{align}
\end{subequations}
where the electromagnetic fields are defined as (cf.\ \cref{eq:gy_flr_electomagfields})
\begin{subequations}
  \label{eq:reduced_flr_fields}
  \begin{align}
    \vc{E}_1^{\effective,\reduced} &\defeq -\nabla \gav{\evalperturbedgc{\phi}_1} - \pard{\gav{\evalperturbedgc{A}_{1,\shortparallel}}}{t} \hb,\label{eq:reduced_flc_electricfield}
    \\
    \vc{B}_1^{\effective,\reduced}
    &\defeq \curl (\gav{\evalperturbedgc{A}_{1,\shortparallel}} \hb)
    ,
  \end{align}
\end{subequations}
and we have defined $\vc{b}_s^{\effective,\reduced}$ and $B_{s,\shortparallel}^{\effective,\reduced}$ analogously to \cref{eq:gy_bstarstar,eq:gy_Bpastar}, i.e.\ by replacing $\vc{B}_{1}^\effective$ by $\vc{B}_{1}^{\effective,\reduced}$.

For the derivation of the field equations we again give Low's action explicitly resulting in (cf.\ \cref{eq:low_split_explicit,eq:low_split_explicit_darwin})
\begin{multline}\label{eq:low_split_reduced_explicit}
  \mathfrak{A}^\reduced(\gy{\vc{Z}}, \phi_1, \Apa)
  = \sum_s \int \gyfzero(\gyzzero) \biggl[
      q_s \roundpar{
          \vc{A}_{0,s}^\effective 
        + \gav{\evalperturbedgc{A}_{1,\shortparallel}} \hb} \bcdot \dot{\gy{\vc{R}}}
      + \frac{m_s \gy{\Mu}}{q_s} \dot{\gy{\Theta}} 
      - \frac{m_s}{2} \gy{U}_\shortparallel^2 
      - \gy{\Mu} B_0
      \\
      - q_s \gav{\evalperturbedgc{\phi}_1}
  \biggr] \jacobiangy^\reduced(\gyzzero, \tzero) \diff^6 \gyzzerod \diff t
  + \sum_s \int \gyfzero \biggl[
      \frac{m_s}{2 B_0^2} \abs{\nabla_\perp (\phi_1 - \gy{u}_\shortparallel \Apa)}^2 
    - \frac{\gy{\mu}}{2 B_0} \abs{\nabla_\perp \Apa}^2
  \biggr] \jacobiangcslab \diff^6 \gy{z} \diff t\\
  - \frac{1}{2 \mu_0} \int \abs{\vc{B}_0 + \nabla_\perp \Apa \times \hb}^2 \diff^3 x \diff t,
\end{multline}
where the Jacobian is now given by $\jacobiangy^\reduced \defeq {B_{s,\shortparallel}^{\effective,\reduced}} / {m_s}$ (cf.\ \cref{eq:gy_jacobian}).
Compared to the action of the proposed gyrokinetic Maxwell model, as given in \cref{eq:low_split_explicit}, we consider the limit of quasi-neutrality (i.e.\ $\epsilon_0 = 0$).

The quasi-neutrality equation (cf.\ \cref{eq:quasi_neutrality,eq:darwin_qn}) and Amp\`ere's law (cf.\ \cref{eq:ampere,eq:darwin_ampere}) are given by
\begin{subequations}
  \label{eq:reduced_fieldeqs}
  \begin{align}
    -\int \vc{\mathcal{P}}_1^\reduced \bcdot \nabla_\perp \cp \diff^3 \gy{r}
      &=
    \sum_s q_s \int \gy{f}_s 
      \gav{\evalperturbedgc{\cp}} 
      \jacobiangy^\reduced \diff^6 \gy{z}
    ,\label{eq:reduced_quasi_neutrality}\\
    \frac{1}{\mu_0} \int \nabla_\perp \Apa \bcdot \nabla_\perp \cp \diff^3 x
    &= 
    \int \vc{\mathcal{M}}_1^\reduced \bcdot (\hb \times \nabla_\perp \cp) \diff^3 \gy{r}
    + \sum_s q_s \int \gy{f}_s \dot{\gy{R}}_\shortparallel
        \gav{\evalperturbedgc{\cp}}
    \jacobiangy^\reduced \diff^6 \gy{z},\label{eq:reduced_ampere}
  \end{align}
\end{subequations}
where the reduced polarisation (cf.\ \cref{eq:polarisation,eq:darwin_polarisation}) and magnetisation (cf.\ \cref{eq:magnetisation,eq:darwin_magnetisation}) are given by
\begin{subequations}
  \begin{align}
    \vc{\mathcal{P}}_1^\reduced &\defeq \sum_s \int \gyfzero \vc{P}_{1,s}^\reduced \jacobiangcslab \diff^3 \gy{u}
    , \quad
    \vc{P}_{1,s}^\reduced \defeq \frac{m_s}{q_s B_0^2} \vc{F}_{1,\perp}^\reduced
    , \quad
    \vc{F}_{1,\perp}^\reduced \defeq -q_s \nabla_\perp (\phi_1 - \gy{u}_\shortparallel \Apa),\label{eq:reduced_polarisation}\\
    \vc{\mathcal{M}}_1^\reduced &\defeq \sum_s \int \gyfzero \vc{M}_{1,s}^\reduced \jacobiangcslab \diff^3 \gy{u}
    , \quad
    \vc{M}_{1,s}^\reduced \defeq
        - \hb \times \roundpar{\gy{u}_\shortparallel \vc{P}_{1,s}^\reduced
      - \frac{\gy{\mu}}{B_0} \nabla_\perp \Apa}
    .\label{eq:reduced_magnetisation}
  \end{align}
\end{subequations}

When considering the EOMs~\eqref{eq:reduced_split_eoms} as well as field equations~\eqref{eq:reduced_fieldeqs} of the reduced model, we find that this model coincides with reduced parallel models from the literature.
We note that this is only due to the choice $(\gparamr, \gparamtheta) = (1, 0)$, which yields the appropriate gyro-averages on both the scalar and vector potential.


\subsubsection{Strong formulation}
In order to be able to interpret the equations more easily, we present the strong formulation of the field equations as follows (cf. \cref{eq:darwin_saddle})
\begin{subequations}\label{eq:field_eqs_parallel}
  \begin{align}
    - \nabla \bcdot \squarepar{\sum_s\frac{ m_s \gy{n}_{0,s}}{B_0^2} \roundpar{\nabla_\perp \phi_1 - \gy{u}_{0,\shortparallel,s} \nabla_\perp \Apa}} &= \evalperturbedgcadjoint{\mathcal{R}}{}^\free, \\
    -\frac{1}{\mu_0} \nabla \bcdot \nabla_\perp \Apa 
    - \nabla \bcdot \roundpar{
      \coeffcupar \nabla_\perp \phi_1 
    - \frac{p_{0,\shortparallel} - p_{0,\perp}}{B_0^2} \nabla_\perp \Apa
  } 
    &=  \evalperturbedgcadjoint{\mathcal{J}}{}^{\free,\reduced}_\shortparallel,
  \end{align}
\end{subequations}
where we recall that $\gy{n}_{0,s}$ and $\gy{u}_{0,\shortparallel,s}$ denote the background particle density and parallel velocity, as defined in \cref{eq:background_density_velocity}.
The gyro-average adjoint of the parallel component of the free current density~$\evalperturbedgcadjoint{{\mathcal{J}}}{}^{\free,\reduced}_\shortparallel$ is different from the one discussed in \cref{sec:macromaxwell_formulation}, and it is defined in a weak sense as (cf.\ \cref{eq:free_current_gavadjoint})
\begin{equation}
  \int \evalperturbedgcadjoint{{\mathcal{J}}}{}^{\free,\reduced}_\shortparallel \cp \diff^3 \gy{r} 
  \defeq 
  \sum_s q_s \int \gy{f}_s \dot{\gy{{R}}}_\shortparallel \gav{\evalperturbedgc{\cp}}  \jacobiangy^\reduced \diff^6 \gy{z}.
\end{equation}

The following identities
\begin{equation}\label{eq:centeredMax_integrate0}
  \int \gyfzero \gy{u}_\shortparallel \jacobiangcslab \diff^3 \gy{u} = 0
  , \quad
  \int \gyfzero \roundpar{m_s \gy{u}_\shortparallel^2 - \gy{\mu} B_0} \jacobiangcslab \diff^3 \gy{u} = 0,
\end{equation}
which hold for a centred Maxwellian background distribution
\begin{equation}\label{eq:maxwellian}
  \gyfzerocm(\gy{\vc{r}}, \gy{u}_\shortparallel, \gy{\mu}) \defeq \frac{\gy{n}_{0,s}(\gy{\vc{r}})}{\sqrt{\upi}^{3} \uths(\gy{\vc{r}})^3}
    \exp{\left(-\frac{m_s \gy{u}_\shortparallel^2 + 2 \gy{\mu} B_0(\gy{\vc{r}})}{m_s \uths(\gy{\vc{r}})^2} \right)},
\end{equation}
result in decoupling the field equations~\eqref{eq:field_eqs_parallel}.
Note that \cref{eq:centeredMax_integrate0}, in physical terms, coincides with the absence of a parallel background current density ($\gy{u}_{0,\shortparallel,s} = 0$) as well as the isotropy of the pressure/temperature ($p_{0,\perp} = p_{0,\shortparallel}$).

\subsection{Symplectic gyrokinetic model from Brizard--Hahm}\label{sec:brizardhahm}
Besides the gauge invariant model described in \citet{burby2019gauge}, there are also gauge variant gyrokinetic models which include $\Adperp$.
In particular, we consider the symplectic model from \citet[Eqs.~(171) and (173) with $(\alpha, \beta) = (1, 1)$]{brizard2007}, which we hereafter refer to as the Brizard--Hahm (BH) model.

\subsubsection{The gyrocentre single-particle Lagrangian}
The symplectic part of the Lagrangian is as given in \cref{eq:gy_gamma_brizard}, whereas the first- and second-order Hamiltonian are derived from \citet[Eqs.~(171) and (173) with $(\alpha, \beta) = (1, 1)$]{brizard2007} resulting in (cf.\ \cref{eq:gy_H1,eq:H1_parallel})
\begin{subequations}
  \begin{equation}\label{eq:H1_bh}
    \gy{H}_{1,s}^\bh \defeq q_s \gav{\evalperturbedgc{\phi}_1} + \Mu \dgav{\evalperturbedgcburby{B}_{1,\shortparallel}}
  \end{equation}
  and  (cf.\ \cref{eq:gy_H2,eq:H2_parallel})
  \begin{equation}\label{eq:H2_bh}
    \gy{H}_{2,s}^{\bh} \defeq 
    - \frac{m_s}{2 B_0^2} \abs{\nabla_\perp (\phi_1 - \gy{U}_\shortparallel A_{1,\shortparallel})}^2
    + \frac{\Mu}{2 B_0} \abs{\nabla_\perp \Apa}^2,
  \end{equation}
\end{subequations}
respectively. 
The derivation, which neglects terms of $\bigO(\varepsilon_\perp^3)$ in $\gy{H}_{2,s}^{\bh}$, can be found in \cref{app:brizard_hahm}.
We note that this result can alternatively be derived from \cref{eq:gy_H2}, when making use of the approximation $\vc{B}_1 \approx \nabla_\perp \Apa \times \hb + (\curl \Adperp)_\shortparallel \hb$ and neglecting ${\partial \vc{A}_{1,\perp}} / {\partial t}$.

Note that the second-order Hamiltonian coincides exactly with that of the parallel-only model: $\gy{H}_{2,s}^{\bh} = \gy{H}_{2,s}^\reduced$.
This is remarkable, as it implies an absence of the polarisation current density as well as an absence of a contribution to the magnetisation from the perpendicular part of the vector potential.

Both the proposed gyrokinetic Maxwell model and the Brizard--Hahm model~\citep{brizard2007} reduce to the parallel-only model when neglecting the perpendicular part of the vector potential.
Note that the same cannot be said about the gauge invariant model from \citet{burby2019gauge}, which coincides with $(\gparamr, \gparamtheta) = (0, 0)$. 
Therein, we find that, e.g.\ $\vc{A}^\effective_1 = \Ad$ such that the vector potential without gyro-averaging appears in the EOMs.

\subsubsection{Principle of least action}
The EOMs are found by substituting the expression for the particle Hamiltonian $\gy{H}^{\mathrm{part},\bh} = \gy{H}_0 + \gy{H}_1^\bh$ into the general form of the EOMs given by \cref{eq:gy_eoms}, where we now have $\vc{A}^{\effective,\bh}_1 \defeq \gav{\evalperturbedgc{\vc{A}}_1}$.
This results in EOMs which have an identical structure as the EOMs of the quasi-neutral gyrokinetic Maxwell model~\eqref{eq:gy_split_eoms}, except that the
electromagnetic fields are defined as (cf.\ \cref{eq:gy_flr_electomagfields,eq:reduced_flr_fields})
\begin{subequations}
  \begin{align}
    \vc{E}_1^{\effective,\bh} &\defeq -\nabla \gav{\evalperturbedgc{\phi}_1} - \pard{\gav{\evalperturbedgc{\vc{A}}_{1}}}{t},\label{eq:bh_flc_electricfield}\\
    \vc{B}_1^{\effective,\bh} &\defeq \curl \gav{\evalperturbedgc{\vc{A}}_1},
  \end{align}
\end{subequations}
and we have defined $\vc{b}_s^{\effective,\bh}$ and $B_{s,\shortparallel}^{\effective,\bh}$ analogously to \cref{eq:gy_bstarstar,eq:gy_Bpastar}.

The explicit expression of Low's action is given by (cf.\ \cref{eq:low_split_explicit,eq:low_split_explicit_darwin,eq:low_split_reduced_explicit})
\begin{multline}\label{eq:low_split_bh_explicit}
  \mathfrak{A}^\bh(\gy{\vc{Z}}, \phi_1, \Ad, \lambda)
  = \sum_s \int \gyfzero(\gyzzero) \biggl[
      q_s \roundpar{
          \vc{A}_{0,s}^\effective 
        + \gav{\evalperturbedgc{\vc{A}}_{1}}} \bcdot \dot{\gy{\vc{R}}}
      + \frac{m_s \gy{\Mu}}{q_s} \dot{\gy{\Theta}} 
      - \frac{m_s}{2} \gy{U}_\shortparallel^2
      - \gy{\Mu} B_0 \\
      - q_s \gav{\evalperturbedgc{\phi}_1} - \Mu \dgav{\evalperturbedgcburby{B}_{1,\shortparallel}}
  \biggr] \jacobiangy^\bh(\gyzzero, \tzero) \diff^6 \gyzzerod \diff t
  + \sum_s \int \gyfzero \biggl[
      \frac{m_s}{2 B_0^2} \abs{\nabla_\perp (\phi_1 - \gy{u}_\shortparallel \Apa)}^2\\
    - \frac{\gy{\mu}}{2 B_0} \abs{\nabla_\perp \Apa}^2
  \biggr] \jacobiangcslab \diff^6 \gy{z} \diff t
  - \frac{1}{2 \mu_0} \int \abs{\vc{B}_0 + \vc{B}_1}^2 \diff^3 x \diff t
  + \int C^\bh(\gy{\vc{Z}}, \phi_1, \Ad, \lambda) \diff^3 x \diff t,
\end{multline}
where a Lagrange multiplier $\lambda$ is introduced, with an associated constraint $C^\bh$, to ensure that a well-posed system of equations is found, as is discussed in more detail in \cref{sec:bh_wellposed}.
Therein, we show that the constraint necessarily depends on all unknowns of our problem, including the characteristics $\gy{\vc{Z}}$.
Hence, when using a variational formulation of the Brizard--Hahm model~\citep{brizard2007} we find that well-posedness of the model implies that the Lagrange multiplier~$\lambda$ affects the EOMs as well as each of the field equations.
We do not show this dependence here explicitly, and for now we ignore the contribution due to the constraint.
Moreover, the Jacobian is now given by $\jacobiangy^\bh \defeq {B_{s,\shortparallel}^{\effective,\bh}} / {m_s}$ (cf.\ \cref{eq:gy_jacobian}).

The action~\eqref{eq:low_split_bh_explicit} results in the same quasi-neutrality equation as found for the reduced parallel-only model, as given by \cref{eq:reduced_quasi_neutrality}.
We find that Amp\`ere's law is now given by
\begin{align}
  \frac{1}{\mu_0} \int (\vc{B}_0 + \vc{B}_1) \bcdot (\curl \ca) \diff^3 x
  = {} & \int \nabla \bcdot (\vc{\mathcal{M}}_1^\reduced \times \hb) \capa \diff^3 \gy{r} \label{eq:ampere_bh} \\ &
  + \sum_s \int \gy{f}_s \squarepar{q_s \dot{\gy{\vc{R}}} \bcdot \gav{\evalperturbedgc{\ca}} - \gy{\mu} \dgav{(\evalperturbedgcburby{\nabla} \times \ca)_\shortparallel}} \jacobiangy^\bh \diff^6 \gy{z}
  . \nonumber 
\end{align}

\subsubsection{Strong formulation}
As usual, we consider the strong formulation of the field equations, this results in the following quasi-neutrality equation and Amp\`ere's law (cf.\ \cref{eq:maxwell,eq:darwin_saddle,eq:field_eqs_parallel})
\begin{subequations}\label{eq:bh_strong}
  \begin{align}
    - \nabla \bcdot \squarepar{\sum_s\frac{ m_s \gy{n}_{0,s}}{B_0^2} \roundpar{\nabla_\perp \phi_1 - \gy{u}_{0,\shortparallel,s} \nabla_\perp \Apa}} &= \evalperturbedgcadjoint{\mathcal{R}}{}^\free, \label{eq:bh_gauss} \\
    \frac{1}{\mu_0} \curl (\curl \vc{A}_1)
    - \hb \nabla \bcdot \roundpar{
      \sum_s\frac{ m_s \gy{n}_{0,s} \gy{u}_{0,\shortparallel,s}}{B_0^2} \nabla_\perp \phi_1 
    - \frac{p_{0,\shortparallel} - p_{0,\perp}}{B_0^2} \nabla_\perp \Apa
    }
    &= 
    \evalperturbedgcadjoint{\vc{\mathcal{J}}}{}^{\free,\bh} \nonumber\\
    - \frac{1}{\mu_0} \curl & \vc{B}_0
    .\label{eq:bh_ampere}
  \end{align}
\end{subequations}

The meaning of the gyro-average adjoint of the free current density~$\evalperturbedgcadjoint{\vc{\mathcal{J}}}{}^{\free,\bh}$ is different from the one discussed in \cref{sec:macromaxwell_formulation}, and it is defined in a weak sense as (cf.\ \cref{eq:free_current_gavadjoint})
\begin{equation}
  \int \evalperturbedgcadjoint{\vc{\mathcal{J}}}{}^{\free,\bh} \bcdot \ca \diff^3 \gy{r} 
  \defeq 
  \sum_s \int \gy{f}_s \squarepar{q_s \dot{\gy{\vc{R}}} \bcdot \gav{\evalperturbedgc{\ca}} - \gy{\mu} \dgav{(\evalperturbedgcburby{\nabla} \times \ca)_\shortparallel}} \jacobiangy^\bh \diff^6 \gy{z}.
\end{equation}
This adjoint of the free current density coincides with the one from the proposed gyrokinetic model, as given by \cref{eq:free_current_gavadjoint}, whenever $\varepsilon_B = 0$ thanks to \cref{eq:gradient_theorem_vector_approx}.

Contrary to the proposed gyrokinetic Maxwell model discussed in strong formulation in \cref{sec:macromaxwell_formulation}, we now fail to recognise the structure of the macroscopic Maxwell's equations in \cref{eq:bh_strong}.
This is primarily due to the absence of the polarisation current density, but also due to the fact that the magnetisation current density $\vc{\mathcal{J}}^{\mathrm{m},\bh}$ cannot be written as the curl of a magnetisation.
Even when neglecting $\bigO(\varepsilon_B)$ contributions this is not possible, in which case we find
\begin{equation}
  \vc{\mathcal{J}}{^\mathrm{m,\bh}} = (\curl \vc{\mathcal{M}}_1^\reduced)_\shortparallel + \bigO(\varepsilon_B).
\end{equation}


\subsubsection{Well-posedness of the field equations}\label{sec:bh_wellposed}
For the symplectic Brizard--Hahm model~\citep{brizard2007} we note that the free current density is defined differently from the free current density of the proposed model~\eqref{eq:free_current_gavadjoint}.
This implies, in particular, that the gyro-average of the free-charge continuity equation is no longer satisfied
\begin{equation}\label{eq:gy_continuity_gavadjoint_bh}
  \int \roundpar{\pard{\evalperturbedgcadjoint{\mathcal{R}}{}^\free}{t} + \nabla \bcdot \evalperturbedgcadjoint{\vc{\mathcal{J}}}{}^{\free,\bh}} \cp \diff^3 \gy{r} 
  = \int \vc{\mathcal{J}}^\free \bcdot \roundpar{\nabla \gav{\evalperturbedgc{\cp}} - \gav{\evalperturbedgc{\nabla} \cp}} \diff^3 \gy{r}
  .
\end{equation}
As the right-hand side does not vanish in general, we find that a result analogous to \cref{eq:gy_continuity_gavadjoint} is not obtained.

Upon inspecting \cref{eq:ampere_bh}, we find that the bound current density is given by
\begin{equation}
  \vc{\mathcal{J}}^{\bound,\bh} 
  = \vc{\mathcal{J}}^{\mathrm{m},\bh}
  = \hb \nabla \bcdot (\vc{\mathcal{M}}_1^\reduced \times \hb),
\end{equation}
which is not divergence free.
Computation of the divergence of Amp\`ere's law~\eqref{eq:bh_ampere} results in the unsatisfied constraint
\begin{equation}\label{eq:total_continuity_bh}
  0 
  = \pard{}{t} (\evalperturbedgcadjoint{\mathcal{R}}{}^\free + \mathcal{R}^{\bound,\bh})
  + \nabla \bcdot (\evalperturbedgcadjoint{\vc{\mathcal{J}}}{}^{\free,\bh} + \vc{\mathcal{J}}^{\bound,\bh} 
  )
  ,
\end{equation}
upon addition of the quasi-neutrality equation~\eqref{eq:bh_gauss} where we let $\mathcal{R}^{\bound,\bh} \defeq - \nabla \bcdot \vc{\mathcal{P}}_1^\reduced$.
In this case, the gyro-average adjoint of the free-charge continuity equation does not vanish, and we are therefore left with a total continuity equation which must be enforced by means of a Lagrange multiplier.
Due to the non-zero right-hand side of \cref{eq:gy_continuity_gavadjoint_bh} we find that the constraint \cref{eq:total_continuity_bh} also depends on the characteristics~$\gy{\vc{Z}}$, which is very undesirable as it implies that the Lagrange multiplier $\lambda$ also affects the EOMs.

Hence, for the symplectic Brizard--Hahm model~\citep{brizard2007} we find that $\lambda \neq 0$ for three independent reasons: the polarisation current density is missing, the magnetisation current density is not the curl of a magnetisation, and the gyro-average adjoint of the free-charge continuity equation does not hold.
Note that a polarisation current density can be included by considering higher-order approximations of the first-order generating function, see e.g.\ \citep{qin1999gyrokinetic}.


\subsection{Linearized models in a slab}\label{sec:slab_susceptibility}
In order to study the dispersive properties of the models under consideration, we consider the case of slab geometry, wherein the background magnetic field $\vc{B}_0$ is assumed to be constant.
In this case the FLR corrected electromagnetic fields exactly coincide with what one would expect (cf.\ \cref{eq:gy_flr_electomagfields_appro})
\begin{equation}
  \vc{B}_1^\effective = \gav{\evalperturbedgc{\vc{B}}_1}
  , \quad
  \vc{E}_1^\effective = \gav{\evalperturbedgc{\vc{E}}_1}
  ,
\end{equation}
as $\varepsilon_B = 0$.
Furthermore, the term that multiplies $\dot{\gy{\vc{R}}}$ in the Amp\`ere--Maxwell law now exactly coincides with $\gav{\evalperturbedgc{\ca}}$ (cf.\ \cref{eq:gradient_theorem_vector_approx}).
We reiterate that each of these three identities holds only because of the specific choice of our gyrocentre coordinate transformation $(\gparamr, \gparamtheta) = (1, 0)$.

\subsubsection{Proposed gyrokinetic Maxwell model}
We make use of the so-called susceptibility tensor in order to study the dispersive properties of the proposed model.
The susceptibility tensor represents the linearised model with a Fourier ansatz.
More precisely, we find that computing the time derivative of the Amp\`ere--Maxwell law~\eqref{eq:maxwell_ampere} results in
\begin{equation}\label{eq:ampere_E}
  -\frac{1}{\mu_0} \curl (\curl \vc{E}_1)
  = 
  \pardd{}{t} \roundpar{\epsilon_0 \vc{E}_1 + \vc{\mathcal{P}}_1} + \curl \pard{\vc{\mathcal{M}}_1}{t} + \pard{\evalperturbedgcadjointalt{\vc{\mathcal{J}}}{}^\free}{t} 
  ,
\end{equation}
where we have substituted \cref{eq:maxwell_displacement,eq:maxwell_magfield,eq:maxwell_faraday}.
By substituting the expressions for the polarisation~\eqref{eq:polarisation}, magnetisation~\eqref{eq:magnetisation}, and the gyrocentre current density, and by repeatedly using Faraday's law~\eqref{eq:Faraday} we can obtain an equation which is expressed entirely in terms of the perturbed electric field $\vc{E}_1$.
We then linearize this equation and substitute the Fourier ansatz
\begin{equation}\label{eq:reduced_fourierE}
  \vc{E}_1 = \fourier{\vc{E}}_1 \euler^{\ii (\vc{k} \bcdot \gy{\vc{r}} - \omega t)},
\end{equation}
which results in an equation of the form \cite[Eq.~(2.37)]{hasegawa1975physics}
\begin{equation}\label{eq:wave_dispersionE}
  \omega^2 (\mat{I}_3 + \gyFIXED{\mat{X}}) \fourier{\vc{E}}_1 + c^2 \vc{k} \times (\vc{k} \times \fourier{\vc{E}}_1) = \vc{0}_3,
\end{equation}
where $\gyFIXED{\mat{X}}$ is referred to as the gyrokinetic susceptibility tensor, $\mat{I}_3$ denotes the $3 \times 3$ identity matrix and $c = \sqrt{1 / (\mu_0 \epsilon_0)}$ denotes the speed of light in vacuum.
Note that $\mat{I}_3 + \gyFIXED{\mat{X}}$ is often referred to as the dielectric tensor.
The susceptibility tensor contains the (linearised) contributions to the Amp\`ere--Maxwell law from the polarisation, magnetisation as well as the gyrocentre current density and reduces the linearised gyrokinetic model to a material property: the permittivity of the `material' is given by $\epsilon_0 (\mat{I}_3 + \gyFIXED{\mat{X}})$.

We follow the discussion from \citet{zonta2021dispersion}, wherein the gyrokinetic susceptibility tensor $\gyFIXED{\mat{X}}^\zlr$ is derived for the drift kinetic model from \citet{burby2019gauge} (i.e.\ the proposed model with $(\gparamr, \gparamtheta) = (0, 0)$ in the ZLR limit $\varepsilon_\perp \rightarrow 0$) and subsequently compared to the ZLR limit \cite[Eq.~(2.159)]{hasegawa1975physics} of the Vlasov--Maxwell susceptibility tensor \cite[Eqs.~(2.42) and (2.43)]{hasegawa1975physics}.

The derivation of the susceptibility tensor can be found in \cref{app:susceptibility}, and the expression for the susceptibility tensor in its most general form can be found in \cref{eq:gy_susceptibility_sub}.
We make two simplifications: first, we consider the ZLR limit of the susceptibility tensor, resulting in the drift kinetic susceptibility tensor given by \cref{eq:gy_susceptibility_zlr}, which coincides exactly with the low-frequency and ZLR limit of the Vlasov--Maxwell susceptibility tensor found in \citet[Eq.~(2.159)]{hasegawa1975physics}.
And second, we consider the use of a centred Maxwellian background particle distribution as defined in \cref{eq:maxwellian}.
Note that a constant background magnetic field $\vc{B}_0$ combined with a centred Maxwellian background distribution with constant density trivially satisfies the equilibrium conditions discussed in \cref{sec:field_equilibria}.
When using the identities given by \cref{eq:centeredMax_integrate0} we find that the resulting susceptibility tensor is given by
\begin{equation}\label{eq:gy_susceptibility_zlr_cm}
  \begin{aligned}
    \frac{\omega^2}{c^2} \gyFIXED{\mat{X}}= {} &
      \sum_s \frac{\beta_{0,s}}{\uths^2} 
      \begin{pmatrix}
        \omega^2 & \ii \omega \omega_{\mathrm{c},s} & 0\\
        -\ii \omega \omega_{\mathrm{c},s} & \omega^2 - \kperp^2 \uths^2 & 0\\
        0 & 0 & 0\\
      \end{pmatrix}\\
    & - \sum_s \frac{\beta_{0,s}}{\gy{n}_{0,s} \uths^2} \int \frac{
      \kpa^2 \gyfzerocm
    }
    {(\kpa \gy{u}_\shortparallel - \omega)^2}
    \begin{pmatrix}
      0 & 0 & 0\\
      0 & \gy{u}_\tau^4 \kperp^2 \frac{1}{4} & \ii \frac{\kperp}{\kpa} \gy{u}_\tau^2 \frac{\omega \omega_{\mathrm{c},s}}{2}\\
      0 & -\ii \gy{u}_\tau^2 \frac{\kperp}{\kpa} \frac{\omega \omega_{\mathrm{c},s}}{2} & \frac{\omega^2 \omega_{\mathrm{c},s}^2}{\kpa^2} 
    \end{pmatrix} \jacobiangcslab \diff^3 \gy{u},
  \end{aligned}
\end{equation}
where we recall that the plasma-$\beta$ is defined in \cref{eq:beta}.

The remaining integrals can be expressed in terms of the plasma dispersion function~$\mathcal{Z}$
\begin{equation}\label{eq:plasma_disp_fun}
  \mathcal{Z}(\zeta) \defeq \frac{1}{\sqrt{\upi}} \int_{-\infty}^{\infty} \frac{\euler^{-u^2}}{u - \zeta} \diff u.
\end{equation}
Substitution of \cref{eq:gy_susceptibility_zlr_cm} in \cref{eq:wave_dispersionE} results in 
\begin{equation}\label{eq:wave_dispersionE_mat}
  \gyFIXED{\mat{D}}\fourier{\vc{E}}_1 = \vc{0}_3
  , \quad
  \gyFIXED{\mat{D}}
  =
    \frac{\omega^2}{c^2} \mat{I}_3
  + \frac{\omega^2}{c^2} \gyFIXED{\mat{X}}
  + \mat{K}^2,
\end{equation}
where the matrix $\mat{K}$ is such that
\begin{equation}\label{eq:Kmatrix}
  \mat{K} \vc{S} = \vc{S} \times \vc{k},
\end{equation}
and the matrix $\gyFIXED{\mat{D}}$ can explicitly be written as
\begin{equation}\label{eq:gy_dispersionmat_zlr_cm}
  \begin{aligned}    
    \gyFIXED{\mat{D}}
    = {} &
    \begin{pmatrix}
      \frac{\omega^2}{c^2} - \kpa^2 & 0 & \kpa \kperp\\
      0 & \frac{\omega^2}{c^2} - k^2 & 0\\
      \kpa \kperp & 0 & \frac{\omega^2}{c^2} - \kperp^2
    \end{pmatrix}
    + \sum_s \frac{\beta_{0,s}}{\uths^2} 
    \begin{pmatrix}
      \omega^2 & \ii \omega \omega_{\mathrm{c},s} & 0\\
      -\ii \omega \omega_{\mathrm{c},s} & \boxed{\omega^2} - \kperp^2 \uths^2 & 0\\
      0 & 0 & 0\\
    \end{pmatrix}\\
    & - \sum_s \frac{\beta_{0,s}}{\uths^2} \mathcal{Z}'({\omega} / (\kpa \uths))
    \begin{pmatrix}
      0 & 0 & 0\\
      0 & \frac{\kperp^2 \uths^2}{2} & \ii \frac{\kperp \omega \omega_{\mathrm{c},s}}{2\kpa}\\
      0 & -\ii \frac{\kperp \omega \omega_{\mathrm{c},s}}{2\kpa} & \frac{\omega^2 \omega_{\mathrm{c},s}^2}{\kpa^2 \uths^2}
    \end{pmatrix}.
  \end{aligned}
\end{equation}
We highlight the boxed ${\omega^2}$ term for the discussion that follows in \cref{sec:susceptibility_darwin}.

Non-trivial solutions to \cref{eq:wave_dispersionE_mat} exist if and only if the dispersion matrix given by \cref{eq:gy_dispersionmat_zlr_cm} is singular, and to this end we implicitly define the dispersion relation $\omega(\vc{k})$ \textit{via}
\begin{equation}
  \det \gyFIXED{\mat{D}}(\omega(\vc{k}), \vc{k}) = 0.
\end{equation}

\subsubsection{Other models}\label{sec:susceptibility_darwin}
It is \textit{a priori} not clear what effect the Darwin approximation has on the dispersive properties of the proposed model, and to this end we also consider the gyrokinetic Darwin susceptibility tensor
for which a derivation can be found in \cref{sec:darwin_susceptibility}.
When using a centred Maxwellian, as we did in \cref{sec:slab_susceptibility}, we find that the susceptibility tensor is given by \cref{eq:gy_dispersionmat_zlr_cm} where the boxed {$\omega^2$} term is removed while the limit of quasi-neutrality yields $c \rightarrow \infty$.

As discussed in \cref{sec:brizardhahm}, the EOMs of the symplectic Brizard--Hahm (BH) model~\citep{brizard2007} coincide with the ones from the proposed gyrokinetic Maxwell model if $\varepsilon_B = 0$, which is what we assume here.
Similarly, we find that the magnetisation term vanishes, as it does in the proposed gyrokinetic model, for the centred Maxwellian background particle distribution function that we consider here.
Hence, the only difference between the proposed model and the Brizard--Hahm model, under the current simplifying assumptions, is the missing polarisation current density.

However, when reconsidering \cref{eq:total_continuity_bh} under the current simplifying assumptions, and while including the same gauge condition used in the quasi-neutral gyrokinetic Darwin model, we find that the compatibility constraint reduces to
\begin{equation}
  \nabla \bcdot \roundpar{\coeffcone \nabla_\perp \lambda}
  = \pard{}{t} \nabla \bcdot \roundpar{\coeffcone \nabla_\perp \phi_1}
  ,
\end{equation}
upon substitution of \cref{eq:bh_gauss}, and it therefore follows that in this case $\lambda = \partial \phi_1 / \partial t$.
This implies that the Lagrange multiplier for the Brizard--Hahm model restores the polarisation current density and thereby coincides with the quasi-neutral gyrokinetic Darwin model.
Note that this only holds under the simplifying assumptions that we use here: the model is linearised, we assume $\varepsilon_B = 0$, use a centred Maxwellian background distribution, and we consider the ZLR limit.


For the parallel-only model we make use of the dispersion relation provided in \cite[Eq.~(40)]{Kleiber_pullback}.

\subsubsection{Comparison of the dispersion relations}
The models are compared by numerically evaluating the dispersion relations (the code can be found in~\cite[{\texttt{examples/gauge\_invariant.ipynb}}]{dispersoncurves}).
We consider electron and ion (deuteron) species, with $T_\ii = T_\euler$, $q_\ii = - q_\euler$ and an electron to ion mass ratio given by $m_\euler / m_\ii = 1 / 3671$.
Furthermore, we non-dimensionalise the frequency and wave vector as
\begin{equation}
  \nondim{\omega} \defeq \frac{\omega}{\uthi \kpa}
  , \quad
  \nondim{\vc{k}} \defeq \varrho_\ii \vc{k}
  .
\end{equation}
In this case, we expect to find a shear and compressional Alfv\'en wave, which have the following frequencies (in the quasi-neutral limit $c \rightarrow \infty$)
\begin{equation}\label{eq:aflven_frequencies}
  \nondim{\omega}_\mathrm{As} \defeq \frac{1}{\sqrt{{\beta_{0}}}}
  , \quad
  \nondim{\omega}_\mathrm{Ac} \defeq \frac{k}{\kpa} \nondim{\omega}_\mathrm{As}
  ,
\end{equation}
respectively.
We note that the shear Alfv\'en frequency is constant w.r.t.\ $\nondim{k}_\perp$, whereas the compressional Alfv\'en frequency increases linearly with increasing wavenumber.
Hence, the compressional Alfv\'en wave is a fast wave.
This is especially true when small perpendicular length scales are considered (turbulence), as in such a case the compressional Alfv\'en frequency becomes comparable to the cyclotron frequency 
\begin{equation}
  \varepsilon_\perp \sim 1, \quad \omega = \omega_{\mathrm{Ac}} 
  \quad \implies \quad 
  \varepsilon_\omega = \frac{\omega_{\mathrm{Ac}}}{\omega_{\mathrm{c},\ii}} = \frac{\nondim{k}}{\sqrt{\beta_{0}}} \sim 1.
\end{equation}

The presence of the compressional Alfv\'en wave in the proposed gyrokinetic model is therefore incompatible with the low-frequency approximation of the first-order generating function~\eqref{eq:gy_S1_approx}, which relies on $\varepsilon_\omega \ll 1$.
This incompatibility seems to suggest that we have made a mistake in the derivation of the proposed gyrokinetic model.
The origin of this issue lies in the low-frequency approximation of the first-order generating function as discussed in \cref{sec:gy_H1_gauge_invariant}, where we have intentionally kept the gauge invariant parts of the right-hand side of \cref{eq:gy_S1_def_approx}.
While this ensures gauge invariance, it neglects the fact that the part of the electric field that comes from the vector potential has a time derivative, and this term would therefore be neglected if we had neglected all $\bigO(\varepsilon_\omega)$ terms.

\begin{figure}
  \centering
  \begin{tikzpicture}[]
\begin{axis}[
  hide axis,
  width=2cm,
  height=2cm,
  legend style={legend columns=-1,
                column sep=1ex, line width=0.5}
]

  \addplot [line width=\tikzlinewidth, color=\parallelcolor, domain=-0.1:0.1,samples=2, draw=none, line width=2pt] {0};
  \addlegendentry{Parallel-only}


  \addplot [line width=\tikzlinewidth, color=\darwincolor, domain=-0.1:0.1,samples=2, draw=none, line width=1.5pt] {0};
  \addlegendentry{QN gyrokinetic Darwin / BH}

  \addplot [line width=\tikzlinewidth, color=\gaugeinvariantcolor, domain=-0.1:0.1,samples=2, draw=none, line width=1pt] {0};
  \addlegendentry{Gyrokinetic Maxwell}

  
\end{axis}
\end{tikzpicture}
  \vspace*{\floatsep}

  \subcaptionbox{The real part.\label{fig:alfven:eval:vary_perp_re}}
  [\twofigwidth]{
    \input{tikz/dispersion_varyperp_re.tex}
  }
  \hfill
  \subcaptionbox{The negative imaginary part.}
  [\twofigwidth]{
    \input{tikz/dispersion_varyperp_im.tex}
  }
  \caption{The dispersion relations for a fixed value of $\nondim{k}_\shortparallel = 2 \times 10^{-3}$ and $\beta_{0} = 10 \%$.
  The black dotted line corresponds to $\nondim\omega = \nondim\omega_\mathrm{As}$, whereas the black dashed line corresponds to $\nondim\omega = \nondim\omega_\mathrm{Ac}$.}\label{fig:alfven:eval:vary_perp}
\end{figure}
In \cref{fig:alfven:eval:vary_perp} we show the dispersion relations for a fixed value of $\nondim{k}_\shortparallel = 2 \times 10^{-3}$ and $\beta_{0} = 10 \%$.
The gyrokinetic Maxwell model with $(\gparamr, \gparamtheta) = (1, 0)$ results in two waves, which have the correct real part of the frequency, whereas the imaginary part has the correct sign which results in damping.
The dispersion relation of the other models result only in the shear Alfv\'en wave, as expected.


\begin{table}
  \centering
  \begin{tabular}{rrrrrr}
    \toprule
    & $R_0$ & $a$ & $\varrho_\mathrm{i}$ & $\nondim{k}_\shortparallel$ & $\nondim{k}_\perp$\\
    \midrule
    AUG & 1.6 m & 0.8 m & $3.4 \times 10^{-3}$ m & $2.1\times 10^{-3}$ & $7.2\times 10^{-3}$\\
    ITER & 6.2 m & 2 m & $2.4 \times 10^{-3}$ m & $3.9\times 10^{-4}$ & $2.0 \times 10^{-3}$\\
    W7-X & 5.5 m & 0.53 m & $4.4 \times 10^{-3}$ m & $7.9\times 10^{-4}$ & $1.4 \times 10^{-2}$\\
    \bottomrule
  \end{tabular}
  \caption{The length scales used for determining the wave vector $\nondim{\vc{k}}$, as obtained from~\citep{Zoni2021}.
  The non-dimensional wavenumbers $\nondim{k}_\shortparallel$ and $\nondim{k}_\perp$ are computed according to \cref{eqn:alfven:eval:wavenumber_comp}.}\label{tab:alfven:eval:tokamak_length_scales}
\end{table}
In order to further explore the parameter space, we use the geometrical parameters of the tokamak fusion devices ASDEX Upgrade (AUG) and ITER as well as the stellarator Wendelstein 7-X (W7-X) to determine the value of the wave vector $\nondim{\vc k}$.
For both the parallel and perpendicular direction we consider the lowest non-trivial wavenumber
\begin{equation}\label{eqn:alfven:eval:wavenumber_comp}
  \nondim{k}_\shortparallel = \frac{\varrho_\ii}{R_0}, \quad \nondim{k}_\perp = \frac{2\upi \varrho_\ii}{a\sqrt{4 + \upi^2}} ,
\end{equation}
where $R_0$ and $a$ denote the major and minor radii of the tokamak, respectively.
The values of $R_0, a, \varrho_\ii$ are shown in \cref{tab:alfven:eval:tokamak_length_scales} and are taken from~\citet{Zoni2021} for the tokamak fusion devices and from \citet{grieger1992modular,klinger2019overview} for the stellarator W7-X.
We moreover show the resulting wavenumbers according to \cref{eqn:alfven:eval:wavenumber_comp}.

\begin{figure}
  \centering
  \begin{tikzpicture}
  \begin{axis}[
      hide axis,
      width=2.cm,
      height=2cm,
      ymin=0.99,
      ymax=1,
  ]
      \addplot [color=\parallelcolor, domain=-0.1:0.1,samples=2, draw=none, line width=2pt] {1};
          \label{plot:dispersion_legend:line1}
      \addplot [color=\gaugeinvariantcolor, domain=-0.1:0.1,samples=2, draw=none, line width=1pt] {1};
          \label{plot:dispersion_legend:line3}
      \addplot [color=\darwincolor, domain=-0.1:0.1,samples=2, draw=none, line width=1.5pt] {1};
          \label{plot:dispersion_legend:line4}
      \addplot [dashed, color=\parallelcolor, domain=-0.1:0.1,samples=2, draw=none, line width=2pt] {1};
          \label{plot:dispersion_legend:line5}
      \addplot [dashed, color=\gaugeinvariantcolor, domain=-0.1:0.1,samples=2, draw=none, line width=1pt] {1};
          \label{plot:dispersion_legend:line7}
      \addplot [dashed, color=\darwincolor, domain=-0.1:0.1,samples=2, draw=none, line width=1.5pt] {1};
          \label{plot:dispersion_legend:line8}
      \addplot [dotted, color=\parallelcolor, domain=-0.1:0.1,samples=2, draw=none, line width=2pt] {1};
          \label{plot:dispersion_legend:line9}
      \addplot [dotted, color=\gaugeinvariantcolor, domain=-0.1:0.1,samples=2, draw=none, line width=1pt] {1};
          \label{plot:dispersion_legend:line11}
      \addplot [dotted, color=\darwincolor, domain=-0.1:0.1,samples=2, draw=none, line width=1.5pt] {1};
          \label{plot:dispersion_legend:line12}

      %
      \coordinate (legend) at (axis description cs:0.5,0.5);
  \end{axis}

  %
  \draw (0,0) node{
      \begin{tabular}{rccc}
          \toprule
             & Parallel-only & QN gyrokinetic Darwin / BH & Gyrokinetic Maxwell\\
          \midrule
          AUG & \ref{plot:dispersion_legend:line1}  & \ref{plot:dispersion_legend:line4} & \ref{plot:dispersion_legend:line3}\\
          ITER & \ref{plot:dispersion_legend:line5} & \ref{plot:dispersion_legend:line8} & \ref{plot:dispersion_legend:line7}\\
          W7-X & \ref{plot:dispersion_legend:line9}  & \ref{plot:dispersion_legend:line12} & \ref{plot:dispersion_legend:line11}\\
          \bottomrule
      \end{tabular}
  };

\end{tikzpicture}
  \vspace*{\floatsep}
  
  \subcaptionbox{The real part.}
  [\twofigwidth]{
    \input{tikz/dispersion_varybeta_re.tex}
  }
  \hfill
  \subcaptionbox{The negative imaginary part.}
  [\twofigwidth]{
    \input{tikz/dispersion_varybeta_im.tex}
  }
  \caption{The dispersion relations for fixed values of $\nondim{k}_\perp$ and $\nondim{k}_\shortparallel$, as determined from \cref{tab:alfven:eval:tokamak_length_scales}.
  Only the shear Alfv\'en wave is shown.}\label{fig:alfven:eval:vary_beta}
\end{figure}
For each of the three models and each of the three machines, we compute the dispersion curve where we vary $\beta_{0}$ and keep the wave vectors fixed.
The results are shown in \cref{fig:alfven:eval:vary_beta}.
We find that the real part of the dispersion curves all overlap and agree with \cref{eq:aflven_frequencies}, which suggests that the shear Alfv\'en frequency depends only on $\beta_{0}$.
When considering the imaginary part of the dispersion curve, we find differences not only between the machines, but also between the models, in particular for larger values of $\beta_{0}$.
The gyrokinetic Maxwell and quasi-neutral gyrokinetic Darwin model agree well, which shows that the removal of the compressional Alfv\'en wave has not altered the shear Alfv\'en wave.
Moreover, we find that the parallel-only model yields a different imaginary part of the shear Alfv\'en frequency.

\subsection{Summary of comparison}
\begin{table}
  \centering
  \begin{tabular}{lccccc}
    \toprule
    & $\gyFIXED{\vc{\gamma}}_{1,\vc{R}}$ & \shortstack{Reduces to \\ parallel-only} & \shortstack{Gauge \\ invariant} & \shortstack{Fast \\ waves} & \shortstack{Compatible \\ equations}\\
    \midrule
    Parallel-only & $q \gav{\evalperturbedgc{A}_{1,\shortparallel}} \hb$ & N/A & N & N & Y\\
    Brizard--Hahm & $q \gav{\evalperturbedgc{\vc{A}}_1}$ & Y & N & \phantom{$^\dagger$}N$^\dagger$ & N\\
    Burby--Brizard & $q \vc{A}_1$ & N & Y & Y/N$^\ast$ & Y/N$^\ast$\\
    \textbf{QN gyrokinetic Darwin} & $q \Ad + q \radgav{\evalperturbedgcburby{\vc{B}}_1 \times \vc{\rho}}$ & Y & Y & N & Y\\
    \textbf{Gyrokinetic Maxwell} & $q \Ad + q \radgav{\evalperturbedgcburby{\vc{B}}_1 \times \vc{\rho}}$ & Y & Y & Y & Y\\
    \bottomrule
  \end{tabular}
  \caption{
    Properties of the different gyrokinetic models under consideration.
    The two models proposed in this paper are in boldface: the gyrokinetic Maxwell model with $(\gparamr, \gparamtheta) = (1, 0)$ and its corresponding quasi-neutral gyrokinetic Darwin approximation.
    $^\dagger$ This can be a `yes (Y)' if the approach from \citet{qin1999gyrokinetic} is followed. 
    $^\ast$ If the polarisation current density is kept then the compressional Alfv\'en wave is present, and the Lagrange multiplier vanishes, but if the polarisation current density is neglected then the compressional Alfv\'en wave is absent, and the Lagrange multiplier is needed to restore the bound-charge continuity equation.
    }\label{tab:compare}
\end{table}
We have compared five different gyrokinetic models, all of which are derived from an action principle.
An overview of the most essential properties is given in \cref{tab:compare}, wherein we have also included the symplectic part of the gyrocentre single-particle phase-space Lagrangian $\gyFIXED{\vc{\gamma}}_{1,\vc{R}}$ -- with all other components equal to zero -- which defines the gyrocentre coordinate transformation.
We summarise the comparison of the models.

The parallel-only model~\citep{Kleiber_pullback}, as described in \cref{sec:reduced_parallel}, neglects the perpendicular part of the vector potential and thereby results in two scalar, well-posed, perpendicular Poisson-type field equations which decouple for a centred Maxwellian background particle distribution function.
The model is not gauge invariant.
Due to the absence of the perpendicular part of the vector potential, we find that the EOMs of the parallel model do not contain a grad-$B_{1,\shortparallel}$ drift.
This model is still the `working horse', and it is widely used for the global simulation of turbulence in fusion devices~\citep{garbet2010gyrokinetic,mishchenko2023global}.
It is favourable if more complex models reduce to this well-known model in the limiting case $\vc{A}_{1,\perp} = \vc{0}_3$, not only because it is reassuring that the traditional parallel-only model is recovered in this limit, but also because it implies that any differences in simulation results are only due to the presence of the perpendicular part of the vector potential when results are compared between the parallel-only model and a more complex gyrokinetic model.
In particular, differences in simulation results cannot be due to a different choice of coordinates.

The symplectic Brizard--Hahm model~\citep{brizard2007} (see \cref{sec:brizardhahm}) contains the full vector potential and results in a curl-curl type Amp\`ere's law which is coupled to the same quasi-neutrality equation as found in the parallel-only model.
This model is also not gauge invariant and reduces to the parallel-only model when neglecting the perpendicular part of the vector potential.
The grad-$B_{1,\shortparallel}$ drift is present in the EOMs; this is the case for each of the gyrokinetic models which have the full vector potential.
Interestingly, the second-order Hamiltonian of the Brizard--Hahm model coincides with that of the parallel-only model.
This implies that this model has no polarisation current density, and the magnetisation effects do not depend on the perpendicular part of the vector potential.
Moreover, the magnetisation current density is not the curl of a magnetisation, and we find that the continuity equation is not satisfied for the gyro-average adjoint of the free charge and current density.
For those reasons the total continuity equation, consisting of the free and bound charge density, is not satisfied, which is a necessary condition for well-posedness of the field equations.
It follows that well-posedness of the Brizard--Hahm model~\citep{brizard2007} requires the introduction of a Lagrange multiplier, which affects both the EOMs and the field equations in an unphysical way.

The gyrokinetic model from \citet{burby2019gauge} has played an essential role in this paper, as it has guided us to the development of a family of gauge-invariant gyrokinetic models. 
Their model coincides with the parameter choice $(\gparamr, \gparamtheta) = (0, 0)$.
In order to suppress the compressional Alfv\'en wave, they propose to neglect the polarisation current density altogether, as it is a higher-order contribution in $\varepsilon_\omega$.
This choice, however, breaks the bound-charge continuity equation and therefore requires the introduction of a Lagrange multiplier to restore it.
Moreover, letting $\gparamr = 0$ results in a model where the symplectic part is essentially drift kinetic and does not reduce to the parallel-only model if the perpendicular part of the vector potential is neglected. 

Instead, we propose to use $(\gparamr, \gparamtheta) = (1, 0)$, which results in a gauge-invariant gyrokinetic model for which all terms in both the EOMs and the field equations can be interpreted from a physical point of view.
Moreover, it yields the smallest coordinate transformation (see the discussion at the end of \cref{sec:gy_H1_gauge_invariant}) while resulting in a gauge invariant model wherein the gyrocentre magnetic moment is an invariant.
This model does reduce to the parallel-only model if the perpendicular part of the vector potential is neglected.
The gyrocentre single-particle Lagrangian and resulting Vlasov--Maxwell action principle of this model are derived and described in detail in \cref{sec:single_particle,sec:vlasov_maxwell}, respectively.
We find that the continuity equation is satisfied for the gyro-average adjoint of the free charge and current density, and, since the model is gauge invariant and derived from an action principle, it follows that the bound-charge continuity equation is satisfied as well.
These are necessary conditions for well-posedness of the field equations, which are satisfied without the need of a Lagrange multiplier.

The presence of the polarisation current density in the proposed gyrokinetic Maxwell model results in a compressional Alfv\'en wave, which is often undesired due to its relatively high frequency.
To this end, the quasi-neutral Darwin approximation can be applied to the proposed gyrokinetic Maxwell model (see \cref{sec:darwin}).
Therein, we consider the limit of quasi-neutrality and remove the part of the polarisation current density which is responsible for the compressional Alfv\'en wave (as we have demonstrated by making use of the dispersion relation), while retaining the compatibility of the field equations as well as gauge invariance of the model.


Finally, we have derived a dispersion relation for each of the models, which has confirmed the expected properties of the models: each model contains the shear Alfv\'en wave and only the proposed gyrokinetic Maxwell model includes the compressional Alfv\'en wave.
In both cases the real part of the frequency agrees well with the theory, whereas the imaginary part is negative and thereby results in damping of the wave.

  \section{Conclusions}\label{sec:conclusion}
Motivated by the need for a more complete gyrokinetic model, wherein the perpendicular part of the vector potential is kept, we have discussed the gyrocentre coordinate transformation in detail.
The purpose of the gyrocentre coordinate transformation is to transform the perturbed guiding-centre single-particle phase-space Lagrangian in such a way that it becomes independent of the gyro-phase.
This results in a reduction of the phase-space dimension from six to five (the gyrocentre position, the parallel velocity and the invariant magnetic moment), and when moreover considering the limit of quasi-neutrality, removes the fastest time scales from the large range of length and time scales present in the Vlasov--Maxwell model.

When gyrokinetic modelling is considered at the level of the Lagrangian, which thereby utilises a variational principle, it is ensured that the gyrokinetic model is structure preserving.
In particular, we find that energy is conserved regardless of the modelling error introduced by the truncated gyrocentre coordinate transformation.
However, an aspect that is often overlooked in gyrokinetics is the property of gauge invariance~\citep{brizard2007}, which ensures that the model is invariant under the gauge transformation, as it is the case for the Vlasov--Maxwell model.
In particular, the traditionally used parallel-only models are not gauge invariant. 
To this end, we have generalised the approach proposed in \citet{burby2019gauge}, wherein a gauge-invariant gyrokinetic model is introduced.
We have derived sufficient conditions on the gyrocentre coordinate transformation which ensure that the resulting gyrocentre single-particle phase-space Lagrangian is gauge invariant.
Despite this additional restriction, the gyrocentre coordinate transformation is by no means uniquely defined, and this approach therefore results in a family of gauge-invariant gyrokinetic models.
The family of models is parametrized by two parameters $\gparamr, \gparamtheta$, where the model from \citet{burby2019gauge} coincides with $(\gparamr, \gparamtheta) = (0, 0)$.
Our derivation is presented by making use of vector calculus, as opposed to the more customarily used formalism of differential geometry.

In an effort to obtain a gyrokinetic model for which each of the equations of motion as well as the field equations can be interpreted from a physical point of view, we have chosen $(\gparamr, \gparamtheta) = (1, 0)$.
This choice leads to the smallest coordinate transformation which results in a gauge invariant model wherein the gyrocentre magnetic moment is an invariant.
We find that the proposed model reduces to the parallel-only model when the perpendicular component of the vector potential is neglected, which does not hold for the gauge invariant model from \citet{burby2019gauge}.
The resulting model has been derived to second-order in the perturbation parameter $\varepsilon_\delta$, and the second-order part of the Lagrangian contains polarisation and magnetisation effects which have a clear physical meaning.
Due to gauge invariance the model can be expressed directly in terms of the electromagnetic fields, rather than the potentials.
The gyrokinetic model thereby results in the macroscopic Maxwell's equations.
Moreover, we have derived equilibrium conditions on the background distribution function and magnetic field which justify the smallness of the perturbation parameter $\varepsilon_\delta$.

We find that the proposed gyrokinetic Maxwell model possesses a magnetisation current density which is the curl of a magnetisation. 
In addition, it has a polarisation current density, and we find that the free-charge continuity equation holds for the gyro-average adjoints of the free charge and current density.
Each of those three properties is essential for showing that the bound-charge continuity equation is naturally satisfied, which is necessary for the well-posedness of the field equations.
Hence, unlike the Brizard--Hahm model~\citep{brizard2007}, we find that the field equations of the proposed gyrokinetic Maxwell model are compatible without the need of a Lagrange multiplier.
A brief summary of the comparison between each of the models under consideration is found in \cref{tab:compare}.

In addition, we have derived the gyrokinetic susceptibility tensor, which covers the material properties of the linearised gyrokinetic model in the macroscopic Maxwell's equations, for each of the models under consideration.
In the ZLR limit, we find that the gyrokinetic susceptibility tensor of the proposed gyrokinetic Maxwell model agrees with that of the Vlasov--Maxwell system.
Moreover, the resulting dispersion relation shows that, besides the usual shear Alfv\'en wave, a fast compressional Alfv\'en wave is present in the proposed gyrokinetic Maxwell model.
Due to the potentially high frequency of this wave we have proposed a quasi-neutral Darwin approximation to the proposed gyrokinetic Maxwell model which successfully removes the fast compressional Alfv\'en wave.
We find that the quasi-neutral gyrokinetic Darwin model is still well-posed and gauge invariant.

In future work, we plan to implement the two proposed models in the gyrokinetic particle-in-cell code EUTERPE~\citep{jost2001euterpe,kleiber2024euterpe} and compare them to the well-established and traditionally used parallel-only model.

    \section*{Acknowledgements}

  This work has been carried out within the framework of the EUROfusion Consortium, funded by the European Union \textit{via} the Euratom Research and Training Programme (Grant Agreement No~101052200 -- EUROfusion). 
  Views and opinions expressed are however those of the author(s) only and do not necessarily reflect those of the European Union or the European Commission. 
  Neither the European Union nor the European Commission can be held responsible for them.

  We thank 
  A.~Bottino, 
  M.~Campos~Pinto, 
  R.~Kleiber, 
  A.~K\"onies, 
  P.~Lauber, 
  O.~Maj, 
  A.~Mishchenko, 
  S.~Possanner, 
  and B.~Scott 
  for helpful discussions. 
  We also thank our anonymous reviewers for their valuable comments which have led to significant improvement of the manuscript and in particular has led to the insight that the quasi-neutral gyrokinetic Darwin model is gauge invariant.

  \appendix
  \setcounter{theorem}{0}
  \renewcommand\gy[1]{\gyFIXED{#1}}
  \section{Inversion of the Lagrange matrix}\label{app:matrix_inverse}
We consider the inversion of the Lagrange matrix
\begin{equation}
  \mat{W} = \begin{pmatrix}
    \mat{W}_{11} & \mat{W}_{12}\\
    -\mat{W}_{12}^\transpose & \mat{W}_{22}
  \end{pmatrix},
\end{equation}
where
\begin{equation}
  \mat{W}_{11} = \begin{pmatrix}
    q \mat{B} & -m \hb\\
    m \hb^\transpose & 0
  \end{pmatrix},\quad
  \mat{W}_{12} = \begin{pmatrix}
    \frac{m}{q} \vc{w} & \vc{0}_3\\
    0 & 0\\
  \end{pmatrix}, \quad
  \mat{W}_{22} = \begin{pmatrix}
    0 & \frac{m}{q} \\
    -\frac{m}{q} & 0\\
  \end{pmatrix}.
\end{equation}
Here $\mat{B}$ is such that
\begin{equation}
  \mat{B} \vc{S} = \vc{S} \times \vc{B}
\end{equation}
for some magnetic field $\vc{B}$. 

The inverse of $\mat{W}_{22}$ is readily given by
\begin{equation}
  \mat{W}_{22}^{-1} = \begin{pmatrix}
    0 & - \frac{q}{m}\\
    \frac{q}{m} & 0\\
  \end{pmatrix}.
\end{equation}
Using the Schur complement
\begin{equation}
  \mat{S} 
  = \mat{W}_{11} + \mat{W}_{12} \mat{W}_{22}^{-1} \mat{W}_{12}^\transpose,
\end{equation}
we find that the inverse of the $2\times 2$ block matrix $\mat{W}$ is given by
\begin{equation}
  \mat{W}^{-1} = \begin{pmatrix}
    \mat{S}^{-1} & - \mat{S}^{-1} \mat{W}_{12} \mat{W}_{22}^{-1}\\
    \mat{W}_{22}^{-1} \mat{W}_{12}^\transpose \mat{S}^{-1} & \mat{W}_{22}^{-1} - \mat{W}_{22}^{-1} \mat{W}_{12}^\transpose \mat{S}^{-1} \mat{W}_{12} \mat{W}_{22}^{-1}
  \end{pmatrix}.
\end{equation}
The expressions for the block matrices result in
\begin{equation}\label{eq:matinverse_S_W11}
  \mat{S} = \mat{W}_{11}
\end{equation}
for which the inverse is given by
\begin{equation}
  \mat{S}^{-1} = \begin{pmatrix}
    -\frac{\mat{B}_0}{q B_0 B_{\shortparallel}} & \frac{\vc{B}}{m B_{\shortparallel}}\\
  -\frac{\vc{B}^\transpose}{m B_{\shortparallel}} & 0
  \end{pmatrix}
\end{equation}
as can be verified by computing the product
\begin{equation}
  \mat{S}\mat{S}^{-1} = \begin{pmatrix}
    q \mat{B} & -m \hb\\
    m \hb^\transpose & 0
  \end{pmatrix} \begin{pmatrix}
    -\frac{\mat{B}_0}{q B_0 B_{\shortparallel}} & \frac{\vc{B}}{m B_{\shortparallel}}\\
  -\frac{\vc{B}^\transpose}{m B_{\shortparallel}} & 0
  \end{pmatrix}
  = \begin{pmatrix}
    \frac{-\mat{B} \mat{B}_0 + \vc{B}_0 \vc{B}^\transpose}{B_0 B_{\shortparallel}} & \frac{q \mat{B}\vc{B}}{m B_{\shortparallel}} \\
    -\frac{m \hb^\transpose \mat{B}_0}{q B_0 B_{\shortparallel}} & 1
  \end{pmatrix}
\end{equation}
and observing that using \cref{eq:matvec_is_timescurl}
\begin{equation}
  \mat{B} \vc{B} = \vc{B} \times \vc{B} = \vc{0}, \quad
  \mat{B}_0^\transpose \hb = \vc{B}_{0} \times \hb = \vc{0}
\end{equation}
as well as
\begin{align}
  \roundpar{-\mat{B} \mat{B}_0 + \vc{B}_0 \vc{B}^\transpose} \vc{S} 
  &= -\mat{B} (\vc{S} \times \vc{B}_0) + \vc{B}_0 (\vc{B} \bcdot \vc{S})\nonumber\\
  &= -(\vc{S} \times \vc{B}_0) \times \vc{B}  + \vc{B}_0 (\vc{B} \bcdot \vc{S})\nonumber\\
  &= B_{\shortparallel} B_0 \vc{S}
\end{align}
for any vector $\vc{S}$, which implies that $-\mat{B} \mat{B}_0 + \vc{B}_0 \vc{B}^\transpose = B_{\shortparallel} B_0 \mat{I}_3$.

Using these intermediate results we are ready to evaluate the blocks of the inverse of $\mat{W}$
\begin{equation}
  (\mat{W}^{-1})_{12} 
  = - \mat{S}^{-1} \mat{W}_{12} \mat{W}_{22}^{-1} 
  = \begin{pmatrix}
    \vc{0}_3 & - \frac{\vc{w} \times \hb}{q B_{\shortparallel}} \\
    0 & -\frac{\vc{B} \bcdot \vc{w}}{m B_{\shortparallel}}\\
  \end{pmatrix}
\end{equation}
and
\begin{equation}
  (\mat{W}^{-1})_{22}  
  = \mat{W}_{22}^{-1} - \mat{W}_{22}^{-1} \mat{W}_{12}^\transpose \mat{S}^{-1} \mat{W}_{12} \mat{W}_{22}^{-1}
  = \begin{pmatrix}
    0 & - \frac{q}{m} \\
    \frac{q}{m} & 0 \\
  \end{pmatrix}
\end{equation}
such that the inverse of the Lagrange matrix is given by
\begin{equation}
  \mat{W}^{-1} = \begin{pmatrix}
    -\frac{\mat{B}_0}{q B_0 B_{\shortparallel}} & \frac{\vc{B}}{m B_{\shortparallel}} & \vc{0}_3 & - \frac{\vc{w} \times \hb}{q B_{\shortparallel}}\\
    -\frac{\vc{B}^\transpose}{m B_{\shortparallel}} & 0 & 0 & -\frac{\vc{B} \bcdot \vc{w}}{m B_{\shortparallel}}\\
    \vc{0}_3^\transpose & 0 & 0 & -\frac{q}{m}\\
    \frac{(\vc{w} \times \hb)^\transpose }{q B_{\shortparallel}} & \frac{\vc{B} \bcdot \vc{w}}{m B_{\shortparallel}} & \frac{q}{m}& 0\\
  \end{pmatrix}.
\end{equation}

  \section{Gyrocentre coordinate transformation}\label{app:transform}
The gyrocentre Lagrangian $\gy{L}$ is defined according to \cref{eq:lie_lagrangian_def}.
The transformation rules for the Hamiltonian and symplectic part follow directly by substituting \cref{eq:lie_coord_transform} into \cref{eq:lie_lagrangian_def} and by subsequently making use of a Taylor series expansion centred around the gyrocentre coordinate $\gy{\vc{Z}}$.
The contribution due to the generating function $\gy{S}$ is omitted here.

For the guiding-centre Hamiltonian this results in
\begin{align}
  \gc{H}(\gc{\vc{Z}})
  = \gc{H}
  + \pard{\gc{H}}{\gy{\vc{Z}}} \bcdot \roundpar{- \gy{\vc{G}}_1 + \frac{1}{2} \pard{\gy{\vc{G}}_1}{\gy{\vc{Z}}} \gy{\vc{G}}_1 - \gy{\vc{G}}_2}
  + \frac{1}{2} \pardd{\gc{H}}{\gy{\vc{Z}}} \bcddot \roundpar{\gy{\vc{G}}_1 \otimes \gy{\vc{G}}_1}
  + \bigO(\varepsilon_\delta^3)\label{eqn:gc_gy_hamtransform_taylor}
  ,
\end{align}
upon substitution of \cref{eq:lie_coord_transform} and using a Taylor-series expansion of $\gc{H}$ centred around the gyrocentre coordinate $\gy{\vc{Z}}$. 
We use the notational convention that an absence of arguments implies evaluation at the gyrocentre coordinate, and we let $\otimes$ denote the outer product of two vectors
\begin{subequations}\label{eq:otimes}
  \begin{equation}
    (\vc{a} \otimes \vc{b})_{ij} = a_i b_j
  \end{equation}
  such that
  \begin{align}
    \mat{M} \bcddot (\vc{a} \otimes \vc{b}) = \sum_{i,j = 1}^{3} \matcmp{M}_{ij} a_i b_j.
  \end{align}
\end{subequations}
This expression can be simplified to
\begin{equation}\label{eq:gc_gy_hamtransform}
  \gc{H}(\gc{\vc{Z}})
  = \gc{H}
  - \pard{\gc{H}}{\gy{\vc{Z}}} \bcdot \roundpar{\gy{\vc{G}}_1 + \gy{\vc{G}}_2}
  + \frac{1}{2} \squarepar{\pard{}{\gy{\vc{Z}}}\roundpar{\pard{\gc{H}}{\gy{\vc{Z}}} \bcdot \gy{\vc{G}}_1}} \bcdot \gy{\vc{G}}_1
  + \bigO(\varepsilon_\delta^3)
  .
\end{equation}

The same approach is followed for the symplectic part of the guiding-centre Lagrangian.
Let $\gc{\Gamma}(\gc{\vc{Z}}) = \gc{\vc{\gamma}}(\gc{\vc{Z}}) \bcdot \dot{\gc{\vc{Z}}}$. 
Substitution of \cref{eq:lie_coord_transform} results in
\begin{align}
  \gc{\Gamma}(\gc{\vc{Z}})
  = {} & \squarepar{
      \gc{\vc{\gamma}} 
    - \pard{\gc{\vc{\gamma}}}{\gy{\vc{Z}}} \roundpar{\gy{\vc{G}} - \frac{1}{2} \pard{\gy{\vc{G}}_1}{\gy{\vc{Z}}} \gy{\vc{G}}_1}
    + \frac{1}{2} \pardd{\gc{\vc{\gamma}}}{\gy{\vc{Z}}} \bcddot \roundpar{\gy{\vc{G}}_1 \otimes \gy{\vc{G}}_1}
  } \nonumber\\ & \bcdot \squarepar{\dot{\gy{\vc{Z}}} - \totd{}{t} \roundpar{\gy{\vc{G}} - \frac{1}{2} \pard{\gy{\vc{G}}_1}{\gy{\vc{Z}}} \gy{\vc{G}}_1}}
  + \bigO(\varepsilon_\delta^3)
  ,\label{eq:gc_gy_symtransform_defsub}
\end{align}
where we write $\gy{\vc{G}} \defeq \gy{\vc{G}}_1 + \gy{\vc{G}}_2$ for brevity, and we use the notational convention that an absence of arguments implies evaluation at the gyrocentre coordinate $\gy{\vc{Z}}$.
We are permitted to add a total derivative, by virtue of considering an action principle; we add the total derivative of
\begin{equation}
  \roundpar{\gc{\vc{\gamma}} - \frac{1}{2} \pard{\gc{\vc{\gamma}}}{\gy{\vc{Z}}} \gy{\vc{G}}} \bcdot \roundpar{\gy{\vc{G}} - \frac{1}{2} \pard{\gy{\vc{G}}_1}{\gy{\vc{Z}}} \gy{\vc{G}}_1}
\end{equation}
resulting in the following symplectic part of the gyrocentre Lagrangian
\begin{align}
  \gy{\vc{\gamma}} 
  = {} & 
      \gc{\vc{\gamma}} 
    + \gc{\mat{W}} \roundpar{\gy{\vc{G}} - \frac{1}{2} \pard{\gy{\vc{G}}_1}{\gy{\vc{Z}}} \gy{\vc{G}}_1}
    + \frac{1}{2} \pardd{\gc{\vc{\gamma}}}{\gy{\vc{Z}}} \bcddot \roundpar{\gy{\vc{G}}_1 \otimes \gy{\vc{G}}_1} \nonumber \\ &
    + \frac{1}{2} \roundpar{\squarepar{\pard{\gy{\vc{G}}_1}{\gy{\vc{Z}}}}^\transpose \pard{\gc{\vc{\gamma}}}{\gy{\vc{Z}}}
    - \squarepar{\pard{}{\gy{\vc{Z}}} \roundpar{\pard{\gc{\vc{\gamma}}_0}{\gy{\vc{Z}}} \gy{\vc{G}}_1}}^\transpose} \gy{\vc{G}}_1
  ,\label{eq:gc_gy_symtransform_defsub_pi}
\end{align}
where the following piece of the symplectic part of the transformed guiding-centre Lagrangian moves to the Hamiltonian part of the gyrocentre Lagrangian (while switching sign)
\begin{equation}\label{eq:gc_gy_symtransform_hampart1}
  \pard{}{t} \roundpar{\gc{\vc{\gamma}}_1
  - \frac{1}{2} \pard{\gc{\vc{\gamma}}_0}{\gy{\vc{Z}}} \gy{\vc{G}}_1} \bcdot \gy{\vc{G}}_1,
\end{equation}
where we have made use of $\partial \gc{\vc{\gamma}}_0 / \partial t = \vc{0}_6$.

To further simplify \cref{eq:gc_gy_symtransform_defsub_pi}, we write it component wise as
\begin{align}
  \gy{{\gamma}}_i
  = {} &
      \gc{{\gamma}}_i 
    + \gc{\matcmp{W}}_{ij} \roundpar{\gy{{G}}_j - \frac{1}{2} \pard{\gy{{G}}_{1,j}}{\gy{{Z}}_k} \gy{{G}}_{1,k}}
    + \frac{1}{2} \roundpar{\frac{\partial^2 \gc{{\gamma}}_i}{\partial \gy{{Z}}_j \partial \gy{{Z}}_k} - \frac{\partial^2\gc{{\gamma}}_{0,j}}{\partial \gy{{Z}}_i \partial \gy{{Z}}_k}} \gy{{G}}_{1,j} \gy{{G}}_{1,k} \nonumber \\ &
    + \frac{1}{2} \gc{\matcmp{W}}_{kj} \pard{\gy{{G}}_{1,j}}{\gy{{Z}}_i} \gy{{G}}_{1,k} 
  ,\label{eq:gc_gy_symtransform_comp_expand}
\end{align}
where we have simplified the expression, and we note that a repeated index implies summation.
Note that
\begin{align}
  \pard{}{\gy{Z}_i} \roundpar{\gc{\matcmp{W}}_{kj} \gy{{G}}_{1,j}} - \pard{}{\gy{Z}_k} \roundpar{\gc{\matcmp{W}}_{ij} \gy{{G}}_{1,j}}
  = {} &
  \roundpar{\frac{\partial^2 \gc{\gamma}_{i}}{\partial \gy{Z}_j \partial \gy{Z}_k} - \frac{\partial^2 \gc{\gamma}_{k}}{\partial \gy{Z}_j \partial \gy{Z}_i}}  \gy{{G}}_{1,j} 
  \nonumber \\ & + \gc{\matcmp{W}}_{kj} \pard{\gy{{G}}_{1,j}}{\gy{Z}_i} - \gc{\matcmp{W}}_{ij} \pard{\gy{{G}}_{1,j}}{\gy{Z}_k},
\end{align}
which allows \cref{eq:gc_gy_symtransform_comp_expand} to be written as
\begin{align}
  \gy{{\gamma}}_i
  = {} &
      \gc{{\gamma}}_i 
    + \gc{\matcmp{W}}_{ij}\gy{{G}}_j
    + \squarepar{\pard{}{\gy{Z}_i} \roundpar{\gc{\matcmp{W}}_{kj} \gy{{G}}_{1,j}} - \pard{}{\gy{Z}_k} \roundpar{\gc{\matcmp{W}}_{ij} \gy{{G}}_{1,j}}} \gy{{G}}_{1,k} 
  .\label{eq:gc_gy_symtransform_comp}
\end{align}

Finally, the gyrocentre Hamiltonian follows from subtracting \cref{eq:gc_gy_symtransform_hampart1} from \cref{eq:gc_gy_hamtransform} resulting in
\begin{equation}
  \gy{H}
  = \gc{H}
  - \pard{\gc{H}}{\gy{\vc{Z}}} \bcdot \roundpar{\gy{\vc{G}}_1 + \gy{\vc{G}}_2}
  + \squarepar{
    - \pard{}{t} \roundpar{\gc{\vc{\gamma}}_1 - \frac{1}{2} \pard{\gc{\vc{\gamma}}_0}{\gy{\vc{Z}}} \gy{\vc{G}}_1}
    + \frac{1}{2} \pard{}{\gy{\vc{Z}}}\roundpar{\pard{\gc{H}}{\gy{\vc{Z}}} \bcdot \gy{\vc{G}}_1}
  } \bcdot \gy{\vc{G}}_1
  .
\end{equation}
  \section{Proof of sufficient condition for gauge invariance}\label{sec:gauge_invariance_proof}
\begin{theorem}[Sufficient condition for gauge invariance]
  The gyrocentre single-particle phase-space Lagrangian (to second-order) is gauge invariant up to a total derivative
  \begin{equation}
    \gy{L} 
    \overset{\text{\cref{eq:gauge_transform}}}{\mapsto}
    \gy{L} + q \roundpar{\nabla \eta \bcdot \dot{\gy{\vc{R}}} + \pard{\eta}{t}}
    = 
    \gy{L} + q \totd{\eta}{t}.
  \end{equation}  
  provided that $\gy{\vc{\gamma}}_1 - \gc{\vc{\gamma}}_1$ and $\gy{\vc{\gamma}}_2$ are gauge invariant.
\end{theorem}
\begin{proof}
  We assume that $\gy{\vc{\gamma}}_1 - \gc{\vc{\gamma}}_1$ is gauge invariant.
  Note that the perturbed guiding-centre Lagrangian~\eqref{eq:gc_lag_pert_burby_split} is gauge invariant (up to a total derivative)
  \begin{equation}
    \gc{L}_1 
    \overset{\text{\cref{eq:gauge_transform}}}{\mapsto}
    \gc{L}_1 + q \roundpar{\nabla \eta \bcdot \dot{\gc{\vc{R}}} + \pard{\eta}{t}}
    = 
    \gc{L}_1 + q \totd{\eta}{t}.
  \end{equation}  
  From \cref{eq:lie2_gamma_transform1,eq:lie2_F_transform1} it follows that the first-order gyrocentre Lagrangian can be written as
  \begin{equation}
    \gy{L}_1 = \gc{L}_1 + \roundpar{\gc{\mat{W}}_0 \gy{\vc{G}}_1 + \pard{\gy{S}_1}{\gy{\vc{Z}}}} \bcdot \dot{\gy{\vc{Z}}} + \pard{\gc{H}_0}{\gy{\vc{Z}}} \bcdot \gy{\vc{G}}_1 + \pard{\gy{S}_1}{t},
  \end{equation}
  from which it follows that $\gy{L}_1$ is gauge invariant up to a total derivative if the first-order generating vector $\gy{\vc{G}}_1$ and first-order generating function $\gy{S}_1$ are gauge invariant.

  Using \cref{eq:gy_effpot,eq:gy_S1_def,eq:gc_lag_pert_burby_split_hamiltonian} we find that
  \begin{equation}\label{eq:gy_S1_pde}
    \pard{\gy{S}_1}{t} + \bracketgc{\gy{S}_1}{\gy{H}_0} 
    = q \wt{\psi}_1 
    = \wt{\gc{H}_1} + \dot{\gc{\vc{Z}}} \bcdot \wt{\roundpar{\gy{\vc{\gamma}}_1 - \gc{\vc{\gamma}}_1}}
    = \wt{\gc{H}_1^\flr} + \dot{\gc{\vc{Z}}} \bcdot \wt{\roundpar{\gy{\vc{\gamma}}_1 - \gc{\vc{\gamma}}_1}},
  \end{equation}
  where we note that $\gc{H}_1^\flr$ is gauge invariant as it is expressed in terms of the electric field, and therefore $\gy{S}_1$ is gauge invariant as $\gy{\vc{\gamma}}_1 - \gc{\vc{\gamma}}_1$ is assumed to be gauge invariant. 
  It follows that the generating vector $\gy{\vc{G}}_1$, by \cref{eq:gy_genvector1}, and thereby also the first-order gyrocentre Lagrangian are gauge invariant.

  We additionally assume that $\gy{\vc{\gamma}}_2$ is gauge invariant.
  The second-order gyrocentre Lagrangian is given by
  \begin{equation}\label{eq:gy_L2}
    \gy{L}_2 = \gy{\vc{\gamma}}_2 \bcdot \dot{\gy{\vc{Z}}} - \gy{H}_2,
  \end{equation}
  where the second-order gyrocentre Hamiltonian is given by \cref{eq:gy_H2_general}.
  In order for the second-order gyrocentre Lagrangian to be gauge invariant, we therefore need to show that the corresponding Hamiltonian is gauge invariant, which, upon inspection of \cref{eq:gy_H2_general}, requires the vector $\vc{T}_1$
   to be gauge invariant.

  The vector $\vc{T}_1$, as given by \cref{eq:gy_Tvector}, can be written as
  \begin{equation}
    \vc{T}_1 
    = \underbrace{\squarepar{\gc{\mat{W}}_1 \dot{\gc{\vc{Z}}} - \pard{\gc{H}_1}{\gy{\vc{Z}}} - \pard{\gc{\vc{\gamma}}_1}{t}}}_{\vc{T}_1^\dagger}
    + \frac{1}{2}\squarepar{\pard{\gc{\vc{\gamma}}_0}{\gy{\vc{Z}}} \pard{\gy{\vc{G}}_1}{t} + (\gy{\mat{W}}_1 - \gc{\mat{W}}_1) \dot{\gc{\vc{Z}}} + \pard{}{\gy{\vc{Z}}}\roundpar{\pard{\gc{H}_0}{\gy{\vc{Z}}} \bcdot \gy{\vc{G}}_1}},
  \end{equation}
  where we note that $\gy{\mat{W}}_1 - \gc{\mat{W}}_1$ is gauge invariant because $\gy{\vc{G}}_1$ is gauge invariant (cf.\ \cref{eq:perturbed_Lagrange}).
  It therefore remains to be shown that $\vc{T}_1^\dagger$ is gauge invariant, which is expressed in terms of the first-order symplectic and Hamiltonian part of guiding-centre Lagrangian.
  We note that the FLR part of the first-order guiding-centre Lagrangian is expressed in terms of the electromagnetic fields (cf.\ \cref{eq:gc_lag_pert_burby_split}) and is therefore gauge invariant.
  Hence, we need only to show that the ZLR part $\vc{T}_1^{\dagger,\zlr}$ is gauge invariant, which is given by
  \begin{align}
    \vc{T}_1^{\dagger,\zlr} 
    = \gc{\mat{W}}_1^\zlr \dot{\gc{\vc{Z}}} - \pard{\gc{H}_1^\zlr}{\gy{\vc{Z}}} - \pard{\gc{\vc{\gamma}}_1^\zlr}{t}
    = \begin{pmatrix}
      q (\vc{E}_1 + \dot{\gc{\vc{R}}} \times \vc{B}_1)\\
      0\\
      0\\
      0\\
    \end{pmatrix}
  \end{align}
  and is therefore gauge invariant.
\end{proof}
  \section{Gyro-averaging identities}\label{app:gyro_identities}

\subsection{Taylor-series expansions}
Let $Q = Q(\vc{r}, t)$ be a smooth function.
We denote by $\mat{H} Q$ the Hessian matrix of second-order partial derivatives of $Q$
\begin{equation}
  (\mat{H} Q)_{ij} = \frac{\partial^2 Q}{\partial r_i \partial r_j},
\end{equation}
which allows us to write the Taylor series expansion of $Q$ centred around $\vc{r}$ as
\begin{align}\label{eq:series_expansion}
  \evalperturbedgc{Q} = Q(\vc{r} + \rho\hrho) = Q + \rho\hrho \bcdot\nabla Q + \frac{\rho^2}{2} (\mat{H} Q) \bcddot (\hrho \otimes \hrho) + \bigO(\rho^3).
\end{align}
Where we recall \cref{eq:otimes}, and note that an absence of arguments implies evaluation at $(\vc{r}, t)$.

Non-dimensionalisation of \cref{eq:series_expansion} results in
\begin{align}\label{eq:series_expansion_nondim}
  \nondim{Q}(\vc{r} + \rho\hrho) = \nondim{Q} + \varepsilon_\perp \hrho \bcdot\nabla \nondim{Q} + \frac{\varepsilon_\perp^2}{2} (\nondim{\mat{H}} \nondim{Q}) \bcddot (\hrho \otimes \hrho) + \bigO(\varepsilon_\perp^3),
\end{align}
where $Q = \units{Q} \nondim{Q}$, and we recall that the non-dimensional perpendicular wavenumber is defined in \cref{eq:eps_perp}.

We note that the $\bigO(\varepsilon_\perp^3)$ term in \cref{eq:series_expansion_nondim} contains a $\hrho \otimes \hrho \otimes \hrho$ term, which yields zero upon computing the gyro-average.
It follows that the gyro-average of $\evalperturbedgc{Q}$ has the following Taylor series expansion
\begin{align}\label{eq:gyro_identity_gav_lwa}
  \gav{\evalperturbedgc{Q}} =
    Q 
  + \frac{\rho^2}{4} (\mat{H} Q) \bcddot (\vc{\hat{e}}_1 \otimes \vc{\hat{e}}_1 + \vc{\hat{e}}_2 \otimes \vc{\hat{e}}_2) 
  + \bigO(\varepsilon_\perp^4),
\end{align}
where we have multiplied \cref{eq:series_expansion_nondim} by $\units{Q}$ and have made use of
\begin{subequations}
  \begin{align}
    \gav{\hrho} &= \vc{0}_3,\\
    \gav{\hrho \otimes \hrho} &= \frac{1}{2}\roundpar{\vc{\hat{e}}_1 \otimes \vc{\hat{e}}_1 + \vc{\hat{e}}_2 \otimes \vc{\hat{e}}_2}.\label{eq:gavrhorho}
  \end{align}
\end{subequations}

\subsection{Stokes's theorem}
Using the definition of the gyro-average, as found in \cref{eq:gyro_average}, we find that
\begin{equation}
    \gav{\evalperturbedgc{S}_\tau} 
  = \frac{1}{2\upi} \int_0^{2\upi} \vc{S}(\gy{\vc{R}} + \vc{\rho}) \bcdot \htau(\gy{\vc{R}}, \gy{\theta}) \diff \gy{\theta}
  = -\frac{1}{2\upi \rho} \int_{\partial D_\rho} \vc{S} \bcdot \vc{\hat{t}} \diff{l}
  ,
\end{equation}
where $\vc{\hat{t}} \perp \hb$ is the counter-clockwise tangent to the boundary of the disk $D_\rho$ centred at $\gy{\vc{R}}$ (i.e.\ the shaded disk shown in \cref{fig:illustrate_gc}), which results in the minus sign.
Using Stokes's theorem and \cref{eq:disc_average}, we then find that (see also \citep{porazik2011gyrokinetic})
\begin{equation}\label{eq:gyro_identity_gav_tau}
  \gav{\evalperturbedgc{S}_\tau}
= -\frac{1}{2\upi \rho} \int_{D_\rho} (\nabla \times \vc{S})_\shortparallel \diff^2 x
= - \frac{\rho}{2} \frac{1}{\upi\rho^2} \int_{D_\rho} (\nabla \times \vc{S})_\shortparallel \diff^2 x 
= -\frac{\rho}{2} \dgav{(\evalperturbedgcburby{\nabla} \times \vc{S})_\shortparallel},
\end{equation}
where $\diff^2 x$ is an infinitesimal area element on the disk.

\subsection{The gradient theorem}
Terms containing a radial average (i.e.\ an integral over $\burby$) can often be interpreted in some alternative way.
For instance, the gradient theorem (or fundamental theorem of calculus for line integrals) results in
\begin{equation}\label{eq:gradient_theorem_scalar}
  \rho \radgav{\hrho \bcdot \evalperturbedgcburby{\nabla} Q} 
  = \frac{1}{2\upi} \int_0^{2\upi}\int_0^1 \vc\rho \bcdot (\nabla Q)(\gy{\vc{R}} + \burby \vc\rho) \diff \burby \diff\gy{\theta}
  = \gav{\evalperturbedgc{Q}} - Q.
\end{equation}
The same identity can be derived for vector fields
\begin{equation}\label{eq:gradient_theorem_vector}
  \rho \radgav{(\evalperturbedgcburby{\nabla} \vc{S})^\transpose \hrho} 
  = \gav{\evalperturbedgc{\vc{S}}} - \vc{S}.
\end{equation}
By making use of $\nabla (\vc{a} \bcdot \vc{b}) = \vc{a} \times (\curl \vc{b}) + \vc{b} \times (\curl \vc{a}) + (\nabla \vc{b})^\transpose \vc{a} + (\nabla \vc{a})^\transpose \vc{b}$, while neglecting $\bigO(\varepsilon_B)$ contributions, we then find that
\begin{equation}\label{eq:gradient_theorem_vector_approx}
    \radgav{\evalperturbedgcburby{\nabla} (\rho S_\rho)}
  + \rho \radgav{(\evalperturbedgcburby{\nabla} \times \vc{S}) \times \hrho}
  + \bigO(\varepsilon_B)
  =
    \gav{\evalperturbedgc{\vc{S}}} - \vc{S}.
\end{equation}

Similarly, we find that application of \cref{eq:gradient_theorem_vector} to the curl of a vector field results in
\begin{equation}
  \rho \radgav{[\evalperturbedgcburby{\nabla} (\curl \vc{S})]^\transpose \hrho} 
  = \gav{\evalperturbedgc{\nabla} \times \vc{S}} - \curl \vc{S}.
\end{equation}
By making use of $\curl(\vc{a} \times \vc{b}) = \vc{a} \nabla \bcdot \vc{b} - \vc{b} \nabla \bcdot \vc{a} + (\nabla \vc{a})^\transpose \vc{b} - (\nabla \vc{b})^\transpose \vc{a}$, while neglecting $\bigO(\varepsilon_B)$ contributions, we then find that
\begin{equation}\label{eq:gradient_theorem_curlvector_approx}
  \rho \radgav{\evalperturbedgcburby{\nabla} \times [(\curl \vc{S}) \times \hrho]} 
  + \bigO(\varepsilon_B)
  = \gav{\evalperturbedgc{\nabla} \times \vc{S}} - \curl \vc{S}.
\end{equation}

  \section{Smallness of the coordinate transformation}\label{app:parameter_motivation}
The choice of the parameter $\gparamr$ is based on the smallness of the gyrocentre coordinate transformation, which is relevant because we truncate the expansion (in the small parameter $\varepsilon_\delta$) at second-order in \cref{eq:lie2_F_transform,eq:lie2_gamma_transform}.
To make this more explicit, we consider a specific contribution of $\gc{H}_1$ to $\gy{H}_3$ which is therefore neglected in the proposed second-order accurate model. 
We consider the following gyro-averaged contribution from \cref{eqn:gc_gy_hamtransform_taylor}
\begin{align}
  \gy{H}_3^\effective = \frac{1}{2} \sgav{\pardd{\gc{H}_1}{\gy{\vc{Z}}} \bcddot \roundpar{\gy{\vc{G}}_1 \otimes \gy{\vc{G}}_1}} = \bigO(\varepsilon_\delta^3),
\end{align}
where we note that the corresponding fluctuating part would be absorbed by the third-order generating function $\gy{S}_3$ if this term were to be included in the proposed model.
By decomposing each term in terms of its mean and fluctuating part, we find that
\begin{align}
  \nonumber
  \gy{H}_3^\effective 
  = {} & \frac{1}{2} \sgav{\pardd{\gc{H}_1}{\gy{\vc{Z}}} \bcddot \roundpar{\wt{\gy{\vc{G}}}_1 \otimes \wt{\gy{\vc{G}}}_1}}
   + \frac{1}{2} \pardd{\gav{\gc{H}_1}}{\gy{\vc{Z}}} \bcddot \roundpar{\gav{\gy{\vc{G}}_1} \otimes \gav{\gy{\vc{G}}_1}} \\
  &+ \sgav{\pardd{\wt{\gc{H}}_1}{\gy{\vc{Z}}} \bcddot \roundpar{\wt{\gy{\vc{G}}}_1 \otimes \gav{\gy{\vc{G}}_1}}},
\end{align}
where we note that the first contribution does not depend on $\gparamr$.
The magnitude of $\gy{H}_3^\effective$ can be bounded from above as follows
\begin{align}
  \abs{\gy{H}_3^\effective} \le 
      \frac{1}{2} \sabs{\pardd{\gav{\gc{H}_1}}{\gy{\vc{Z}}}}_\mathrm{F} \abs{\gav{\gy{\vc{G}}_1}}^2
    + \sabs{\sgav{\pardd{\wt{\gc{H}}_1}{\gy{\vc{Z}}} \wt{\gy{\vc{G}}}_1}} \abs{\gav{\gy{\vc{G}}_1}} + \ldots,
\end{align}
where have omitted the contribution that does not depend on $\gparamr$ (now denoted by $\ldots$), and by making use of the triangle and Cauchy--Schwarz inequalities, substituting \cref{eq:gc_lag_pert_burby_split_hamiltonian}, and by letting $\abs{\cdot}_\mathrm{F}$ denote the Frobenius norm for which $\abs{\vc{S} \otimes \vc{S}}_\mathrm{F} = \abs{\vc{S}}^2$.

As the fluctuating part of the first-order generating vector $\wt{\gy{\vc{G}}}_1$ is independent of the choice of $\gparamr$, we can minimise the upper bound of the magnitude of the neglected term~$\gy{H}_3^\effective$ by minimising the Euclidean norm (squared) of the gyro-average of the first-order generating vector
\begin{equation}
  \abs{\gav{\gy{\vc{G}}_1}}^2 = \frac{\gy{\Mu}^2}{B_0^2} \dgav{\evalperturbedgcburby{B}_{1,\shortparallel}}^2
    + (1 - \gparamr)^2 \roundpar{
          \frac{\radgav{\evalperturbedgcburby{B}_{1,\shortparallel}}^2}{(B_{0,\shortparallel}^\effective)^2}
        + \frac{q^2}{m^2} \radgav{\evalperturbedgcburby{B}_{1,\tau}}^2
    },
\end{equation}
which is achieved by letting $\gparamr = 1$.
This choice yields a second-order accurate (in terms of $\varepsilon_\delta$) gyrocentre Lagrangian for which the upper bound of the truncation error is minimised w.r.t. the contributions from the first-order generating vector, under the constraint that the resulting gyrocentre magnetic moment is an invariant (i.e.\ letting $\gparamtheta = 0$).

  \section{Approximation of the second-order Hamiltonian}\label{app:H20}
We consider \cref{eq:gy_H2_general_}, with the first-order generating vector approximated by \cref{eq:gy_G10} and $\vc{T}_{1}$ is as defined by \cref{eq:gy_Tvector}, which we repeat here for convenience
\begin{equation}\label{eq:gy_Tvector_split}
  \vc{T}_1 = 
    \underbrace{\frac{1}{2} (\gc{\mat{W}}_1 + \gy{\mat{W}}_1) \dot{\gc{\vc{Z}}}}_{\circled{A}}
  - \underbrace{\pard{}{t} \roundpar{\gc{\vc{\gamma}}_1
  - \frac{1}{2} \pard{\gc{\vc{\gamma}}_0}{\gy{\vc{Z}}} \gy{\vc{G}}_1}}_{\circled{B}}
  - \underbrace{\pard{}{\gy{\vc{Z}}}\roundpar{\gc{H}_1 - \frac{1}{2} \pard{\gc{H}_0}{\gy{\vc{Z}}} \bcdot \gy{\vc{G}}_1}}_{\circled{C}}.
\end{equation}
In this section of the appendix, we use the following shorthand notation
\newcommand\paramsum[2]{#1^{\dagger}}
\begin{equation}
  \paramsum{Q}{R} \defeq \frac{1}{2} (Q + \gav{Q}).
\end{equation}
For the approximation of $\gy{H}_2$ we omit all derivatives of the background magnetic field as well as of the perturbed electromagnetic fields.

Using \cref{eq:gc_lag_pert_burby_split_gamma10,eq:gy_gamma_param} we find that
\begin{equation}
  \frac{1}{2} (\gc{\vc{\gamma}}_{1} + \gy{\vc{\gamma}}_{1}) =
  \begin{pmatrix}
    q\vc{A}_1 + q \rho \int_0^1 \paramsum{(\evalperturbedgcburby{\vc{B}}_1 \times \hrho)}{R} \diff \burby \\
    0\\
    0\\
    - \frac{q \rho^2}{2} \int_0^1 \burby (\evalperturbedgcburby{B}_{1,\shortparallel}) \diff \burby\\
  \end{pmatrix},
\end{equation}
from which it follows that
\begin{equation}
  \frac{1}{2} (\gc{\mat{W}}_{1} + \gy{\mat{W}}_{1}) =
  \int_0^1 \begin{pmatrix}
    q \mat{B}_1 & \vc{0}_3 & -\frac{q \rho}{2 \gy{\Mu}} \paramsum{(\evalperturbedgcburby{\vc{B}}_1 \times \hrho)}{R} & -\frac{q \rho}{2} \evalperturbedgcburby{\vc{B}}_1 \times \htau\\
    \vc{0}_3^\transpose & 0 & 0 & 0\\
    \frac{q \rho}{2 \gy{\Mu}} \squarepar{\paramsum{(\evalperturbedgcburby{\vc{B}}_1 \times \hrho)}{R}}^\transpose & 0 & 0 & -\frac{q \rho^2}{2\gy{\Mu}} \burby (\evalperturbedgcburby{B}_{1,\shortparallel})\\
    \frac{q \rho}{2} \roundpar{ \evalperturbedgcburby{\vc{B}}_1 \times \htau}^\transpose & 0 & \frac{q \rho^2}{2\gy{\Mu}} \burby (\evalperturbedgcburby{B}_{1,\shortparallel}) & 0\\
  \end{pmatrix} \diff\burby,
\end{equation}
where the matrix $\mat{B}_1$ is defined analogously to \cref{eq:matvec_is_timescurl}. 
We have omitted all derivatives of the background magnetic field as well as of the perturbed electromagnetic fields.
Subsequent evaluation of the matrix-vector product with $\dot{\gc{\vc{Z}}}$ (while neglecting $\bigO(\varepsilon_B)$ contributions), where the guiding-centre EOMs are given by \cref{eq:gc_eoms}, results in
\begin{equation}\label{eq:gy_H2_W1_times_Zdot}
  \circled{A} 
  = \frac{1}{2} (\gc{\mat{W}}_{1} + \gy{\mat{W}}_{1}) \dot{\gc{\vc{Z}}} 
  = \int_0^1 \begin{pmatrix}
    q \gc{U}_\shortparallel \hb \times \vc{B}_1 - \frac{q \rho \omega_\mathrm{c}}{2} \evalperturbedgcburby{\vc{B}}_1 \times \htau\\
    0\\
    \frac{q \rho \gc{U}_\shortparallel}{2 \gy{\Mu}} \paramsum{(\evalperturbedgcburby{\vc{B}}_1 \times \hrho)}{R} \bcdot \hb -\frac{\omega_\mathrm{c} q \rho^2}{2\gy{\Mu}} \burby (\evalperturbedgcburby{B}_{1,\shortparallel})\\
    \frac{q \rho \gc{U}_\shortparallel}{2} (\evalperturbedgcburby{\vc{B}}_1 \times \htau) \bcdot \hb\\
  \end{pmatrix} \diff \burby.
\end{equation}

The second term in $\vc{T}_1$ is given by
\begin{equation}\label{eq:gy_T1_B}
  \circled{B} =  
  \begin{pmatrix}
    q\pard{\vc{A}_1}{t}\\
    0\\
    0\\
    0\\
  \end{pmatrix}
\end{equation}
by making use of \cref{eq:gc_lag_pert_burby_split_gamma10,eq:gc_lag,eq:gy_G10} as well as Faraday's law~\eqref{eq:Faraday}.
The third term in $\vc{T}_1$ involves the average of the first-order guiding-centre and gyrocentre Hamiltonians
\begin{equation}
  \frac{1}{2} (\gc{H}_1 + \gy{H}_1) = 
    q \phi_1
  -  \int_0^1 \squarepar{\frac{q \rho}{2} (\evalperturbedgcburby{E}_{1,\rho} + \gav{\evalperturbedgcburby{E}_{1,\rho}})
  - \gy{\Mu}  \burby \gav{\evalperturbedgcburby{B}_{1,\shortparallel}}} \diff\burby,
\end{equation}
which follows from substituting \cref{eq:gc_lag_pert_burby_split_hamiltonian,eq:gy_H1}.
This results in
\begin{equation}\label{eq:gy_H2_partialderivativesH1}
  \circled{C} 
  = \int_0^1 \begin{pmatrix}
    q \nabla \phi_1\\
    0\\
    - \frac{q \rho}{4 \gy{\Mu}} (\evalperturbedgcburby{E}_{1,\rho} + \gav{\evalperturbedgcburby{E}_{1,\rho}}) 
    + \burby \gav{\evalperturbedgcburby{B}_{1,\shortparallel}}\\
    - \frac{q \rho}{2} \evalperturbedgcburby{E}_{1,\tau} \\
  \end{pmatrix} \diff \burby,
\end{equation}
where we have again neglected all derivatives of the background and perturbed electromagnetic fields.

When combining \cref{eq:gy_H2_W1_times_Zdot,eq:gy_T1_B,eq:gy_H2_partialderivativesH1} we find that $\vc{T}_1$ is given by
\begin{equation}\label{eq:gy_H2_T1}
  \vc{T}_1 = 
  \int_0^1 \begin{pmatrix}
      \vc{F}_1 - \frac{q \rho \omega_\mathrm{c}}{2} \evalperturbedgcburby{\vc{B}}_1 \times \htau\\
      0\\
      \frac{\rho}{4 \gy{\Mu}} (\evalperturbedgcburby{F}_{1,\rho} + \gav{\evalperturbedgcburby{F}_{1,\rho}})
    - \burby (\evalperturbedgcburby{B}_{1,\shortparallel} + \gav{\evalperturbedgcburby{B}_{1,\shortparallel}})\\
      \frac{\rho}{2} \evalperturbedgcburby{F}_{1,\tau} \\
  \end{pmatrix} \diff \burby,
\end{equation}
where we have made use of the definition of the Lorentz force $\vc{F}_1$ as given by \cref{eq:gc_Lorentz}.
We are ready to evaluate the second-order gyrocentre Hamiltonian, as expressed in \cref{eq:gy_H2_general_}, by substituting \cref{eq:gy_G10,eq:gy_H2_T1}
\begin{align}
  \gy{H}_{2} = {} & 
    \sgav{\roundpar{\vc{F}_1 - \frac{q \rho \omega_\mathrm{c}}{2} \int_0^1 \evalperturbedgcburby{\vc{B}}_1 \times \htau \diff \burby} \bcdot \roundpar{-\frac{\hb}{B_{0,\shortparallel}^\effective} \times \int_0^1 \wt{\evalperturbedgcburby{\vc{B}}_1 \times \vc\rho} \diff\burby + \frac{\rho B_{1,\rho}}{B_0} \hb}}\nonumber\\ &
  - \sgav{\int_0^1 \squarepar{\frac{\rho}{4 \gy{\Mu}} (\evalperturbedgcburby{F}_{1,\rho} + \gav{\evalperturbedgcburby{F}_{1,\rho}}) - \burby (\evalperturbedgcburby{B}_{1,\shortparallel} + \gav{\evalperturbedgcburby{B}_{1,\shortparallel}}) }\diff \burby \squarepar{\frac{q^2 \rho^2}{m} \int_0^1 \burby \evalperturbedgcburby{B}_{1,\shortparallel} \diff\burby - \frac{\rho}{B_0} F_{1,\rho}}}\nonumber\\ &
  + \sgav{\squarepar{\frac{\rho}{2} \int_0^1 \evalperturbedgcburby{F}_{1,\tau} \diff \burby} \squarepar{-\frac{\rho}{2 B_0 \gy{\Mu}} F_{1,\tau}}}.
\end{align}

We consider the ZLR limit of all the terms.
That is, we make use of the approximation $\evalperturbedgcburby{Q} \approx Q$ resulting in
\begin{equation}
  \begin{aligned}
    \gy{H}_{2} = {} & 
      \frac{\rho}{B_0}\sgav{\squarepar{\vc{F}_1 - \frac{q \rho \omega_\mathrm{c}}{2} {\vc{B}}_1 \times \htau} \bcdot \squarepar{\hb \times \roundpar{\gav{{\vc{B}}_1 \times \hrho} - {\vc{B}}_1 \times \hrho} + B_{1,\rho} \hb}}\\
      & + \frac{1}{2} \sgav{\squarepar{\frac{\rho}{2 \gy{\Mu}} ({F}_{1,\rho} + \gav{{F}_{1,\rho}}) - {B}_{1,\shortparallel}} \squarepar{-\frac{q^2 \rho^2}{2 m} {B}_{1,\shortparallel} - \frac{\rho}{B_0} F_{1,\rho}}}
       - \frac{\rho^2}{4 B_0 \gy{\Mu}} \sgav{F_{1,\tau}^2}
      ,
  \end{aligned}
\end{equation}
where we have moreover approximated $B_{0,\shortparallel}^\effective \approx B_0$.
When subsequently making use of $\gav{\htau} = \gav{\hrho} = \vc{0}_3$ as well as 
\begin{align}
  \gav{S_\rho^2} 
  = \gav{S_\tau^2} 
  = (\vc{S} \otimes \vc{S}) \bcddot \gav{\hrho \otimes \hrho} 
  = \frac{1}{2} (\vc{S} \otimes \vc{S}) \bcddot (\vc{\hat{e}}_1 \otimes \vc{\hat{e}}_1 + \vc{\hat{e}}_2 \otimes \vc{\hat{e}}_2)
  = \frac{1}{2} \abs{\vc{S}_\perp}^2,
\end{align}
we find that
\begin{equation}\label{eq:gy_H2_gcF10}
  \gy{H}_{2} = 
      \frac{\gy{\Mu}}{2 B_0} \abs{\vc{B}_{1,\perp}}^2
    - \frac{m}{2 q^2 B_0^2} \abs{\vc{F}_{1,\perp}}^2
    .
\end{equation}

  \section{Computation of the Jacobian}\label{app:jacobian}
\renewcommand\gy[1]{\gyFIXED{#1}}
The aim is to compute the Jacobian of the transformation from the physical coordinates $\tilde{\vc{Z}} = (\vc{R}, \vc{U})$ to the gyrocentre coordinates $\gy{\vc{Z}}$.
To this end, we write the Lagrangian in physical coordinates $\tilde{\vc{Z}}$ as
\begin{equation}\label{eq:tilde_lagrangian}
  \tilde{L} = \tilde{\vc{\gamma}} \bcdot \dot{\tilde{\vc{Z}}} - \tilde{H}.
\end{equation}
Here the Hamiltonian $\tilde{H}$ is defined such that, when the physical Lagrangian is transformed to gyrocentre coordinates, we find
\begin{equation}\label{eq:tilde_eq_gy}
  \tilde{L}(\tilde{\vc{Z}}) = \gy{L}(\gy{\vc{Z}}).
\end{equation}
Letting $\tilde{\vc{Z}} = \tilde{\vc{Z}}(\gy{\vc{Z}})$, we find that
\begin{equation}
  \dot{\tilde{\vc{Z}}} = \pard{\tilde{\vc{Z}}}{\gy{\vc{Z}}} \dot{\gy{\vc{Z}}}
\end{equation}
such that \cref{eq:tilde_lagrangian,eq:tilde_eq_gy} imply
\begin{equation}
  \tilde{\vc{\gamma}} \bcdot \roundpar{\pard{\tilde{\vc{Z}}}{\gy{\vc{Z}}} \dot{\gy{\vc{Z}}}} = \gy{\vc{\gamma}} \bcdot \dot{\gy{\vc{Z}}}
  \quad \implies \quad
  \gy{\vc{\gamma}} = \roundpar{\pard{\tilde{\vc{Z}}}{\gy{\vc{Z}}}}^\transpose \tilde{\vc{\gamma}}.
\end{equation}
This results in the following expression for the gyrocentre Lagrange matrix 
\begin{equation}
  \gy{\mat{W}} 
  = \roundpar{\pard{\gy{\vc{\gamma}}}{\gy{\vc{Z}}}}^\transpose - \pard{\gy{\vc{\gamma}}}{\gy{\vc{Z}}}
  = \roundpar{\pard{\tilde{\vc{\gamma}}}{\gy{\vc{Z}}}}^\transpose \pard{\tilde{\vc{Z}}}{\gy{\vc{Z}}}  - \roundpar{\pard{\tilde{\vc{Z}}}{\gy{\vc{Z}}}}^\transpose \pard{\tilde{\vc{\gamma}}}{\gy{\vc{Z}}} 
  = \roundpar{\pard{\tilde{\vc{Z}}}{\gy{\vc{Z}}}}^\transpose \tilde{\mat{W}} \pard{\tilde{\vc{Z}}}{\gy{\vc{Z}}},
\end{equation}
from which it follows that
\begin{equation}
  \det \pard{\tilde{\vc{Z}}}{\gy{\vc{Z}}} = \sqrt{\frac{\det \gy{\mat{W}}}{\det \tilde{\mat{W}}}}.
\end{equation}

We note that
\begin{equation}
  \det \tilde{\mat{W}} 
  = \det \begin{pmatrix}
    q \mat{B} & -m\mat{I}_3\\
    m\mat{I}_3 & \mat{0}_3 \\
  \end{pmatrix}
  = m^6
\end{equation}
and
\begin{equation}
  \det \gy{\mat{W}} 
  = \det \gy{\mat{S}} \det \gy{\mat{W}}_{22} 
  = \det \gy{\mat{W}}_{11} \frac{m^2}{q^2}
\end{equation}
by the Schur complement determinant formula, where we use the block structure of the Lagrange matrix as discussed in \cref{app:matrix_inverse}.
Finally, we note that direct computation shows that
\begin{equation}
  \det \gy{\mat{W}}_{11} 
  = \det \begin{pmatrix}
    q \mat{B}^\effective & -m \hb\\
    m \hb^\transpose & 0
  \end{pmatrix}
  = (q m B_\shortparallel^\effective)^2
\end{equation}
such that
\begin{equation}
  \det \pard{\tilde{\vc{Z}}}{\gy{\vc{Z}}} = \frac{B_\shortparallel^\effective}{m}.
\end{equation}

  \section{Proof of Liouville's theorem}\label{app:liouville}
\begin{theorem}[Gyrocentre Liouville theorem]
  The phase-space volume is conserved:
  \begin{equation}\label{eq:app_gy_liouville}
    \pard{\jacobiangy}{t} + \nabla \bcdot (\jacobiangy \dot{\gy{\vc{R}}}) + \pard{}{\gy{u}_\shortparallel} (\jacobiangy \dot{\gy{U}}_\shortparallel) 
    = 0.
  \end{equation}
  Furthermore, integrals of the form \cref{eq:integral_physical_to_gyrocentre} can be expressed in terms of the initial phase-space coordinates in the following way
  \begin{equation}\label{eq:app_gy_liouville_integral_equality}
    \int \gy{f}_s \mathcal{F} \jacobiangy \diff^6 \gy{z}
    =
    \int \gyfzero(\gyzzero) \mathcal{F}(\gy{\vc{Z}}(t; \gyzzero, \tzero)) \jacobiangy(\gyzzero, \tzero) \diff^6 \gyzzerod
    ,
  \end{equation}
  where an absence of arguments implies evaluation at ($\gy{\vc{z}}, t$).
\end{theorem}
\begin{proof}
  The gyrocentre EOMs~\eqref{eq:gy_eoms} imply that
  \begin{subequations}
    \begin{align}
      \pard{B_{s,\shortparallel}^\effective}{t} &= \pard{B^\effective_{1,\shortparallel}}{t},\\
      \nabla \bcdot (B_{s,\shortparallel}^\effective \dot{\gy{\vc{R}}}) &= \frac{\vc{B}_s^\effective}{m_s} \bcdot \nabla \pard{\gy{H}_s}{\gy{U}_\shortparallel} + \frac{1}{q_s}\roundpar{\nabla \gy{H}_s + q_s \pard{\vc{A}^\effective_1}{t}} \bcdot (\curl \hb) - \pard{B^\effective_{1,\shortparallel}}{t},\\
      \pard{}{\gy{U}_\shortparallel} (B_{s,\shortparallel}^\effective \dot{\gy{U}}_\shortparallel) &= -\frac{1}{q_s} (\curl \hb) \bcdot \roundpar{\nabla \gy{H}_s + q_s \pard{\vc{A}^\effective_1}{t}} - \frac{\vc{B}_s^\effective}{m_s} \bcdot \pard{}{\gy{U}_\shortparallel} \nabla \gy{H}_s,
    \end{align}
  \end{subequations}
  from which \cref{eq:app_gy_liouville} follows by simply adding the contributions.

  We consider the coordinate transformation $\gy{\vc{z}} = \gy{\vc{Z}}(t; \gyzzero, \tzero) \mapsto \gyzzero$ applied to the left-hand side of \cref{eq:app_gy_liouville_integral_equality} resulting in
  \begin{multline}\label{eq:liouville_intermediate}
    \int \gy{f}_s(\gy{\vc{z}}, t) \mathcal{F}(\gy{\vc{z}}, t) \jacobiangy(\gy{\vc{z}}, t) \diff^6 \gy{z}
    = \\
    \int \gyfzero(\gyzzero) \mathcal{F}(\gy{\vc{Z}}(t; \gyzzero, \tzero)) \jacobiangy(\gy{\vc{Z}}(t; \gyzzero, \tzero), t) \mathfrak{H}(\gy{\vc{Z}}(t; \gyzzero, \tzero), t) \diff^6 \gyzzerod
  \end{multline}
  by defining the Jacobian
  \begin{equation}
    \mathfrak{H} = \det \pard{\gy{\vc{Z}}}{\gyzzero}
  \end{equation}
  and by making use of \cref{eq:gy_distfun_def}.
  Therefore, we must show that
  \begin{equation}
    \totd{}{t} (\jacobiangy \mathfrak{H}) = 0,
  \end{equation}
  where we note that the product $\jacobiangy \mathfrak{H}$ is the Jacobian of the coordinate transformation from initial gyrocentre coordinates to physical coordinates.
  
  We note that \cref{eq:app_gy_liouville} implies
  \begin{equation}\label{eq_gy_liouville_res0}
    \dotjacobiangy = - \jacobiangy \pard{}{\gy{\vc{Z}}} \bcdot \dot{\gy{\vc{Z}}}.
  \end{equation}
  In order to proceed, we make use of (for a proof we refer to \cite[Proposition~1.1]{bouchut2000kinetic})
  \begin{equation}
      \dot{\mathfrak{H}} 
    = \mathfrak{H} \pard{}{\gy{\vc{Z}}} \bcdot \dot{\gy{\vc{Z}}}.
  \end{equation}
  When combining this result with \cref{eq_gy_liouville_res0}, we find
  \begin{equation}
    \totd{}{t} (\jacobiangy \mathfrak{H})
    = \dotjacobiangy \mathfrak{H} + \jacobiangy \dot{\mathfrak{H}}
    = \dotjacobiangy \mathfrak{H} + \jacobiangy \mathfrak{H} \pard{}{\gy{\vc{Z}}} \bcdot \dot{\gy{\vc{Z}}}
    = \dotjacobiangy \mathfrak{H} - \mathfrak{H} \dotjacobiangy
    = 0
    .
  \end{equation}
  It follows that
  \begin{equation}
    \jacobiangy \mathfrak{H} 
    = \jacobiangy(\gyzzero, \tzero) \mathfrak{H}(\gyzzero, \tzero)
    = \jacobiangy(\gyzzero, \tzero),
  \end{equation}
  which, upon substitution in \cref{eq:liouville_intermediate}, results in \cref{eq:app_gy_liouville_integral_equality}.
\end{proof}
  \section{Proof of local energy conservation}\label{app:poynting}
\begin{theorem}[Local energy conservation]
  The kinetic energy density~\eqref{eq:energy_kin_density} satisfies
  \begin{equation}\label{eq:app_energy_kin_evolution}
    \pard{\gy{\mathcal{K}}}{t} + \nabla \bcdot \roundpar{\sum_s \int \gy{f}_s \dot{\gy{\vc{R}}} \gy{K}_s \jacobiangy \diff^3 \gy{u}} = \evalperturbedgcadjointalt{\vc{\mathcal{J}}}{}^\free \bcdot \vc{E},
  \end{equation}
  whereas the potential energy density~\eqref{eq:energy_pot} satisfies Poynting's theorem
  \begin{equation}\label{eq:app_energy_pot_evolution}
    \pard{\gy{\mathcal{U}}}{t} + \nabla \bcdot (\vc{E} \times \vc{\mathcal{H}}) = - \evalperturbedgcadjointalt{\vc{\mathcal{J}}}{}^\free \bcdot \vc{E}.
  \end{equation}
  The magnetizing field $\vc{\mathcal{H}}$ and free current density $\evalperturbedgcadjointalt{\vc{\mathcal{J}}}{}^\free$ are defined in Eqs.~\eqref{eq:maxwell_magfield} and \eqref{eq:free_current_gavadjoint}, respectively.
  It follows that the following local energy conservation law holds
  \begin{equation}\label{eq:app_energy_tot_evolution}
    \pard{}{t} (\gy{\mathcal{K}} + \gy{\mathcal{U}}) + \nabla \bcdot \roundpar{\vc{E} \times \vc{\mathcal{H}} + \sum_s \int \gy{f}_s \dot{\gy{\vc{R}}} \gy{K}_s \jacobiangy \diff^3 \gy{u}} = 0.
  \end{equation}
\end{theorem}

\begin{proof}
  A direct computation of the partial derivative of the kinetic energy density~\eqref{eq:energy_kin_density} per species results in
  \begin{align}
    \pard{\gy{\mathcal{K}}_s}{t}
    ={} & \int \pard{}{t} (\gy{f}_s \jacobiangy) \gy{K}_s \diff^3 \gy{u}
    - \int \gy{f}_s \gy{\mu} \dgav{(\evalperturbedgcburby{\nabla} \times \vc{E}_{1})_\shortparallel} \jacobiangy \diff^3 \gy{u} \nonumber\\
    ={} & - \nabla \bcdot \int \gy{f}_s \dot{\gy{\vc{R}}} \gy{K}_s \jacobiangy \diff^3 \gy{u} 
    + \int \gy{f}_s \gy{\mu} \dot{\gy{\vc{R}}} \bcdot \nabla (B_0 + \dgav{\evalperturbedgcburby{B}_{1,\shortparallel}}) \jacobiangy \diff^3 \gy{u} \nonumber\\ &
    + m_s \int \gy{f}_s \dot{\gy{U}}_\shortparallel \gy{u}_\shortparallel \jacobiangy \diff^3 \gy{u}
    - \int \gy{f}_s \gy{\mu} \dgav{(\evalperturbedgcburby{\nabla} \times \vc{E}_{1})_\shortparallel} \jacobiangy \diff^3 \gy{u} \nonumber\\
    ={} & - \nabla \bcdot \int \gy{f}_s \dot{\gy{\vc{R}}} \gy{K}_s \jacobiangy \diff^3 \gy{u}
    + \frac{1}{m_s} \int \gy{f}_s \gy{\mu} \roundpar{\gy{u}_\shortparallel \vc{B}_s^\effective 
    - \hb \times \vc{E}^\effective_1} \bcdot \nabla (B_0 + \dgav{\evalperturbedgcburby{B}_{1,\shortparallel}}) \diff^3 \gy{u} \nonumber\\ &
    + \frac{1}{m_s} \int \gy{f}_s  \vc{B}_s^\effective \bcdot \squarepar{
      q_s \vc{E}^\effective_1
    - \gy{\mu} \nabla (B_0 + \dgav{\evalperturbedgcburby{B}_{1,\shortparallel}})} \gy{u}_\shortparallel \diff^3 \gy{u}
    - \int \gy{f}_s \gy{\mu} \dgav{(\evalperturbedgcburby{\nabla} \times \vc{E}_{1})_\shortparallel} \jacobiangy \diff^3 \gy{u} \nonumber\\
    ={} & - \nabla \bcdot \int \gy{f}_s \dot{\gy{\vc{R}}} \gy{K}_s \jacobiangy \diff^3 \gy{u} 
    + q_s \int \gy{f}_s \dot{\gy{\vc{R}}} \bcdot \vc{E}^\effective_1 \jacobiangy \diff^3 \gy{u} 
    - \int \gy{f}_s \gy{\mu} \dgav{(\evalperturbedgcburby{\nabla} \times \vc{E}_{1})_\shortparallel} \jacobiangy \diff^3 \gy{u},
  \end{align}
  upon substitution of Faraday's law~\eqref{eq:Faraday}, the conservative form of the Vlasov equation~\eqref{eq:gy_vlasov_conservative} and the gyrocentre EOMs~\eqref{eq:gy_split_eoms}.
  It follows that the total kinetic energy evolves as \cref{eq:app_energy_kin_evolution} upon substitution of the definition of the free current density~\eqref{eq:free_current_gavadjoint} and summation over the species $s$.

  For computing the partial time derivative of the potential energy density~\eqref{eq:energy_pot} we note that
  \begin{equation}\label{eq:pot_evo_part1}
    \pard{}{t} (\vc{\mathcal{D}} \bcdot \vc{E}) = 
      2\pard{\vc{\mathcal{D}}}{t} \bcdot \vc{E} 
    + \sum_s \frac{m_s}{B_0^2} \int \gyfzero \gy{u}_\shortparallel \squarepar{(\hb \times \vc{B}_1) \bcdot \pard{\vc{E}}{t} - \roundpar{\hb \times \pard{\vc{B}_1}{t}} \bcdot \vc{E}} \jacobiangcslab \diff^3 \gy{u} 
  \end{equation}
  and similarly
  \begin{equation}\label{eq:pot_evo_part2}
    \pard{}{t} (\vc{\mathcal{H}} \bcdot \vc{B}) =
      2\vc{\mathcal{H}} \bcdot \pard{\vc{B}}{t}
    + \pard{\vc{B}}{t} \bcdot \vc{\mathcal{M}}_1 - \pard{\vc{\mathcal{M}}_1}{t} \bcdot \vc{B}.
  \end{equation}
  Upon substitution of the strong form of the Amp\`ere--Maxwell law~\eqref{eq:maxwell_ampere} as well as Faraday's law~\eqref{eq:maxwell_faraday} we find that
  \begin{equation}
    \pard{\vc{\mathcal{D}}}{t} \bcdot \vc{E} + \vc{\mathcal{H}} \bcdot \pard{\vc{B}}{t} 
    = (\curl \vc{\mathcal{H}} - \evalperturbedgcadjointalt{\vc{\mathcal{J}}}{}^\free) \bcdot \vc{E} - \vc{\mathcal{H}} \bcdot (\curl \vc{E})
    = - \nabla \bcdot (\vc{E} \times \vc{\mathcal{H}}) - \evalperturbedgcadjointalt{\vc{\mathcal{J}}}{}^\free \bcdot \vc{E}.
  \end{equation}
  It remains to be shown that the two remaining terms from \cref{eq:pot_evo_part1,eq:pot_evo_part2} vanish.
  The sum of the two terms is written as
  \begin{align}
    \sum_s \frac{m_s}{B_0^2} \int \gyfzero \Biggl( &
      \gy{u}_\shortparallel \squarepar{(\hb \times \vc{B}_1) \bcdot \pard{\vc{E}_1}{t} - \roundpar{\hb \times \pard{\vc{B}_1}{t}} \bcdot \vc{E}_1} \nonumber\\ &
      - \pard{\vc{B}_1}{t} \bcdot \squarepar{\gy{u}_\shortparallel \hb \times \vc{E}_{1,\perp} + \roundpar{\frac{\gy{\mu} B_0}{m_s} - \gy{u}_\shortparallel^2} \vc{B}_{1,\perp}} \nonumber \\ &
      + \pard{}{t} \squarepar{\gy{u}_\shortparallel \hb \times \vc{E}_{1,\perp} + \roundpar{\frac{\gy{\mu} B_0}{m_s} - \gy{u}_\shortparallel^2} \vc{B}_{1,\perp}} \bcdot \vc{B}_1
    \biggr) \jacobiangcslab \diff^3 \gy{u}
  \end{align}
  upon substitution of the definition of the magnetisation and polarisation as defined in Eqs.~\eqref{eq:magnetisation} and~\eqref{eq:polarisation}, respectively.
  Indeed, the sum vanishes and therefore \cref{eq:app_energy_pot_evolution} also holds.
\end{proof}
  \section{Derivation of the Brizard--Hahm Hamiltonian}\label{app:brizard_hahm}
The aim is to compute the ZLR approximation of the second-order Hamiltonian as found in \cite[Eq.~(173)]{brizard2007}.
Their first- and second-order Hamiltonian are given by
\begin{subequations}
  \begin{equation}
    \gy{H}_1^\bh = q \gav{{\psi}_1^\bh}
    , \quad
    {\psi}_1^\bh \defeq \evalperturbedgc{\phi}_1 - \gy{U}_\shortparallel \wt{A}_{1,\shortparallel} - \gy{U}_\tau \evalperturbedgc{A}_{1,\tau}
  \end{equation}
  and
  \begin{equation}
    \gy{H}_2^\bh = - \frac{q}{2} \gav{\bracketgc{{\gy{S}}_1^\bh}{\wt{\psi}_1^\bh}} + \frac{q^2}{2m} \squarepar{\gav{\abs{\evalperturbedgc{\vc{A}}_{1,\perp}}^2} + \gav{(\wt{A}_{1,\shortparallel})^2}} + \frac{q}{B_0} \gav{\evalperturbedgc{\vc{A}}_{1,\perp}} \bcdot (\hb \times \nabla \gav{{\psi}_1^\bh}),
  \end{equation}
\end{subequations}
respectively. 
The first-order generating function $\gy{S}_1^\bh$ again satisfies \cref{eq:gy_S1_def_approx} to zeroth-order in $\varepsilon_\omega$.

\subsection{ZLR approximation of the Poisson bracket}
We use the following shorthand notation for the components of a vector field $\vc{S}$ in terms of the local coordinates $\vc{\hat{e}}_1, \vc{\hat{e}}_2$
\begin{equation}
  S_{\hat{i}} \defeq \vc{S} \bcdot \vc{\hat{e}}_i
\end{equation}
as well as the corresponding directional derivative of a scalar function
\begin{equation}
  \hatdd{i} Q \defeq \vc{\hat{e}}_i \bcdot \nabla Q.
\end{equation}

Provided with the first-order generating function $\gy{S}_1^\bh$, we want to approximate the leading order term in $\varepsilon_\perp$ of the gyro-averaged guiding-centre Poisson bracket~\eqref{eq:gc_bracket} while neglecting $\bigO(\varepsilon_\omega)$ and $\bigO(\varepsilon_B)$ terms. 
This results in
\begin{equation}\label{eq:H2_poissonbracket_term_tmp}
  \gav{\bracketgc{{\gy{S}}_1^\bh}{\wt{\psi}_1^\bh}} = 
  \underbrace{\frac{q}{B_0} \sgav{\pard{}{\gy{\Mu}} (\wt{\psi}_1^\bh)^2}}_{\circled{A}}
  - \underbrace{\frac{\hb}{q B_0}\bcdot \sgav{\nabla_\perp {\gy{S}}_1^\bh \times \nabla_\perp \wt{\psi}_1^\bh}}_{\circled{B}}
  + \bigO(\varepsilon_\perp^3).
\end{equation}

Application of \cref{eq:gyro_identity_gav_lwa} to $\wt{\psi}_1^\bh$ results in
\begin{align}\label{eq:psit_smallLR}
  \wt{\psi}_1^\bh
  = {} & 
  - \frac{B_0^2}{m} \vc{\rho} \bcdot \vc{Q}_{1}
  - \rho \Vtau \hrho \bcdot \nabla_\perp {A_{1,\tau}}
  - \frac{\rho \Vtau}{2} (\curl \Adperp)_\shortparallel
  - \frac{\rho^2 \Vtau}{2} (\hrho \otimes \hrho) \bcddot (\mat{H} {A_{1,\tau}})
  \nonumber\\ & 
  + \frac{\rho^2}{2} (\hrho \otimes \hrho) \bcddot [\mat{H} (\phid - \gy{U}_\shortparallel \Apa)]
  - \frac{\rho^2}{2} \nabla_\perp^2 (\phid - \gy{U}_\shortparallel \Apa)
  + \bigO(\varepsilon_\perp^3),
\end{align}
which shows that the first term can be written as
\begin{multline}
  \circled{A}
  = {} \frac{q}{B_0} \pard{}{\gy{\Mu}}  
    \Biggl\langle  
        \underbrace{\roundpar{\frac{B_0^2}{m} \vc{\rho} \bcdot \vc{Q}_{1}}^2}_{\circled{A$_1$}}
        + \underbrace{\rho^2 \Vtausqr \hrho \bcdot \nabla_\perp \roundpar{\htau \bcdot \Adperp} (\curl \Adperp)_\shortparallel}_{\circled{A$_3$}}\\
      + \underbrace{\frac{\rho^2 \Vtausqr}{4} [(\curl \Adperp)_\shortparallel]^2}_{\circled{A$_2$}}
       + \underbrace{\squarepar{\rho \Vtau \hrho \bcdot \nabla_\perp \roundpar{\htau \bcdot \Adperp}}^2}_{\circled{A$_4$}}
      + \underbrace{\rho^2 \Vtausqr A_{1,\tau} (\hrho \otimes \hrho) \bcddot (\mat{H} {A_{1,\tau}})}_{\circled{A$_5$}}
  \Biggr\rangle,
\end{multline}
where we made use of the fact that integration over the interval $[0, 2\upi]$ of odd powers of cosine and/or sine functions yields zero, and we have defined the vector $\vc{Q}_{1}$ as
\begin{equation}\label{eq:polarisation_bh}
  \vc{Q}_{1} \defeq -\frac{m}{B_0^2} \squarepar{\nabla_\perp (\phi_1 - \gy{U}_\shortparallel A_{1,\shortparallel}) + \omega_{\mathrm{c}} \Adperp \times \hb}.
\end{equation}
Moreover, we used the fact that the $\Adperp$ term in $\vc{Q}_{1}$ is the only $\bigO(\varepsilon_\perp^0)$ term of $\wt{\psi}_1^\bh$, and therefore the only term that remains upon multiplication by the $\rho^2 (\mat{H} A_{1,\tau})$ term, when neglecting $\bigO(\varepsilon_\perp^3)$ terms and after integration over $\theta$.
Recall that the outer product is defined in \cref{eq:otimes}.

The first contribution can be evaluated as follows
\begin{equation}
  \circled{A$_1$} = \frac{B_0^2}{q m} \abs{\vc{Q}_{1}}^2,
\end{equation}
where we made use of \cref{eq:gavrhorho}.
The second contribution can trivially be evaluated as
\begin{equation}
  \circled{A$_2$} = \frac{2 \gy{\Mu}}{q B_0} [(\curl \Adperp)_\shortparallel]^2,
\end{equation}
whereas the third contribution yields
\begin{equation}
  \circled{A$_3$} 
  = \frac{4 \gy{\Mu}}{q B_0} (\curl \Adperp)_\shortparallel \roundpar{\hatdd{2} A_{1,\hat{1}} - \hatdd{1} A_{1,\hat{2}}}
  = -\frac{4 \gy{\Mu}}{q B_0} (\curl \Adperp)_\shortparallel^2
\end{equation}
by making use of
\begin{equation}
  \gav{\hrho \otimes \htau} 
  = \frac{1}{2} (\vc{\hat{e}}_2 \otimes \vc{\hat{e}}_1 - \vc{\hat{e}}_1 \otimes \vc{\hat{e}}_2).
\end{equation}
The fourth contribution can be written as
\begin{equation}
  \circled{A$_4$} = \frac{\gy{\Mu}}{q B_0}
  \roundpar{
    \abs{\nabla_\perp \Adperp}^2
    + 2 [(\curl \Adperp)_\shortparallel]^2
    - 2 \det \mat{G}_\perp
  },
\end{equation}
where we make use of
\begin{equation}\label{eq:sin4_cos4}
  \gav{\sin^4\theta} = \gav{\cos^4\theta} = \frac{3}{8}
  , \quad
  \gav{\sin^2\theta \cos^2\theta} = \frac{1}{8}
\end{equation}
and define
\begin{equation}
  \mat{G}_\perp \defeq \begin{pmatrix}
    \hatdd{1} A_{1,\hat{1}} & \hatdd{2} A_{1,\hat{1}}\\
    \hatdd{1} A_{1,\hat{2}} & \hatdd{2} A_{1,\hat{2}}
  \end{pmatrix}.
\end{equation}
Finally, the fifth contribution can be written as 
\begin{equation}
  \circled{A$_5$} = \frac{\gy{\Mu}}{q B_0} \squarepar{
      \Adperp \bcdot \nabla_\perp^2 \Adperp
    - 2 (\Adperp \times \nabla_\perp [(\curl \Adperp)_\shortparallel])_\shortparallel
  }.
\end{equation}

Since the $\circled{B}$ term in \cref{eq:H2_poissonbracket_term_tmp} already has a factor $\varepsilon_\perp^2$, we need to only keep the zeroth order term of $\wt{\psi}_1^\bh$ (cf.\ \cref{eq:psit_smallLR}) for obtaining an $\bigO(\varepsilon_\perp^3)$ approximation
\begin{equation}
  \wt{\psi}_1^\bh = \Vtau A_{1,\tau} + \bigO(\varepsilon_\perp)
\end{equation}
and similarly for $\gy{S}_1^\bh$ for the computation of the second term in \cref{eq:H2_poissonbracket_term_tmp} resulting in
\begin{equation}\label{eq:bracket_other_contribution}
  \circled{B} 
  = -\frac{\Vtausqr}{B_0 \omega_{\mathrm{c}}} 
    \hb \bcdot \roundpar{\nabla_\perp A_{1,\hat{1}} \times \nabla_\perp A_{1,\hat{2}}} + \bigO(\varepsilon_\perp^3)
  = -\frac{2 \gy{\Mu}}{q B_0} \det \mat{G}_\perp + \bigO(\varepsilon_\perp^3)
  .
\end{equation}

By making use of
\begin{equation}
  \frac{1}{2} \nabla_\perp^2 \abs{\Adperp}^2 = \abs{\nabla_\perp \Adperp}^2 + \Adperp \bcdot \nabla_\perp^2 \Adperp,
\end{equation}
we then find that the contribution to $\gy{H}_2^\bh$ due to the first-order generating function can be approximated by (neglecting $\bigO(\varepsilon_\perp^3)$ terms)
\begin{equation}
  \gav{\bracketgc{{\gy{S}}_1^\bh}{\wt{\psi}_1^\bh}} = 
    \frac{B_0^2}{q m} \abs{\vc{Q}_{1}}^2
  + \frac{\gy{\Mu}}{q B_0} \squarepar{
      \frac{1}{2} \nabla_\perp^2 \abs{\Adperp}^2
    - 2 \hb \bcdot (\Adperp \times \nabla \squarepar{(\curl \Adperp)_\shortparallel})
  }.
\end{equation}
Note that in principle a high-frequency approximation of the first-order generating function~$\gy{S}_1^\bh$ can also be considered, such as in the work of \citep{qin1999gyrokinetic}.

\subsection{ZLR approximation of the Hamiltonian}
The remaining gyro-averaged terms in $\gy{H}_2^\bh$ can be approximated in the ZLR limit as follows
\begin{align}
  \frac{q^2}{2m} \gav{(\wt{A}_{1,\shortparallel})^2} 
  &= \frac{\gy{\Mu}}{2 B_0} \abs{\nabla_\perp \Apa}^2 + \bigO(\varepsilon_\perp^4),\label{eq:fluc_mag_parallel_approx}\\
  \frac{q^2}{2m} \gav{\abs{\evalperturbedgc{\vc{A}}_{1,\perp}}^2} &= \frac{q^2}{2m} \abs{\Adperp}^2 + \frac{\gy{\Mu}}{4 B_0} \nabla_\perp^2 \abs{\Adperp}^2 + \bigO(\varepsilon_\perp^4),\label{eq:fluc_mag_perp_approx}\\
  \gav{\evalperturbedgc{\vc{A}}_{1,\perp}} &= \vc{A}_{1,\perp} + \frac{\rho^2}{4} \nabla_\perp^2 \vc{A}_{1,\perp} + \bigO(\varepsilon_\perp^4)
\end{align}
by making use of \cref{eq:gyro_identity_gav_lwa}.
Moreover, we note that
\begin{equation}
  \gav{\evalperturbedgc{\psi}_1^\bh} = 
  \phi_1 - \gy{U}_\shortparallel A_{1,\shortparallel} + \frac{\rho^2}{4} \nabla_\perp^2 (\phi_1 - \gy{U}_\shortparallel A_{1,\shortparallel}) + \frac{\gy{\Mu}}{q} (\nabla \times \vc{A}_1)_\shortparallel + \bigO(\varepsilon_\perp^3),
\end{equation}
by making use of \cref{eq:gyro_identity_gav_tau,eq:gyro_identity_gav_lwa}.
Combining the ZLR approximations results in the following ZLR approximation of the second-order Hamiltonian
\begin{equation}
  \gy{H}_2^{\bh} = - \frac{B_0^2}{2 m} \abs{\vc{Q}_{1}}^2
  + \frac{\gy{\Mu}}{2 B_0} \abs{\nabla_\perp \Apa}^2 
  + \frac{q^2}{2m} \abs{\Adperp}^2
  + \frac{q}{B_0} \vc{A}_{1,\perp} \bcdot [\hb \times \nabla_\perp (\phi_1 - \gy{U}_\shortparallel A_{1,\shortparallel})]
  ,
\end{equation}
which can be simplified to \cref{eq:H2_bh}.

  \section{Well-posedness of the saddle-point problem}\label{app:saddle_point}
Here we briefly, and rather informally, discuss the well-posedness  of the saddle-point problem given by \cref{eq:darwin_saddle} under the simplifying assumptions of a constant background magnetic field, an isotropic pressure $p_{0,\perp} = p_{0,\shortparallel}$ as well as a spatially constant coefficient $\mathcal{C}(1)$.
Therefore, the following system of equations are considered
\begin{subequations}\label{eq:saddle_point}
  \begin{align}
    \curl (\curl \vc{A}) - \nabla_\perp \lambda &= \vc{\mathcal{J}},\label{eq:saddle_point_curl}\\
    \nabla_\perp \bcdot \vc{A} &= 0,\label{eq:saddle_point_div}
  \end{align}
\end{subequations}
where $\nabla \bcdot \vc{\mathcal{J}} = 0$ and with appropriate boundary conditions.
Assume that $\lambda = 0$ and that $\vc{A}$ satisfies 
\begin{subequations}\label{eq:saddle_point_altform}
  \begin{align}
    -\nabla_\perp^2 A_\shortparallel &= \mathcal{J}_\shortparallel,\label{eq:saddle_point_parallel}\\
    -\nabla^2 \vc{A}_\perp &= \vc{\mathcal{J}}_\perp - \nabla_\perp (\hb \bcdot \nabla A_\shortparallel),\label{eq:saddle_point_perp}
  \end{align}
\end{subequations}
then we can show that $(\lambda, \vc{A})$ also satisfy \cref{eq:saddle_point} provided that $\varepsilon_B = 0$.
As each of the PDEs in \cref{eq:saddle_point_altform} is well-posed provided with appropriate boundary conditions, we find that \cref{eq:saddle_point} is well-posed as well.

The equivalence can be shown as follows.
Assume that $\vc{A}$ solves \cref{eq:saddle_point_altform}.
Computing the divergence of \cref{eq:saddle_point_perp} results in
\begin{equation}
  \nabla^2 (\nabla_\perp \bcdot \vc{A}) 
  = \nabla_\perp \bcdot \vc{\mathcal{J}} - \hb \bcdot \nabla (\nabla_\perp^2 A_\shortparallel)
  = \nabla_\perp \bcdot \vc{\mathcal{J}} + \hb \bcdot \nabla (\mathcal{J}_\shortparallel)
  = \nabla \bcdot \vc{\mathcal{J}}
  = 0
\end{equation}
upon substitution of \cref{eq:saddle_point_parallel}.
If homogenous Dirichlet boundary conditions are `included' in the Laplace operator, then this implies that \cref{eq:saddle_point_div} holds.
Finally, we add $\hb$ times \cref{eq:saddle_point_parallel} to \cref{eq:saddle_point_perp} resulting in
\begin{equation}
  \hb (\hb \bcdot \nabla)^2 A_\shortparallel - \nabla^2 \vc{A}
  = \vc{\mathcal{J}} - \nabla_\perp (\hb \bcdot \nabla A_\shortparallel).
\end{equation}
When rearranging the terms and by making use of \cref{eq:saddle_point_div}, we find that
\begin{equation}
  \nabla (\nabla \bcdot \vc{A}) - \nabla^2 \vc{A}
  = \vc{\mathcal{J}},
\end{equation}
which shows that $\vc{A}$ satisfies \cref{eq:saddle_point_curl}.
  \section{Susceptibility tensor in a slab}\label{app:susceptibility}
For the computation of the susceptibility tensor we require a linearization of the proposed model under a Fourier ansatz.
In particular, we require the Fourier component of the linearised distribution function, the free current density, the polarisation current density, and finally the magnetisation current density.
Once each of these Fourier components is known, we combine the results into the gyrokinetic susceptibility tensor, and we compute the drift kinetic limit.

\subsection{The Fourier component of the linearised distribution function}
We need a closed-form expression for $\gyfone$, which can be obtained upon substitution of the Fourier ansatz
\begin{equation}\label{eq:fourier_f1}
  \gyfone = \fourier{\gyfone}(\gy{u}_\shortparallel, \gy{\mu}) \euler^{\ii (\vc{k} \bcdot \gy{\vc{r}} - \omega t)}
\end{equation}
into the linearised Vlasov equation (cf.\ \cref{eq:gy_vlasov_bracket})
\begin{equation}
    \pard{\gyfone}{t} 
  + \gy{u}_\shortparallel \hb \bcdot \nabla \gyfone 
    - \frac{1}{m_s} \hb \bcdot \roundpar{\mu \nabla \dgav{\evalperturbedgcburby{B}_{1,\shortparallel}}
    - q_s \gav{\evalperturbedgc{\vc{E}}_1}} \pard{\gyfzero}{\gy{u}_\shortparallel} 
  = 0,
\end{equation}
where we have made use of
\begin{equation}\label{eq:gy_eomR_linpart}
  (B_{s,\shortparallel}^\effective \dot{\gy{\vc{R}}})_0 = \vc{B}_0 \gy{u}_\shortparallel
  , \quad
  (B_{s,\shortparallel}^\effective \dot{\gy{\vc{R}}})_1 = 
      \gy{u}_\shortparallel \gav{\evalperturbedgc{\vc{B}}_{1}}
      + \roundpar{
          \gav{\evalperturbedgc{\vc{E}}_1}
        - \frac{\mu}{q_s} \nabla \dgav{\evalperturbedgcburby{B}_{1,\shortparallel}}
       } \times \hb
\end{equation}
using the gyrocentre EOM~\eqref{eq:gy_split_eoms_R_subbstarstar} and \cref{eq:gy_genvecpot,eq:flrcorrected_B1}
as well as 
\begin{equation}
  (B_{s,\shortparallel}^\effective \dot{\gy{U}}_\shortparallel)_0 = 0
  , \quad
  (B_{s,\shortparallel}^\effective \dot{\gy{U}}_\shortparallel)_1 = -\frac{1}{m} 
      \vc{B}_0 
    \bcdot \roundpar{\mu \nabla \dgav{\evalperturbedgcburby{B}_{1,\shortparallel}}
  - q_s \gav{\evalperturbedgc{\vc{E}}_1}},
\end{equation}
which follows from substituting \cref{eq:gy_bstarstar,eq:gy_gradHpart} into the gyrocentre EOM~\eqref{eq:gy_eoms_Vpa} and having assumed $\nabla \gyfzero = \vc{0}_3$.
Subsequent substitution of the Fourier ansatzes~\eqref{eq:fourier_f1} and Eq.~\eqref{eq:reduced_fourierE} (and similarly for the magnetic field) results in
\begin{equation}\label{eq:lin_vlasov_fourier_gav}
  - \ii \omega \fourier{\gyfone}
  + \ii \gy{u}_\shortparallel \hb \bcdot \vc{k} \fourier{\gyfone} 
    - \frac{1}{m_s} \hb \bcdot \roundpar{\ii \mu \vc{k} \fourier{\dgav{\evalperturbedgcburby{B}_{1,\shortparallel}}}
    - q_s \fourier{\gav{\evalperturbedgc{\vc{E}}_1}}} \pard{\gyfzero}{\gy{u}_\shortparallel} 
  = 0.
\end{equation}

In order to proceed, we must express the Fourier component of the gyro-average of a function in terms of the Fourier component of that function itself.
We write the wave vector $\vc k$ in terms of its parallel and perpendicular components as
\begin{equation}\label{eq:wavevector}
  \vc k = \kpa \hb + \kperp(\vc{\hat{e}}_1 \cos \alpha + \vc{\hat{e}}_2 \sin \alpha)
\end{equation}
for some angle $\alpha$.
It follows that the gyro-average of a function $Q$
\begin{equation}
  Q(\vc{r}) = \int \fourier{Q}(\vc{k}) \euler^{\ii\vc{k} \bcdot \vc{r}} \diff^3 k
\end{equation} 
can be expressed in terms of the Fourier component in the following way
\begin{equation}
  \gav{\evalperturbedgc{Q}}(\vc{r}) 
  = \frac{1}{2\upi} \int \fourier{Q} \euler^{\ii\vc{k} \bcdot \vc{r}} \int_0^{2\upi} \euler^{\ii \kperp \rho \cos \theta} \diff \theta \diff^3 k
  = \int \fourier{Q} \euler^{\ii\vc{k} \bcdot \vc{r}} J_0 \diff^3 k\label{eq:gav_bessel0}
\end{equation} 
and therefore
\begin{equation}\label{eq:fourier_gav}
  \fourier{\gav{\evalperturbedgc{Q}}} = J_0 \fourier{Q},
\end{equation}
where $J_n$ denotes the $n$-th order Bessel function of the first kind evaluated at $\kperp \rho$
\begin{equation}\label{eq:bessel_n}
  J_n = \frac{1}{2\upi \ii^n} \int_0^{2\upi} \euler^{\ii \kperp \rho \cos \theta} \euler^{\ii \theta n} \diff \theta.
\end{equation}
A similar computation for the disc average, \cref{eq:disc_average}, yields
\begin{equation}\label{eq:fourier_dgav}
  \fourier{\dgav{\evalperturbedgcburby{Q}}} 
  = 2 \int_0^1 \burby J_0(\burby \kperp \rho) \diff \burby \, \fourier{Q}
  = \frac{2 J_1}{\kperp \rho} \fourier{Q}.
\end{equation}

\newcommand\pertfvec{\vc{f}_{s}}
\begin{equation}\label{eq:lin_vlasov_fourier}
  \fourier{\gyfone} = \pertfvec \bcdot \fourier{\vc{E}}_1
  , \quad
  \pertfvec
  \defeq
  \roundpar{
      J_0 \hb
    - \ii \frac{2 \gy{\mu} \kpa J_1}{\omega q_s \kperp \rho} \hb \times \vc{k}}
  \frac{
    \omega_{\mathrm{c},s} \pard{\gyfzero}{\gy{u}_\shortparallel}
  }
  {\ii B_0 (\omega - \kpa \gy{u}_\shortparallel)}
  ,
\end{equation}
where we have made use of Faraday's law~\eqref{eq:Faraday} in Fourier space
\begin{equation}\label{eq:Faraday_fourier}
  \fourier{\vc{B}}_1 = \frac{1}{\omega} \vc{k} \times \fourier{\vc{E}}_1.
\end{equation}

\subsection{The Fourier component of the linearised gyrocentre free-current density}\label{sec:gyrocentre_current_fourier}
We write the gyrocentre free-current density~\eqref{eq:free_current_gavadjoint} as
\begin{equation}
  \evalperturbedgcadjointalt{\vc{\mathcal{J}}}{}^\free = \sum_s \evalperturbedgcadjoint{\vc{\mathcal{J}}}{}^\free_{s}
  , \quad
  \evalperturbedgcadjoint{\vc{\mathcal{J}}}{}^\free_s \defeq \int \squarepar{
        q_s \gav{\evalperturbedgcadjoint{\gy{f}_s \jacobiangy  \dot{\gy{\vc{R}}}}}
      - \gy{\mu} \dgav{\evalperturbedgcadjoint{\curl (\gy{f}_s \jacobiangy \hb)}}
    } \diff^3 \gy{u},
\end{equation}
where we have made use of \cref{eq:gradient_theorem_vector_approx} with $\varepsilon_B = 0$ and define the gyro-average adjoint~$\gav{\evalperturbedgcadjoint{Q}}$ as (and similarly for the disc average)
\begin{equation}
  \int \gav{\evalperturbedgcadjoint{Q}} \cp \diff^3 \gy{r}
  \defeq
  \int Q \gav{\evalperturbedgc{\cp}} \diff^6 \gy{z}
\end{equation}
for all suitable test functions $\cp$.

Linearization of the gyrocentre free-current density~\eqref{eq:free_current_gavadjoint}, as is required for computing the susceptibility tensor, results in
\begin{align}\label{eq:djdt_tmp}
  \evalperturbedgcadjoint{\vc{\mathcal{J}}}{}^\free_{1,s} = {} & 
    \frac{q_s}{m_s} \int \biggl[
      \gyfzero \roundpar{ \gav{\evalperturbedgcadjoint{(B_{s,\shortparallel}^\effective \dot{\gy{\vc{R}}})_1}}
    - \frac{\gy{\mu}}{q_s} \dgav{\evalperturbedgcadjoint{\curl \vc{B}_{1,s,\shortparallel}^\effective}}} \nonumber \\ &
    + \gav{\evalperturbedgcadjoint{\gyfone}} \vc{B}_0 \gy{u}_\shortparallel
    - \frac{\gy{\mu}}{q_s} \dgav{\evalperturbedgcadjoint{\curl (\gyfone \vc{B}_0)}} \biggr] \diff^3 \gy{u}
  ,
\end{align}
where we have substituted $(B_{s,\shortparallel}^\effective \dot{\gy{\vc{R}}})_0 = B_0 \gy{u}_\shortparallel \hb$ as follows from \cref{eq:gy_split_eoms_R}.

We need the Fourier component of the adjoint of the gyro-average for evaluating \cref{eq:djdt_tmp}.
The gyro-average adjoint on $\Omega = \mathbb{R}^3$, with constant $\vc{B}_0$, is given by
\begin{equation}
  \int Q \cp(\gy{\vc{r}} + \rho(\gy{\mu}) \hrho(\gy{\theta})) \diff^6 \gy{z}
  = 
  \int Q(\gy{\vc{r}} - \rho(\gy{\mu}) \hrho(\gy{\theta})) \cp \diff^6 \gy{z}
  \quad \implies \quad
  \gav{\evalperturbedgcadjoint{Q}} = \gav{Q(\gy{\vc{r}} - \vc{\rho})},
\end{equation}
where we have made use of the coordinate transformation
\begin{equation}
  \gy{\vc{r}} \mapsto \gy{\vc{r}} - \rho(\gy{\mu}) \hrho(\gy{\theta}),
\end{equation}
which has unit Jacobian.
It follows that the corresponding Fourier component is given by
\begin{equation}\label{eq:fourier_gavadjoint}
  \fourier{\gav{\evalperturbedgcadjoint{Q}}} = J_0(-\kperp \rho) \fourier{Q} = J_0 \fourier{Q}
\end{equation}
by the symmetry of the zeroth-order Bessel function of the first kind.
A similar result holds for the adjoint of the disc average
\begin{equation}
  \dgav{\evalperturbedgcadjoint{Q}} = \dgav{Q(\gy{\vc{r}} - \burby \vc{\rho})}
  \quad \implies \quad
  \fourier{\dgav{\evalperturbedgcadjoint{Q}}} = \frac{2 J_1}{\kperp \rho} \fourier{Q}
  .
\end{equation}

It follows that \cref{eq:djdt_tmp}, in terms of Fourier components, can be expressed as
\begin{align}
  \fourier{\evalperturbedgcadjoint{\vc{\mathcal{J}}}}{}^\free_{1,s} = {} &
    \frac{q_s}{m_s} \int \biggl[
      \gyfzero \roundpar{
          J_0 \fourier{(B_{s,\shortparallel}^\effective \dot{\gy{\vc{R}}})_1} 
        + \ii \frac{\gy{\mu}}{q_s} \frac{2 J_1}{\kperp \rho} \fourier{B_{1,s,\shortparallel}^\effective} \hb \times \vc{k}
      } \nonumber \\ &
    + \fourier{\gyfone} \roundpar{
          J_0 \vc{B}_0 \gy{u}_\shortparallel 
        + \ii \frac{\gy{\mu}}{q_s} \frac{2 J_1}{\kperp \rho} \vc{B}_0 \times \vc{k}
      }
    \biggr]  \diff^3 \gy{u}\label{eq:djdt_tmp_fourier}
\end{align}
for which we find that \cref{eq:gy_eomR_linpart} can be written in terms of Fourier components as follows
\newcommand\Rdotmat{\mat{R}_s}
\begin{equation}\label{eq:gy_eomR_linpart_fourier}
  \fourier{(B_{s,\shortparallel}^\effective \dot{\gy{\vc{R}}})_1} = \Rdotmat \fourier{\vc{E}}_1
  , \quad
  \Rdotmat \defeq 
    J_0 \roundpar{
      - \frac{\gy{u}_\shortparallel}{\omega} \mat{K}
      + \frac{1}{B_0} \mat{B}_0
    }
    + \ii \frac{2 \gy{\mu} J_1}{\omega q_s \kperp \rho} (\vc{k} \times \hb) (\vc{k} \times \hb)^\transpose
  ,
\end{equation}
where the matrix $\mat{K}$ is such that \cref{eq:Kmatrix} holds.
Moreover, the Fourier transform of $B_{1,s,\shortparallel}^\effective$ is given by
\begin{equation}\label{eq:B1sparallel_fourier}
  \fourier{B_{1,s,\shortparallel}^\effective} 
  = J_0 \fourier{B}_{1,\shortparallel}
  = \frac{J_0}{\omega} (\hb \times \vc{k}) \bcdot \fourier{\vc{E}}_1
\end{equation}
by making use of \cref{eq:Faraday_fourier}.

When substituting \cref{eq:fourier_gavadjoint,eq:B1sparallel_fourier,eq:gy_eomR_linpart_fourier} into \cref{eq:djdt_tmp_fourier} we find
\newcommand\pertJmat{\mat{J}}
\begin{equation}\label{eq:fourier_current}
  \fourier{\evalperturbedgcadjoint{\vc{\mathcal{J}}}}{}^\free_1 = \pertJmat \fourier{\vc{E}}_1,
\end{equation}
where
\begin{align}
  \pertJmat \defeq {} &  \sum_s \frac{q_s}{m_s} \int \biggl[ \gyfzero \roundpar{
      J_0 \Rdotmat  
    + \ii \frac{\gy{\mu}}{q_s} \frac{2 J_1}{\kperp \rho} \frac{J_0}{\omega} (\hb \times \vc{k}) (\hb \times \vc{k})^\transpose} 
    \nonumber \\ &
    + \roundpar{
        J_0 \vc{B}_0 \gy{u}_\shortparallel 
      + \ii \frac{\gy{\mu}}{q_s} \frac{2 J_1}{\kperp \rho} \vc{B}_0 \times \vc{k}
      } \pertfvec^\transpose
    \biggr] \diff^3 \gy{u}
  .\label{eq:fourier_currentmat}
\end{align}

\subsection{The Fourier component of the polarisation and magnetisation}
Computing the Fourier component of the polarisation~\eqref{eq:polarisation}, while substituting Faraday's law~\eqref{eq:Faraday_fourier}, results in
\newcommand\polarisationmat{\mat{P}}
\begin{equation}\label{eq:fourier_polarisationmat}
  \fourier{\vc{\mathcal{P}}}_1 = \polarisationmat \fourier{\vc{E}}_{1}
  , \quad 
  \polarisationmat \defeq \sum_s
  \int \frac{\gyfzero}{B_0} \squarepar{
      \mat{\Pi}_\perp 
    + \frac{\gy{u}_\shortparallel}{\omega} (\vc{k} \hb^\transpose - \mat{I}_3 \kpa)
  } \diff^3 \gy{u},
\end{equation}
where we have made use of the perpendicular projection as defined in \cref{eq:perpendicular_projection}.
For the magnetisation~\eqref{eq:magnetisation} we find
\newcommand\magnetisationmat{\mat{M}}
\begin{equation}\label{eq:fourier_magnetisationmat}   
    \fourier{\vc{\mathcal{M}}}_1 =  
    \magnetisationmat \fourier{\vc{E}}_1
    , \quad 
    \magnetisationmat \defeq
    - \sum_s \int \frac{\gyfzero}{m_s} \squarepar{
        \frac{m_s \gy{u}_\shortparallel}{B_0} \roundpar{-\frac{1}{B_0} \mat{B}_0 \mat{\Pi}_\perp + \frac{\gy{u}_\shortparallel}{\omega} \mat{\Pi}_\perp \mat{K}}
      - \frac{\gy{\mu}}{\omega} \mat{\Pi}_\perp \mat{K} 
    } \diff^3 \gy{u}
    .
\end{equation}

\subsection{The gyrokinetic susceptibility tensor}
We consider the Fourier component of the linearization of \cref{eq:ampere_E}
\begin{equation}
  -\omega^2 \roundpar{\epsilon_0 \fourier{\vc{E}}_1 + \fourier{\vc{\mathcal{P}}}_1} 
  =
    \frac{1}{\mu_0} \vc{k} \times (\vc{k} \times \fourier{\vc{E}}_1) 
  - \omega \vc{k} \times \fourier{\vc{\mathcal{M}}}_1 
  + \ii\omega \fourier{\evalperturbedgcadjoint{\vc{\mathcal{J}}}}{}^\free_1
\end{equation}
and therefore, when comparing to \cref{eq:wave_dispersionE}, we find that the gyrokinetic susceptibility tensor $\gyFIXED{\mat{X}}$ is given by
\begin{equation}\label{eq:gy_susceptibility}
  \epsilon_0 \gyFIXED{\mat{X}} = 
    \polarisationmat
  + \frac{1}{ \omega} \mat{K} \magnetisationmat
  + \ii \frac{1}{\omega} \pertJmat,
\end{equation}
where the polarisation, magnetisation and current matrices are given by Eqs.~\eqref{eq:fourier_polarisationmat}, \eqref{eq:fourier_magnetisationmat} and \eqref{eq:fourier_currentmat}, respectively.

To facilitate the comparison with the results from \citep{hasegawa1975physics,zonta2021dispersion}, we choose the following parallel and perpendicular directions
\begin{equation}\label{eq:par_perp_dirs}
  \hb = \vc{\hat{e}}_z
  , \quad
  \vc{k} = \kperp \vc{\hat{e}}_x + \kpa \vc{\hat{e}}_z.
\end{equation}
When substituting this into \cref{eq:fourier_currentmat,eq:fourier_polarisationmat,eq:fourier_magnetisationmat,eq:gy_susceptibility} we find
\begin{equation}\label{eq:gy_susceptibility_sub}
  \begin{aligned}    
    \gyFIXED{\mat{X}} = {} &
      \sum_s \frac{\omega_{\mathrm{p},s}^2}{\omega \omega_{\mathrm{c},s}^2} \int \frac{\gyfzero}{\gy{n}_{0,s}}
      \begin{pmatrix}
        \omega - 2\kpa \gy{u}_\shortparallel & 0 & \kperp \gy{u}_\shortparallel\\
        0 & \omega - 2\kpa \gy{u}_\shortparallel & 0\\
        \kperp \gy{u}_\shortparallel & 0 & 0\\
      \end{pmatrix} 
      \jacobiangcslab \diff^3 \gy{u}\\
    & - \sum_s \frac{\omega_{\mathrm{p},s}^2}{\omega^2 \omega_{\mathrm{c},s}^2} \int \frac{\gyfzero}{\gy{n}_{0,s}}
    \roundpar{\gy{u}_\shortparallel^2 - \frac{\gy{u}_\tau^2}{2}} 
    \jacobiangcslab \diff^3 \gy{u} \begin{pmatrix}
      - \kpa^2 & 0 & \kperp \kpa\\
     0 &- \kpa^2 & 0\\
      \kpa \kperp & 0 & -\kperp^2 
   \end{pmatrix}\\
    & + \ii \sum_s \frac{\omega_{\mathrm{p},s}^2}{\omega^2\omega_{\mathrm{c},s}} \int 
    \frac{\gyfzero}{\gy{n}_{0,s}} J_0^2
    \begin{pmatrix}
      0 & \omega - \kpa \gy{u}_\shortparallel & 0\\
      \kpa \gy{u}_\shortparallel - \omega & \ii \frac{2\kperp \gy{u}_\tau J_1}{J_0} & -\kperp \gy{u}_\shortparallel\\
      0 & \kperp \gy{u}_\shortparallel & 0
    \end{pmatrix} \jacobiangcslab \diff^3 \gy{u} \\
    & - \sum_s \frac{\omega_{\mathrm{p},s}^2}{\omega^2} \int \frac{
      \kpa^2 \gyfzero
    }
    {\gy{n}_{0,s} (\kpa \gy{u}_\shortparallel - \omega)^2}
    \begin{pmatrix}
      0 & 0 & 0\\
      0 & \gy{u}_\tau^2 J_1^2 & \ii \frac{\omega}{\kpa} \gy{u}_\tau J_0 J_1\\
      0 & -\ii \frac{\omega}{\kpa} \gy{u}_\tau  J_0 J_1 & \frac{\omega^2}{\kpa^2} J_0^2
    \end{pmatrix} \jacobiangcslab \diff^3 \gy{u},
  \end{aligned}
\end{equation}
where $\jacobiangcslab = B_0 / m_s$, and the plasma frequency is given by
\begin{equation}\label{eq:plasma_frequency}
  \omega_{\mathrm{p},s} = q_s \sqrt{\frac{\gy{n}_{0,s}}{\epsilon_0 m_s}}.
\end{equation}
In deriving \cref{eq:gy_susceptibility_sub} we have made use of partial integration w.r.t.\ $\gy{u}_\shortparallel$ and assume that $\gyfzero$ vanishes at $\gy{u}_\shortparallel = \pm \infty$.


\subsection{Drift kinetic susceptibility tensor}\label{app:susceptibility:zlr}
In order to compare with the results from \citep{zonta2021dispersion}, we ignore FLR effects by assuming $\kperp \rho \ll 1$ such that we may approximate the Bessel functions as follows
\begin{equation}
  J_0 \approx 1
  , \quad
  J_1 \approx \frac{\kperp \rho}{2}.
\end{equation}
This results in the following ZLR limit of \cref{eq:gy_susceptibility_sub}
\begin{equation}\label{eq:gy_susceptibility_zlr}
  \begin{aligned}    
    \gyFIXED{\mat{X}}^\zlr = {} & 
      \sum_s \frac{\omega_{\mathrm{p},s}^2}{\omega \omega_{\mathrm{c},s}^2} \int \frac{\gyfzero}{\gy{n}_{0,s}}
      \begin{pmatrix}
        \omega - 2\kpa \gy{u}_\shortparallel & 0 & \kperp \gy{u}_\shortparallel\\
        0 & \omega - 2\kpa \gy{u}_\shortparallel & 0\\
        \kperp \gy{u}_\shortparallel & 0 & 0\\
      \end{pmatrix} 
      \jacobiangcslab \diff^3 \gy{u}\\
    & - \sum_s \frac{\omega_{\mathrm{p},s}^2}{\omega^2 \omega_{\mathrm{c},s}^2} \int \frac{\gyfzero}{\gy{n}_{0,s}}
    \roundpar{\gy{u}_\shortparallel^2 - \frac{\gy{u}_\tau^2}{2}} 
    \jacobiangcslab \diff^3 \gy{u} \begin{pmatrix}
      - \kpa^2 & 0 & \kpa \kperp\\
     0 &- \kpa^2 & 0\\
      \kpa \kperp & 0 & -\kperp^2 
   \end{pmatrix}\\
    & + \ii \sum_s \frac{\omega_{\mathrm{p},s}^2}{\omega^2 \omega_{\mathrm{c},s}^2} \int 
    \frac{\gyfzero \jacobiangcslab}{\gy{n}_{0,s}}
    \begin{pmatrix}
      0 & \omega_{\mathrm{c},s} (\omega - \kpa \gy{u}_\shortparallel) & 0\\
      -\omega_{\mathrm{c},s} (\omega - \kpa \gy{u}_\shortparallel) & \ii \kperp^2 \gy{u}_\tau^2 & - \omega_{\mathrm{c},s} \kperp \gy{u}_\shortparallel\\
      0 & \omega_{\mathrm{c},s} \kperp \gy{u}_\shortparallel & 0
    \end{pmatrix} \diff^3 \gy{u} \\
    & - \sum_s \frac{\omega_{\mathrm{p},s}^2}{\omega^2} \int \frac{
      \kpa^2 \gyfzero
    }
    {\gy{n}_{0,s} (\kpa \gy{u}_\shortparallel - \omega)^2}
    \begin{pmatrix}
      0 & 0 & 0\\
      0 & \gy{u}_\tau^4 \kperp^2 \frac{1}{4 \omega_{\mathrm{c},s}^2} & \ii \frac{\kperp}{\kpa} \gy{u}_\tau^2 \frac{\omega}{2 \omega_{\mathrm{c},s}}\\
      0 & -\ii \gy{u}_\tau^2 \frac{\kperp}{\kpa} \frac{\omega}{2 \omega_{\mathrm{c},s}} & \frac{\omega^2}{\kpa^2} 
    \end{pmatrix} \jacobiangcslab \diff^3 \gy{u}.
  \end{aligned}
\end{equation}
It can be verified that this agrees with \cite[Eq.~(2.159)]{hasegawa1975physics}.

\subsection{Gyrokinetic Darwin susceptibility tensor}\label{sec:darwin_susceptibility}
This is very similar to the gauge-invariant gyrokinetic model, except that we must express the additional term in the Darwin polarisation~\eqref{eq:darwin_polarisation} in terms of the electric field $\fourier{\vc{E}}_1$.
That is, we must express the perpendicular part of the vector potential in terms of the electromagnetic fields.

From the definition of the electric field~\eqref{eq:electromagnetic_fields} we find that
\begin{equation}
  \fourier{\vc{E}}_1 = -\ii \vc{k} \fourier{\phi}_1 + \ii \omega \fourier{\vc{A}}_1
  \quad \implies \quad
  \vc{k}_\perp \bcdot \fourier{\vc{E}}_1 = -\ii \kperp^2 \fourier{\phi}_1,
\end{equation}
after having substituted the Fourier ansatz as well as the constraint~\eqref{eq:darwin_constraint_saddle}, $\vc{k}_\perp \bcdot \fourier{\vc{A}}_1 = 0$.
It follows that the perpendicular part of the vector potential is given by
\begin{equation}
  \fourier{\vc{A}}_{1,\perp} = \frac{\fourier{\vc{E}}_{1,\perp} \kperp^2 - \vc{k}_\perp (\vc{k}_\perp \bcdot \fourier{\vc{E}}_1)}{\ii \omega \kperp^2}.
\end{equation}

Using the specific parallel and perpendicular directions given by \cref{eq:par_perp_dirs} we find
\begin{equation}\label{eq:darwin_A_as_E_xz}
  \fourier{\vc{A}}_{1,\perp} 
  = 
  \mat{A}_\perp \fourier{\vc{E}}_1
  , \quad
  \mat{A}_\perp = 
  \begin{pmatrix}
    0 & 0 & 0\\
    0 & -\frac{\ii}{ \omega} & 0\\
    0 & 0 & 0
  \end{pmatrix}
\end{equation}
such that when considering \cref{eq:gy_susceptibility,eq:darwin_polarisation} we find that the gyrokinetic Darwin susceptibility tensor is given by
\begin{equation}
  \gyFIXED{\mat{X}}^\darwin = \gyFIXED{\mat{X}} - \ii \omega \sum_s \frac{\omega_{\mathrm{p},s}^2}{\omega_{\mathrm{c},s}^2} \mat{A}_\perp,
\end{equation}
where we have moreover substituted the definition of the plasma frequency~\eqref{eq:plasma_frequency}.

  \bibliography{mixvar}

\begin{thebibliography}{52}
\expandafter\ifx\csname natexlab\endcsname\relax\def\natexlab#1{#1}\fi
\def\au#1{#1} \def\ed#1{#1} \def\yr#1{#1}\def\at#1{#1}\def\jt#1{\textit{#1}}
  \def\bt#1{#1}\def\bvol#1{\textbf{#1}} \def\vol#1{#1} \def\pg#1{#1}
  \def\publ#1{#1}\def\arxiv#1{#1}\def\org#1{#1}\def\st#1{\textit{#1}}

\bibitem[Bittencourt(2004)]{bittencourt2004}
{\sc \au{Bittencourt, J.~A.}} \yr{2004} {\em Fundamentals of Plasma Physics\/},
  3rd edn.  \publ{Springer New York}.

\bibitem[Bottino \& Sonnendr\"ucker(2015)]{bottino}
{\sc \au{Bottino, A.} \& \au{Sonnendr\"ucker, E.}} \yr{2015}  \at{{M}onte
  {C}arlo particle-in-cell methods for the simulation of the
  {V}lasov--{M}axwell gyrokinetic equations}.  \jt{Journal of Plasma Physics}
  \bvol{81},  \pg{435810501+39}.

\bibitem[Bouchut {\em et~al.\/}(2000)Bouchut, Golse \&
  Pulvirenti]{bouchut2000kinetic}
{\sc \au{Bouchut, F.}, \au{Golse, F.} \& \au{Pulvirenti, M.}} \yr{2000}
  \at{Kinetic equations and asymptotic theory}.  \bt{In {\em Series in applied
  mathematics\/} (ed. \ed{L.~Desvillettes \& B.~Perthame})}.
  \publ{Gauthiers-Villars, Paris}.

\bibitem[Brizard(1990)]{brizard_PhD1990}
{\sc \au{Brizard, A.~J.}} \yr{1990}  \at{Nonlinear gyrokinetic tokamak
  physics}. PhD thesis, Princeton University.

\bibitem[Brizard(2021{\natexlab{{\em a\/}}})]{brizard2021exact}
{\sc \au{Brizard, A.~J.}} \yr{2021{\natexlab{{\em a\/}}}}  \at{Exact
  conservation laws for gauge-free electromagnetic gyrokinetic equations}.
  \jt{Journal of Plasma Physics}  \bvol{87}~(3),  \pg{905870307}.

\bibitem[Brizard(2021{\natexlab{{\em b\/}}})]{brizard2021hamiltonian}
{\sc \au{Brizard, A.~J.}} \yr{2021{\natexlab{{\em b\/}}}}  \at{Hamiltonian
  structure of a gauge-free gyrokinetic {V}lasov--{M}axwell model}.
  \jt{Physics of Plasmas}  \bvol{28}~(12),  \pg{122107}.

\bibitem[Brizard \& Hahm(2007)]{brizard2007}
{\sc \au{Brizard, A.~J.} \& \au{Hahm, T.~S.}} \yr{2007}  \at{Foundations of
  nonlinear gyrokinetic theory}.  \jt{Reviews of Modern Physics}  \bvol{79},
  \pg{421--468}.

\bibitem[Burby \& Brizard(2019)]{burby2019gauge}
{\sc \au{Burby, J.~W.} \& \au{Brizard, A.~J.}} \yr{2019}  \at{Gauge-free
  electromagnetic gyrokinetic theory}.  \jt{Physics Letters A}
  \bvol{383}~(18),  \pg{2172--2175}.

\bibitem[Burby {\em et~al.\/}(2015)Burby, Brizard, Morrison \&
  Qin]{burby2015hamiltonian}
{\sc \au{Burby, J.~W.}, \au{Brizard, A.~J.}, \au{Morrison, P.~J.} \& \au{Qin,
  H.}} \yr{2015}  \at{Hamiltonian gyrokinetic {V}lasov--{M}axwell system}.
  \jt{Physics Letters A}  \bvol{379}~(36),  \pg{2073--2077}.

\bibitem[Cary \& Littlejohn(1983)]{cary1983}
{\sc \au{Cary, J.~R.} \& \au{Littlejohn, R.~G.}} \yr{1983}  \at{Noncanonical
  {H}amiltonian mechanics and its application to magnetic field line flow}.
  \jt{Annals of Physics}  \bvol{151}~(1),  \pg{1--34}.

\bibitem[Chen {\em et~al.\/}(2021)Chen, Chen, Zonca \&
  Lin]{chen2021gyrokinetic}
{\sc \au{Chen, L.}, \au{Chen, H.}, \au{Zonca, F.} \& \au{Lin, Y.}} \yr{2021}
  \at{A gyrokinetic simulation model for low frequency electromagnetic
  fluctuations in magnetized plasmas}.  \jt{Science China Physics, Mechanics \&
  Astronomy}  \bvol{64}~(4),  \pg{245211}.

\bibitem[Chen \& Zonca(2016)]{Chen_Zonka_2016}
{\sc \au{Chen, L.} \& \au{Zonca, F.}} \yr{2016}  \at{Physics of {A}lfv\'en
  waves and energetic particles in burning plasmas}.  \jt{Reviews of Modern
  Physics}  \bvol{88},  \pg{015008}.

\bibitem[Degond \& Raviart(1992)]{DegondRaviart}
{\sc \au{Degond, P.} \& \au{Raviart, P.~A.}} \yr{1992}  \at{An analysis of the
  {D}arwin model of approximation to {M}axwell's equations}.  \jt{Forum
  Mathematicum}  \bvol{4}~(Jahresband),  \pg{13--44}.

\bibitem[Dragt \& Finn(1976)]{dragtfinn1976}
{\sc \au{Dragt, A.~J.} \& \au{Finn, J.~M.}} \yr{1976}  \at{Lie series and
  invariant functions for analytic symplectic maps}.  \jt{Journal of
  Mathematical Physics}  \bvol{17}~(12),  \pg{2215--2227}.

\bibitem[Dudkovskaia {\em et~al.\/}(2023)Dudkovskaia, Wilson, Connor, Dickinson
  \& Parra]{Dudkovskaia_2023}
{\sc \au{Dudkovskaia, A.~V.}, \au{Wilson, H.~R.}, \au{Connor, J.~W.},
  \au{Dickinson, D.} \& \au{Parra, F.~I.}} \yr{2023}  \at{Nonlinear second
  order electromagnetic gyrokinetic theory for a tokamak plasma}.  \jt{Plasma
  Physics and Controlled Fusion}  \bvol{65}~(4),  \pg{045010}.

\bibitem[Fisher(1971)]{fisher1971_dipoles}
{\sc \au{Fisher, G.~P.}} \yr{1971}  \at{The {E}lectric {D}ipole {M}oment of a
  {M}oving {M}agnetic {D}ipole}.  \jt{American Journal of Physics}
  \bvol{39}~(12),  \pg{1528--1533}.

\bibitem[Frieman \& Chen(1982)]{Frieman1982}
{\sc \au{Frieman, E.~A.} \& \au{Chen, L.}} \yr{1982}  \at{Nonlinear gyrokinetic
  equations for low-frequency electromagnetic waves in general plasma
  equilibria}.  \jt{Physics of Fluids}  \bvol{25}~(3),  \pg{502--508}.

\bibitem[Garbet {\em et~al.\/}(2010)Garbet, Idomura, Villard \&
  Watanabe]{garbet2010gyrokinetic}
{\sc \au{Garbet, X.}, \au{Idomura, Y.}, \au{Villard, L.} \& \au{Watanabe,
  T.~H.}} \yr{2010}  \at{Gyrokinetic simulations of turbulent transport}.
  \jt{Nuclear Fusion}  \bvol{50}~(4),  \pg{043002}.

\bibitem[Grad(1966)]{grad1966}
{\sc \au{Grad, H.}} \yr{1966}  \at{Variational principle for a guiding-center
  plasma}.  \jt{Physics of Fluids}  \bvol{9}~(2),  \pg{225--251}.

\bibitem[Grad(1967)]{grad1967}
{\sc \au{Grad, H.}} \yr{1967}  \at{Toroidal containment of a plasma}.
  \jt{Physics of Fluids}  \bvol{10}~(1),  \pg{137--154}.

\bibitem[Grad \& Rubin(1958)]{grad1958hydromagnetic}
{\sc \au{Grad, H.} \& \au{Rubin, H.}} \yr{1958}  \at{Hydromagnetic equilibria
  and force-free fields}.  \jt{Journal of Nuclear Energy}  \bvol{7}~(3--4),
  \pg{284--285}.

\bibitem[Grieger {\em et~al.\/}(1992)Grieger, Beidler, Harmeyer, Lotz,
  Kisslinger, Merkel, N\"uhrenberg, Rau, Strumberger \&
  Wobig]{grieger1992modular}
{\sc \au{Grieger, G.}, \au{Beidler, C.}, \au{Harmeyer, E.}, \au{Lotz, W.},
  \au{Kisslinger, J.}, \au{Merkel, P.}, \au{N\"uhrenberg, J.}, \au{Rau, F.},
  \au{Strumberger, E.} \& \au{Wobig, H.}} \yr{1992}  \at{Modular stellarator
  reactors and plans for {W}endelstein 7-{X}}.  \jt{Fusion Technology}
  \bvol{21}~(3P2B),  \pg{1767--1778}.

\bibitem[Hahm(1988)]{hahm1988nonlinear}
{\sc \au{Hahm, T.~S.}} \yr{1988}  \at{Nonlinear gyrokinetic equations for
  tokamak microturbulence}.  \jt{Physics of Fluids}  \bvol{31}~(9),
  \pg{2670--2673}.

\bibitem[Hasegawa(1975)]{hasegawa1975physics}
{\sc \au{Hasegawa, A.}} \yr{1975}  \at{Plasma instabilities and non-linear
  effects}.  \bt{In {\em Physics and chemistry in space\/} (ed. \ed{J.~G.
  Roederer \& J.~T. Wasson})}.  \publ{Springer-Verlag New York}.

\bibitem[Hindenlang {\em et~al.\/}(2019)Hindenlang, Maj, Strumberger, Rampp \&
  Sonnendr\"ucker]{hindenlang2019gvec}
{\sc \au{Hindenlang, F.}, \au{Maj, O.}, \au{Strumberger, E.}, \au{Rampp, M.} \&
  \au{Sonnendr\"ucker, E.}} \yr{2019} {GVEC}: {A} newly developed {3D} ideal
  {MHD} {G}alerkin {V}ariational {E}quilibrium {C}ode. Annual Meeting of the
  Simons Collaboration on Hidden Symmetries and Fusion Energy.

\bibitem[Hirshman \& Whitson(1983)]{VMEC1983}
{\sc \au{Hirshman, S.~P.} \& \au{Whitson, J.~C.}} \yr{1983}
  \at{Steepest-descent moment method for three-dimensional magnetohydrodynamic
  equilibria}.  \jt{Physics of Fluids}  \bvol{26}~(12),  \pg{3553--3568}.

\bibitem[Hirvijoki {\em et~al.\/}(2020)Hirvijoki, Burby, Pfefferl\'e \&
  Brizard]{hirvijoki2020energy}
{\sc \au{Hirvijoki, E.}, \au{Burby, J.~W.}, \au{Pfefferl\'e, D.} \&
  \au{Brizard, A.~J.}} \yr{2020}  \at{Energy and momentum conservation in the
  {E}uler--{P}oincar\'e formulation of local {V}lasov--{M}axwell-type systems}.
   \jt{Journal of Physics A: Mathematical and Theoretical}  \bvol{53}~(23),
  \pg{235204}.

\bibitem[Hnizdo(2012)]{Hnizdo_2012}
{\sc \au{Hnizdo, V.}} \yr{2012}  \at{Magnetic dipole moment of a moving
  electric dipole}.  \jt{American Journal of Physics}  \bvol{80},
  \pg{645--647}.

\bibitem[Jost {\em et~al.\/}(2001)Jost, Tran, Cooper, Villard \&
  Appert]{jost2001euterpe}
{\sc \au{Jost, G.}, \au{Tran, T.~M.}, \au{Cooper, W.~A.}, \au{Villard, L.} \&
  \au{Appert, K.}} \yr{2001}  \at{Global linear gyrokinetic simulations in
  quasi-symmetric configurations}.  \jt{Physics of Plasmas}  \bvol{8}~(7),
  \pg{3321--3333}.

\bibitem[Kleiber {\em et~al.\/}(2024)Kleiber, Borchardt, Hatzky, K\"onies,
  Leyh, Mishchenko, Riemann, Slaby, Garc\'ia-Rega\~na, S\'anchez \&
  Cole]{kleiber2024euterpe}
{\sc \au{Kleiber, R.}, \au{Borchardt, M.}, \au{Hatzky, R.}, \au{K\"onies, A.},
  \au{Leyh, H.}, \au{Mishchenko, A.}, \au{Riemann, J.}, \au{Slaby, C.},
  \au{Garc\'ia-Rega\~na, J.~M.}, \au{S\'anchez, E.} \& \au{Cole, M.}} \yr{2024}
   \at{{EUTERPE}: A global gyrokinetic code for stellarator geometry}.
  \jt{Computer Physics Communications}  \bvol{295},  \pg{109013}.

\bibitem[Kleiber {\em et~al.\/}(2016)Kleiber, Hatzky, K\"onies, Mishchenko \&
  Sonnendr\"ucker]{Kleiber_pullback}
{\sc \au{Kleiber, R.}, \au{Hatzky, R.}, \au{K\"onies, A.}, \au{Mishchenko, A.}
  \& \au{Sonnendr\"ucker, E.}} \yr{2016}  \at{An explicit large time step
  particle-in-cell scheme for nonlinear gyrokinetic simulations in the
  electromagnetic regime}.  \jt{Physics of Plasmas}  \bvol{23},
  \pg{032501+12}.

\bibitem[Klinger {\em et~al.\/}(2019)Klinger, Andreeva, Bozhenkov, Brandt,
  Burhenn, Buttensch\"on, Fuchert, Geiger, Grulke, Laqua {\em
  et~al.\/}]{klinger2019overview}
{\sc \au{Klinger, T.}, \au{Andreeva, T.}, \au{Bozhenkov, S.}, \au{Brandt, C.},
  \au{Burhenn, R.}, \au{Buttensch\"on, B.}, \au{Fuchert, G.}, \au{Geiger, B.},
  \au{Grulke, O.}, \au{Laqua, H.~P.} \& \au{others}} \yr{2019}  \at{Overview of
  first {W}endelstein 7-{X} high-performance operation}.  \jt{Nuclear Fusion}
  \bvol{59}~(11),  \pg{112004}.

\bibitem[Kraus {\em et~al.\/}(2017)Kraus, Kormann, Morrison \&
  Sonnendr\"ucker]{gempic}
{\sc \au{Kraus, M.}, \au{Kormann, K.}, \au{Morrison, P.~J.} \&
  \au{Sonnendr\"ucker, E.}} \yr{2017}  \at{{GEMPIC}: {G}eometric
  electromagnetic particle-in-cell methods}.  \jt{Journal of Plasma Physics}
  \bvol{83},  \pg{905830401+51}.

\bibitem[Littlejohn(1982)]{littlejohn1982}
{\sc \au{Littlejohn, R.~G.}} \yr{1982}  \at{Hamiltonian perturbation theory in
  noncanonical coordinates}.  \jt{Journal of Mathematical Physics}
  \bvol{23}~(5),  \pg{742--747}.

\bibitem[Littlejohn(1983)]{littlejohn1983}
{\sc \au{Littlejohn, R.~G.}} \yr{1983}  \at{Variational principles of guiding
  centre motion}.  \jt{Journal of Plasma Physics}  \bvol{21},  \pg{111--125}.

\bibitem[Low(1958)]{low}
{\sc \au{Low, F.~E.}} \yr{1958}  \at{A {L}agrangian formulation of the
  {B}oltzmann--{V}lasov equation for plasmas}.  \jt{Proceedings of the Royal
  Society of London A.}  \bvol{248}~(1253),  \pg{282--287}.

\bibitem[McMillan(2023)]{mcmillan2023relationship}
{\sc \au{McMillan, B.~F.}} \yr{2023}  \at{Relationship between drift kinetics,
  gyrokinetics and magnetohydrodynamics in the long-wavelength limit}.
  \jt{Journal of Plasma Physics}  \bvol{89}~(1),  \pg{905890115}.

\bibitem[Miloshevich \& Burby(2021)]{Miloshevich_Burby_2021}
{\sc \au{Miloshevich, G.} \& \au{Burby, J.~W.}} \yr{2021}  \at{Hamiltonian
  reduction of {V}lasov--{M}axwell to a dark slow manifold}.  \jt{Journal of
  Plasma Physics}  \bvol{87}~(3),  \pg{835870301}.

\bibitem[Mishchenko {\em et~al.\/}(2023)Mishchenko, Borchardt, Hatzky, Kleiber,
  K\"onies, N\"uhrenberg, Xanthopoulos, Roberg-Clark \&
  Plunk]{mishchenko2023global}
{\sc \au{Mishchenko, A.}, \au{Borchardt, M.}, \au{Hatzky, R.}, \au{Kleiber,
  R.}, \au{K\"onies, A.}, \au{N\"uhrenberg, C.}, \au{Xanthopoulos, P.},
  \au{Roberg-Clark, G.} \& \au{Plunk, G.~G.}} \yr{2023}  \at{Global gyrokinetic
  simulations of electromagnetic turbulence in stellarator plasmas}.
  \jt{Journal of Plasma Physics}  \bvol{89}~(3),  \pg{955890304}.

\bibitem[Noether(1918)]{Noether1918}
{\sc \au{Noether, E.}} \yr{1918}  \at{Invariante {V}ariationsprobleme}.
  \jt{Nachrichten von der Gesellschaft der Wissenschaften zu G\"ottingen,
  Mathematisch-Physikalische Klasse}  \bvol{1918},  \pg{235--257}.

\bibitem[Novikau {\em et~al.\/}(2021)Novikau, Biancalani, Bottino, {Di Siena},
  Lauber, Poli, Lanti, Villard, Ohana \& Briguglio]{NOVIKAU2021107032}
{\sc \au{Novikau, I.}, \au{Biancalani, A.}, \au{Bottino, A.}, \au{{Di Siena},
  A.}, \au{Lauber, P.}, \au{Poli, E.}, \au{Lanti, E.}, \au{Villard, L.},
  \au{Ohana, N.} \& \au{Briguglio, S.}} \yr{2021}  \at{Implementation of energy
  transfer technique in {ORB5} to study collisionless wave-particle
  interactions in phase-space}.  \jt{Computer Physics Communications}
  \bvol{262},  \pg{107032}.

\bibitem[Parra \& Calvo(2011)]{parra2011phase}
{\sc \au{Parra, F.~I.} \& \au{Calvo, I.}} \yr{2011}  \at{Phase-space
  {L}agrangian derivation of electrostatic gyrokinetics in general geometry}.
  \jt{Plasma Physics and Controlled Fusion}  \bvol{53}~(4),  \pg{045001}.

\bibitem[Peifeng {\em et~al.\/}(2021)Peifeng, Hong \&
  Jianyuan]{peifeng2021discovering}
{\sc \au{Peifeng, F.}, \au{Hong, Q.} \& \au{Jianyuan, X.}} \yr{2021}
  \at{Discovering exact, gauge-invariant, local energy--momentum conservation
  laws for the electromagnetic gyrokinetic system by high-order field theory on
  heterogeneous manifolds}.  \jt{Plasma Science and Technology}
  \bvol{23}~(10),  \pg{105103}.

\bibitem[Porazik \& Lin(2011)]{porazik2011gyrokinetic}
{\sc \au{Porazik, P.} \& \au{Lin, Z.}} \yr{2011}  \at{Gyrokinetic simulation of
  magnetic compressional modes in general geometry}.  \jt{Communications in
  Computational Physics}  \bvol{10}~(4),  \pg{899--911}.

\bibitem[Qin(2005)]{qin2005short}
{\sc \au{Qin, H.}} \yr{2005}  \at{A short introduction to general gyrokinetic
  theory}.  \bt{In {\em Topics in kinetic theory\/} (ed. \ed{T.~Passot,
  C.~Sulem \& P.~L. Sulem})}.  \publ{American Mathematical Society}.

\bibitem[Qin {\em et~al.\/}(2000)Qin, Tang \& Lee]{qin2000gyrocenter}
{\sc \au{Qin, H.}, \au{Tang, W.~M.} \& \au{Lee, W.~W.}} \yr{2000}
  \at{Gyrocenter-gauge kinetic theory}.  \jt{Physics of Plasmas}
  \bvol{7}~(11),  \pg{4433--4445}.

\bibitem[Qin {\em et~al.\/}(1999)Qin, Tang, Lee \& Rewoldt]{qin1999gyrokinetic}
{\sc \au{Qin, H.}, \au{Tang, W.~M.}, \au{Lee, W.~W.} \& \au{Rewoldt, G.}}
  \yr{1999}  \at{Gyrokinetic perpendicular dynamics}.  \jt{Physics of Plasmas}
  \bvol{6}~(5),  \pg{1575--1588}.

\bibitem[Remmerswaal(2023)]{dispersoncurves}
{\sc \au{Remmerswaal, R.}} \yr{2023} {D}ispersion{C}urves.jl: A small package
  for plotting dispersion curves.
  \url{https://gitlab.mpcdf.mpg.de/rwr/dispersioncurves.jl}.

\bibitem[Sugama(2000)]{sugama}
{\sc \au{Sugama, H.}} \yr{2000}  \at{Gyrokinetic field theory}.  \jt{Physics of
  Plasmas}  \bvol{7},  \pg{466--480}.

\bibitem[Sugama {\em et~al.\/}(2018)Sugama, Nunami, Satake \&
  Watanabe]{sugama2018eulerian}
{\sc \au{Sugama, H.}, \au{Nunami, M.}, \au{Satake, S.} \& \au{Watanabe, T.~H.}}
  \yr{2018}  \at{Eulerian variational formulations and momentum conservation
  laws for kinetic plasma systems}.  \jt{Physics of Plasmas}  \bvol{25}~(10),
  \pg{102506}.

\bibitem[Zoni \& Possanner(2021)]{Zoni2021}
{\sc \au{Zoni, E.} \& \au{Possanner, S.}} \yr{2021} On the accuracy of
  gyrokinetic equations in fusion applications.  \bt{In {\em Recent Advances in
  Kinetic Equations and Applications\/} (ed. \ed{F.~Salvarani})},  \pg{pp.
  367--393}.  \publ{Springer International Publishing}.

\bibitem[Zonta {\em et~al.\/}(2021)Zonta, Iorio, Burby, Liu \&
  Hirvijoki]{zonta2021dispersion}
{\sc \au{Zonta, F.}, \au{Iorio, R.}, \au{Burby, J.~W.}, \au{Liu, C.} \&
  \au{Hirvijoki, E.}} \yr{2021}  \at{Dispersion relation for gauge-free
  electromagnetic drift kinetics}.  \jt{Physics of Plasmas}  \bvol{28}~(9),
  \pg{092504}.

\end{thebibliography}
  \bibliographystyle{jpp}

\end{document}